%% file: THE_diphoton_paper.tex
\documentclass[article,12pt]{article} 
\pdfoutput=1
\usepackage{color}
\usepackage{jheppub}
\usepackage[utf8]{inputenc}
\usepackage{feynmp}
\usepackage{amsmath}          
\usepackage{amsfonts}         
\usepackage{amssymb}
\usepackage{slashed}
\usepackage{array}
\usepackage{graphicx}   
\usepackage[T1]{fontenc}       
\usepackage{xspace}
\usepackage{longtable}
\usepackage{pdflscape}
\usepackage[capitalise, english]{cleveref}
\usepackage{siunitx}
\usepackage{booktabs}
\usepackage{soul}
\usepackage[normalem]{ulem}
\usepackage[toc,page]{appendix}

\newcommand{\nocontentsline}[3]{}
\newcommand{\tocless}[2]{\bgroup\let\addcontentsline=\nocontentsline#1{#2}\egroup}

\newcommand*{\TakeFourierOrnament}[1]{{%
\fontencoding{U}\fontfamily{futs}\selectfont\char#1}}
\newcommand*{\danger}{\TakeFourierOrnament{66}}

\newcolumntype{L}[1]{>{\raggedright\let\newline\\\arraybackslash\hspace{0pt}}m{#1}}
\newcolumntype{C}[1]{>{\centering\let\newline\\\arraybackslash\hspace{0pt}}m{#1}}
\newcolumntype{R}[1]{>{\raggedleft\let\newline\\\arraybackslash\hspace{0pt}}m{#1}}

\newcommand{\DRbar}{{\ensuremath{\overline{\mathrm{DR}}}}}
\newcommand{\MSbar}{{\ensuremath{\overline{\mathrm{MS}}}}}
\newcommand\SARAH{{\tt SARAH}\xspace}

\newcommand\MG{{\tt MadGraph}\xspace}

\newcommand{\MGv}[1]{\MG~{\texttt{#1}}\xspace}
\newcommand\MGNLO{{\tt MG5\_aMC@NLO}\xspace}
\newcommand\FeynArts{{\tt FeynArts}\xspace}

\newcommand\FlavorKit{{\tt FlavorKit}\xspace}
\newcommand\FormCalc{{\tt FormCalc}\xspace}
\newcommand\CalcHep{{\tt CalcHep}\xspace}
\newcommand\CompHep{{\tt CompHep}\xspace}

\newcommand\WHIZARD{{\tt WHIZARD}\xspace}
\newcommand\OMEGA{{\tt O'Mega}\xspace}
\newcommand\Sherpa{{\tt Sherpa}\xspace}
\newcommand\SPheno{{\tt SPheno}\xspace}
\newcommand\Herwig{{\tt Herwig++}\xspace}
\newcommand\FS{{\tt FlexibleSUSY}\xspace}
\newcommand\Vevacious{{\tt Vevacious}\xspace}
\newcommand\MO{{\tt MicrOmegas}\xspace}
\newcommand\HB{{\tt HiggsBounds}\xspace}
\newcommand\HS{{\tt HiggsSignals}\xspace}

\newcommand\UFO{{\tt UFO}\xspace}

\newcommand\Susyno{{\tt Susyno}\xspace}
\newcommand\Mathematica{{\tt Mathematica}\xspace}

\newcommand{\Fortran}{\texttt{Fortran}\xspace}
\newcommand{\GeV}{\text{GeV}\xspace}
\newcommand{\TeV}{\text{TeV}\xspace}

\newcommand{\BLSSM}{\textit{B-L}-SSM\xspace}
\newcommand{\code}[1]{\lstinline[basicstyle=\ttfamily]|#1|}
\newcommand{\be}{\begin{equation}}
\newcommand{\ee}{\end{equation}}
\newcommand{\ba}{\begin{array}}
\newcommand{\ea}{\end{array}}

\def\zTwo{$\mathbb{Z}_2$}
\def\zThree{$\mathbb{Z}_3$}
\def\ThreeTwoOne{$\mathrm{SU(3)_c} \times \mathrm{SU(2)_L} \times \mathrm{U(1)_Y}$}
\def\ThreeThreeOne{$\mathrm{SU(3)_c} \times \mathrm{SU(3)_L} \times \mathrm{U(1)_\mathcal{X}}$}
\def\hc{\text{h.c.}}
\def\one{\ensuremath{\mathbf{1}}}
\def\two{\ensuremath{\mathbf{2}}}
\def\three{\ensuremath{\mathbf{3}}}
\def\threeS{\ensuremath{\mathbf{\bar 3}}}
\def\five{\ensuremath{\mathbf{5}}}
\def\fiveb{\ensuremath{\mathbf{\bar 5}}}
\def\seven{\ensuremath{\mathbf{7}}}
\def\eight{\ensuremath{\mathbf{8}}}
\def\ten{\ensuremath{\mathbf{10}}}
\def\tenb{\ensuremath{\mathbf{\bar {10}}}}

\newcounter{para}

\newcommand\modelparalabel[2]{{\bf \bigskip\par\noindent\refstepcounter{para}\label{#2}\thepara\space #1 }}
\numberwithin{para}{subsubsection}
\crefname{para}{Sec.}{Secs.}

\newenvironment{itemize*}%
  {\begin{itemize}%
    \setlength{\itemsep}{0pt}%
    \setlength{\parskip}{0pt}}%
  {\end{itemize}}

\usepackage{enumitem}
\usepackage{nopageno}
\usepackage{enumitem}
\usepackage{listings}

-lmtk10

\lstdefinestyle{mathematica}{
        basicstyle=\ttfamily\mdseries,
	language=bash,
	frame=false,
	xleftmargin=.25in}   

\lstdefinestyle{terminal}{
        basicstyle=\ttfamily,
	language=bash,
        prebreak={\textbackslash},
	frame=lines,
	xleftmargin=.5in,
        numbers=none}
        
\lstdefinestyle{file}{
        basicstyle=\ttfamily\mdseries,
	language=bash,
	frame=shadowbox,
        numbers=left,   
        numberstyle=\tiny} 
        
\lstdefinestyle{mssm}{
        basicstyle=\ttfamily\mdseries,
	language=bash,
	frame=shadowbox,
        numbers=left,  
        stringstyle=\color{gray},  
        title=\hspace{14.5cm}MSSM,
        numberstyle=\tiny}     
        
\lstdefinestyle{blssm}{
        basicstyle=\ttfamily\mdseries,
	language=bash,
	frame=shadowbox,
        title=\hspace{14.2cm}\BLSSM,	
        numbers=left,   
        numberstyle=\tiny}
        
\lstdefinestyle{both}{
        basicstyle=\ttfamily\mdseries,
	language=bash,
	frame=shadowbox,
        title=\hspace{13cm}{MSSM,\,\BLSSM},		
        numbers=left,   
        numberstyle=\tiny}  

\newcounter{magicrownumbers}

\newcommand{\AddrCERN}{%
Theoretical Physics Department, CERN, Geneva, Switzerland
}
\newcommand{\AddrValencia}{%
Instituto de F\'{\i}sica Corpuscular (CSIC-Universitat de Val\`{e}ncia), Apdo. 22085, E-46071 Valencia, Spain.
}
\newcommand{\AddrBonn}{%
Bethe Center for Theoretical Physics \& Physikalisches Institut der 
Universit\"at Bonn, \\
Nussallee 12, 53115 Bonn, Germany
}
\newcommand{\AddrDESY}{%
Deutsches Elektronen-Synchrotron DESY, 22607 Hamburg, Germany
}
\newcommand{\AddrTAU}{%
Raymond and Beverly Sackler School of Physics and Astronomy, \\
 Tel-Aviv University, Tel-Aviv 69978, Israel
}
\newcommand{\AddrMonash}{%
ARC Centre of Excellence for Particle Physics at the Terascale, \\
School of Physics, Monash University, Melbourne, Victoria 3800, Australia
}
\newcommand{\AddrAdelaide}{%
ARC Centre of Excellence for Particle Physics at the Terascale, \\
Department of Physics, The University of Adelaide, Adelaide, South Australia
5005, Australia
}
\newcommand{\AddrCPPM}{%
CPPM, Aix-Marseille Universit\'e, CNRS-IN2P3, UMR 7346, 163 avenue de Luminy, 13288 Marseille Cedex 9, France
}
\newcommand{\AddrLPTHE}{%
LPTHE, UMR 7589, CNRS  and  Universit\'e Pierre et Marie Curie, Sorbonne Universit\'es, 75252 Paris Cedex 05, France
}

\preprint{ADP-16-8/T963 \\ \hspace*{\fill} BONN-TH-2016-01\\ \hspace*{\fill} CERN-TH-2016-033 \\ \hspace*{\fill}COEPP-MN-16-4 \\ \hspace*{\fill}DESY 16-025\\ \hspace*{\fill}IFIC/16-09}

\title{Precision tools and models to narrow in on the 750 GeV diphoton resonance}

\author[a]{Florian Staub,}\emailAdd{florian.staub@cern.ch}
\author[b]{Peter Athron,} \emailAdd{peter.athron@coepp.org.au}
\author[c]{Lorenzo Basso,} \emailAdd{basso@cppm.in2p3.fr}
\author[d]{Mark D. Goodsell,} \emailAdd{goodsell@lpthe.jussieu.fr}
\author[e]{Dylan Harries,} \emailAdd{dylan.harries@adelaide.edu.au}
\author[f]{Manuel E. Krauss,} \emailAdd{mkrauss@th.physik.uni-bonn.de}
\author[f]{Kilian Nickel,} \emailAdd{nickel@th.physik.uni-bonn.de}
\author[f]{Toby Opferkuch,} \emailAdd{toby@th.physik.uni-bonn.de}
\author[g]{Lorenzo Ubaldi,} \emailAdd{ubaldi.physics@gmail.com}
\author[h]{Avelino Vicente,} \emailAdd{avelino.vicente@ific.uv.es}
\author[i]{Alexander Voigt} \emailAdd{alexander.voigt@desy.de}

\affiliation[a]{\AddrCERN}
\affiliation[b]{\AddrMonash}
\affiliation[c]{\AddrCPPM}
\affiliation[d]{\AddrLPTHE}
\affiliation[e]{\AddrAdelaide}
\affiliation[f]{\AddrBonn}
\affiliation[g]{\AddrTAU}
\affiliation[h]{\AddrValencia}
\affiliation[i]{\AddrDESY}

\abstract{ The hints for a new resonance at 750~GeV from ATLAS and CMS
  have triggered a significant amount of attention.  Since the
  simplest extensions of the standard model cannot accommodate the
  observation, many alternatives have been considered to explain the
  excess.  Here we focus on several proposed renormalisable
  weakly-coupled models and revisit results given in the literature.
  We point out that physically important subtleties are often missed
  or neglected. To facilitate the study of the excess we have created
  a collection of 40 model files, selected from recent literature, for
  the \Mathematica package \SARAH.  With \SARAH one can generate files
  to perform numerical studies using the tailor-made spectrum
  generators \FS and \SPheno. These have been extended to
  automatically include crucial higher order corrections to the
  diphoton and digluon decay rates for both CP-even and CP-odd
  scalars. Additionally, we have extended the {\tt UFO} and {\tt
    CalcHep} interfaces of \SARAH, to pass the precise information
  about the effective vertices from the spectrum generator to a
  Monte-Carlo tool.  Finally, as an example to demonstrate the power
  of the entire setup, we present a new supersymmetric model that
  accommodates the diphoton excess, explicitly demonstrating how a
  large width can be obtained. We explicitly show several steps in
  detail to elucidate the use of these public tools in the precision
  study of this model. }

\begin{document}
\maketitle

\input{tex/introduction}
\input{tex/motivation}

\input{tex/sarah}

\input{tex/sg}
\input{tex/models}

\input{tex/example}
\input{tex/summary}

\section*{Acknowledgements}
We thank Martin Winkler for beta-testing and valuable impact, and Michael
Spira for helpful discussions.  Mark
Goodsell acknowledges support from Agence Nationale de Recherche grant
ANR-15-CE31-0002 ``HiggsAutomator,'' and would like to thank Pietro
Slavich and Luc Darm\'e for interesting discussions.  Avelino Vicente
acknowledges financial support from the ``Juan de la Cierva'' program
(27-13-463B- 731) funded by the Spanish MINECO as well as from the
Spanish grants FPA2014-58183-P, Multidark CSD2009-00064 and
SEV-2014-0398 (MINECO), FPA2011-22975 and PROMETEOII/2014/084
(Generalitat Valenciana), and is grateful to Wei-Chih Huang for
discussions about the GTHDM model.  Manuel E.\ Krauss is supported by
the BMBF grant 00160287 and thanks Cesar Bonilla for useful
discussions on left-right models.  Lorenzo Basso acknowledges support
by the OCEVU Labex (ANR-11-LABX- 0060) and the A*MIDEX project
(ANR-11-IDEX-0001-02), funded by the “Investissements d’Avenir” French
government program managed by the ANR. The work by Peter Athron is in
part supported by the ARC Centre of Excellence for Particle Physics at
the Tera-scale.  Dylan Harries is supported by the University of
Adelaide and the ARC Centre of Excellence for Particle Physics at the
Tera-scale.  Florian Staub thanks Alfredo Urbano for discussions about
the Georgi-Machacek model and for testing the model file.

\begin{appendix}
\input{tex/appendix_sarah}
\end{appendix}

\bibliographystyle{JHEP} 
\bibliography{diphoton}
\end{document}

%% file: tex/introduction.tex
\section{Introduction}
The first data from the 13~TeV run of the Large Hadron Collider (LHC) contained a surprise: ATLAS and CMS reported a resonance at about 750~GeV in the diphoton channel with local significances of $3.9 \sigma$ and $2.6 \sigma$, respectively \cite{AtlasDiphoton,CMS:2015dxe}. When including the look-elsewhere-effect, the deviations from standard model (SM) expectations drop to $2.3 \sigma$ and $1.2 \sigma$. 

This possible signal caused a lot of excitement, as it is the largest deviation from the SM which has been seen by both experiments. Such an excitement, in turn, led to an avalanche of papers in a very short time which analysed the excess from various perspectives
 \cite{Abel:2016pyc,
 Agrawal:2015dbf,
 Ahmed:2015uqt,
 Allanach:2015ixl,
 Alexander:2016uli,
 Altmannshofer:2015xfo,
 Aloni:2015mxa,
 Alves:2015jgx,
 An:2015cgp,
 Anchordoqui:2015jxc,
 Antipin:2015kgh,
 Aparicio:2016iwr,
 Arbelaez:2016mhg,
 Mambrini:2015wyu,
 Arun:2016ela,
 Arun:2015ubr,
 Aydemir:2016qqj,
 Backovic:2015fnp,
 Badziak:2015zez,
 Bae:2016xni,
 Bai:2015nbs,
 Bardhan:2015hcr,
 Barducci:2015gtd,
 Barrie:2016ntq,
 Bauer:2015boy,
 Becirevic:2015fmu,
Bellazzini:2015nxw,
Belyaev:2015hgo,
Ben-Dayan:2016gxw,
Benbrik:2015fyz,
Berlin:2016hqw,
Bernon:2015abk,
Berthier:2015vbb,
Bertuzzo:2016fmv,
Bhattacharya:2016lyg,
Bi:2015lcf,
Bi:2015uqd,
Bian:2015kjt,
Bizot:2015qqo,
Dasgupta:2015pbr,
Borah:2016uoi,
Boucenna:2015pav,
Buckley:2016mbr,
Buttazzo:2015txu,
Cai:2015hzc,
Cao:2015twy,
Cao:2016udb,
Cao:2015apa,
Cao:2015xjz,
Cao:2016cok,
Cao:2015scs,
Cao:2015pto,
Carpenter:2015ucu,
Casas:2015blx,
Chakrabortty:2015hff,
Chakraborty:2015gyj,
Chakraborty:2015jvs,
Chala:2015cev,
Chang:2015sdy,
Chang:2015bzc,
Chao:2016aer,
Chao:2016mtn,
Chao:2015nac,
Chao:2015nsm,
Chao:2015ttq,
Cheung:2015cug,
Chiang:2015tqz,
Chiang:2016ydx,
Cho:2015nxy,
Chway:2015lzg,
Cline:2015msi,
Cox:2015ckc,
Craig:2015lra,
Csaki:2016raa,
Csaki:2015vek,
Curtin:2015jcv,
Cvetic:2015vit,
D'Eramo:2016mgv,
Danielsson:2016nyy,
Laperashvili:2016cah,
Das:2015enc,
Davis:2016hlw,
Davoudiasl:2015cuo,
Kim:2015ron,
Kim:2015ksf,
Dasgupta:2016wxw,
deBlas:2015hlv,
Delaunay:2016zmu,
Demidov:2015zqn,
Deppisch:2016scs,
Dev:2015vjd,
Dev:2015isx,
Dey:2015bur,
Dhuria:2015ufo,
DiChiara:2015vdm,
Ding:2016udc,
Ding:2016ldt,
Ding:2015rxx,
Djouadi:2016eyy,
Dorsner:2016ypw,
Draper:2016fsr,
Dutta:2015wqh,
Dutta:2016jqn,
Ellis:2015oso,
Ellwanger:2016qax,
Fabbrichesi:2016alj,
Falkowski:2015swt,
Faraggi:2016xnm,
Feng:2015wil,
Fichet:2015vvy,
Fichet:2016pvq,
Franceschini:2015kwy,
Franzosi:2016wtl,
Frugiuele:2016rii,
Gabrielli:2015dhk,
Gao:2015igz,
Ge:2016xcq,
Geng:2016xin,
Ghorbani:2016jdq,
Ghoshal:2016jyj,
Giddings:2016sfr,
Godunov:2016kqn,
Goertz:2015nkp,
Gross:2016ioi,
Gu:2015lxj,
Gupta:2015zzs,
Hall:2015xds,
Hamada:2015skp,
Hamada:2016vwk,
Han:2015cty,
Han:2015yjk,
Han:2015dlp,
Han:2015qqj,
Han:2016bus,
Han:2016bvl,
Han:2016fli,
Harigaya:2016pnu,
Harigaya:2015ezk,
Harland-Lang:2016qjy,
Harland-Lang:2016apc,
Hati:2016thk,
Heckman:2015kqk,
Hektor:2016uth,
Hernandez:2015hrt,
Hernandez:2015ywg,
Hernandez:2016rbi,
Higaki:2015jag,
Huang:2015evq,
Huang:2015rkj,
Huang:2015svl,
Ibanez:2015uok,
Ito:2016zkz,
Jiang:2015oms,
Jung:2015etr,
Kanemura:2015vcb,
Kanemura:2015bli,
Kaneta:2015qpf,
Kang:2015roj,
Karozas:2016hcp,
Kavanagh:2016pso,
Kawamura:2016idj,
Kim:2015xyn,
King:2016wep,
Knapen:2015dap,
Ko:2016wce,
Ko:2016lai,
Kobakhidze:2015ldh,
Kulkarni:2015gzu,
Li:2015jwd,
Li:2016xcj,
Liao:2015tow,
Liu:2015yec,
Low:2015qho,
Low:2015qep,
Luo:2015yio,
Martin:2016byf,
Martinez:2015kmn,
Martini:2016ahj,
Marzola:2015xbh,
Matsuzaki:2015che,
Salvio:2015jgu,
McDermott:2015sck,
Megias:2015ory,
Modak:2016ung,
Molinaro:2015cwg,
Moretti:2015pbj,
Murphy:2015kag,
Nakai:2015ptz,
No:2015bsn,
Nomura:2016rjf,
Nomura:2016seu,
Nomura:2016fzs,
Okada:2016rav,
Palti:2016kew,
Palle:2015vch,
Park:2015ysf,
Patel:2015ulo,
Pelaggi:2015knk,
Petersson:2015mkr,
Pilaftsis:2015ycr,
Potter:2016psi,
Sahin:2016lda,
Salvio:2016hnf,
Son:2015vfl,
Stolarski:2016dpa,
Tang:2015eko,
Wang:2015omi,
Wang:2015kuj,
Yu:2016lof,
Zhang:2016pip,
Zhang:2015uuo,
Zhang:2016xei}.

It is hard to explain the excess within the most commonly considered frameworks for physics beyond the standard model (BSM), like two-Higgs-doublet models (THDM) or the minimal supersymmetric standard model (MSSM) \cite{Angelescu:2015uiz}, to mention a couple of well-known examples. Thus, many alternative ideas for BSM models have been considered, some of which lack a deep theoretical motivation and are rather aimed at just providing a decent fit to the diphoton bump. Most of the papers in the avalanche were written quickly, some in a few hours, many in a few days, so the analyses of the new models are likely to have shortcomings. Some effects could be missed in the first attempt and some statements might not hold at a second glance. Indeed we have found a wide range of mistakes or unjustified assumptions, which represented the main motivation that prompted this work.

Now that the dust has settled following the stampede caused by the presentation of the ATLAS and CMS data, the time has come for a more detailed and careful study of the proposed ideas. In the past few years several tools have been developed which can be very helpful in this respect. In the context of renormalisable models, the Mathematica package \SARAH \cite{Staub:2008uz,Staub:2009bi,Staub:2010jh,Staub:2012pb,Staub:2013tta,Staub:2015kfa} offers all features for the precise study of a new model: it calculates all tree-level properties of the model (mass, tadpoles, vertices), the one-loop corrections to all masses as well as the two-loop renormalisation group equations, and it can be interfaced with the spectrum generators \SPheno \cite{Porod:2003um,Porod:2011nf} and \FS \cite{Athron:2014yba}. These codes, in turn, can be used for a numerical analysis of any model, which can compete with the precision of state-of-the-art spectrum generators dedicated just to the MSSM and NMSSM \cite{Staub:2015aea}. The RGEs are solved numerically at the two-loop level and the mass spectrum is calculated at one loop.  Both codes have the option to include the known two-loop corrections \cite{Degrassi:2001yf,Brignole:2001jy,Dedes:2002dy,Brignole:2002bz,Dedes:2003km,Degrassi:2009yq} to the Higgs masses in the MSSM and NMSSM, which may, depending on the model, provide a good approximation of the dominant corrections. \SPheno  can also calculate the full two-loop corrections to the Higgs masses in the gaugeless limit at zero external momentum\cite{Goodsell:2014bna,Goodsell:2015ira}. \FS has an extension to calculate the two-loop Higgs mass corrections using the complementary effective field theory approach, which is to be released very soon.  \SPheno makes predictions for important flavour observables, which have been not yet implemented in \FS. Of particular importance for the current study is that \SPheno and \FS calculate the effective vertices for the diphoton and digluon couplings of the scalars, which can then be used by Monte-Carlo (MC) tools like \CalcHep \cite{Pukhov:2004ca,Boos:1994xb} or \MG \cite{Alwall:2011uj,Alwall:2014hca}.  Other numerical tools like \MO \cite{Belanger:2014hqa}, \HB \cite{Bechtle:2008jh,Bechtle:2011sb}, \HS \cite{Bechtle:2013xfa} or \Vevacious \cite{Camargo-Molina:2013qva} can easily be included in the framework.

These powerful packages provide a way to get a thorough understanding of the new models. The main goal of this work is to support the model builders and encourage them to use these tools. We provide several details about the features of the packages in the spirit of making this paper self-contained and bringing the reader unfamiliar with the tools to the level of knowledge necessary to use them. More information can be found in the manuals of each package. 
We have created a database of diphoton models in \SARAH, by implementing 40 among those proposed in recent literature, which is now available to all interested researchers. For each model we have written model files to interface \SARAH with \SPheno and \FS.  \\

Although in each case we have tried to check very carefully that we implement the model which has been proposed in the literature, it is of course possible that some details have been missed.  The original authors of these models are encouraged to check the implementation themselves to satisfy that what we have implemented really does correspond to the model they proposed.  In the description of some of the models we state cases where the model has problems or where we find difficulties for the proposed solution. This helps inform potential users about what they may see when running these models through the tools we are discussing here. {\it However especially in these cases we encourage the original authors to check what we have written and let us know if they disagree with any claim we make.} \\

This paper is long but modular, and each section is to a large extent self contained, so the reader can easily jump to the section of greater interest. We have structured it as follows:
\begin{itemize}
\item In \cref{sec:motivation}, we give a list of common mistakes we have found in the literature and emphasise how they are easily avoided by using the tools. This provides the main motivation of this paper and we hope that other model builders will also see the necessity of using these packages.
\item In \cref{sec:sarah}, we discuss the basic features of \SARAH, describing how one can use it to extract all analytical properties of the model and how to generate model files or source code for the numerical tools.
\item In \cref{sec:sg}, we give an introduction to \SPheno and \FS and explain how these codes can be easily interfaced with MC tools. We also discuss at some length the implementation of the diphoton and digluon effective vertices.
\item In \cref{sec:models}, we give an overview of the models which we have implemented in \SARAH\   and briefly discuss their main features. 
\item In \cref{sec:example}, we provide an explicit example of how to quickly work out the details of a model, analytically with \SARAH and numerically with the other tools. For this purpose we extended a natural supersymmetric (SUSY) model  to accommodate the 750 GeV resonance.
\item We conclude in \cref{sec:summary}.
\item We provide three appendices containing frequently asked questions (FAQs) aimed at further clarifying the use of the packages.
\end{itemize}

%% file: tex/motivation.tex
\section{Motivation}
\label{sec:motivation}
Precision studies in high energy physics have reached a high level of automation.
There are publicly available tools to perform Monte-Carlo
studies at LO or NLO \cite{Bahr:2008pv,Bellm:2015jjp,Sjostrand:2006za,Sjostrand:2014zea,Gleisberg:2008ta,Hoeche:2012ft}, many spectrum generators \cite{Porod:2003um,Porod:2011nf,Allanach:2001kg,Allanach:2013kza,Allanach:2009bv,Allanach:2011de,Allanach:2014nba,Allanach:2016rxd,Djouadi:2002ze,Heinemeyer:1998yj,Drechsel:2016jdg,Ellwanger:2006rn,Baglio:2013iia,Vega:2015fna} for the calculation of pole masses 
including important higher order corrections, codes dedicated only to Higgs \cite{Djouadi:1997yw,Frisch:2010gw,Ellwanger:2004xm,Ellwanger:2005dv} or sparticle decays \cite{Muhlleitner:2003vg,Hlucha:2011yk,Das:2011dg}, and codes to check flavour \cite{Degrassi:2007kj,Mahmoudi:2008tp,Rosiek:2010ug,Crivellin:2012jv,Lee:2012wa} or other precision observables \cite{Athron:2015rva}. This machinery has been used in the past mainly for detailed
studies of some promising BSM candidates, like the MSSM, NMSSM or
variants of THDMs. There are two main reasons why these tools are
usually the preferred method to study these models: (i) it has been
shown that there can be large differences between the exact numerical
results and the analytic approximations; (ii) writing private routines for
specific calculations is not only time consuming but also error
prone. On the other hand, the number of tools available to study the new ideas
proposed to explain the diphoton excess is still limited. Of these tools, many are not yet widely used largely due to the community's reluctance in adopting new codes.
However, we think it is beneficial to adopt this new generation of generic tools like \SARAH.

We noticed that
several studies done in the context of the 750~GeV excess have
overlooked important subtleties in some models, neglected important
higher order corrections, or made many simplifying
assumptions which are difficult to justify.
Using generic software tools in this context can help address these issues:
many simplifications will no longer be necessary and important higher
order corrections can be taken into account in a consistent manner.
In order to illustrate this we comment, in the
following subsections, on several issues we became aware of when
revisiting some of the results in the literature. 

\subsection{Calculation of the diphoton and digluon widths}
\subsubsection{The diphoton and digluon rates beyond leading order}
\label{sec:motivation_rates}
A precise calculation of the diphoton rate is of crucial importance. In the validation process of this work, we identified several results in the literature that deviate, often by an order of magnitude or more, in comparison to our results \cite{Chao:2015nac,Pilaftsis:2015ycr,Martinez:2015kmn}.
Additionally we observed that in many cases there are important subtleties which we think are highly relevant.

First of all, the choice of the renormalisation scale of the running couplings appearing
in the calculation.  The majority of recent studies use the
electromagnetic coupling at the scale of the decaying
particle. However, one should rather use $\alpha_{em} (0)$, i.e.\ the Thompson limit (see for
instance Refs.~\cite{Spira:1995rr,Hartmann:2015oia}), in order to keep the NLO corrections under control. 
Taking this into account
already amounts to an $\mathcal O (10~\%)$ change of the diphoton rate
 compared to many studies in the literature. 
In addition, as we will discuss in \cref{sec:motivation_constraints_BR}, an important prediction of  
a model is the ratio $\text{Br}(S\to
gg)/\text{Br}(S\to \gamma\gamma)$, where $S$ is the $\SI{750}{\GeV}$ scalar resonance. It is well known that the digluon channel
receives large QCD corrections. If one neglects these corrections
the ratio  will be severely underestimated.
\begin{figure}[htbp]
\begin{center}
\includegraphics{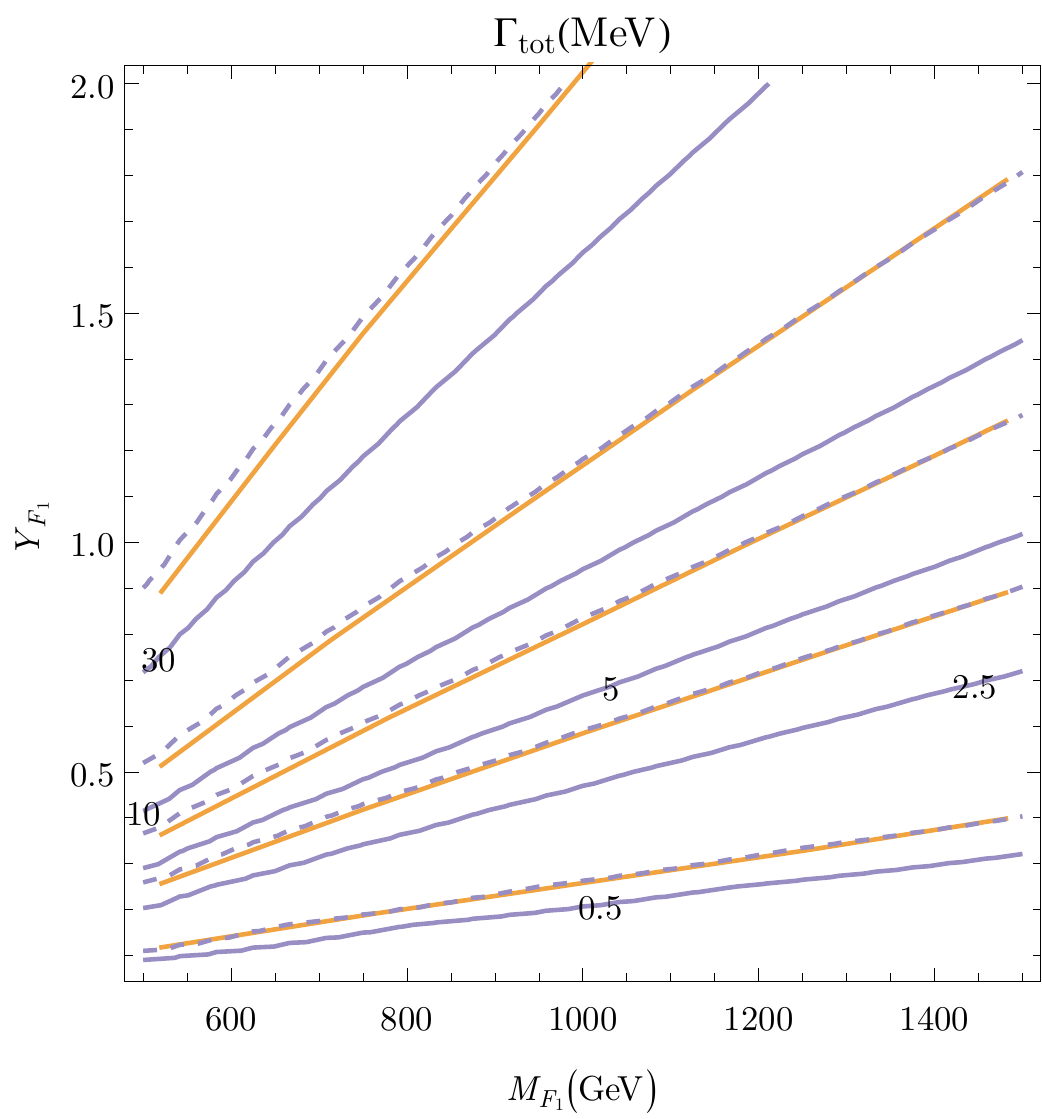} 
\end{center}
\caption{The approximate total width (sum of the diphoton and digluon channels) of $S$ as a function of the coupling $Y_{F_1}$ and
  the mass $m_{F_1}$ of the vector-like particle $F_1$, calculated
  using \texttt{SPheno} (blue) at LO (dashed) and NLO (solid). The
  orange contours are the results of the LO calculation from
  Ref.~\cite{Knapen:2015dap}. Here we assume a single generation of
  vector-like quarks.}
\label{fig:TotWidthF1ToyModel}
\end{figure}
\begin{table}[htbp]
\centering
\begin{tabular}{c c c c c}
\toprule
 Model &  & Br($gg/\gamma\gamma$) & $\Gamma_{S\to gg}$ [$\SI{}{\MeV}$] & $\Gamma_{S\to \gamma\gamma}$ [$\SI{}{\MeV}$] \\
 \midrule
$\Psi_{F_1}$ &Ref.~\cite{Knapen:2015dap} LO & 11.62/- & 6.74/- & 0.58/- \\
&\texttt{SPheno} LO & 13.47/12.22 & 6.78/14.27 & 0.50/1.17 \\
&\texttt{SPheno} NLO & 23.27/20.27 & 11.04/23.71 & 0.47/1.17 \\ \midrule
$\Psi_{F_2}$ &Ref.~\cite{Knapen:2015dap} LO & 24.42/- & 15.14/- & 0.62/- \\
&\texttt{SPheno} LO & 28.32/25.70 & 15.26/32.12 & 0.54/1.25 \\
&\texttt{SPheno} NLO & 48.93/42.67 & 24.85/52.34 & 0.51/1.25 \\\midrule
$\Psi_{F_3}$ &Ref.~\cite{Knapen:2015dap} LO & 33.80/- & 6.76/- & 0.20/- \\
&\texttt{SPheno} LO & 39.20/35.56 & 6.78/14.27 & 0.17/0.40 \\
&\texttt{SPheno} NLO & 67.72/59.06 & 11.04/23.71 & 0.16/0.40 \\\midrule
$\Psi_{F_4}$ &Ref.~\cite{Knapen:2015dap} LO & 49.84/- & 14.95/- & 0.30/- \\
&\texttt{SPheno} LO & 57.80/52.44 & 15.26/32.12 & 0.26/0.61 \\
&\texttt{SPheno} NLO & 99.85/87.09 & 24.85/53.34 & 0.25/0.61 \\\midrule
$\Psi_{F_5}$ &Ref.~\cite{Knapen:2015dap} LO & 150.0/- & 1.50/- & \SI{10.0E-3}{}/- \\
&\texttt{SPheno} LO & 177.0/160.6 & 1.70/3.57 & \SI{9.58E-3}{}/\SI{22.22E-3}{} \\
&\texttt{SPheno} NLO & 305.8/266.7 & 2.76/5.93 & \SI{9.03E-3}{}/\SI{22.22E-3}{} \\\midrule
$\Psi_{F_6}$ &Ref.~\cite{Knapen:2015dap} LO & 390.0/- & 7.80/- & \SI{2.00E-2}{}/- \\
&\texttt{SPheno} LO & 453.2/411.1 & 6.78/14.27 & \SI{1.50E-2}{}/\SI{3.47E-2}{} \\
&\texttt{SPheno} NLO & 782.8/682.8 & 11.04/23.71 & \SI{1.41E-2}{}/\SI{3.47E-2}{} \\
\bottomrule
\end{tabular}
\caption{Branching fraction ratio, as well as the partial decay widths for the digluon and diphoton channels for the toy model (\cref{sec:toymodelsec}) containing only the relevant vector-like fermion pair $\Psi_{F_i}$. The above values are for the benchmark points $Y_{F_i}=1$ and $m_{F_i}=\SI{1}{\TeV}$, where the values are for a CP-even/CP-odd scalar resonances, respectively. The \SPheno NLO calculation includes $\text{N}^3$LO corrections for the digluon channel, while the diphoton decay width is calculated at NLO and LO for a CP-even and odd scalar respectively.}
\label{tab:SSMVLevenValidateTable}
\end{table}

To demonstrate these effects we show in \cref{fig:TotWidthF1ToyModel}
the total decay width\footnote{Here, the total width is simply the sum of the diphoton and digluon channels ignoring small contributions from other sub-dominant channels.} of the singlet $S$ as a function of the mass
$M_{F_1}$ and coupling $Y_{F_1}$ for a simple toy model containing
only the vector-like fermions $\Psi_{F_1}$, presented in
\cref{sec:toymodelsec}.  \cref{tab:SSMVLevenValidateTable} contains
benchmark points for the partial widths of the digluon and diphoton
channels as well as the ratio of these two channels for both CP-even
and CP-odd scalar resonances. This table contains the LO calculations
performed using \SPheno as a comparison to results previously shown in
the literature \cite{Knapen:2015dap}. We also show the partial widths
including NLO corrections for the diphoton channel\footnote{NLO corrections in the case of a CP-odd vanish in the limit $m_f \gg m_S$, see \cref{sec:diphotoncalc} for more detail.} and
N${}^3$LO QCD corrections to the gluon fusion production as implemented in
\cref{sec:diphotoncalc}. The discrepancies between the LO calculations
arise purely through the choice of the renormalisation scale for the
gauge couplings. However, the NLO results clearly emphasise that loop
corrections at the considered mass scales are the dominant source of
errors. To our knowledge, these uncertainties have thus far not
received a sufficiently careful treatment in the literature; we give
further discussion of this (and the remaining uncertainty in the
\SARAH calculation) in \cref{subsec:accuracy}.

\subsubsection{Constraints on a large diphoton width}
In order to explain the measured signal, one needs a large diphoton rate of 
$\Gamma(S\to \gamma\gamma)/M_S \simeq 10^{-6}$ assuming a narrow width for $S$, while for a large width one requires
 $\Gamma(S\to \gamma\gamma)/M_S \simeq 10^{-4}$ \cite{Franceschini:2015kwy}.
In weakly-coupled models there are three different possibilities to obtain such a large width:
\begin{enumerate}
 \item Assuming a large Yukawa-like coupling between the resonance and charged fermions
 \item Assuming a large cubic coupling between the resonance and charged scalars
 \item Using a large multiplicity and/or a large electric charge for the scalars and/or fermions in the loop
\end{enumerate}
However, all three possibilities are also constrained by very fundamental considerations, which we briefly summarise 
in the following.

\modelparalabel{Large couplings to fermions}{sec:motivation_perturbation}
A common idea to explain the diphoton excess is the presence of
vector-like states which enhance the loop-induced coupling of a
neutral scalar to two photons or two gluons.  This led some authors
to consider Yukawa-like couplings of the scalar to the vector-like fermions larger than
$\sqrt{4 \pi}$, which is clearly beyond the perturbative regime\footnote{This diphoton excess could be triggered 
by strong interactions. Of course, in this case one cannot use perturbative methods to understand it.}.
Nevertheless, a one-loop calculation is used in these analyses to obtain predictions for the partial 
widths \cite{Wang:2015omi}, despite being in a non-perturbative region of parameter space. \\
Moreover, even if the couplings are chosen to be within the perturbative regime at the scale $Q=\SI{750}{\GeV}$, they can quickly grow 
at higher energies. This issue of a Landau pole has been already discussed to some extent in the literature 
\cite{Franceschini:2015kwy,Goertz:2015nkp,Son:2015vfl,Bertuzzo:2016fmv,Bae:2016xni,Salvio:2016hnf}, 
and one should ensure that the model does not break down at unrealistic small scales.

\begin{figure}[hbt]
\includegraphics[width=0.49\linewidth]{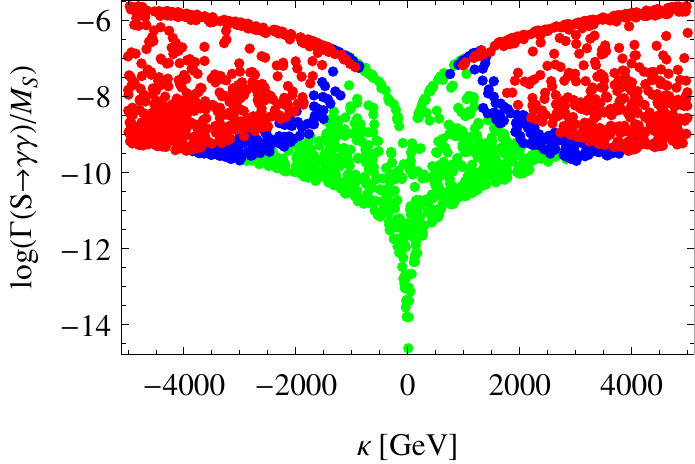} \hfill 
\includegraphics[width=0.49\linewidth]{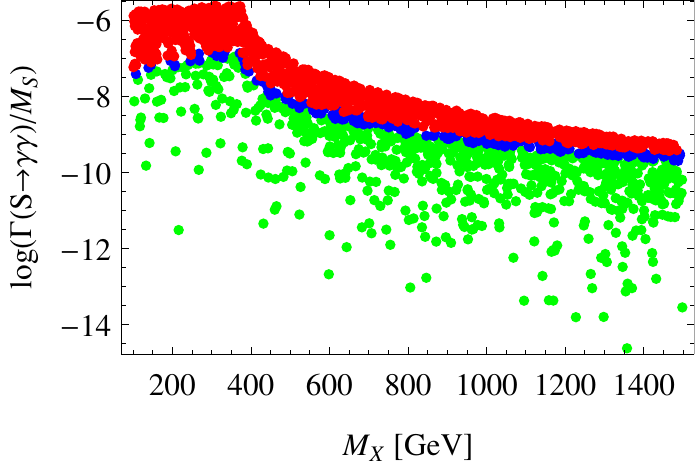} 
\caption{$\Gamma(S\to \gamma\gamma)/M_S$ as a function of $\kappa$
  (left) and $M_X$ (right). Green points have a stable vacuum,
  blue points have a meta-stable but long-lived vacuum, while for the red ones it decays in a short time, in comparison to cosmological time scales, with a survival probability below 10\%. The hypercharge of $X$ was
  set to $Y=1$.}
\label{fig:VacuumX}
\end{figure}

\modelparalabel{Large couplings to scalars}{}
One possibility to circumvent large Yukawa couplings
is to introduce charged scalars, which give large loop contributions
to the diphoton/digluon decay. A large cubic coupling between the
charged scalar and the $750$~GeV one does not lead to a Landau pole
for the dimensionless couplings because of dimensional
reasons. However, it is known that large cubic couplings can
destabilize the scalar potential: if they are too large, the
electroweak vacuum could tunnel into a deeper vacuum where 
$U(1)_{\text{em}}$ gauge invariance is spontaneously broken, depending
on the considered scenario.  The simplest example with such a
scenario is the SM, extended by a real scalar $S$ and a complex scalar
$X$ with hypercharge $Y$. The scalar potential of this example is
\begin{equation}
  V \supset  \kappa \, S |X|^2 + \frac12 M_S S^2 + M_X |X|^2 + \cdots .
\end{equation}
In Fig.~\ref{fig:VacuumX} the dependence of the
diphoton partial width as a function of $\kappa$ and $M_X$ is shown, and the
stability of the electroweak potential as well as the life-time of its
ground state is checked with \Vevacious and {\tt CosmoTransitions}.
For more details about vacuum stability in the presence of large
scalar cubic terms, we refer to Ref.~\cite{Salvio:2016hnf}. 
 The
overall conclusion of \cite{Salvio:2016hnf} is that the maximal
possible diphoton width, even when allowing for a meta-stable but
sufficiently long-lived electroweak vacuum, is not much larger than in
the case of vector-like fermions when requiring that the model is
perturbative up to the Planck scale.
  It is therefore essential to
perform these checks when studying a model that predicts large cubic
scalar couplings.

\begin{figure}[hbt]
\centering
\includegraphics[width=0.6\linewidth]{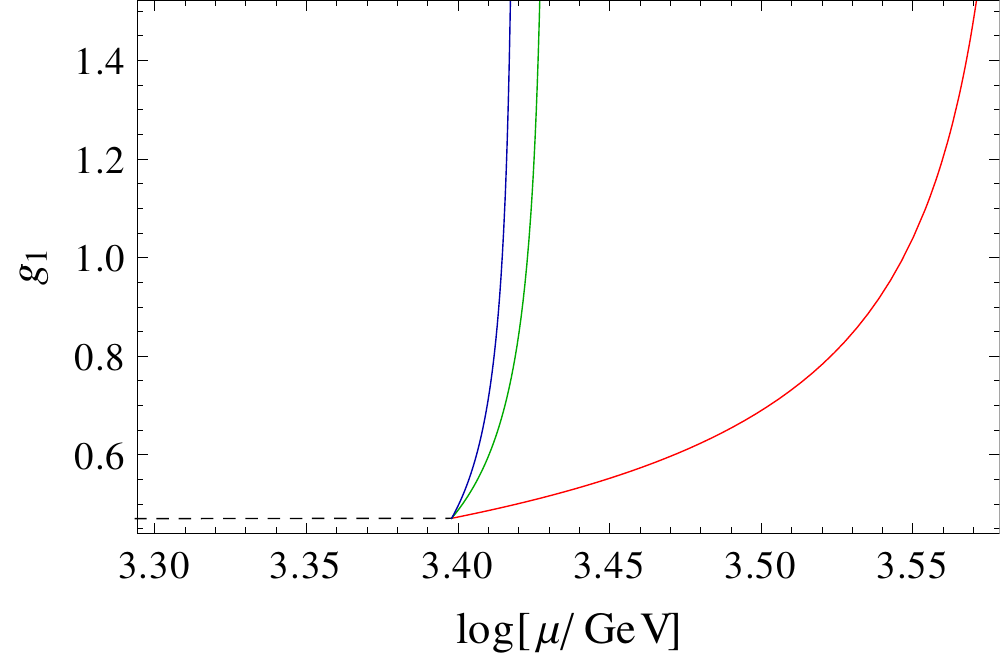}
\caption{Running of the $U(1)_Y$ gauge coupling, $g_1$, in the model
  presented in \cite{Nomura:2016fzs} for $N_k = 1000$ (red), $N_k = 6000$
  (green) and $N_k = 9000$ (blue). The black dashed line corresponds
  to the SM running below $\mu_{NP} = 2.5$ TeV.}
\label{fig:running}
\end{figure}

\modelparalabel{Large multiplicities}{} 
To circumvent large Yukawa or cubic couplings, other models require a large number of generations of
new BSM fields and/or large electric charges.   
 As a consequence the running of the $U(1)_Y$ gauge coupling,
  $g_1$, gets strongly enhanced  well below the Planck scale. 
   Moreover, even before
reaching the Landau pole, the model develops large (eventually
non-perturbative) gauge couplings. This implies an enhancement of
Drell-Yan processes at the LHC, with current data already setting
stringent constraints and potentially excluding some of the models
proposed to explain the diphoton excess \cite{Gross:2016ioi,Goertz:2016iwa}.
For general studies on running effects in the
  context of the diphoton excess see
  \cite{Franceschini:2015kwy,Goertz:2015nkp,Son:2015vfl,Bertuzzo:2016fmv,Bae:2016xni}.
We briefly discuss dramatic examples of this class of models
proposed in Refs.~\cite{Kanemura:2015bli} and \cite{Nomura:2016fzs}, which feature
approximately $\sim\,100$ and $6000$--$9000$ generations of doubly-charged
scalar fields respectively. In the model of Ref.
\cite{Nomura:2016fzs} the SM particle content is
enlarged by a vector-like doubly-charged fermion
$E$, a Majorana fermion $N_R$, a singlet scalar $S$, a singly-charged
scalar $h^+$ and $N_k$ generations of the doubly-charged scalar field
$k^{++}$.  At the one-loop level the
running of $g_1$ is governed by the renormalisation group equation (RGE)
\begin{equation}
\frac{d g_1}{dt} = \beta_{g_1} = \frac{1}{16 \pi^2} \beta_{g_1}^{(1)} \, ,
\end{equation}
where $t = \log \mu$, $\mu$ being the renormalisation scale, and
\begin{equation} \label{eq:betaf}
\beta_{g_1}^{(1)} = \frac{g_1^3}{10} \, \left( 75 + 8 \, N_k \right)
\end{equation}
is the one-loop $\beta$ function.  It is clear from
\cref{eq:betaf} that a large value of $N_k$ necessarily leads to
a very steep increase of $g_1$ with the renormalisation scale, soon
reaching a Landau pole. This is shown in Fig.~\ref{fig:running},
obtained by setting the masses of all the charged BSM states to
$\mu_{\rm NP} = 2.5$ TeV, which is already the largest mass considered in this 
analysis.
 The running up to $\mu_{\rm NP}$ is governed by the SM RGEs,
and the result for $g_1$ is displayed with a black dashed line. For
scales above $\mu_{\rm NP} = 2.5$ TeV, the contributions from BSM fields
become effective.  \cref{fig:running} shows that a Landau pole can be
reached at relatively low energies once we allow for such large values of
$N_k$. In fact, for $N_k = 9000$, we find that a Landau pole appears
already at $\mu \simeq 2.6$ TeV. In this specific example the appearance of a Landau
pole below $10^{16}$ GeV is unavoidable as soon as $N_k > 10$, as
shown in \cref{tab:Landau}.
\begin{table}[!t]
\centering
\begin{tabular}{rr}
\toprule
$N_k$ & $\mu_{\text{Landau}}$ \\
\midrule
$10$ & $2 \cdot 10^{13}$ GeV \\
$100$ & $1.2 \cdot 10^5$ GeV \\
$1000$ & $3.8$ TeV \\
$6000$ & $2.7$ TeV \\
$9000$ & $2.6$ TeV \\
\bottomrule
\end{tabular}
\caption{Energy scale at which a Landau pole in $g_1$ appears as a
  function of $N_k$ in the model of Ref.~\cite{Nomura:2016fzs}.}
\label{tab:Landau}
\end{table}

\subsubsection{How do the tools help?}\label{diphoton_howdotoolshelp}
The tools which we describe in more detail in the following sections
can help to address all the above issues:
\begin{enumerate}
 \item \FS and \SPheno can calculate the diphoton and digluon rate
   including important higher order corrections.
 \item Using the effective vertices calculated by \FS /\SPheno and the
   interface to \CalcHep or \MG, the gluon-fusion production
   cross-section of the 750~GeV mediator can be calculated numerically
   and one does not have to rely on analytical (and sometimes
   erroneous\footnote{It is straightforward to see that the analytical
     estimate of the production cross section in Eq.~(10) of
     Ref.~\cite{Pilaftsis:2015ycr} 
     is wrong by orders of magnitude:
     consider the production of a SM-like scalar $H$ with
     $m_H=\SI{750}{\GeV}$ via top-loops. Then, the factor $h_F^2
     m_t^2/m_F^2$ drops out and one obtains $\sigma=1440$~pb which is
     too large by roughly three orders of magnitude
     \cite{Dittmaier:2011ti}. The authors of Ref.~\cite{Martinez:2015kmn} (which originally
     made use of this analytic estimate) have revised their results in an updated version of their
     paper.})  approximations.
 \item \SARAH calculates the RGEs for a model which can be used to
   check for the presence of Landau pole.
 \item \Vevacious can be used to check the stability of the scalar
   potential.
\end{enumerate}

\subsection{Properties of the 750~GeV scalar}
\subsubsection{Mixing with the SM Higgs}\label{sect:motivation:mixing}
It is often assumed that $S$, although it is a CP-even scalar, does
not mix with the SM-like Higgs $h$. However, if this is done in a very
ad hoc way and not motivated by any symmetry, this assumption will not
hold when radiative corrections are taken into account. To see this,
one can consider, for example, the scalar potential
\begin{align}
  \begin{split}
    V &= \frac12 M_S S^2 + M_X |X|^2 + \mu^2 |H|^2 + \kappa S |X|^2\\
    &\phantom{={}}+ \lambda_{S} S^4 + \lambda_{SX} S^2 X^2 +
    \lambda_{HX} |H|^2 |X|^2 + \lambda |H|^4 \, ,
  \end{split}
\end{align}
where $H$ is the SM Higgs $SU(2)_L$ doublet, which contains the SM Higgs $h$.
This potential in principle has all ingredients to get a large diphoton decay
of $S$ via a loop involving the charged scalar $X$. Note, however, that
the potentially dangerous term $\kappa_H \, S |H|^2$ has been
omitted. One can see immediately that this term would be
generated radiatively by the diagram below.
\begin{center}
 \includegraphics[width=0.3\linewidth]{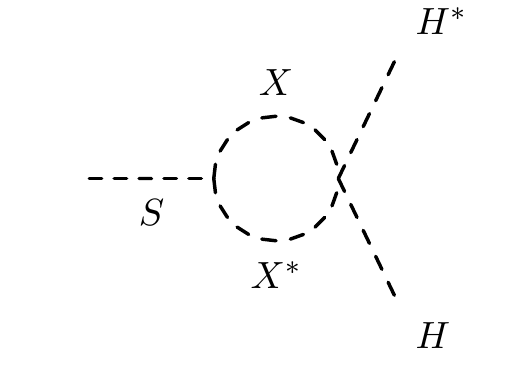} 
\end{center}
Note that it is also not possible to circumvent this decay by
forbidding the $\lambda_{HX}$ term: since $H$ and $X$ are charged
under $SU(2)_L \times U(1)_Y$, also the $\lambda_{HX} |H|^2 |X|^2$
term would be generated radiatively via diagrams like
\begin{center}
 \includegraphics[width=0.3\linewidth]{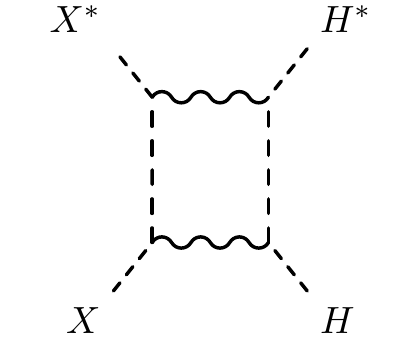} 
\end{center}
Such a mixing has important consequences since it opens the decay
channels $S\to h h$ and $S\to ZZ$, $S\to W^+ W^-$ at tree-level, which
are tightly constrained.

\begin{figure}[hbt]
\centering
 \includegraphics[width=0.5\linewidth]{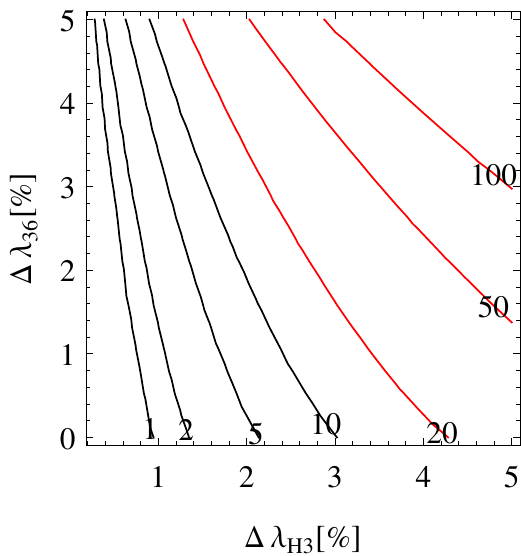}
 \caption{The impact on small deviations from the parameter relations
   assumed in Ref.~\cite{Modak:2016ung}.  Specifically, the y-axis axis
   represents the deviation in the coupling $\lambda_{36}$, while the
   x-axis represents deviations in the coupling $\lambda_{H3}$.  The
   contour lines show the ratio $\text{Br}(S\to W^+
   W^-)/\text{Br}(S\to \gamma \gamma)$. The red line indicates where
   the model would already be in conflict with current collider
   limits.}
 \label{fig:ComparisonBLVLtuning}
\end{figure}

Another possibility is that all terms allowed by symmetries are taken
into account, but very special relations among them are imposed like
in Ref.~\cite{Modak:2016ung}. When these relations hold, the above-mentioned 
tree-level decays in SM particles would cancel. However, as
long as there is no symmetry behind these relations, they will not be
invariant under RGE running. Therefore, immediately the question arises
how large the tuning among the parameters must be to have a point that
fulfils all constraints.  To illustrate this issue, we make small
  variations in the couplings $\lambda_{H3}$ and $\lambda_{36}$, which
  cause non-vanishing tree-level couplings between $S$ and the massive
  vector bosons, and check for which size of the deviations the
  condition $\text{Br}(S\to W^+ W^-)/\text{Br}(S\to \gamma \gamma) <
  20$ holds. The result is shown in
  \cref{fig:ComparisonBLVLtuning}. Here, the diphoton rate was
  maximized by setting the masses of the vector-like fields to
  $375$~GeV and using a Yukawa coupling of $O(1)$.
In principle, one could try to check what this means for the scale
dependence of these ratios by calculating the RGEs.  However, this
cannot really be done for this setup since one obtains the following
condition from the relations which have been imposed: $\lambda_{03} =
f_Y^2 \frac{M_S^2}{M_F^2}$, i.e. $\lambda_{03} = 4 f_Y^2$ is needed to
maximize the diphoton branching ratio. Thus, $f_Y$ of $\mathcal{O}(1)$ 
immediately leads to a huge quartic coupling.

Thus, in general, it is very difficult to justify the assumption that the
$750$~GeV particles do not mix with the SM-like Higgs if there is no
fundamental symmetry to forbid this mixing. However, this can already
be forbidden using the CP symmetry: the mentioned problems can be
circumvented in models where the diphoton excess stems from a CP-odd
particle. While, in the case of a CP-even particle, it is crucial to include
the mixing effects and to check at least how large the tuning in
parameters must be. 

\subsubsection{To VEV or not to VEV?}
\label{sect:motivation:vev}
\begin{figure}[hbt]
\centering
\parbox{0.2\linewidth}{
\includegraphics[width=\linewidth]{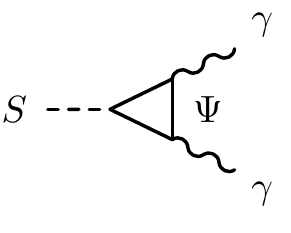}}
\parbox{0.2\linewidth}{\centering \bf $\Rightarrow$}
\parbox{0.2\linewidth}{
\includegraphics[width=\linewidth]{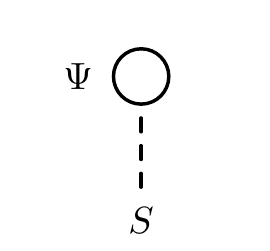}}
\caption{Tadpole terms for $S$ generated at one-loop level.}
\label{fig:TadSloop}
\end{figure}

The possibility that the new scalar receives a vacuum expectation value
(VEV) is also often neglected. However, as we have just
discussed, it often occurs that a $H$--$S$ mixing will be induced, at
least radiatively, in many models. Such radiative effects would
immediately lead to a non-zero VEV for the new scalar. Even in
cases where there is a symmetry which prevents a mixing with the SM
Higgs, the \SI{750}{\GeV} particle will still receive a VEV if it is a
CP-even scalar. This arises due to the introduced couplings to charged particles
which are necessary to allow diphoton and digluon decays. More specifically, these introduced
couplings will generate one-loop tadpole diagrams for $S$ as shown in
\cref{fig:TadSloop}.
Thus, the tadpole equation reads at the one-loop level
\begin{equation}
  \frac{\partial V^{(1L)}}{\partial v_S} = T^{(1L)} = T^{(T)} + \delta T = 0\,,
\end{equation}
where $T^{(T)}$ is the tree-level tadpole, given by
\begin{equation}
  \frac{\partial V^{(T)}}{\partial v_S} = 
  T^{(T)} = c_1 v_S + c_2 v^2_S  + c_3 v_S^3 = 0\,.
\end{equation}
Here, we have parametrised the tree-level expression so that the general form has the solution $v_S=0$. One finds in general that the one-loop corrections are
\begin{equation}
 \delta T = 
   \begin{cases}
     \kappa A(M^2_X) & \quad \text{for a scalar loop}\,,   \\
     2 Y M_\Psi A(M^2_\Psi) & \quad \text{for a fermion loop}\,,   \\
   \end{cases}
\end{equation}
with $A(x^2) = \frac{1}{16 \pi^2} x^2 \left[1 + \log
  (Q^2/x^2)\right]$. Taking $M_\Psi$, $\kappa$, $M_X$ of the order
1~TeV, results in a VEV which is naturally of order
$1\,\text{TeV}^3/(16 \pi^2 c_1)$. As a result, the simplifying assumption that
$v_S$ vanishes is in general hard to justify. Therefore, it is
important to check how the conclusions made about the model
depend on this assumption. Here, the tools discussed in the following sections
can really help, as including the non-vanishing $v_S$ is no more difficult than assuming the VEV vanishes.

\subsubsection{Additional decay channels}
\label{sec:motivation_constraints_BR}
Many analyses concentrate only on the decay $S\to \gamma \gamma$ and
completely neglect other potential decay channels. However, there are stringent 
constraints on the branching ratios of $S$ into other SM final states, which are summarised in 
\cref{tab:BRconstraints}.
\begin{table}[hbt]
\centering 
\begin{tabular}{c c c c c c c c c c c c c}
\toprule  
$e^+e^- + \mu^+ \mu^-$ & $\tau^+ \tau^-$ & $Z\gamma$ & $ZZ$ & $Zh$ & $hh$ &$W^+W^-$ & $t\bar t$ & $b \bar b$ & $jj$ & inv. \\
\hline 
0.6 & 6 & 6 & 6 & 10 &  20 & 20 & 300 & 500 & 1300 & 400 \\
\bottomrule
\end{tabular}
\caption{Limits on $\Gamma(S\to X)/\Gamma(S\to\gamma\gamma)$ assuming a production of $S$ via 
gluon fusion or heavy quarks. Values are taken from  Ref.~\cite{Franceschini:2015kwy}.}
\label{tab:BRconstraints}
\end{table}
Thus,
any model which tries to explain the excess via additional coloured
states in the loop must necessarily worry about limits from dijet searches
 \cite{Aad:2014aqa}. Therefore, an accurate calculation of the digluon decay rate 
 is a necessity.
As an example that illustrates why both additional channels and the diphoton/digluon width calculation are important we consider the model presented in Refs.~\cite{Cao:2015twy,Ding:2016ldt} and considered in more detail here in \cref{sec:scalarOctet}. 

This model extends the SM with a singlet and a scalar $SU(2)$-doublet colour octet. As an approximation the ratio of the singlet decays to gluons and to photons is 
\begin{align}
\frac{\Gamma (S \rightarrow gg)}{\Gamma (S \rightarrow \gamma \gamma)} \simeq \frac{9}{2} \frac{\alpha_s^2}{\alpha^2}.
\end{align}
In \cite{Cao:2015twy} this is quoted as $\simeq 750$; before any NLO corrections are applied, we find $700$. However, once we include all of the N$^3$LO corrections this is enhanced to $1150$, near the bound for constraints on dijet production at $8$ TeV and significantly squeezing the parameter space of the model. \\

Additionally in many works we observed that potential decay channels of the resonance were missed.
 For instance in Ref.
 \cite{Fabbrichesi:2016alj}, the authors, who considered the
 Georgi-Machacek model \cite{Georgi:1985nv}, missed the decay of the scalar into $W^\pm
 H^\mp$, which can be the
 dominant mode when kinematically
 allowed. 

\subsubsection{How do the tools help?}  
Many of the assumptions which we criticised were made to keep the study simple. However, 
when using the public tools presented in the next two section, there is no need for these simplifying assumptions:
\begin{enumerate}
 \item \SARAH automatically calculates all expressions for the masses and vertices, no matter how complicated they are. 
 \item \FS and \SPheno give numerical predictions for the mass spectrum and the mixing among all states 
  including higher order corrections. 
\end{enumerate}
Thus, for the user the study becomes no more difficult when he/she drops all simplifying assumptions but considers the 
model in full generality. Moreover, there is no chance to miss important effects in the decays of the new scalar:
\begin{enumerate}
 \item As outlined above and described in detail in \cref{sec:calculationDecays}, \FS and \SPheno calculate the diphoton and digluon rate very accurately
 \item \SPheno does calculate all other two body decays\footnote{Even three-body decays into another scalar and two fermions can be calculated with \SPheno.} of the scalar. This makes it impossible to miss any channel.
\end{enumerate}

\subsection{Considering a full model}
\subsubsection{Additional constraints in a full model}
There are several studies which extend an already existing model by
vector-like states and then assume that this part of the model is
decoupled from the rest. When this assumption is made it is clear that the
results from toy models, with the minimal particle content
will be reproduced. However, it is often not clear if this decoupling
can be done without invoking specific structures in the choice of
parameters, and if these assumptions hold at the loop level.
  
On the other hand, if model-specific features are used to explain the
diphoton excess, it is likely that there will be important
constraints on the model coming from other sectors. For instance, there might be
bounds from flavour observables, dark matter, Higgs searches, neutrino
mixing, electroweak precision observables, searches for BSM particles
at colliders, and so on.  All of that has to be checked to be sure
that any benchmark point presented is indeed a valid explanation for
all observations.  Such a wide range of constraints is much easier to
address by making exhaustive use of tools which provide a high level
of automation.

\subsubsection{Theoretical uncertainties of other predictions}
Even if the attempts are made to include the effects of the new states 
on other sectors of the model,
it is important to be aware that there are large uncertainties
involved in certain calculations. If the level of uncertainty is
underestimated, this can have an impact on what is inferred from
the calculation.  The large uncertainty in a LO calculation of the
diphoton and digluon rate has already been addressed in
\cref{sec:motivation_rates}. However, there are also other important
loop corrections especially in SUSY models: the accurate calculation
of the Higgs mass is a long lasting endeavour where for the simplest SUSY
models even the dominant three-loop corrections are partially tackled
\cite{Kant:2010tf}. The current ball-park of the remaining uncertainty
is estimated to be 3~GeV.

\begin{figure}[h!]
\center
 \includegraphics[width=0.6\linewidth]{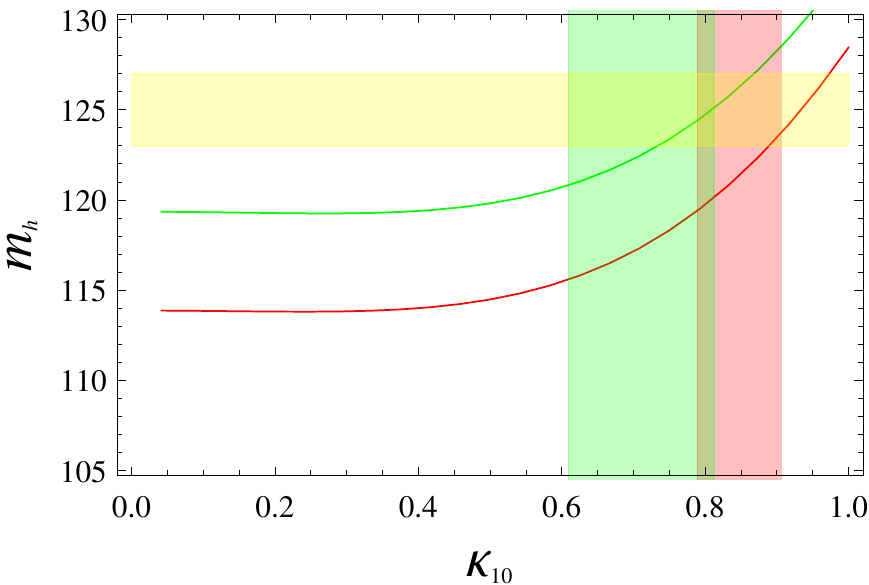}
 \caption{Comparison of the two-loop Higgs mass calculation of
   Ref.~\cite{Dutta:2016jqn} with the results obtained by \SPheno. The
   parameters are those of Fig.~7 in Ref.~\cite{Dutta:2016jqn} and we
   fixed $X_{\kappa_{10}}=0$. The lines are the results from \SPheno,
   while the green and red shades areas are the ranges of
   $\kappa_{10}$ which predict $m_h=[123,127]$~GeV according to
   Ref.~\cite{Dutta:2016jqn}. Red takes $X_t =4$ and green $X_t=2$,
   where $X_t$ is defined in Eq. (9) of the reference.}
 \label{fig:ComparisonNMSSMLV}
\end{figure}

However, it is clear that the MSSM cannot explain the excess, hence it
must be extended.  A common choice is to add additional pairs of
vector-like superfields together with a gauge singlet, see
\cref{sec:example}. These new fields can also be used to increase the SM-like Higgs mass.
However, this will in general
also increase the theoretical uncertainty in the Higgs mass
prediction, because these new corrections are not calculated with
the same precision as the MSSM corrections. For instance,
Ref.~\cite{Dutta:2016jqn} has taken into account the effect of the new
states on the SM-like Higgs. There, they use a one-loop effective 
potential approach considering the new Yukawa couplings to be $\mathcal{O}(1)$
or below, while also including the dominant two-loop corrections
from the stop quark. They assumed that including these corrections is sufficient in order to achieve an
uncertainty of 2~GeV in the Higgs mass prediction. One can compare
their results encoded in Fig.~7 of Ref.~\cite{Dutta:2016jqn} with a
calculation including, in addition to the corrections taken into
account in the paper, momentum dependence and electroweak
corrections at the one-loop level, as well as the additional two-loop
corrections arising from all the newly introduced states. These corrections can be important, as was shown for instance in
Ref.~\cite{Nickel:2015dna}. The result of the comparison is shown in
Fig.~\ref{fig:ComparisonNMSSMLV}.
We find a similar behaviour, but still there are several GeV
difference between both calculations. For $\kappa_{10}=0.8$ and
$X_t=4$, the point would be within the interesting range for
$m_h=[123,127]$~GeV, while the more sophisticated calculation predicts
a mass below 120~GeV. Thus, the assumed uncertainty of 2~GeV in
Ref.~\cite{Dutta:2016jqn}, which would even be optimistic in the MSSM,
is completely unrealistic without including all the aforementioned higher
order corrections. 

\subsubsection{How do the tools help?}
The tools help to ensure that one really considers all aspects of a full model:
\begin{enumerate}
 \item All masses of the model are calculated with high accuracy: \FS and \SPheno include the full. 
 one-loop correction to all fields in a model, while \SPheno covers even the dominant two-loop corrections introduced by adding new states.
 \item \SPheno makes predictions for all important flavour observables in the model.
 \item A link to \MO provides the possibility to calculate the dark matter relic density.
 \item The interface to \HB and \HS offers the possibility to check all constraints from Higgs searches and to 
 check if the results for the SM-like Higgs can be reproduced.
\end{enumerate}

%% file: tex/sarah.tex
\section{Using \SARAH to understand a model}
\label{sec:sarah}
\subsection{General}
One of the reasons that makes high energy particle physics an
exciting field is the vast amount of experimental data available. When
proposing a model one first has to check its self consistency,
checking for instance the particle mass spectrum and vacuum stability
requirements.  Then it has to be tested against data related to
collider searches, flavour observables, dark matter observations and
Higgs measurements. A lot of effort has been devoted to developing an
arsenal of specific tools to explore these quantities with high
precision for specific classes of models, such as the MSSM, the THDM
and the NMSSM to some extent. However, none of them, in their simplest
versions, can explain the diphoton excess.  For the time being there
is no specific model which is clearly preferred over others as an
explanation of the excess, as reflected by the large variety of models
that different authors have proposed, and it would be impractical to
repeat the process of developing a code for each one of them.  In the
absence of a dedicated tool, the alternative is often to resort to
approximations or just to leading order expressions, as described in
the previous section, in which case the analysis (in particular for
more complicated models) is of limited value.

Luckily, a dedicated powerful tool already exists.  It is the
\Mathematica package \SARAH
\cite{Staub:2008uz,Staub:2009bi,Staub:2010jh,Staub:2012pb,Staub:2013tta,Staub:2015kfa},
which can perform the most advanced quantum field theory computations
and apply them generically to any given model. \SARAH has been
optimised for an easy, fast, and exhaustive study of renormalisable
BSM models: not only can it calculate all the relevant quantities
within a given model analytically -- it also provides routines to
export the derived information in order to use it for numerical
calculations with dedicated tools. \SARAH can be used for SUSY and
non-SUSY models to write model files for \CalcHep/\CompHep
\cite{Pukhov:2004ca,Boos:1994xb}, \FeynArts/\FormCalc
\cite{Hahn:2000kx,Hahn:2009bf}, \WHIZARD/\OMEGA
\cite{Kilian:2007gr,Moretti:2001zz}, as well as in the \UFO format
\cite{Degrande:2011ua} which can be handled for instance by \MGv{5}
\cite{Alwall:2011uj}, {\tt GoSam} \cite{Cullen:2011ac}, {\tt Herwig++}
\cite{Gieseke:2003hm,Gieseke:2006ga,Bellm:2013hwb}, and {\tt Sherpa}
\cite{Gleisberg:2003xi,Gleisberg:2008ta,Hoche:2014kca}.  The modules
created by \SARAH for \SPheno calculate the full one-loop and
partially two-loop corrected mass spectrum, branching ratios and decay
widths of all states, and many flavour and precision observables
thanks to the \FlavorKit \cite{Porod:2014xia} functionality. Moreover,
an easy link to \HB \cite{Bechtle:2008jh,Bechtle:2011sb} and \HS
\cite{Bechtle:2013xfa} exists. One can as well use another tailor-made
spectrum generator, \FS\ \cite{Athron:2014yba}, which can handle both
SUSY and non-SUSY models generated with \SARAH.  Finally, \SARAH can
also produce model files for \Vevacious \cite{Camargo-Molina:2013qva},
which is a tool dedicated to studying vacuum stability.

\subsection{Models supported by \SARAH}
\SARAH is optimised for handling a wide range of SUSY and non-SUSY
models. The basic idea of \SARAH is to give the user the possibility
to implement models in an easy, compact and straightforward way. The
user simply has to declare symmetries, particle content, and the
(super)potential. All steps to derive the full Lagrangian for the
gauge eigenstates from this information are then fully
automatised. But this is not the final aim: usually one is interested
in the mass eigenstates after spontaneous gauge symmetry breaking. To
perform the necessary rotations to the new eigenstates, the user has
to provide additional information: (i) definition of the fields which
develop VEVs, and (ii) definitions of the fields which mix. Using this
information, all necessary re-definitions and field rotations are
performed by \SARAH. In addition, gauge fixing terms for the new
eigenstates are derived and ghost interactions are added. Plenty of
information can be derived by \SARAH for all states in the model, as
explained in \cref{sec:analytical}. Before moving into it, we
elaborate on the kind of models and features supported by \SARAH.

\subsection{Supported models and features}
\label{sec:supported_models}

\subsubsection{Global symmetries}
\SARAH can handle an arbitrary number of global symmetries which are
either $Z_N$ or $U(1)$ symmetries.  In a SUSY model one can also
impose a continuous $R$-symmetry, $U(1)_R$, at the level of the
superpotential. However, \SARAH will then generate automatically the
scalar potential including soft trilinear terms that do not respect
the $U(1)_R$.  The user has two options: either add ``{\tt AddTterms =
  False}'' to the \SARAH model file to forbid the automatic generation
of the soft trilinear terms, or set such terms to zero manually within
the input files for the spectrum generators.  \SARAH can also handle
approximate symmetries: if the user specifies terms in the
Lagrangian/superpotential which violate a global symmetry, the terms
will not be forbidden but a warning will be printed.

\subsubsection{Gauge sector}
\paragraph*{Gauge groups} \SARAH is not restricted to the SM gauge
sector, but many more gauge groups can be defined. To improve the
power in dealing with gauge groups, \SARAH has linked routines from
the \Mathematica package \Susyno \cite{Fonseca:2011sy}. \SARAH,
together with \Susyno, takes care of all group-theoretical
calculations: it calculates the Dynkin and Casimir invariants, derives
the needed representation matrices as well as the generalised
Clebsch-Gordan coefficients. For all Abelian groups one can also give
a GUT normalisation; these factors usually arise from
considerations about embedding a model in a larger symmetry group such as
$SU(5)$ or $SO(10)$.

\paragraph*{Gauge kinetic mixing} 
With more than one Abelian gauge group, terms of the form
\begin{equation}
\label{eq:off}
\mathcal{L} = -\frac{1}{4} \kappa F_{\mu\nu}^A F^{B,\mu\nu}, \hspace{1cm} A\neq B \, ,
\end{equation}
are allowed, where $F^{\mu\nu}$ are the field strength tensors of two different
Abelian groups $A$, $B$ \cite{Holdom:1985ag}. $\kappa$ is in general
an $n \times n$ matrix if $n$ Abelian groups are present. \SARAH fully
includes the effects of kinetic mixing for any number of
Abelian groups by performing a rotation to bring
the field strength to a diagonal form. The kinetic mixing is then absorbed into
the covariant derivatives, which take the form
\begin{equation}
  D_\mu \phi = \left(\partial_\mu - i \sum_{x,y} Q_\phi^x g_{xy}   V^\mu_y \right)\phi.
\end{equation}
The indices $x,y$ run over all $U(1)$ groups, and $g_{xy}$ are the
entries of the gauge coupling matrix $G$ which now also contains mixed gauge couplings. Gauge kinetic mixing is not only included
in the interactions with vector bosons, but also in the derivation of
the $D$-terms and in the gaugino interactions for SUSY models.

\subsubsection{Matter sector} 
\label{sec:mattersector}
One can define up to 99 matter fields in a single model in
\SARAH. Each one of them can come with an arbitrary number of
generations and can transform as any irreducible representation with
respect to the defined gauge groups.

\paragraph*{Supersymmetric models}
The matter interactions in SUSY models are usually fixed by the
superpotential and the soft SUSY-breaking terms. \SARAH takes as input
the renormalisable terms in the superpotential
\begin{equation}
\label{eq:W}
 W = c_L L_i \hat \phi_i + c_M M^{ij} \hat \phi_i \hat \phi_j + c_T Y^{ijk} \hat \phi_i \hat \phi_j \hat \phi_k 
\end{equation}
which the user has to write in the model file,
and automatically generates the corresponding soft-breaking terms
\begin{equation}
  L_{SB,W} =  c_L t_i  \phi_i + c_M B^{ij}  \phi_i \phi_j + c_T T^{ijk}  \phi_i  \phi_j  \phi_k + \text{h.c.}
\end{equation}
$c_L$, $c_M$, $c_T$ are real coefficients, while the linear, bilinear, and trilinear
 parameters are treated
by default in the most general way by taking them as complex tensors
of appropriate order and dimension. If identical fields are involved
in the same coupling, \SARAH also derives the symmetry properties for
the parameter.

In recent years models with Dirac gauginos have been largely explored.
They feature mass terms of the form $m^{\hat \phi_i A}_D \lambda_A
\psi_i $, where $\lambda_A$ is a gaugino and $\psi_i$ the fermionic
component of a chiral superfield $\hat \phi_i$ in the adjoint
representation of the gauge group $A$. In addition, there are new
$D$-term couplings (see e.g. \cite{Benakli:2011vb}). To generate Dirac
mass terms for all the gauginos, these models always come with an
extended matter sector, including at least one singlet, one triplet
under $SU(2)$, and one octet under $SU(3)$. Furthermore, these models
generate new structures in the RGEs \cite{Goodsell:2012fm}. All these
are fully supported in \SARAH.

\paragraph*{Non-Supersymmetric models}
For non-supersymmetric models, \SARAH supports all general, renormalisable  Lagrangians of the form
\begin{equation}
L =  m_{ij}^2  \phi_i \phi_j + \frac{1}{3} \kappa_{ijk}  \phi_i  \phi_j  \phi_k + \frac{1}{4} \lambda \phi_i \phi_j \phi_k \phi_l + M^F_{ij} \psi_i \psi_j + Y_{ijk} \phi_i \psi_i \psi_j
\end{equation}
for scalars $\phi_i$, and Weyl fermions $\psi_j$. The Lagrangian needs to be defined
by the user in the model file. Note, that we have omitted a tadpole term, $t \phi$, 
for a gauge singlet, as it can always be absorbed in a shift of $\phi$. 

\subsection{Checks of implemented models}
\label{sec:checks}
After the initialisation of a model, \SARAH provides functions to
check its (self-)con\-sis\-ten\-cy:
\begin{itemize}
 \item Check for gauge anomalies, and mixed gauge/gravity anomalies;
 \item Check for Witten anomalies \cite{Witten:1982fp};
 \item Check if all terms in the (super)potential are in agreement with all global and gauge symmetries;
 \item Check if terms allowed by all symmetries are missing in the (super)potential;
 \item Check if additional mass eigenstates can mix in principle;
 \item Check if all mass matrices are irreducible. 
\end{itemize}
In addition, \SARAH performs other formal checks. For instance, it checks
if the number of Particle Data
Group numbers for a given family of particles fits to the number of
generations for each particle class, if \LaTeX\ names are defined for
all particles and parameters, and if the position in a Les Houches
spectrum file is defined for all parameters.

\subsection{Analytical calculations performed by \SARAH}
\label{sec:analytical}
The full power of \SARAH can be unleashed on exhaustive numerical
analyses via the dedicated interfaces to other tools, discussed in the
next sections. However, since \SARAH itself is a \Mathematica package,
it is capable of many analytical computations beyond the derivation of
the Lagrangian. Here we list some of them.

\subsubsection{Tree-level properties}
\paragraph{Tadpole equations} During the evaluation of a model, \SARAH calculates `on the fly' all the minimisation conditions of the tree-level potential, the so-called
tadpole equations.

\paragraph{Masses} \SARAH calculates the mass matrices for the states which are rotated to mass eigenstates. In addition, it calculates the masses of states for which no field rotation has taken place. 

\paragraph{Vertices} \SARAH has functions to extract in an efficient way all tree-level vertices from the Lagrangian. These vertices are saved in different \Mathematica arrays according to their generic type. 

\subsubsection{Renormalization group equations}
\label{sec:RGEs}
\SARAH calculates the full two-loop RGEs in SUSY and non-SUSY models including the full CP and flavour structure. For this purpose, it makes use of the most sophisticated generic calculations available in the literature.
\paragraph*{SUSY RGEs} The calculation of the SUSY RGEs is mainly based on Ref.~\cite{Martin:1993zk}, which
however did not cover all possible subtleties which can appear in SUSY models. \SARAH has implemented also results from more recent literature: 
\begin{itemize}
 \item In the case of multiple $U(1)$ gauge groups, gauge-kinetic mixing can arise if the groups are not orthogonal. Substitution rules to translate the results of Ref.~\cite{Martin:1993zk} to those including gauge kinetic mixing were presented in Ref.~\cite{Fonseca:2011vn} and are used by \SARAH; 
 \item The calculation of the RGEs in the presence of Dirac gauginos is based on Ref.~\cite{Goodsell:2012fm};
 \item The results of Refs.~\cite{Sperling:2013eva,Sperling:2013xqa} are used to obtain the gauge dependence in the running of the VEVs.
\end{itemize}

\paragraph*{Non-SUSY RGEs} \SARAH uses the expressions of Refs.~\cite{Machacek:1983tz,Machacek:1983fi,Machacek:1984zw,Luo:2002ti} for the calculation of the RGEs in a general quantum field theory. These results are completed by Ref.~\cite{Fonseca:2013bua} to cover gauge kinetic mixing and again by Refs.~\cite{Sperling:2013eva,Sperling:2013xqa} to include the gauge-dependence of the running VEVs also in the non-SUSY case.

\subsubsection{One- and two-loop corrections to tadpoles and self-energies}
\label{sec:loopcorrections}
\paragraph{One-loop corrections} \SARAH calculates the analytical expressions for the one-loop corrections to the tadpoles and the one-loop self-energies for all the particles. For states which are a mixture of several gauge eigenstates, the self-energy matrices are calculated. The calculations are performed in the $\DRbar$ scheme using 't Hooft gauge for SUSY models. In the case of non-SUSY models \SARAH switches to the $\MSbar$ scheme. 

\paragraph{Two-loop corrections} It is even possible to go beyond one loop with \SARAH and to calculate two-loop contributions to the self-energies of real scalars. There are two equivalent approaches implemented in the \SPheno interface of \SARAH to perform these calculations: an effective potential approach and a diagrammatic approach with vanishing external momenta. More details about these calculations are given in \cref{sec:TwoLoop}.

%% file: tex/sg.tex
\section{Spectrum calculation, Monte-Carlo studies, and more}
\label{sec:sg}
As mentioned in the last section, \SARAH can use the analytical information derived about a model and pass it to other tools. We give in the following an overview about the different possibilities.

\subsection{\SPheno}
\SARAH writes Fortran source code
for \SPheno \cite{Porod:2003um,Porod:2011nf} using the derived information about the 
mass matrices, tadpole equations, vertices, loop corrections and RGEs for the given model. 
With this code the user gets a fully functional spectrum generator for the model of their choice. 
The features of a spectrum generator created in this way are 
\begin{itemize}
\item Full {two-loop running} of all parameters
\item One-loop corrections to all masses
\item {Two-loop} corrections to {Higgs} masses
\item Complete {one-loop thresholds} at $M_Z$
\item Calculation of {flavour} and {precision observables} at full one-loop level
\item Calculation of {decay widths} and {branching ratios} for two-- and three body decays
\item Interface to {\tt HiggsBounds} and {\tt HiggsSignals}
\item Estimate of electroweak {Fine-Tuning}
\item Prediction for LHC cross sections for all neutral scalars
\end{itemize}

\subsubsection{Mass calculation with \SPheno}
\paragraph{Threshold corrections}
For a precise calculation of the masses it is necessary to have an accurate input for all parameters which enter the calculation. In general, the running SM parameters depend on the masses of the BSM states. This is due to the influence of the threshold scales, required to match the running parameters to the measured ones. The routines generated by \SARAH perform a full one-loop matching in the given model to calculate the SM gauge and Yukawa couplings. This matching takes the constraints from the CKM matrix into account even if there are additional states which mix with the SM quarks.

\paragraph{One-loop shifts to pole masses}
The one-loop mass spectrum is calculated from the information about the one-loop self-energies and tadpole equations. The procedure is a generalisation of the one explained in detail for the MSSM in Ref.~\cite{Pierce:1996zz}. The main features are 
\begin{itemize}
 \item Any one-loop contribution in a given model to all fermions, scalars and vector bosons is included 
 \item The full $p^2$ dependence in the loop integrals is included
 \item An iterative procedure is applied to find the on-shell masses $m(p^2=m^2)$. 
\end{itemize}

\paragraph{Two-loop shifts to Higgs pole masses}
\label{sec:TwoLoop}
\SARAH can also generate Fortran code to calculate the two-loop corrections to the masses of the CP-even scalar states with \SPheno\footnote{At  the moment, these calculations are just done in the \DRbar\ scheme, but can be also provided for \MSbar\ if necessary.}. The same approximations usually taken for the MSSM are also applied here: (i) all calculations are performed in the gaugeless limit, i.e.\ the electroweak contributions are dropped, and  (ii) the momentum dependence is neglected. Using these routines, the theoretical uncertainty in the Higgs mass prediction for many models has been shrunk to the level of the MSSM. In general, there are two different techniques to calculate the two-loop corrections with \SARAH--\SPheno:
\begin{itemize}
 \item {\bf Effective potential calculation} \cite{Goodsell:2014bna}: \SARAH makes use of the generic two-loop results for the effective potential given in Ref.~\cite{Martin:2001vx}. To get the values for the two-loop self-energies and two-loop tadpoles, the derivatives of the potential with respect to the VEVs are taken numerically, as proposed in Ref.~\cite{Martin:2002wn}.
 \item {\bf Diagrammatic calculation} \cite{Goodsell:2015ira}: A fully diagrammatic calculation for two-loop contributions to scalar self-energies with \SARAH--\SPheno became available with Ref.~\cite{Goodsell:2015ira}. The advantage of the diagrammatic approach is that no numerical derivation is needed. This is now the default calculation. 
\end{itemize}
The implementation of the two independent approaches provides a good possibility to double check results. It has been shown that these generic calculations can provide important two-loop corrections for the NMSSM which are not included in dedicated spectrum generators for the NMSSM \cite{Goodsell:2014pla,Staub:2015aea}.

\subsubsection{Decay widths and branching ratios}
\SPheno modules created by \SARAH calculate all two-body decays for fermion and scalar states as well as for the additional gauge bosons. In addition, the three-body decays of a fermion into three other fermions, and of a scalar into another scalar plus two fermions are included.

In the scalar sector, possible decays into two particles are calculated at tree level. In case of two quarks in the final state, the dominant QCD corrections due to gluons are included \cite{Spira:1995rr}. In addition scalar decays into final states with off-shell gauge bosons ($Z Z^*$, $W W^*$) are included.

The loop-induced decays into two photons and gluons are calculated up to $\text{N}^3$LO. More details about this are given in \cref{sec:diphotoncalc}.

\subsubsection{Flavour observables}
The \SPheno modules written by \SARAH contain out-of-the-box routines to calculate many quark and lepton flavour violating observables:
\begin{itemize}
 \item Lepton flavour violation:
\begin{itemize}
  \item Br($\ell_i \to \ell_j \gamma$), Br($\ell_i \to \ell_j \ell_k \ell_k$), Br($Z\to \ell_i \ell_j$)
  \item CR($\mu-e,N$) { (N=Al,Ti,Sr,Sb,Au,Pb)}, Br($\tau\to \ell P$) {(with $P$=$\pi$, $\eta$,$\eta'$)}
 \end{itemize} 
 \item Quark flavour violation:
\begin{itemize}
  \item Br($B\to X_s\gamma$), Br($B_{s,d}^0 \to \ell \bar{\ell}$), Br($B \to s \ell \bar{\ell}$), Br($K \to \mu \nu$)
  \item Br($B \to q \nu\nu$), Br($K^+ \to \pi^+ \nu\nu$), Br($K_L \to \pi^0 \nu\nu$)
  \item $\Delta M_{B_s,B_d}$, $\Delta M_K$, $\epsilon_K$, Br($B \to K \mu \bar{\mu}$)
  \item Br($B\to \ell \nu$), Br($D_s \to \ell \nu$)
 \end{itemize} 
\end{itemize}
The calculation is based on the \FlavorKit functionality \cite{Porod:2014xia}, which makes use of the chain \FeynArts--\SARAH--\SPheno. This provides a full one-loop calculation in a given model. In addition, this interface can be used to derive Wilson coefficients for new operators, and to calculate new observables with \SPheno using the implemented coefficients. 

\subsubsection{Fine-Tuning}
A widely used measure for the electroweak fine-tuning was proposed in Refs.~\cite{Ellis:1986yg, Barbieri:1987fn} 
\begin{equation} 
\label{eq:measure}
\Delta_{FT} \equiv \max {\text{Abs}}\big[\Delta _{\alpha}\big],\qquad \Delta _{\alpha}\equiv \frac{\partial \ln
  M_Z^{2}}{\partial \ln \alpha} = \frac{\alpha}{M_Z^2}\frac{\partial M_Z^2}{\partial \alpha} \;.
\end{equation}
Here, $\alpha$ is a set of independent parameters, and $\Delta_\alpha^{-1}$ is a measure of the accuracy to which the parameters $\alpha$ must be tuned to obtain the correct VEV. The user can choose the set of parameters $\alpha$ in \SARAH, and \SPheno numerically calculates $\Delta_{FT}$ for that choice using the full two-loop RGEs from the GUT or SUSY breaking scale down to the electroweak scale.

\subsubsection{Production cross-sections}
\label{sec:XS}
\SPheno provides an estimate for the production cross sections of all neutral scalars within a certain mass range: values for gluon-fusion and vector-boson fusion are obtained for $m\in [50,1000]$~GeV, while associated production with $W$, $Z$ and $t$ are considered for $m\in [50,300]$~GeV. The results are obtained by re-weighting the SM cross sections with the effective coupling of the scalar in the model to SM states normalised to the SM values. For 7 and 8~TeV the SM cross-sections provided by the Higgs cross section working group are used, while for 13, 14 and 100~TeV the cross sections have been calculated with {\tt SusHi 1.5.0} \cite{Harlander:2012pb}.

\subsection{\FS}

\FS\ is a Mathematica package which uses the \SARAH-generated
expressions for the mass matrices, self-energies, tadpole equations,
vertices and RGEs to create a C++ spectrum generator for both SUSY and
non-SUSY models.  The spectrum generators created with \FS have the
following features:
\begin{itemize}
\item full two-loop running of all parameters
\item three-loop running of all parameters in the SM and MSSM, except
  for the VEVs
\item calculation of the pole mass spectrum at the full one-loop level
\item partial two-loop corrections to the Higgs masses in the SM,
  SplitMSSM, MSSM, NMSSM, UMSSM and E$_6$SSM and partial three-loop
  corrections to the Higgs mass in the SplitMSSM
\item complete one-loop and partial two-loop threshold corrections at
  the scale $Q = M_Z$ or $Q = M_t$
\item calculation of the $h\gamma\gamma$ and $hgg$
  effective couplings at $\text{N}^3$LO
\item an interface to {\tt GM2Calc} \cite{Athron:2015rva} in MSSM models
  without sfermion flavour violation   
\end{itemize}
\FS\ aims to generate spectrum generators which are modular such that
components can be easily reused or replaced.  This means that it is
quite easy to re-use the precision calculations in \FS spectrum
generators for other purposes or add additional routines.

\subsubsection{Mass calculation with \FS}

\paragraph{Determination of the gauge couplings, Yukawa couplings and
  the Standard Model VEV}

\FS\ generates routines allowing for a one-loop calculation of the
$SU(3)_c\times SU(2)_L\times U(1)_Y$ gauge couplings, if they exist in
the model under consideration.  Furthermore, the Yukawa couplings
corresponding to the Standard Model fermions can be calculated in the
considered model at the full one-loop level from the known fermion
masses.  To determine the top quark Yukawa coupling, two-loop Standard
Model QCD corrections are also added.  \FS\ also performs a
complete one-loop calculation of the running Z and W masses at the
low-energy scale in the considered model, which can be used to
determine the running Standard Model-like vacuum expectation value,
$v$.

\paragraph{One-loop shifts to pole masses}

\FS\ by default performs a full one-loop $\MSbar$/$\DRbar'$\ calculation to
determine the pole masses of all particles in the model, similar to
the procedure presented in Ref.~\cite{Pierce:1996zz} for the MSSM.
Thereby, it makes use of the one-loop self-energies and tadpole
diagrams generated by \SARAH, taking the full momentum-dependence into
account.  To tune the spectrum generator, the user can choose from
three different precision levels, which differ in the way two-loop
momentum dependent terms are treated.  By default, the Higgs masses
are calculated with highest precision, where an iteration over the
momentum is performed to determine the pole mass $M_h$ at $p^2 = M_h^2$.

\paragraph{Two-loop and three-loop shifts to Higgs pole masses}

\FS\ allows the user to add certain predefined two-loop corrections to
the Higgs masses in some specific models: In MSSM-like models the
two-loop corrections of the order $O((\alpha_t + \alpha_b)^2 +
\alpha_t \alpha_s + \alpha_b \alpha_s + \alpha_\tau^2)$ from
\cite{Degrassi:2001yf,Brignole:2001jy,Dedes:2002dy,Brignole:2002bz,Dedes:2003km}
can be added to the two CP-even and one CP-odd Higgs boson.  In
NMSSM-like models with three CP-even and two CP-odd Higgs bosons
the two-loop corrections $O(\alpha_t \alpha_s + \alpha_b \alpha_s)$
from \cite{Degrassi:2009yq} plus the MSSM-like contributions
$O((\alpha_t + \alpha_b)^2 + \alpha_\tau^2)$
\cite{Brignole:2001jy,Dedes:2003km} can be added.  In SM-like models
with one physical Higgs singlet the two-loop corrections $O(\alpha_t
\alpha_s + \alpha_t^2)$ from Refs.~\cite{Degrassi:2012ry} can be added
to the self-energy.  In the split-MSSM the three-loop gluino
contribution $O(\alpha_t \alpha_s^2)$ from Ref.~\cite{Benakli:2013msa}
can be added.

\subsubsection{Decay widths and branching ratios}
\label{sec:calculationDecays}

Like \SPheno, \FS\ can calculate the loop-induced Higgs decays into two
photons and two gluons up to NNNLO, see \cref{sec:diphotoncalc}.

\subsection{Mass spectrum calculation: SUSY vs. Non-SUSY}
We have outlined that \FS and \SPheno can include the radiative
corrections to all particles up to the two-loop level in the $\DRbar'$
scheme. These corrections are included by default for supersymmetric
models. It is known that loop corrections, in particular to the Higgs
mass, are crucial. Typically the $\DRbar'$ and on-shell calculations
are in good agreement. Consequently, the remaining difference between
both calculations is often a good estimate for the theoretical
uncertainty.

The treatment of non-supersymmetric models in \FS and \SPheno is very
similar to the treatment of supersymmetric models.  The main
difference is, that in non-supersymmetric models the parameters are
defined in the \MSbar\ scheme, while in supersymmetric ones the
parameters are defined in the $\DRbar'$\ scheme.  In this paper
we perform only tree-level mass calculations (if not stated
otherwise), in which the definition of the renormalisation scheme is
irrelevant.  Thus, in the mass spectrum calculations performed in the
following, one is allowed to use input parameters which are defined in
the on-shell scheme.
This is for instance the standard approach in the large majority of
studies of the THDM: there are in general enough free parameters to
perform a full on-shell renormalisation keeping all masses and mixing
angles fixed. We find that the one-loop corrections in the $\MSbar$
scheme can give huge corrections to the tree-level masses in nearly
all models presented in the following. Therefore large fine-tuning of
the parameters is necessary once the loop corrections are taken into account.
A detailed analysis using a full
on-shell renormalisation scheme is possible for each model, but is
beyond the scope of this work. Of course, for models where it turns
out to be unavoidable that shifts in the masses and mixings appear at
the loop-level, the user can simply turn on the loop corrections in
\FS and \SPheno via a flag in the Les Houches input file.

\subsection{Calculation of the effective diphoton and digluon vertices in \SPheno and \FS}
\label{sec:diphotoncalc}
For the calculation of the partial width of a neutral scalar $\Phi$ decaying into two gluons or two photons we follow closely \cite{Spira:1995rr} for the LO and NLO contributions. The partial widths at LO are given by 
\begin{align}
 \Gamma(\Phi \to \gamma \gamma)_{\rm LO} &= \frac{G_F \alpha^2(0) m_\Phi^3}{128 \sqrt{2} \pi^3} \Bigg|\sum_f N^f_c Q_f^2 r^\Phi_f A_f(\tau_f) + \sum_s N^s_c r^\Phi_s Q_s^2 A_s(\tau_s)  \nonumber \\
 &\hspace{5cm} + \sum_v N^v_c r^\Phi_v Q_v^2 A_v(\tau_v)  \Bigg|^2, \\
 \Gamma(\Phi \to g g)_{\rm LO} &= \frac{G_F \alpha_s^2(\mu) m_\Phi^3}{36 \sqrt{2} \pi^3} \Bigg|\sum_f \frac{3}{2} D_2^f r^\Phi_f  A_f(\tau_f) + \sum_s \frac{3}{2} D_2^s r^\Phi_s A_s(\tau_s)  \nonumber \\
 &\hspace{5cm} + \sum_v \frac{3}{2} D_2^v r^\Phi_v A_v(\tau_v)  \Bigg|^2.
\end{align}
Here, the sums are over all fermions $f$, scalars $s$ and vector bosons $v$ which are charged or coloured and which couple to the scalar $\Phi$. $Q$ is the electromagnetic charges of the fields, $N_c$ are the colour factors and $D_2$ is the quadratic Dynkin index of the colour representation which is normalised to $\frac12$ for the fundamental representation. We note that the electromagnetic fine structure constant $\alpha$ must be taken at the scale $\mu = 0$, since the final state photons are real \cite{Djouadi:2005gi}. In contrast, $\alpha_s$ is evaluated at $\mu = m_\Phi$ which, for the case of interest here, is $\mu = 750$ GeV. $r^\Phi_i$ are the so-called reduced couplings, the ratios of the couplings of the scalar $\Phi$ to the particle $i$ normalised to SM values. These are calculated as
\begin{align}
r^\Phi_f &=  \frac{v}{2 M_f} (C^L_{\bar f f \Phi}+C^R_{\bar f f \Phi}), \label{eq:rPhif} \\
r^\Phi_s &=  \frac{v}{2 M^2_s} C_{s s^* \Phi},\\
r^\Phi_v &=  -\frac{v}{2 M^2_v} C_{v v^* \Phi}.
\end{align}
Here, $v$ is the electroweak VEV and $C$ are the couplings between the scalar and the different fields with mass $M_i$ ($i=f,s,v$). Furthermore,
\begin{equation}
 \tau_x = \frac{m_\Phi^2}{4 m_x^2}
\end{equation}
holds and the loop functions are given by
\begin{align}
 A_f &= 2 (\tau + (\tau -1) f(\tau))/\tau^2, \\
 A_s &=  -(\tau-f(\tau))/\tau^2, \\
 A_v &= -(2 \tau^2  + 3\tau  + 3 (2 \tau -1) f(\tau) )\tau^2,
\end{align}
with 
\begin{equation}
  f(\tau) = \begin{cases}
    \text{arcsin}^2 \sqrt{\tau} \hspace{1cm} &\text{for} \,\, \tau \le 1,\\
    -\frac{1}{4}\left(\log \frac{1+\sqrt{1-\tau^{-1}}}{1-\sqrt{1-\tau^{-1}}} -i\pi \right)^2 &\text{for} \,\, \tau > 1.
  \end{cases}
\end{equation}
For a pure pseudo-scalar state only fermions contribute, i.e.\ the LO widths read
\begin{align}
 \Gamma(A \to \gamma \gamma)_{\rm LO} &= \frac{G_F \alpha^2 m_A^3}{32 \sqrt{2} \pi^3} \left|\sum_f N^f_c Q_f^2 r^A_f A^A_f(\tau_f)  \right|^2, \\
 \Gamma(A \to g g)_{\rm LO} &= \frac{G_F \alpha_s^2 m_A^3}{36 \sqrt{2} \pi^3} \left|\sum_f 3 D_2^f r^A_f  A^A_f(\tau_f)  \right|^2,
\end{align}
where 
\begin{equation}
  A^A_f= f(\tau)/\tau \, ,
\end{equation}
and $r^A_f$ takes the same form as $r^\Phi_f$ in \cref{eq:rPhif}, simply replacing $C^{L,R}_{\bar f f \Phi}$ by $C^{L,R}_{\bar f f A}$.

\begin{figure}[hbt]
\includegraphics[width=0.49\linewidth]{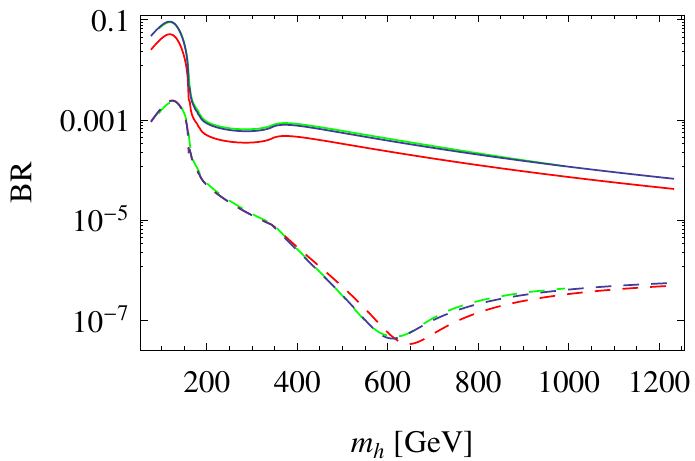} \hfill 
\includegraphics[width=0.49\linewidth]{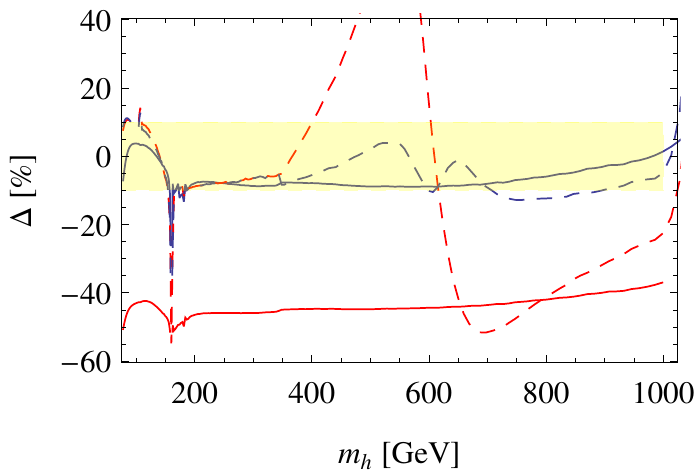}
\caption{On the left: comparison  of $\text{Br}(h\to gg)$  (full lines)  and $\text{Br}(h\to \gamma\gamma)$ (dashed lines) as calculated by \SPheno at LO (red) and including higher order corrections as described in the text (blue). The green line shows the values of the Higgs cross section working group. On the right: relative difference for diphoton (dashed lines) and digluon (full lines) at LO (red) and including higher order corrections (blue). }
\label{fig:ComparisonYP1}
\end{figure}

\begin{figure}[hbt]
\centering
\includegraphics[width=0.49\linewidth]{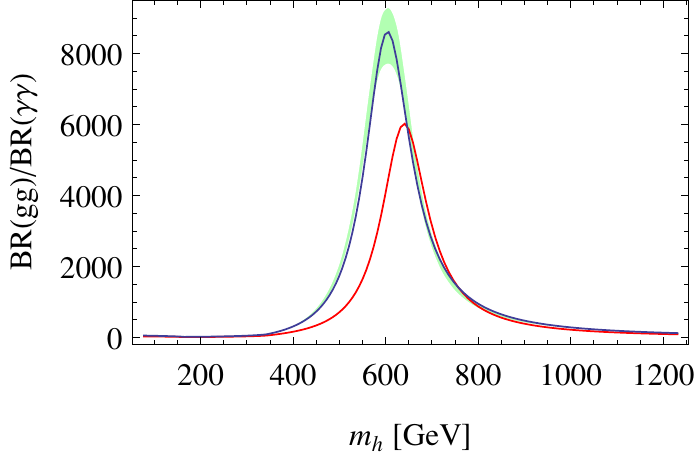}  \hfill
\includegraphics[width=0.49\linewidth]{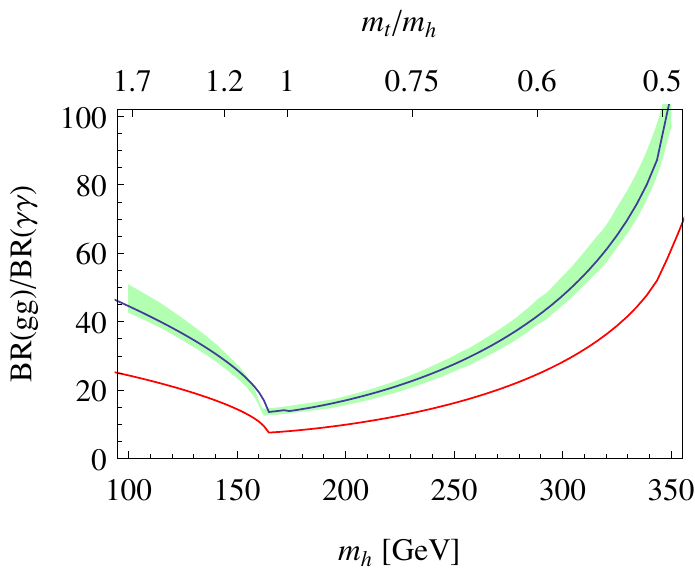}
\caption{$\text{Br}(h\to gg)/\text{Br}(h\to \gamma\gamma)$ as calculated by \SPheno at LO (red) and including higher order corrections as described in the text. The green band shows the values of the Higgs cross section working group including a 10\% uncertainty. On the right we zoom into the range $m_h \in [0.5,2]~m_t$}
\label{fig:ComparisonYP2}
\end{figure}

These formulae are used by \SPheno and \FS to calculate the full LO contributions of any CP-even or odd scalar present in a model including all possible loop contributions at the scale $\mu = m_\Phi$. However, it is well known, that higher order corrections are important. Therefore, NLO, NNLO and even $\text{N}^3$LO corrections from the SM are adapted and used for any model under study. In case of heavy colour fermionic triplets, the included corrections for the diphoton decay are  
\begin{align}
 r^\Phi_f  &\to r_f \left(1 - \frac{\alpha_s}{\pi} \right), \\
 r^\Phi_s &\to r_s \left(1+\frac{8 \, \alpha_s}{3 \pi} \right).
\end{align}
These expressions are obtained in the limit $\tau_f \to 0$ and thus applied only when $m_\Phi < m_f$. $r^A_f$ does not receive any corrections in this limit. 
For the case $m_\Phi > 100 m_f$, we have included the NLO corrections in the light quark limit given by \cite{Spira:1995rr}
\begin{equation}
 r^X_f \to r^X_f \left(1+  \frac{\alpha_s}{\pi} \left[-\frac{2}{3} \log 4\tau +  \frac{1}{18} \left(\pi^2 -\log^2 4\tau\right)  + 2\log \left( \frac{\mu_{\text{NLO}}^2}{m_f^2} \right) + 
 i \frac{\pi}{3}\left(\frac{1}{3} \log 4\tau  +2  \right) 
 \right] \right)
\end{equation}
for $X=\Phi,A$. $\mu_{\text{NLO}}$ is the renormalisation scale used for these NLO corrections, chosen to be $\mu_{\text{NLO}}= m_\Phi/2$\,. In the intermediate range of $100 m_f > m_\Phi > 2 m_f$, no closed expressions for the NLO correction exist. Our approach in this range was to extract the numerical values of the corrections from {\tt HDECAY} \cite{Djouadi:1997yw} and to fit them. 
For the digluon decay rate, the corrections up to N${}^3$LO are included and parametrised by
\begin{equation}
\Gamma(X \to g g) = \Gamma(X \to g g)_{\rm LO} \left(1 + C_X^{\rm NLO} + C_X^{\rm NNLO} + C_X^{\rm \text{N}^3LO} \right)\,,
\end{equation}
with \cite{Spira:1995rr,Kramer:1996iq,Chetyrkin:1997iv,Chetyrkin:2005ia,Schroder:2005hy,Baikov:2006ch,Baglio:2013iia}
\begin{align}
 C_\Phi^{\rm NLO} &= \left(\frac{95}{4} - \frac76 N_F \right) \frac{\alpha_s}{\pi}\,, \\
 C_\Phi^{\rm NNLO} &= \Bigg(370.196 + (-47.1864 + 0.90177 N_F) N_F \nonumber \\
  &\phantom{={}} \hspace{1cm} + (2.375 +  0.666667 N_F)\log \frac{m_\Phi^2}{m_t^2}\Bigg) \frac{\alpha^2_s}{\pi^2}\,, \\\
 C_\Phi^{\rm \text{N}^3LO} &= \left(467.684 + 122.441 \log \frac{m_\Phi^2}{m_t^2} + 10.941 \left(\log \frac{m_\Phi^2}{m_t^2}\right)^2 \right)  \frac{\alpha^3_s}{\pi^3}\,,
\label{EQ:NLOdiphoton}\end{align}
and
\begin{align}
 C_A^{\rm NLO} &= \left(\frac{97}{4} - \frac76 N_F \right) \frac{\alpha_s}{\pi}\,, \\
 C_A^{\rm NNLO} &= \left(171.544 + 5 \log \frac{m_\Phi^2}{m_t^2}\right) \frac{\alpha^2_s}{\pi^2} 
\end{align}
For pseudoscalar we include only corrections up to NNLO as the   $\rm \text{N}^3LO$ are not known for CP-odd scalars. \\
In order to check the accuracy of our implementation, we compared the results obtained with \SARAH--\SPheno for the SM Higgs boson decays with the ones given in the CERN yellow pages \cite{Heinemeyer:2013tqa}. In \cref{fig:ComparisonYP1} we show the results for the Higgs branching ratios into two photons and two gluons with and without the inclusion of higher order corrections. 
One sees that good agreement is generally found when including higher order corrections. In \cref{fig:ComparisonYP2} we show the ratio $\text{Br}(h\to gg)/\text{Br}(h\to \gamma\gamma)$ and compare it again with the recommended numbers by the Higgs cross section working group \cite{Heinemeyer:2013tqa}. Allowing for a 10\% uncertainty, we find that our calculation including higher order corrections agrees with the expectations, while the LO calculation predicts a ratio which is  over a wide range much too small. The important range to look at is actually not the one with $m_h \sim$~750~GeV because this corresponds to a large ratio of the scalar mass compared to the top mass. Important for most diphoton models is the range where the scalar mass is smaller than twice the quark mass. In this mass range we find that the NLO corrections are crucial and can change the ratio of the diphoton and digluon rate up to a factor of 2.
We also note that if we had used $\alpha(m_h)$ instead of $\alpha(0)$ in the LO calculation, the difference would have been even larger, with a diphoton rate overestimated by a factor $(\alpha(m_h)/\alpha(0))^2\simeq (137/124)^2 \simeq 1.22$.

\subsection{Monte-Carlo studies}
\SARAH writes all necessary files to implement a new model in different MC tools. First, a short summary of the output formats is given . Then, it is described how parameter values between \SPheno/\FS and the MC codes can be exchanged. 
\subsubsection{Output of models files}
\paragraph{CalcHep}
The model files for \CalcHep \cite{Pukhov:2004ca,Boos:1994xb} can also
be used with \MO \cite{Belanger:2014hqa} to perform both collider and
dark matter studies. \CalcHep is able to read the numerical values of
the masses and mixing matrices from a SLHA spectrum file based on the
\texttt{SLHA+} functionality \cite{Belanger:2010st}. \SARAH makes use
of this functionality to generate model files by default in a way that
they automatically \emph{expect} to find the input values in spectrum
files written by a \SARAH-generated \SPheno version or by
\FS. However, other choices are possible: the parameters can be given
via the {\tt vars} file or tree-level expressions can be calculated
internally by \CalcHep.

\paragraph{UFO format}
\SARAH can also write model files in UFO format
\cite{Degrande:2011ua}. This is particularly useful to implement new
models in \MGNLO and  perform collider studies. The UFO format is
also supported by other tools like {\tt GoSam} \cite{Cullen:2011ac},
{\tt Aloha} \cite{deAquino:2011ub}, \texttt{Herwig++}
\cite{Gieseke:2003hm,Gieseke:2006ga} and \texttt{MadAnalysis 5}
\cite{Conte:2012fm}. Moreover, the spectrum file written by \SPheno or
by \FS can be directly used as parameter card in \MG.

\paragraph{\WHIZARD/\OMEGA}
\SARAH writes all necessary files to implement a model in \WHIZARD and
\OMEGA \cite{Kilian:2007gr,Moretti:2001zz}. Since the SLHA reader of
\WHIZARD is at the moment restricted to the MSSM and the NMSSM,
\SPheno versions generated by \SARAH can write all information about the
spectrum and parameters in an additional file in the \WHIZARD
specific format. This file can then be read by \WHIZARD. Currently,
the handling of general Lorentz structures in \WHIZARD and the support
of the UFO format are under development. This will provide the
possibility to use \WHIZARD with the calculated diphoton and
digluon vertices as explained in the following.

\subsubsection{Interplay \SARAH--Spectrum-Generator--MC-Tool} 
The tool chains \SARAH--\SPheno/\FS--MC-Tools have one very appealing feature: the implementation of a model in the spectrum generator (\SPheno or \FS) as well as in a MC tool is based on just one single implementation of the model in \SARAH. Thus, the user does not need to worry that the codes might use different conventions to define the model. In addition, \SPheno also provides all widths for the particles so that this information can be used by the MC-Tool to save time. 

\subsubsection{Effective diphoton and digluon vertices for neutral scalars}
The effective diphoton and digluon vertices calculated by \SPheno or \FS are directly available in the UFO model files and the \CalcHep model files: \SARAH includes the effective vertices for all neutral scalars to two photons and two gluons, and the numerical values for these vertices are read from the spectrum file generated with \SPheno or \FS. For this purpose, a new block {\tt EFFHIGGSCOUPLINGS} is included in these files, which contains the values for the effective couplings including all corrections outlined in \cref{sec:diphotoncalc}. \\
It is important to mention that these effective couplings correspond to the decay of the scalar; if we use \CalcHep or \MG to compute the decay $\Phi \rightarrow gg$ then the value matches (as closely as possible) the NNNLO value, which includes \emph{real} emission processes such as $\Phi \rightarrow g g g$. Therefore, the corrections at NLO and beyond for $\Phi \to gg$ are not the same as $pp \to \Phi$ via gluon fusion \cite{Djouadi:2005gi}; the full NNNLO production cross-section includes all processes $g g \rightarrow \Phi + \mathrm{jet}$ and is therefore described by a different $k$-factor to the decay. This $k$-factor can for instance be obtained via 
\begin{equation}
 k = c_{\Phi gg} \cdot \frac{\sigma_{\rm SM}(pp  \to H(M_{\Phi})+ \mathrm{jet})}{\sigma_{\rm MC}(pp  \to \Phi)} 
\end{equation}
where $c_{\Phi gg}$ is the ratio squared of the effective coupling between $\Phi$ and two gluons at LO in the considered model and the SM. These values can for instance be read off by the block {\tt HiggsBoundsInputHiggsCouplingsBosons} in the \SPheno spectrum file. $\sigma_{\rm SM}(pp  \to H(M_{\Phi}))$ is the cross section for a SM-like Higgs with mass $M_\Phi$. This value can be calculated for instance with {\tt Higlu} \cite{Spira:1995mt} or {\tt Sushi} \cite{Harlander:2012pb} for the considered center-of-mass energy. \SPheno also provides values for $c_{\Phi gg} \cdot \sigma_{\rm SM}(pp  \to H(M_{\Phi}))$ for the most common energies in the blocks {\tt HiggsLHCX} ({\tt X}=7,8,13,14)  and {\tt HiggsFCC}, see also \cref{sec:XS}. 

\begin{figure}
\begin{center}
\includegraphics[width=0.5\textwidth]{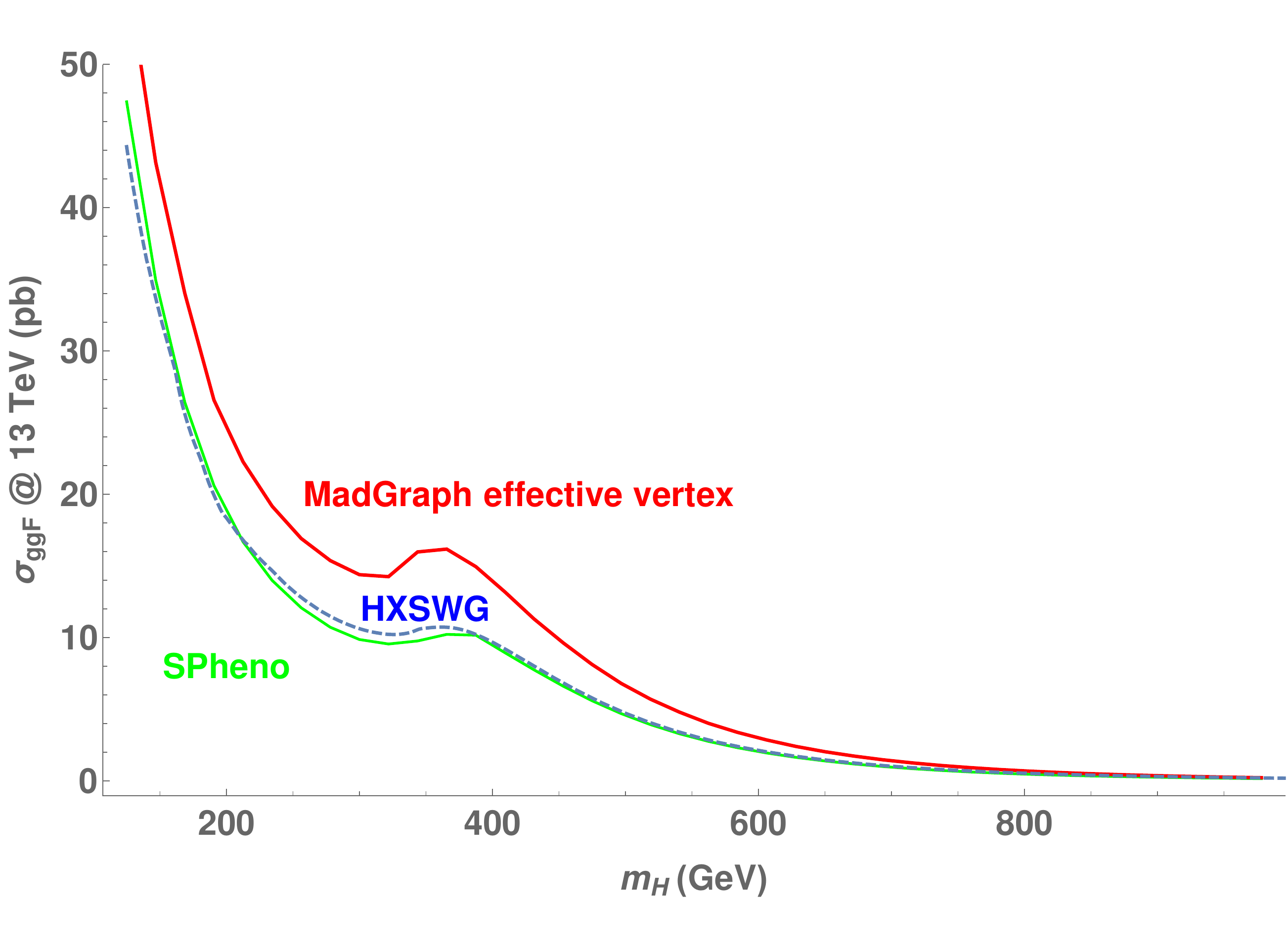}\includegraphics[width=0.5\textwidth]{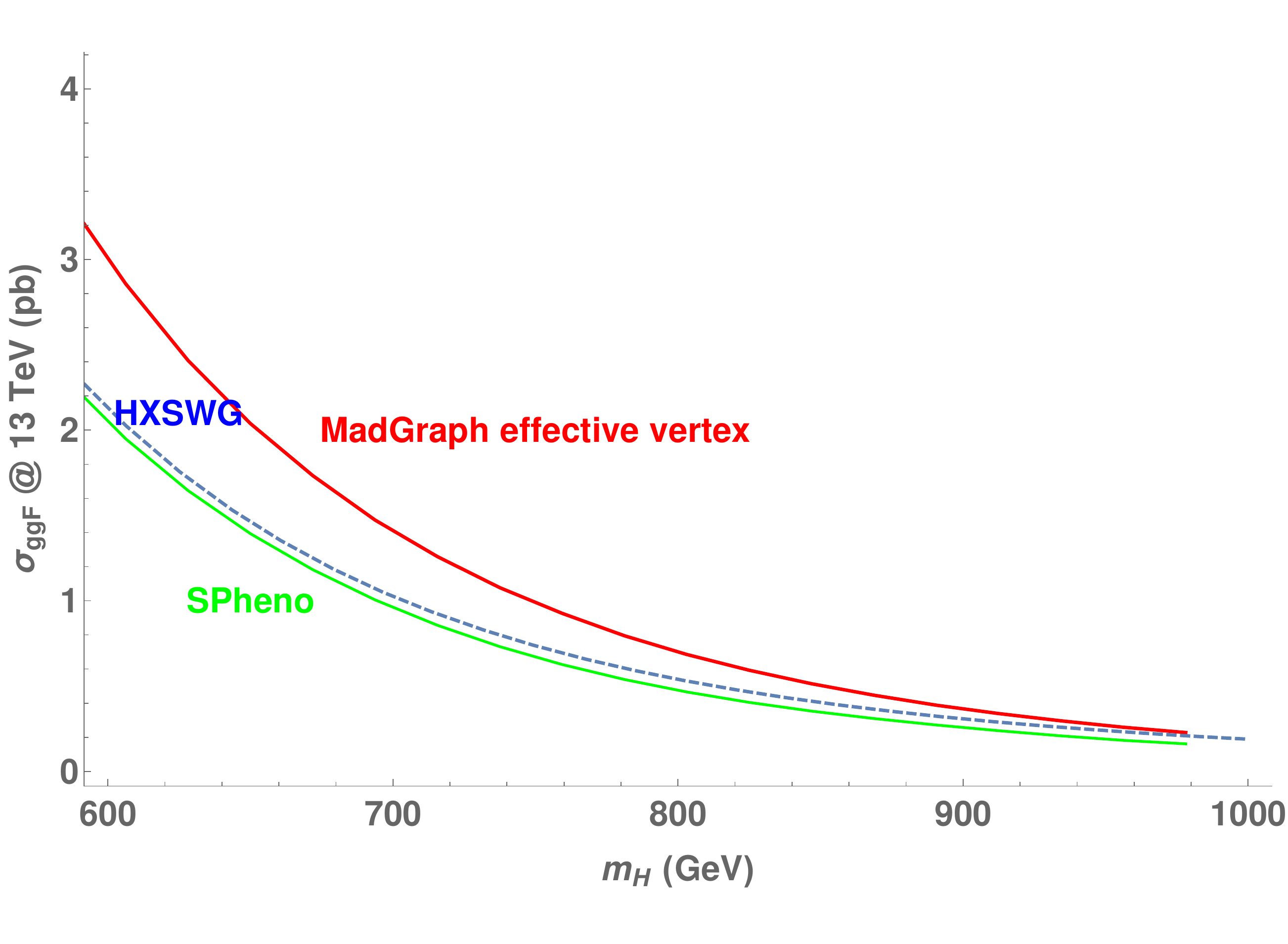}
\caption{Comparisons of the total Higgs production cross-section via gluon fusion in the Standard Model as a function of the Higgs mass, computed using the \SPheno output from \SARAH, the Higgs cross-section working group data, and in \MG using our effective vertex.}\label{FIG:CompareSigmaSM}
\end{center}
\end{figure}

On the other hand, this approach is not entirely appropriate for more refined collider analyses where the user would like to actually include, for example, a hard jet in the final state (without the full loop corrections to the effective vertex this is not an infra-red safe quantity). In this case, we note that around $750$ GeV the effective vertex output by \SARAH gives a fairly accurate result -- to within 30\% -- of the total production cross-section, at least in the Standard Model, when we compute $\sigma_{SM} ( g g \rightarrow \Phi + \mathrm{jet})$ using \MG and the standard cuts on momenta. This is illustrated in figure \ref{FIG:CompareSigmaSM}.

\subsection{Checking Higgs constraints}
\HB \cite{Bechtle:2008jh,Bechtle:2011sb,Bechtle:2013wla} and \HS \cite{Bechtle:2013xfa} are dedicated tools to study the Higgs properties of a given parameter point in a particular model. While \HB checks the parameter point against exclusion limits from Higgs searches, \HS gives a $\chi^2$ value to express how  well the point reproduces the Higgs measurements. In general, \HB and \HS can handle different inputs: either the cross sections for all necessary processes can be given at the parton or hadron level, or the effective couplings of the Higgs states to all SM particles are taken as input. In addition, the masses and widths of all neutral as well as charged Higgs states are always needed. 
\SPheno provides all input for the effective coupling approach. The information is distributed in the SLHA spectrum file and in separated files (called {\tt MH\_GammaTot.dat}, {\tt  MHplus\_GammaTot.dat}, {\tt BR\_H\_NP.dat}, {\tt BR\_Hplus.dat}, {\tt BR\_t.dat}, {\tt effC.dat}). While SLHA files can be used with \HB for models with up to five neutral scalars, the separated files can even be used with up to 99 neutral and 99 charged scalars.

\subsection{Checking the Vacuum stability}
\Vevacious  is a tool to check for the global minimum of the one-loop effective potential allowing for a particular set of non-zero VEVs. For this purpose, \Vevacious first finds all tree-level minima using {\tt HOM4PS2} \cite{lee2008hom4ps}. Afterwards, it minimizes the one-loop effective potential starting from these minima using {\tt minuit}  \cite{James:1975dr}. If the input minimum turns out not to be the global one, life-time of meta-stable vacua can be calculated using {\tt CosmoTransitions} \cite{Wainwright:2011kj}.\\ 
\Vevacious takes the tadpole equations, the polynomial part of the  scalar potential and all mass 
matrices as input. All of this information has to be expressed including all VEVs which should be tested. That means that, in case of supersymmetric models, in
order to check for charge and colour-breaking minima, both the stop and the stau
VEVs must be taken into account everywhere in the scalar potential, the tadpole equations and the mass matrices. Moreover, the possible mixing of all states triggered by the new VEVs must be included. To take care of all that, the corresponding input files can be generated by \SARAH.

\subsection{Accuracy of the diphoton calculation}
\label{subsec:accuracy}

Before concluding this section, we should draw the reader's attention
to the question of how accurate the results are from \SARAH\ in
combination with \SPheno and \FS. While every possible correction has
been included, there are still some irreducible sources of
uncertainty, as we shall discuss below.

\subsubsection{Loop corrections to $ZZ, WW, Z\gamma$ production}

So far in \SARAH, loop-level decays are only computed for processes
where the tree-level process is absent. This is to avoid the technical
issues of infra-red divergences. If the particle that explains the
$750$ GeV excess is a scalar, then it must mix with the Higgs and
acquire tree-level couplings to the $Z$ and $W$ bosons, and these are
fully taken into account.  However, due to the existence of such
terms, the loop corrections to the decays into $Z$s and $W$s are more
complicated and are therefore not yet available in \SARAH.

To estimate the uncertainty incurred by their absence, let us assume
that our $750$ GeV resonance $S$ couples to the $U(1)_Y$ and $SU(2)_L$
gauge bosons via the effective operators $S B_{\mu \nu} B^{\mu \nu}$
and $S W_{\mu \nu} W^{\mu \nu}$.  If we can neglect the tree-level
contributions to the decays and assume that the dominant contribution
originates from a set of particles in the loops, which have the
hypercharge $Y$ and the dynkin index $D_2(W)$ and dimension of the $SU(2)$ representation $d_2$, then the decay widths
are approximately given by
\begin{align}
  \frac{\Gamma (S \rightarrow ZZ)}{\Gamma (S \rightarrow \gamma \gamma
    )} \simeq \frac{( \frac{D_2}{t_W^2} + t_W^2 d_2 Y^2 )^2}{(d_2 Y^2 + D_2)^2} ,\qquad& \frac{\Gamma (S \rightarrow
    Z\gamma)}{\Gamma (S \rightarrow \gamma \gamma )}
  \simeq \frac{2}{t_W^2} \frac{( D_2 - t_W^2 d_2 Y^2)^2}{(d_2 Y^2 + D_2)^2  }, \nonumber\\
  \frac{\Gamma (S \rightarrow WW)}{\Gamma (S \rightarrow \gamma \gamma
    )} \simeq& \frac{2 D_2^2 \mathrm{cosec}^4 \theta_W}{(d_2 Y^2 + D_2)^2  }.
\end{align}
where we abbreviated $t_W$ for $\tan \theta_W$.
Put together, the uncertainty that we find for the decay
$S\to\gamma\gamma$ reads
\begin{align}
  \frac{\delta \Gamma ( S \rightarrow \mathrm{anything})
  }{\Gamma ( S \rightarrow \mathrm{anything})} \simeq& \bigg[
  \frac{55 D_2^2 -2d_2 Y^2 D_2 + 0.69 d_2^2 Y^4}{(d_2 Y^2 + D_2)^2} \bigg] \times \mathrm{Br} ( S \rightarrow \gamma \gamma ).
\end{align}
The factor in square brackets is therefore largest for fields that
only couple to $SU(2)_L$ gauge bosons, giving a factor of $\sim 55$, and for $SU(2)$ doublets with hypercharge $1/2$ it is $13$, although the former case yields too many $W$ bosons (the limit from run $1$  searches is $\frac{\Gamma (S \rightarrow WW)}{\Gamma (S \rightarrow \gamma \gamma
    )} \lesssim 20 $).  
Thus, provided that $\text{Br} ( S \rightarrow \gamma \gamma )
\lesssim 10^{-3}$, the relative uncertainty is guaranteed to be less than $10\%$. In such cases,
the proportional error in the total width transfers directly into the proportional error 
in the total cross-section:
\begin{align}
\frac{\delta \sigma (pp \rightarrow S \rightarrow \gamma \gamma)}{ \sigma (pp \rightarrow S \rightarrow \gamma \gamma)} \simeq -  \frac{\delta \Gamma ( S \rightarrow \mathrm{anything})
  }{\Gamma ( S \rightarrow \mathrm{anything})}
\end{align}
 On the other hand, for models where the dominant decay channel of the singlet 
is into gluons, it is not possible to have $\text{Br} ( S \rightarrow \gamma \gamma )
 \lesssim 10^{-3}$ without violating constraints from dijet production, and the reader should
be careful about the possible errors incurred. Fortunately, provided that the loop particles have
a hypercharge the error is much smaller, for example in the case that $D_2 = 0$ the coefficient above is less than one, thus giving an error of $\sim10^{-3}$ for $\text{Br} ( S \rightarrow \gamma \gamma ) = 10^{-3}$ .

\subsubsection{BSM NLO corrections}

\begin{figure}
\begin{center}
\includegraphics[width=0.5\textwidth]{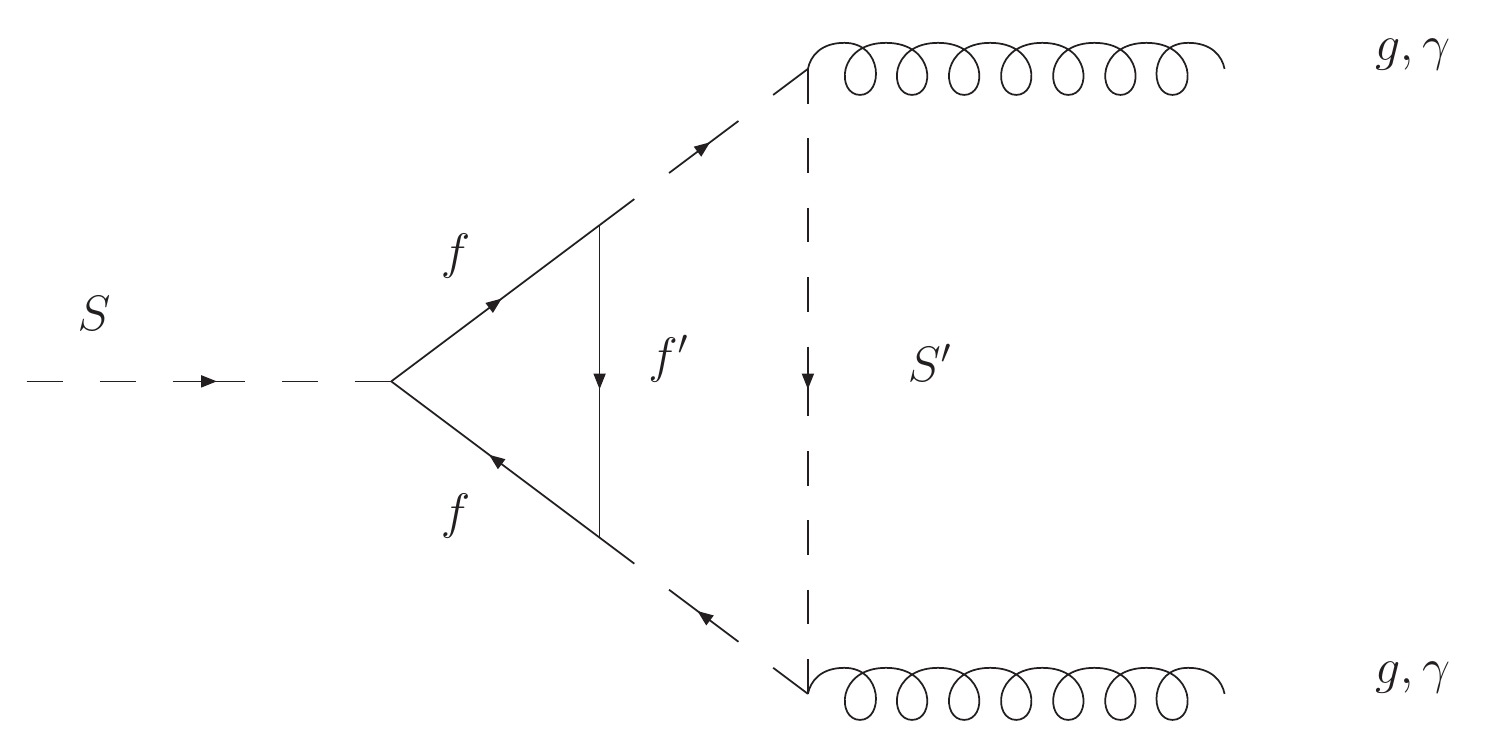}\includegraphics[width=0.5\textwidth]{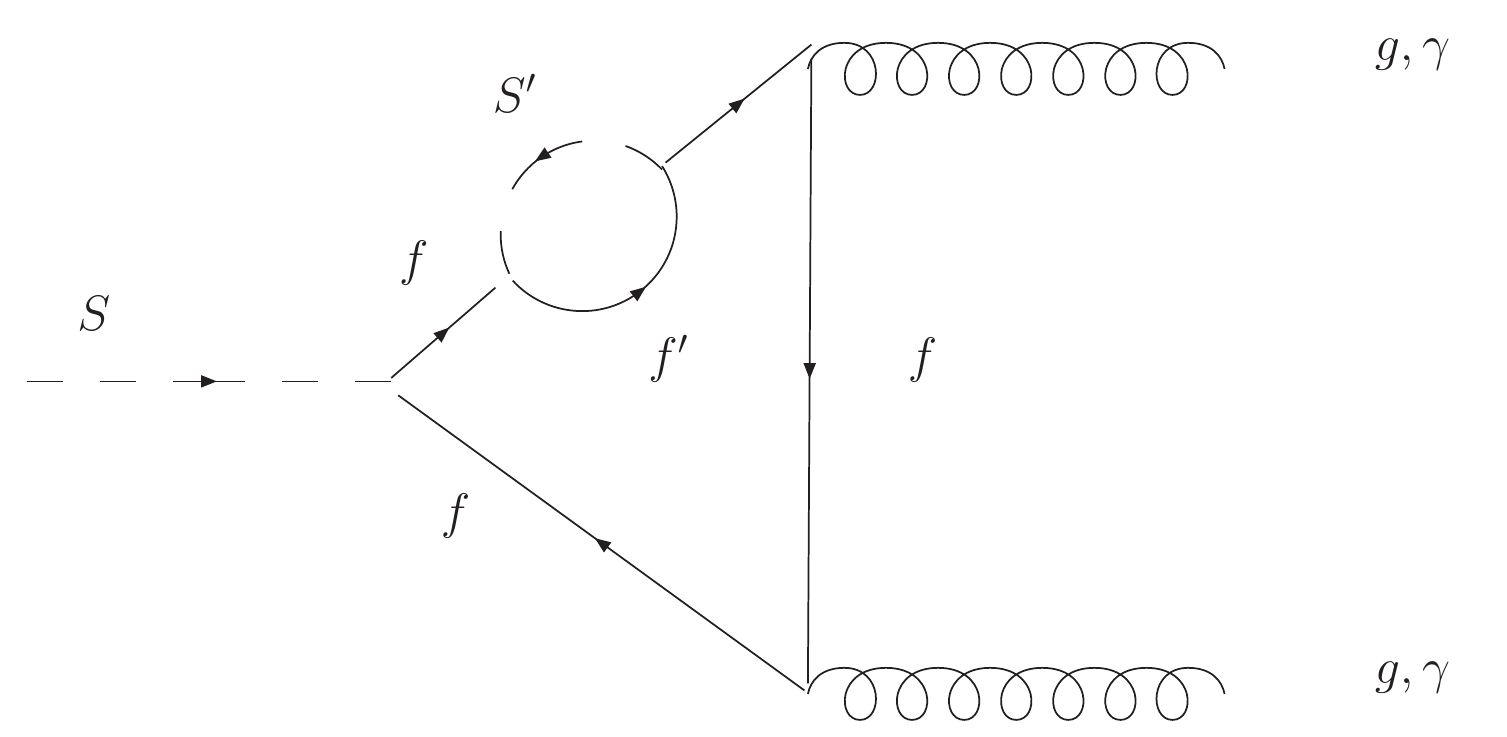} 
\caption{Examples of potentially important NLO corrections.}
\label{FIG:2LDECAY}
\end{center}
\end{figure}

As discussed above, \SARAH includes the leading-order computation of the diphoton and digluon decay amplitudes including the effects of all Standard-Model and Beyond-the-Standard-Model particles in the loops. Furthermore, it also includes the leading-log corrections to the digluon rate at NLO, NNLO  and $\text{N}^3$LO order in $\alpha_s$ \emph{in the Standard Model}, and some NLO corrections due to diagrams with an extra gluon to both the digluon and diphoton rates. However, the NLO corrections are absent for all other particles, which in the case of large Yukawa couplings or hierarchies could be sizeable. Two examples of such a diagrams are given in \cref{FIG:2LDECAY}; in the context of supersymmetric theories, particularly important are diagrams involving the gluino, which (if it is a Majorana particle) would not couple to a singlet at leading order --  naively their contribution is

\begin{align}
\frac{\delta \Gamma ( S \rightarrow gg/\gamma \gamma ) }
{\Gamma ( S \rightarrow gg/\gamma \gamma)}\sim& \frac{\alpha_s}{\pi} \log \frac{m_{\tilde{g}}^2}{\mu_{NLO}^2} \underset{m_{\tilde{g}} = 2\ \mathrm{TeV}}{\longrightarrow}  \sim 10 \ \% ,
\end{align}
although as we shall discuss below this can be (potentially
significantly) an underestimate.

\subsubsection{Tree vs pole masses in loops}

For consistency of the perturbative series and technical expediency,
the masses inside loops (to calculate pole masses and loop decay
amplitudes) are $\overline{\text{MS}}$ or
$\overline{\text{DR}}^\prime$ parameters, not the pole masses of
observed particles. The difference between calculations performed in
this scheme and the on-shell scheme are at two-loop order, and so is
generally small. However, in particular when there are large
hierarchies or Yukawa couplings in a model, there can be a large
difference between the Lagrangian parameters and the pole masses, and
therefore a large discrepancy between the loop amplitudes calculated
from these. In principle, this should be accounted for by including
higher-order corrections such as the right-hand diagram in
\cref{FIG:2LDECAY} -- but applying such a correction to each
propagator in the loop would actually correspond to a four-loop
diagram. The effect of using the pole mass instead is to essentially
resum part of these diagrams, which as is well known is relevant in
the case of large hierarchies of masses -- and so should give a more
accurate result in that case.

If we define 
\begin{align}
\delta m^2 \equiv (m^2)^{\overline{\text{MS}}/\overline{\text{DR}}^\prime} - (m^2)^{\text{on-shell}} 
\end{align}
then for one dominant particle $p$ in the loop, we can estimate the uncertainty as 
\begin{equation}
\frac{\delta \Gamma ( S \rightarrow gg/\gamma \gamma ) }
{\Gamma ( S \rightarrow gg/\gamma \gamma)} \sim \left\{ \begin{array}{cl} - \frac{2\delta m^2}{m^2}\big( \frac{A_p^\prime}{A_p} + 1 \big), & p=s,v, \\
- \frac{4\delta m}{m}\big( \frac{A_p^\prime}{A_p} + \frac{1}{2} \big), & p=f, \end{array}\right.
\end{equation}
where the factor of $1$ or $1/2$ assumes that the couplings $C_{\overline{p}p\Phi}$ do not depend upon the mass $m_p$ (but the prefactor in $r_p^\Phi$ therefore does). For most values of $m_p$ the loop functions are slowly changing (only peaking around $\tau_p =1$) so we will have a proportional uncertainty in the result of $ \frac{2\delta m_{s,v}^2}{m_{s,v}^2}$  or $ \frac{2\delta m_f}{m_f}$. As an example, in supersymmetric theories the soft masses of coloured scalars $\tilde{S}^\prime$ acquire a significant decrease from gluino loops:
\begin{align}
\delta m_{\tilde{S}^\prime}^2 \simeq \frac{C_2(\tilde{S}^\prime) \alpha_s}{\pi} m_{\tilde{g}}^2 \log \frac{m_{\tilde{g}}^2}{\mu_{NLO}^2} .
\end{align}
If the scalar is a colour triplet with a pole mass of $800$ GeV, then
for $2$ TeV gluinos and the $\overline{\text{DR}}^\prime$ mass is
$\sim 1100$ GeV; but $\frac{\delta
  m_{\tilde{S}^\prime}^2}{m_{\tilde{S}^\prime}^2} \sim 1!$ This
corresponds to a shift of a factor of two in the amplitude, and, if
the scalar dominates the total amplitude, a factor of four in $ \Gamma
( S \rightarrow gg/\gamma \gamma ) $; in fact in this case \SARAH
would be potentially \emph{underestimating} the diphoton rate. This is
a relatively mild example regarding this excess: given that the vast
majority of models proposed to explain the diphoton signature contain
large Yukawa couplings and many new particles, there is a significant
potential for large values of $\frac{\delta m^2}{m^2} $, about which
the user should be careful. It is worth noting that this is an effect
that would not be significant in the (N)MSSM, where the Higgs
couplings to photons/gluons are dominated by the top quark (and, for
photons, the W bosons) whose masses are protected by chiral symmetry
from large shifts: this issue is a novelty for the $750$ GeV excess.

For non-supersymmetric models, due to the fact that (almost) every
parameter point is essentially fine-tuned, we have not calculated loop
corrections to the masses by default, and this issue does not arise in
the same way. The user is then free to regard the result as involving
the pole masses of particles instead, if they so desire -- the issue
then becomes one of tuning the potentially large corrections to the
other input parameters.

%% file: tex/models.tex
\section{Models}
\label{sec:models}
A large variety of models has been proposed to explain the diphoton excess at \SI{750}{\GeV}. We have selected and implemented several possible models in \SARAH. Our selection is not exhaustive, but we have tried to implement a sufficient cross-section which are representative of many of the ideas put forward in the context of renormalisable models. These are the ones that \SARAH can handle. Their description is organised in the subsections that follow. 
Before we turn to this discussion we first want to mention other proposals which we do not deal with in this paper.
   
Many authors \cite{Alves:2015jgx, Backovic:2015fnp, Barducci:2015gtd,
  Bellazzini:2015nxw, Berthier:2015vbb, Bi:2015uqd,
  Chakrabortty:2015hff, D'Eramo:2016mgv, deBlas:2015hlv,
  Fichet:2015vvy, Franceschini:2015kwy, Ghorbani:2016jdq,
  Huang:2015svl, Huang:2015evq, Kanemura:2015vcb, Low:2015qho,
  Mambrini:2015wyu, Murphy:2015kag, Park:2015ysf, Sahin:2016lda} have
studied the excess with effective (non-renormalisable) models, which
is sensible given that there are thus far no other striking hints of
new physics at the LHC. As more data becomes available and the
evidence for new physics becomes more substantial, one might want to
UV complete these models, at which point the tools we are advertising
become relevant and necessary.  Other authors \cite{Bai:2015nbs,
  Barrie:2016ntq, Belyaev:2015hgo, Bian:2015kjt, Chiang:2015tqz,
  Cline:2015msi, Curtin:2015jcv, Harigaya:2015ezk, Harigaya:2016pnu,
  Kim:2015ron, Liao:2015tow, Low:2015qep, Matsuzaki:2015che,
  Molinaro:2015cwg, Nakai:2015ptz, No:2015bsn, Son:2015vfl} considered
strongly coupled models, in which the resonance is a composite
state. This possibility would be favoured by a large width of the
resonance, as first indications seem to suggest.  Another possibility
is to interpret the signal in the context of extra-dimensional models
\cite{Abel:2016pyc, Ahmed:2015uqt, Arun:2015ubr, Arun:2016ela,
  Bardhan:2015hcr, Cai:2015hzc, Cox:2015ckc, Davoudiasl:2015cuo,
  Geng:2016xin, Han:2015cty, Megias:2015ory}, with the resonance being
a scalar, a graviton, a dilaton, or a radion, depending on the
scenario. However, some of these interpretations are in tension with
the non-observation of this resonance in other channels. In
Supersymmetry, the scalar partner of the goldstino could provide an
explanation to the diphoton signal
\cite{Casas:2015blx,Petersson:2015mkr,Demidov:2015zqn,Ding:2016udc}. Other
ideas, slightly more exotic, include: a model with a space-time
varying electromagnetic coupling constant \cite{Danielsson:2016nyy},
Gluinonia \cite{Potter:2016psi}, Squarkonium/Diquarkonium
\cite{Luo:2015yio}, flavons \cite{Hernandez:2016rbi}, axions in
various incarnations \cite{Higaki:2015jag, Kim:2015xyn,
  Agrawal:2015dbf, Ben-Dayan:2016gxw, Pilaftsis:2015ycr,
  Aparicio:2016iwr}, a natural Coleman-Weinberg theory
\cite{Antipin:2015kgh,Marzola:2015xbh}, radiative neutrino mass models
\cite{Kanemura:2015bli,Nomura:2016fzs,Nomura:2016seu}, and
string-inspired models \cite{Anchordoqui:2015jxc, Heckman:2015kqk,
  Cvetic:2015vit, Faraggi:2016xnm, Ibanez:2015uok}. \\

We turn now to the weakly coupled models, and discuss the ones which
we implemented in \SARAH. All model files are available for download
at
\begin{center}
{\tt \url{http://sarah.hepforge.org/Diphoton_Models.tar.gz}}
\end{center}
and an overview of all implemented models is given in
Tabs.~\ref{tab:models1} and \ref{tab:models2}. In case of questions,
comments or bug reports concerning these models, please, send an
e-mail to {\tt diphoton-tools$@$cern.ch} which includes all authors.

\begin{table}[hbt]
\hspace{-1cm}
\begin{tabular}{c C{6cm} c c c}
\toprule
Model & Name & Section & Ref. &   \\
\midrule
\multicolumn{5}{c}{\bf Toy models with vector-like fermions} \\
CP-even singlet & {\tt SM+VL/CPevenS} & \ref{model:toy} &   & \\
CP-odd singlet & {\tt SM+VL/CPoddS} & \ref{model:toy} &   & \\
Complex singlet & {\tt SM+VL/complexS} & \ref{model:toy} &   & \\
\hline 
\multicolumn{5}{c}{\bf Models based on the SM gauge-group} \\
Portal dark matter & {\tt SM+VL/PortalDM} & \ref{sec:phiPortal} & \cite{Han:2015yjk,Chao:2015ttq}  & \\
Scalar octet & {\tt SM-S-Octet} &  \ref{sec:scalarOctet} & \cite{Cao:2015twy,Ding:2016ldt} & \danger${}^{(1)}$ \\
$SU(2)$ triplet quark model & \tt{SM+VL/TripletQuarks} & \ref{sec:TripletQuarks} & \cite{Benbrik:2015fyz} \\
Single scalar leptoquark & \tt{LQ/ScalarLeptoquarks} & \ref{sec:ScalarLeptoquarks} & \cite{Bauer:2015boy} \\
Two scalar leptoquarks & \tt{LQ/TwoScalarLeptoquarks} & \ref{sec:TwoScalarLeptoquark} & \cite{Chao:2015nac} & \danger${}^{(3)}$ \\
Georgi-Machacek model & \tt{Georgi-Machacek} & \ref{sec:Georgi-Machacek}  & \cite{Fabbrichesi:2016alj,Chiang:2016ydx} \\
THDM w. colour triplet & \tt{THDM+VL/min-3}  & \ref{sec:minTHDM} &   \cite{Bizot:2015qqo}  \\
THDM w. colour octet & \tt{THDM+VL/min-8}  & \ref{sec:minTHDM} &   \cite{Bizot:2015qqo}  \\
THDM-I w. exotic fermions   & \tt{THDM+VL/Type-I-VL} & \ref{sec:THDMexotic} & \cite{Angelescu:2015uiz,Jung:2015etr} \\
THDM-II w. exotic fermions   & \tt{THDM+VL/Type-II-VL} & \ref{sec:THDMexotic} & \cite{Angelescu:2015uiz,Jung:2015etr} \\
THDM-I w. SM-like fermions   & \tt{THDM+VL/Type-I-SM-like-VL} & \ref{sec:THDMVL} &\cite{Badziak:2015zez} \\
THDM-II w. SM-like fermions   & \tt{THDM+VL/Type-II-SM-like-VL} & \ref{sec:THDMVL} & \cite{Badziak:2015zez} \\
THDM w. scalar septuplet & \tt{THDM/ScalarSeptuplet} & \ref{sec:THDMseptuplet} & \cite{Han:2015qqj,Moretti:2015pbj} \\
\bottomrule
\end{tabular}
\caption{Part I of the overview of proposed models to explain the
  diphoton excess which are now available in \SARAH. The warning
  (\danger) shows that we found serious problems with the model during
  the implementation. The reasons are as follows. (1): the model is in
  conflict with limits from $S\to jj$; (2): we changed the quantum
  numbers and/or the potential because the original model had charge
  violating interactions; (3): we find disagreement with the diphoton rate as calculated in the original reference. For simplicity, we used the abbreviations {\tt LQ} for
  {\tt LeptoQuarks} and {\tt U1Ex} for {\tt U1Extensions} .}
\label{tab:models1}
\end{table}

\begin{table}[hbt]
\hspace{-1cm}
\begin{tabular}{c C{5cm} c c c}
\toprule
Model & Name & Section & Ref. &   \\
\midrule
\multicolumn{5}{c}{\bf $U(1)$ Extensions} \\
Dark $U(1)'$ & \tt{U1Ex/darkU1} & \ref{subsubsec:darkU1}  & \cite{Ko:2016wce} \\
Hidden $U(1)$&   \tt{U1Ex/hiddenU1}  & \ref{sec:hiddenU1} & \cite{Das:2015enc} \\
Simple $U(1)$ &  \tt{U1Ex/simpleU1} & \ref{sec:simpleU1} & \cite{Chang:2015bzc} \\
Scotogenic $U(1)$ & \tt{U1Ex/scotoU1} & \ref{sec:scotoU1}  & \cite{Yu:2016lof} & \danger${}^{(2)}$ \\
Unconventional $U(1)_{B-L}$  & \tt{U1Ex/BL-VL} & \ref{sec:BLVL} & \cite{Modak:2016ung} \\
Sample of $U(1)'$ & \tt{U1Ex/VLsample} & \ref{sec:U1sample}    & \cite{Chao:2015nsm} \\
flavour-nonuniversal charges & \tt{U1Ex/nonUniversalU1}   & \ref{sec:nonUniU1}  & \cite{Martinez:2015kmn} & \\
Leptophobic $U(1)$ & \tt{U1Ex/U1Leptophobic} & \ref{sec:U1leptophobic}   &  \cite{Ko:2016lai} & \danger${}^{(1)}$ \\
$Z'$ mimicking a scalar resonance & \tt{U1Ex/trickingLY} & \ref{sec:trickingLY} & \cite{Chala:2015cev} \\
\hline
\multicolumn{5}{c}{\bf Non-abelian gauge-group extensions of the SM} \\
LR without bidoublets & \tt{LRmodels/LR-VL} & \ref{sec:LRVL} & \cite{Dasgupta:2015pbr,Deppisch:2016scs,Dev:2015vjd} & \danger${}^{(2)}$\\
LR  with $U(1)_L \times U(1)_R$ & \tt{LRmodels/LRLR} & \ref{sec:LRLR}  & \cite{Cao:2015xjz} & \danger${}^{(2)}$\\
LR with triplets & \tt{LRmodels/tripletLR} & \ref{sec:tripletLR} & \cite{Berlin:2016hqw} \\
Dark LR & \tt{LRmodels/darkLR} & \ref{sec:darkLR} & \cite{Dey:2015bur} \\
331 model without exotic charges & \tt{331/v1} & \ref{sec:331v1} & \cite{Boucenna:2015pav}  \\
331 model with exotic charges & \tt{331/v2} & \ref{sec:331v2} & \cite{Cao:2015scs}  \\
Gauged THDM & \tt{GTHDM} &  \ref{sec:gTHDM} & \cite{Huang:2015rkj} & \\
\hline
\multicolumn{5}{c}{\bf Supersymmetric models}\\
NMSSM with vectorlike top & \tt{NMSSM+VL/VLtop} & \ref{sec:NMSSMVLtop} & \cite{Wang:2015omi} & \danger${}^{(1)}$ \\
NMSSM with {\bf 5}'s & \tt{NMSSM+VL/5plets} & \ref{sec:NMSSM5} & \cite{Dutta:2016jqn,Tang:2015eko,Hall:2015xds} \\
NMSSM with {\bf 10}'s& \tt{NMSSM+VL/10plets} &  \ref{sec:NMSSM10} & \cite{Dutta:2016jqn,Tang:2015eko,Hall:2015xds} \\
NMSSM with {\bf 5}'s \& {\bf 10}'s & \tt{NMSSM+VL/10plets} &  \ref{sec:NMSSM15} & \cite{Hall:2015xds} \\
NMSSM with {\bf 5}'s and $R$pV & \tt{NMSSM+VL/5plets+RpV} & \ref{sec:NMSSMRpV} &  \cite{Dutta:2016jqn} \\
Broken MRSSM & \tt{brokenMRSSM} & \ref{sec:bMRSSM} &  \cite{Chakraborty:2015gyj} \\
$U(1)^\prime$-extended MSSM & \tt{MSSM+U1prime-VL}   & \ref{sec:U1pMSSM} & \cite{Jiang:2015oms,An:2012vp} \\
$E_6$  with extra $U(1)$ & \tt{E6MSSMalt} & \ref{sec:E6MSSM} & \cite{Chao:2016mtn} \\
\bottomrule
\end{tabular}
\caption{Part II of the overview of proposed models to explain the
  diphoton excess which are now available in \SARAH. The warning
  (\danger) shows that we found serious problems with the model during
  the implementation. The reasons are as follows. (1):
  non-perturbative couplings needed to explain diphoton excess; (2):
  we changed the quantum numbers and/or the potential because the
  original model had charge violating interactions; (3): we find disagreement with the diphoton rate as calculated in the original reference.}
\label{tab:models2}
\end{table}

\subsection{Toy models}
\label{sec:toymodelsec}
The simplest ideas proposed to explain the diphoton excess extend the
SM by a scalar singlet and vector-like fermions, which serve the
purpose of enhancing the diphoton rate, and -- in the case of coloured
states -- also the production via gluon fusion. An enhancement of
gluon fusion seems to be necessary because a production of the
resonance purely by photon fusion is in some tension with 8~TeV data:
the increase in the cross section from 8 to 13~TeV is just a factor 2,
while a factor of 5 would be needed to make the results from LHC run-I
and II compatible. To explore the multitude of possibilities we first
consider toy models: they do not contain all possible couplings of the
vector-like fermions to SM matter, but are rather engineered to allow
one to easily explore the effects of different representations of
vector-like matter on the diphoton rate and on the relevant partial
widths. Then, from section~\ref{subsec:SMgauge} on, we consider
complete models, containing all the operators consistent with both
field content and symmetries.

\paragraph{Toy models with vector-like fermions}
\label{model:toy}
\begin{itemize}
\item {\bf Reference:} \cite{Knapen:2015dap,Falkowski:2015swt,Han:2015dlp,Han:2015yjk} 
\item {\bf Model names:}\\
\texttt{SM+VL/CPevenS} \\
\texttt{SM+VL/CPoddS} \\
\texttt{SM+VL/complexS}
\end{itemize}

To begin with we categorise the toy models according to the type of the involved scalar singlet. There are three possibilities: (i) the singlet is a real CP-even scalar, (ii) real CP-odd, or (iii) a complex scalar. Each case is considered in separate \texttt{SARAH} model files, where we introduce all possible representations of vector-like fermions. These possibilities, following Tables 3 and 4 of Ref. \cite{Knapen:2015dap}, are shown below in Table \ref{tab:SSM-VL}. This allows one to study combinations of fermion representations or individual choices by giving unwanted fermion representations a mass high enough to effectively decouple them from the model~\footnote{This option has to be used carefully when including loop corrections to the mass spectrum.}. All mixings between the extra fermions and SM fermions are neglected through the assumption of a discrete $\mathbb{Z}_2$ symmetry. Of course in a realistic model the mixings have to be taken into account, as they allow the necessary decays of the coloured vector-like fermions into SM particles. 

\begin{table}[h]
\centering
\begin{tabular}{c c c c c c c}
\toprule
Field & Gen. & $SU(3)_C$ &  $SU(2)_L$ & $U(1)_Y$ & $\mathbb{Z}_2$ &  Ref.\\
\midrule
$S$ & 1 & 1 & 1 & 0 & $+$ \\ \midrule
$\Psi_{F_1}$ & 1 & $\three$ & $\two$ & $\frac{7}{6}$ & $-$ &\\
$\Psi_{F_2}$ & 1 & $\three$ & $\three$ & $\frac{2}{3}$ & $-$ &  \\
$\Psi_{F_3}$ & 1 & $\three$ & $\two$ & $-\frac{5}{6}$ & $-$ & \\
$\Psi_{F_4}$ & 1 & $\three$ & $\three$ & $-\frac{1}{3}$ & $-$ & \\
$\Psi_{F_5}$ & 1 & $\three$ & $1$ & $\frac{2}{3}$  & $-$ & \cite{Falkowski:2015swt,Han:2015dlp}\\
$\Psi_{F_6}$ & 1 & $\three$ & $\two$ & $\frac{1}{6}$ & $-$ &  \\
$\Psi_{F_7}$ & 1 & $\three$ & $1$ & $-\frac{1}{3}$ &  $-$ &\cite{Han:2015dlp}  \\
$\Psi_{F_8}$ & 1 & $1$ & $1$ & $1$ &  $-$ & \\
$\Psi_{F_9}$ & 1 & $1$ & $\two$ & $-\frac{3}{2}$& $-$ &  \\
$\Psi_{F_{10}}$ & 1 & $1$ & $\three$ & $1$ &  $-$ &\\
$\Psi_{F_{11}}$ & 1 & $1$ & $\two$ & $-\frac{1}{2}$& $-$ &  \\
$\Psi_{F_{12}}$ & 1 & $1$ & $\three$ & $0$ &  $-$ &\\
$\Psi_{F_{13}}$ & 1 & $\three$ & $1$ & $\frac{5}{3}$ & $-$ & \cite{Han:2015yjk}  \\
\bottomrule
\end{tabular}
\caption{Extra particle content of the toy models. $S$ is either the CP-even, CP-odd or complex scalar. The various fermions $\Psi_{F_i}\equiv \Psi_{F_{iL}}$ each come with a right-handed partner $\Psi_{F_{iR}}$ with opposite quantum numbers. These models are based on the collection given in Ref. \cite{Knapen:2015dap}, while the last column contains other works where fermions in these specific representations are used. All SM particles have charge `$+$' under the additional $\mathbb{Z}_2$ symmetry.}
\label{tab:SSM-VL}
\end{table}

We write the scalar potentials for the three different types of scalars as
\begin{subequations}
\begin{align}
 V_{\rm even} &= \frac12 m_S^2 S^2 + \frac 14\lambda_S S^4 - \mu^2 |H|^2 + \frac 12 \lambda_H |H|^4 + \frac12\lambda_{HS} S^2 |H|^2 \notag\\
 &+ \kappa_{HS} S |H|^2 + \frac 13 \kappa_S S^3 \,, \\
  V_{\rm odd} &= \frac12m_S^2 |S|^2 + \frac 14\lambda_S |S|^4 - \mu^2 |H|^2 + \frac 12 \lambda_H |H|^4 + \frac12\lambda_{HS} |S|^2 |H|^2 \notag\\
  &+ \left(i \kappa_{HS} S |H|^2 + i \frac 13 \kappa_S S|S|^2 + \hc \right) \,, \\
   V_{\rm complex} &=  m_S^2 |S|^2 + \frac 12\lambda_S |S|^4 - \mu^2 |H|^2 + \frac 12 \lambda_H |H|^4 + \lambda_{HS} |S|^2 |H|^2 \notag\\
   &+ \left(\kappa_{HS} S |H|^2 + \frac 13 \kappa_S S|S|^2 + \hc \right)\,.
\end{align}
\end{subequations}
The Yukawa interactions are given by
\begin{align}
 - \mathcal{L}_Y &= \mathcal{L}^{\rm SM}_Y +\sum_j\left( m_{F_j} \overline{\Psi_{F_jL}}\Psi_{F_jR} +  Y_{F_j} S\, \overline{\Psi_{F_jL}}\Psi_{F_jR} \right)+ \text{h.c.} \,.
\end{align}
In the Lagrangian above one should substitute the expression for the relevant scalar field
\begin{subequations}
\begin{align}
S_{\rm even} &= v_S + \phi_S \,, \qquad\quad \text{where}\quad \langle S \rangle = v_S\,,\\
S_{\rm odd} &= i\sigma_S \,,\\
S_{\rm complex} &= \frac{1}{\sqrt{2}} \left(v_S + \phi_S + i \sigma_S\right)\,.
\end{align}
\end{subequations}
Note that imposing CP conservation forces $\kappa_{HS}$ and $\kappa_S$ to vanish in the CP-odd potential. For both the CP-even and complex singlet models the CP-even component $\phi_S$ mixes with the neutral Higgs field $\phi_h$ at tree-level if $\kappa_{HS} \neq 0$. As discussed in \cref{sect:motivation:mixing}, even if one sets $\kappa_{HS} = 0$ mixing between the CP-even states is induced at the loop level.

\subsection{Models based on the SM gauge-group} \label{subsec:SMgauge}
We now turn our attention to complete models that have been proposed to explain the \SI{750}{\GeV} diphoton excess. To begin with we consider models that are based on the SM gauge group, with or without additional global symmetries. We divide the possible models into two main categories: (i) models with a SM-like Higgs sector and (ii) Two-Higgs-doublet type models.
\subsubsection{Singlet extensions with vector-like fermions}
\input{tex/ModelCategories/GSMmodels}
\subsubsection{Two-Higgs doublet models}
\input{tex/ModelCategories/THDmodels}

\subsection{$U(1)$ extensions of the SM}
\input{tex/ModelCategories/U1models}

\subsection{Non-abelian gauge-group extensions of the SM}
\subsubsection{Left-right symmetric models}
\input{tex/ModelCategories/LRmodels}
\subsubsection{331 models}
\input{tex/ModelCategories/331models}
\subsubsection{Other models}
\input{tex/ModelCategories/OtherBSMmodels}

\subsection{Supersymmetric models}
\input{tex/ModelCategories/SUSYmodels}

%% file: tex/ModelCategories/GSMmodels.tex
\modelparalabel{Portal dark matter model}{sec:phiPortal}

\begin{itemize}
\item {\bf Reference:} \cite{Han:2015yjk,Chao:2015ttq} 
\item {\bf Model name:} \texttt{SM+VL/PortalDM}
\end{itemize}
This model proposes that the resonance is produced by a $\SI{750}{\GeV}$ real scalar singlet $S$, with the diphoton rate boosted through the introduction of vector-like quarks coupling to the new scalar singlet. In this model we have three possible options for the representation of the new vector-like matter. These choices are: (i) the addition of a vector-like up-type quark pair $t^\prime_{L/R}$ \cite{Chao:2015ttq}, (ii) in addition to the vector-like up-type quark pair, a vector-like quark doublet pair $Q^\prime_{L/R}$ is introduced \cite{Chao:2015ttq} and finally, (iii) the addition of only the vector-like pair $X_{L/R}$, also triplet under $SU(3)_C$ but with exotic hypercharge \cite{Han:2015yjk}.

The model also introduces a new real scalar singlet $S_{\rm DM}$ and an additional fermionic singlet $\Psi_{\rm DM}$ as DM candidates, with a {\zTwo} symmetry to stabilise them. The particle content beyond the SM fields is given in Table \ref{tab:phiPortal:tab}.
In order to avoid mixing with the SM quarks, the fields $t^\prime_{L/R}$ and $Q^\prime_{L/R}$ are also odd under the {\zTwo}.

The user can choose between the three model types by setting the couplings of the unwanted fields to zero and choosing their masses to be very large (for example, $10^{12}$ GeV) to decouple them. Originally, the fermionic dark matter is absent from the models described in \cite{Chao:2015ttq}. The exact settings are given below.

\begin{table}[h]
\centering
\begin{tabular}{c c c c c c }
\toprule
Field & Gen. & $SU(3)_C$ &  $SU(2)_L$ & $U(1)_Y$ & $\mathbb{Z}_2$ \\
\midrule
$S$ & 1 & $\one$ & $\one$ & $0$ & $+$ \\
$S_{\rm DM}$ & 1 & $\one$ & $\one$ & $0$ & $-$ \\ \midrule
$\Psi_{\rm DM}$ & 1 & $\one$ & $\one$ & $0$ & $-$ \\
$X$ & 1 & $\three$ & $\one$ & $\frac{5}{3}$ & $-$ \\
$t^\prime$ & 1 & $\three$ & $\one$ & $\frac{2}{3}$ & $-$ \\
$Q^\prime$ & 1 & $\three$ & $\two$ & $\frac{1}{6}$ & $-$ \\
\bottomrule
\end{tabular}
\caption{Extra particle content of the portal DM model. The top/bottom part of the table corresponds to the new scalar/fermionic degrees of freedom. All additional fermionic degrees of freedom are vector-like fermions.}
\label{tab:phiPortal:tab}
\end{table}
The scalar potential for these models reads
\begin{align}
V &= -\mu^2 |H|^2 + \frac{1}{2} \lambda_H \,|H|^4 + \frac{1}{2} M_S^2 \,S^2 + \frac{1}{3} \kappa_S \,S^3 + \frac{1}{4} \lambda_S \,S^4 
+ \frac{1}{2} M_{S_{\rm DM}}^2 \,S_{\rm DM}^2 + \frac{1}{4} \lambda_{S_{\rm DM}} \,S_{\rm DM}^4 \notag\\ 
&+ \kappa_{HS} \,|H|^2 S + \lambda_{HS}\,|H|^2 S^2
+ \lambda_{HS_{\rm DM}} \,|H|^2 S_{\rm DM}^2 +\kappa_{SS_{\rm DM}} S S_{\rm DM}^2 + \lambda_{SS_{\rm DM}} S^2 S_{\rm DM}^2 \,,
\end{align}
whereas the three model variants lead to three distinct forms for the Yukawa interactions, given by: 
\begin{subequations}
\begin{align}
- \mathcal L_Y^{\text{I}} &= \mathcal{L}_Y^{\rm SM} + \left( m_{t^\prime}+ Y_{St^\prime}S\right) \,\overline{t^\prime}_L\, t^\prime_R + Y_{S_{\rm DM} t^\prime} S_{\rm DM} \overline{t^\prime_L} u_R+\hc \,,\\
- \mathcal L_Y^{\text{II}} &= \mathcal L_Y^{\text{I}}+\left( m_{Q^\prime}+Y_{SQ^\prime} S \right) \overline{Q^\prime}_L Q^\prime_R \notag\\ 
&+ Y_{Q^\prime_1} \overline{Q^\prime}_L H t^\prime_R + Y_{Q^\prime_2} \overline{Q^\prime}_R \widetilde{H} t^\prime_L + Y_{S_{\rm DM} Q^\prime} S_{\rm DM} \overline{Q_L} Q^\prime_R+\hc\,,\\
- \mathcal L_Y^{\text{III}} &= \mathcal{L}_Y^{\rm SM} + \left( m_{\rm DM} +\kappa S \right)\overline{\Psi_{\rm DM}}_L {\Psi_{\rm DM}}_R + \left( m_X + Y_X\, S \right) \overline{X}_L X_R + {\rm h.c.}\,,
\end{align}
\end{subequations}
where $\widetilde{H}=i \sigma_2 H^*$. 
In the model variants I and II, the vectorlike quarks decay into SM quarks and the scalar dark matter candidate $S_{\rm DM}$
via the couplings $Y_{S_{\rm DM} Q^\prime}$ and $Y_{S_{\rm DM} t^\prime}$. In model III, in turn, $X$ is stable at the level of the Lagrangian and could only decay through higher-dimensional operators which are not included here.

The symmetry breaking pattern of the models is that of the SM, where the neutral component of the Higgs field acquires a VEV, plus the VEV of the scalar singlet $S$
\begin{align}
S &= v_S + \phi_S \,, \qquad \text{where} \quad \langle S \rangle = v_S\,.
\end{align}
In general, $\phi_S$ mixes with the SM Higgs. 

As mentioned previously, the user can choose between the three different models through the following parameter choices:
\begin{itemize}
\item {\bf Model I:} $Y_{SQ^\prime}=Y_{Q^\prime_i}=Y_{S_{\rm DM} Q^\prime}=Y_X=0$ and $m_{Q^\prime}=m_X=\SI{E12}{\GeV}$
\item {\bf Model II:} $Y_X=0$ and $m_X=\SI{E12}{\GeV}$
\item {\bf Model III:} $Y_{SQ^\prime}=Y_{Q^\prime_i}=Y_{S_{\rm DM} Q^\prime}=Y_{St^\prime}=Y_{S_{\rm DM}}=0$ and $m_{Q^\prime}=m_{t^\prime}=\SI{E12}{\GeV}$
\end{itemize}

\modelparalabel{Scalar octet extension}{sec:scalarOctet}

\begin{itemize}
\item {\bf Reference:} \cite{Cao:2015twy,Ding:2016ldt} 
\item {\bf Model name:}  \texttt{SM-S-Octet}
\end{itemize}

A charged scalar colour octet $O$ coupled to a scalar singlet $S$ was proposed in Refs. ~\cite{Cao:2015twy,Ding:2016ldt}. Here the singlet is the \SI{750}{\GeV} candidate, while the octet enters the loops that contribute to the generation of the couplings of the singlet to the gauge bosons.
While Ref.~\cite{Ding:2016ldt} considers a toy model involving only the term $S\,|O|^2$, Ref.~\cite{Cao:2015twy} takes the singlet extended Manohar-Wise model \cite{Manohar:2006ga}. For the \SARAH implementation we have used the full model. However, since the cubic and quartic terms in $O$ do not play a significant role, they are turned off by default in the \SARAH model file. 
\begin{table}[h]
\centering
\begin{tabular}{c c c c c }
\toprule
Field & Gen. & $SU(3)_C$ &  $SU(2)_L$ & $U(1)_Y$  \\
\midrule
$S$ & 1 & $\one$ & $\one$ & $0$ \\
$O$ & 1 & $\mathbf{8}$ & $\two$ & $\frac12$ \\
\bottomrule
\end{tabular}
\caption{Extra scalar field content of the octet extended SM.}
\label{tab:SOct}
\end{table}

The extra particle content with respect to the SM is a real singlet $S$ and a scalar color octet $O$ which is also charged under $SU(2)_L \times U(1)_Y$, see \cref{tab:SOct}. The isospin components of $O$ are
\begin{equation}
 O^A = \left( \begin{array}{c} O^{+\,A} \\ O^{0\,A} \end{array}\right),
\end{equation}
where $A=1,\dots,8$ is the adjoint colour index. The full scalar potential reads
\begin{align}
 V &= \frac12 m_S^2 S^2 + \lambda_S S^4 - \mu^2 |H|^2 + \lambda_H |H|^4 + \kappa_1 S^2 |H|^2 + 2 m_O^2 \text{Tr}(O^\dagger O) + \kappa_2 S^2 \text{Tr}(O^\dagger O) \nonumber \\
 & + \lambda_1 |H|^2 \text{Tr}(O^\dagger O) + \lambda_2 H^\dagger_i H_j \text{Tr}(O^\dagger_j O_i) + \lambda_6 \text{Tr}(O^\dagger O O^\dagger O) + \lambda_7 \text{Tr}(O^\dagger_i O_j  O^\dagger_j O_i ) \nonumber \\
 & + \lambda_8 \text{Tr}(O^\dagger O)^2 + \lambda_9 \text{Tr}(O^\dagger_i O_j) \text{Tr}(O^\dagger_j O_i) + \lambda_{10} \text{Tr}(O_i O_j) \text{Tr}(O^\dagger_i O^\dagger_j) + \lambda_{11}(O_i O_j O^\dagger_j O^\dagger_i) \nonumber \\
 & + \left(\lambda_3 H^\dagger_i H^\dagger_j \text{Tr}(O_i O_j) + \lambda_4 H^\dagger_i \text{Tr}(O^\dagger_j O_j O_i) + \lambda_5 H^\dagger_i \text{Tr}(O^\dagger_j O_i O_j) + \hc\right)\,.
\end{align}
Electroweak symmetry-breaking (EWSB) is driven by the VEV of the
neutral component of the SM Higgs doublet, which can be decomposed as
\begin{equation} \label{eq:SMVEV}
H^0 = \frac{1}{\sqrt{2}}\left(v + \phi_H + i\, \sigma_H \right)\,.
\end{equation}
Here $\phi_H \equiv h$ is the Higgs boson, to be identified with the
$125$ GeV state discovered at the LHC. Similarly, the singlet $S$
receives a VEV, and the neutral component of the octet is split into
its CP-even and CP-odd eigenstates:
\begin{equation} \label{eq:SMSOctet:VEVs}
S = v_S + \phi_S\,, \,\hspace{1cm} O^0 \to \frac{1}{\sqrt{2}}\left(O^R + i\, O^I \right)\,.
\end{equation}

We will now briefly discuss the parameter space of the model in order
to justify our choice of input parameters. First, we consider the
tadpole equations, which can be automatically derived by \SARAH. Their
solution for $\mu^2$ and $\kappa_1$ is
\begin{align}
\mu^2 =& - \frac{1}{v^2} ( \lambda_H v^4 - m_S^2 v_S^2 - 4 \lambda_S v_S^4) \, , \nonumber\\
\kappa_1 =& -\frac{1}{v^2}(m_S^2 + 4 \lambda_S v_S^2) \, .
\end{align}
The tree-level mass matrix for the CP-even neutral scalars in the
$\left(\phi_H, \phi_S\right)$ basis is given by
\begin{align}
\mathcal{M}^2 =&\left( \begin{array}{cc} \mu^2 + 3 \lambda_H v^2 + \kappa_1 v_S^2  & 2 \kappa_1 v v_S \\
2 \kappa_1 v v_S  & m_S^2 + \kappa_1 v^2 + 12 \lambda_S v_S^2  \end{array}\right) \nonumber \\
 =& \left( \begin{array}{cc} 2 \lambda_H v^2 & - \frac{2v_S}{v} (m_S^2 + 4\lambda_S v_S^2) \\
- \frac{2v_S}{v} (m_S^2 + 4\lambda_S v_S^2) & 8 \lambda_S v_S^2 \end{array}\right).
\end{align}
We note that, in general, there is singlet-doublet mixing. There are two reasons to consider a small singlet-doublet mixing angle, $\theta$. First, the stringent constraints derived from Higgs physics measurements, and second, the required suppressed decay widths into Higgses, W's and Z's in order to fit the diphoton signal -- indeed in \cite{Cao:2015twy} values of $\sim 10^{-2}$ were found to be required. If we have a small mixing angle, then we can write
\begin{align}
\mathcal{M}^2 \sim \left( \begin{array}{cc} m_h^2  & s_\theta c_\theta (m_h^2 - m_{750}^2)  \\
s_\theta c_\theta (m_h^2 - m_{750}^2) & m_{750}^2 \end{array}\right)\,.
\end{align}
This implies $\lambda_S > 0$, but also
\begin{align}
\mu^2 \simeq & - \frac{1}{2} m_h^2 + \frac{v_S^2}{v^2}(m_S^2 + \frac{1}{2} m_{750}^2) \,.
\end{align}
However, we also have $v_S^2 \sim m_{750}^2/8\lambda_S$, and so 
\begin{align}
\mu^2 \simeq & - \frac{1}{2} m_h^2 + \frac{1.2}{\lambda_S}(m_S^2 + \frac{1}{2} m_{750}^2) \,.
\end{align}
We thus require a tachyonic $m_S^2$ for the SM Higgs mass condition:
\begin{align}
m_S^2 \simeq& - \frac{1}{2} m_{750}^2 +  \frac{\lambda_S}{1.2}(\mu^2 +   \frac{1}{2} m_h^2) \lesssim  - (500\ \mathrm{GeV})^2 
\end{align}
where in the last step we have taken $\lambda_S = 1.2$, a rather large value. If we want $\kappa_1 \sim -1$ then we require $m_S^2 \sim - (600\ \mathrm{GeV})^2$. On the other hand, from the second tadpole equation we have
\begin{align}
m_S^2 =& - \kappa_1 v^2 - \frac{1}{2} m_{750}^2,
\end{align}
which, if we require $|\kappa_1| < 2$, gives 
\begin{align}
- (630\ \mathrm{GeV})^2 \le m_S^2 \le  - (400\ \mathrm{GeV})^2 \,,
\end{align}
so putting these together we find the narrow window
\begin{align}
- (630\ \mathrm{GeV})^2 \le m_S^2 \le  - (500\ \mathrm{GeV})^2 \,.
\end{align}

{\bf Alternative implementation in \SARAH}\\

The above discussion suggests to use a different choice for the input parameters of the model in our \SARAH implementation: ideally we would like the particle masses, the mixing and only dimensionless couplings to be the inputs. We shall take the input parameters to be
\begin{align}
m_h, m_{750}, s_\theta, \lambda_S.
\end{align}
In terms of these the other parameters are determined to be
\begin{align}
\lambda_H =& \frac{c_\theta^2 m_h^2 + s_\theta^2 m_{750}^2}{2v^2},  \qquad v_S^2 = \frac{c_\theta^2 m_{750}^2 + s_\theta^2 m_h^2 }{8\lambda_S} \nonumber\\
 m_S^2 =& - \kappa_1 v^2 - \frac{1}{2} m_{750}^2, \qquad \kappa_1 v v_S = s_\theta c_\theta (m_h^2 - m_{750}^2) \nonumber \\
\rightarrow \kappa_1 =& \frac{\sqrt{2 \lambda_S} s_\theta c_\theta (m_h^2 - m_{750}^2) }{v \sqrt{(c_\theta^2 m_{750}^2 + s_\theta^2 m_h^2)}} \simeq - 4.3 \times s_\theta \sqrt{\lambda_S}
\end{align}
The exact version of these equations is implemented in \SARAH and can be selected using the {\tt InputFile$\rightarrow$"SPheno\_diphoton.m"} option in {\tt MakeAll} or {\tt MakeSPheno}. \\

{\bf Octet masses}\\

One further input is taken in \cite{Cao:2015twy}: the physical mass of
the octet scalars.  These are given in terms of the Lagrangian
parameters as:
\begin{align}
m_{O^0_r}^2 =&m_O^2 +  \kappa_2 v_S^2 + \frac{v^2}{2} (\lambda_1  + \lambda_2 + 2 \, \mathrm{Re}(\lambda_3)) \nonumber\\
m_{O^0_i}^2 =&m_O^2 +  \kappa_2 v_S^2 + \frac{v^2}{2} (\lambda_1  + \lambda_2 - 2 \, \mathrm{Re}(\lambda_3)) \nonumber\\
m_{O^+}^2 =&m_O^2 +  \kappa_2 v_S^2   + \frac{1}{2} \lambda_1 v^2 
\end{align}
The values of $\lambda_i$ are taken to be small and equal in order for the octets to have similar masses, but since this is not the general case, we do not impose this choice in \SARAH. The choice in that paper does however hide the possibility of tachyonic $m_O^2$ (and hence possible charge/colour breaking minima) -- indeed, if we insist that $m_O^2 > 0$ we have a lower bound on the masses of
\begin{align}
m_{O^{0,+}}^2 > & \frac{\kappa_2}{8\lambda_S } m_{750}^2 .
\end{align}
Clearly this is violated for $m_{O^{0,+}} = 600$ GeV when $\kappa_2 \sim 1, \lambda_S \ll 1$. On the other hand, this does not guarantee a problem. 

The desired vacuum has energy
\begin{align}
V_0=&\frac{m_S^2v_S^2 }{2} - \frac{ \lambda_H}{4} v^4+ \lambda_S v_S^4 \nonumber\\
\simeq& - \frac{1}{8} v^2 m_h^2  - v_S^2( \frac{1}{2} \kappa_1 v^2 + \frac{1}{4} m_{750}^2 - \frac{m_{750}^2}{8}) \nonumber\\
\simeq& - \frac{1}{8} v^2 m_h^2 - \frac{m_{750}^2}{8\lambda_S} ( - 2 s_\theta \sqrt{\lambda_S} + \frac{m_{750}^2}{8}) 
\end{align}
If we instead concentrate on the potential terms containing the octets, where only one component develops a VEV, we find 
\begin{align}
V(O^R) =& \frac{1}{2}(O^R)^2  \bigg[ m_O^2 + \frac{1}{8}(\lambda_9 + \lambda_{10}+ \frac{1}{9} \lambda_6+ \frac{1}{9} \lambda_7 +  \frac{1}{9} \lambda_{11}) (O^R)^2 \bigg] \nonumber \\
V(O^I) =& \frac{1}{2}(O^I)^2  \bigg[ m_O^2 + \frac{1}{8}(\lambda_9 + \lambda_{10}+ \frac{1}{9} \lambda_6+ \frac{1}{9} \lambda_7 +  \frac{1}{9} \lambda_{11}) (O^I)^2 \bigg] \nonumber \\
V(O^+) =&  |O^+|^2  \bigg[ m_O^2 + \frac{1}{4}(\lambda_9 + \lambda_{10} +  \frac{1}{9} \lambda_6+ \frac{1}{9} \lambda_7 +  \frac{1}{9} \lambda_{11}) |O^+|^2 \bigg] 
\end{align}
Arranging for the additional minimum of the potential to be higher than the colour-breaking one then places a \emph{lower} bound on the octet self-couplings, but for the phenomenology of the diphoton excess -- when we neglect loop corrections to the mass of the octet -- they play no other role.\\

{\bf Comments on fitting the excess}\\

\begin{figure}
\begin{center}
\includegraphics[width=0.5\textwidth]{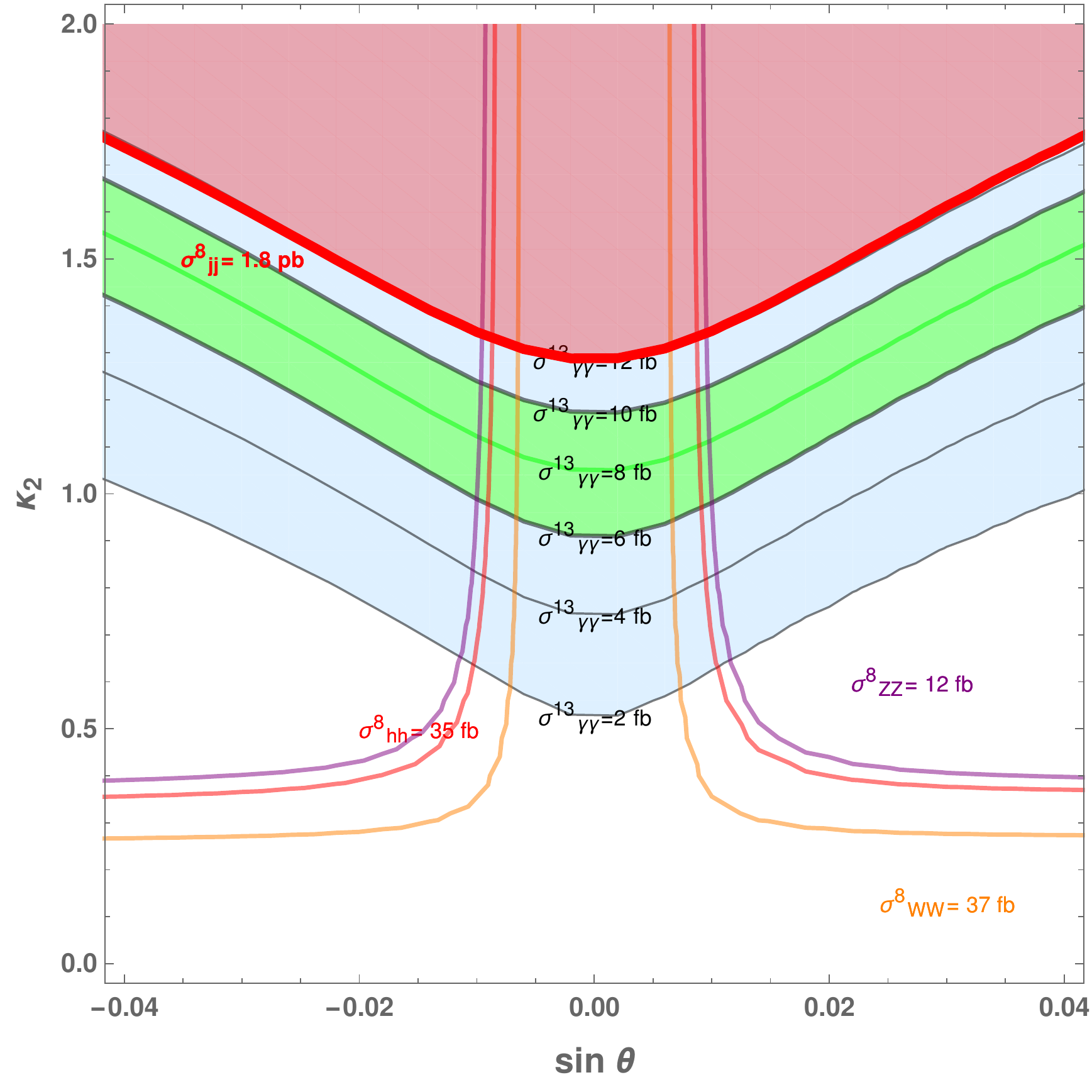}\includegraphics[width=0.5\textwidth]{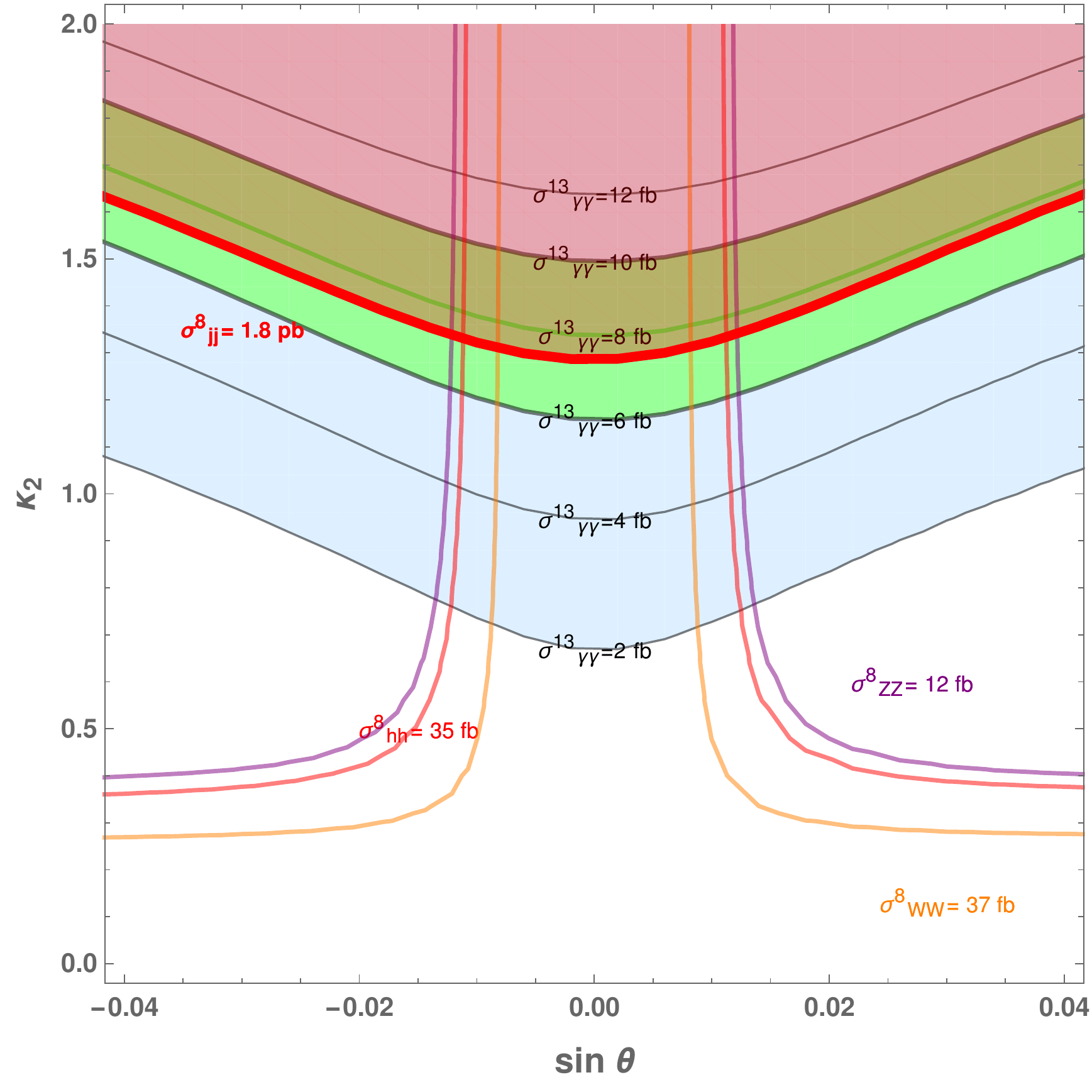}
\caption{Scan over sine of Higgs-singlet mixing angle $\theta$ and $\kappa_2$ for octet masses of $600$ GeV, $\lambda_S = 0.07$ corresponding to $v_S \simeq 1000$ GeV. The contours show the $750$ GeV resonance production cross-sections $\sigma^X_{YY}$ at energy $X$ TeV decaying into channel $YY$. On the left plot, only leading order contributions to the decays are used; on the right, all corrections up to N$^3$LO available in \SARAH are included. }\label{FIG:SMSOCT}
\end{center}
\end{figure} 

In \cite{Cao:2015twy} the authors find that the diphoton excess can be easily fit with octets at $600$ or $1000$ GeV and $\kappa_2 \sim 1.5$ or $4.5$, respectively. The scenario involves merely the simplifying assumption $\lambda_1 = \lambda_2 = \lambda_3$ so that the octets are of approximately equal mass. The ratio between the digluon and diphoton decay rates is then
\begin{align}
\frac{\Gamma (S \rightarrow gg)}{\Gamma (S \rightarrow \gamma \gamma)} \simeq \frac{9}{2} \frac{\alpha_s^2}{\alpha^2}.
\end{align}
In \cite{Cao:2015twy} this is quoted as $\simeq 715$. In \SARAH, before any NLO corrections are applied, the running of the Standard Model gauge couplings yields $\alpha_s (750\ \mathrm{GeV}) = 0.091 $ and we use $\alpha(0) \simeq 137^{-1}$, giving a ratio of $700$, in good agreement. However, when we include corrections up to N$^3$LO, this ratio rises to  $1150$, putting the model near the boundary of exclusion due to dijet production at $8$ TeV. These differences are illustrated in plots produced from \SARAH/\SPheno in figure \ref{FIG:SMSOCT}. To produce these plots, all branching ratios/widths are calculated in \SPheno, as is the production cross-section of the resonance  at $8$ TeV. To calculate $13$ TeV cross-sections the $8$ TeV cross-section was rescaled by the parton luminosity factor for gluons of $4.693$.

\modelparalabel{Vector-like $SU(2)$ triplet quark model}{sec:TripletQuarks}
\begin{itemize}
\item {\bf Reference:} \cite{Benbrik:2015fyz}
\item {\bf Model name:}  \texttt{SM+VL/TripletQuarks}
\end{itemize}

The model introduced in \cite{Benbrik:2015fyz} considers a singlet scalar $S$ as the candidate for the \SI{750}{\GeV} resonance. In order to produce the singlet via gluon fusion at the LHC, the model is further extended with the introduction of vector-like quarks, triplet under $SU(2)_L$. Moreover, the singlet is charged under a {\zTwo} parity, although this is softly broken by the vector-like quark mass terms.

\begin{table}[h]
\centering
\begin{tabular}{c c c c c c}
\toprule
Field & Gen. & $SU(3)_C$ &  $SU(2)_L$ & $U(1)_Y$ & $\mathbb{Z}_2$ \\
\midrule
$S$ & 1 & $\one$ & $\one$ & $0$ & $-$ \\
\midrule
$F_{1L}$ & 1 & $\three$ & $\three$ & $\frac{2}{3}$ & $-$ \\
$F_{1R}$ & 1 & $\three$ & $\three$ & $\frac{2}{3}$ & $+$ \\
$F_{2L}$ & 1 & $\three$ & $\three$ & $-\frac{1}{3}$ & $-$ \\
$F_{2R}$ & 1 & $\three$ & $\three$ & $-\frac{1}{3}$ & $+$ \\
\bottomrule
\end{tabular}
\caption{Extra scalar/fermionic degrees of freedom shown in the top/bottom. All SM particles are even under the imposed discrete symmetry.}
\label{tab:VLTQcontent}
\end{table}

This model is based on the SM gauge symmetry, extended with a {\zTwo} parity. The fermionic and scalar particle content is summarized in \cref{tab:VLTQcontent}. The vector-like $SU(2)_L$ triplet quarks can be expressed in $2 \times 2$ matrix notation as
\begin{equation} \label{eq:frep}
F_1 = \left( \begin{array}{cc}
U_1/\sqrt{2} & X \\
D_1 & - U_1/\sqrt{2}
\end{array} \right) \,,\, \qquad
F_2 = \left( \begin{array}{cc}
D_2/\sqrt{2} & U_2 \\
Y & - D_2/\sqrt{2}
\end{array} \right) \,.\,
\end{equation}
Here we see that the $U_{1,2}$ and $D_{1,2}$ components have the same electric charges as the SM up- and down-type quarks, respectively. The components $X$ and $Y$ are exotic coloured states with electric charges $5/3$ and $-4/3$, respectively, and thus they do not mix.

The Yukawa Lagrangian of the model can be written as
\begin{align} \label{eq:VLTQ:lag}
-\mathcal{L}_Y &= \mathcal{L}_Y^{\text{SM}} + Y_1 \, \overline{Q_L} \, F_{1\,R} \, \widetilde{H} + Y_2 \, \overline{Q_L} \, F_{2\,R} \, H \notag\\
 &+\left( m_{F_1}+ Y_{F_1 S} S\right) \overline{F_{1}} \, F_{1} +\left( m_{F_2} + Y_{F_2 S} S\right)  \overline{F_{2}} \, F_{2} \, + \hc\,,
\end{align}
whereas the scalar potential is given by
\begin{equation}
\label{eq:V}
V = -\mu^2 \, |H|^2 + \lambda_H \, |H|^4 + m_S^2 \, S^2 + \frac{1}{4}\lambda_S \, S^4 + \frac{1}{3}\lambda_{HS} \, |H|^2 S^2  \, .
\end{equation}
We note that the vector-like mass terms for $F_1$ and $F_2$ softly break the {\zTwo} parity. We assume that the SM Higgs doublet obtains a VEV while the introduced singlet does not, hence
\begin{align} \label{eq:TripletQuarks:VEVs}
H = \frac{1}{\sqrt{2}} \, \left( \begin{array}{c}
0 \\
v + \phi_H + i \, \sigma_H
\end{array} \right) \,, \qquad 
S &= \phi_S \,.
\end{align}
The {\zTwo} discrete symmetry, together with $\langle S \rangle = 0$ ensured by the symmetry, implies that no mixing between the SM Higgs boson $\phi_H$ and the singlet $S$ appears at tree-level. 

\modelparalabel{Single scalar leptoquark model}{sec:ScalarLeptoquarks}
\begin{itemize}
\item {\bf Reference:} \cite{Bauer:2015boy}
\item {\bf Model name:}  \texttt{Leptoquarks/ScalarLeptoquarks}
\end{itemize}
The model introduced in Ref. \cite{Bauer:2015boy} considers a singlet
scalar, $S$, as candidate for the $750$ GeV resonance. It is based on
the scalar leptoquark model in Ref. \cite{Bauer:2015knc}, with the
addition of a vector-like fermion multiplet $\chi$ transforming in an
a priori unspecified representation of $SU(2)_L$. The simplest model,
which the authors find to work, requires $\chi$ to be a $SU(2)_L$
triplet.  The model is based on the gauge symmetries of the SM,
augmented with either a discrete or gauge symmetry necessary for the
DM sector. For the sake of simplicity we choose to realise the model
using only a discrete {\zTwo} symmetry. The particle content beyond
the SM is shown in \cref{tab:ScalarLeptoQuarkscontent}.

\begin{table}[h]
\centering
\begin{tabular}{ c c c c c c}
\toprule
Field & Gen. & $SU(3)_C$ &  $SU(2)_L$ & $U(1)_Y$ & $\mathbb{Z}_2$ \\
\midrule 
$S$ & 1 & \one & \one & $0$ & $+$ \\
$\phi$ & 1 & \three & \one & $-\frac{1}{3}$ & $+$ \\ \midrule
$\chi$ & $N_\chi$ & \one & \three & $Y_\chi$ & $-$ \\ \bottomrule
\end{tabular}
\caption{Additional scalar and fermionic particles beyond the SM shown in the top and bottom respectively. Note that all SM fields are neutral under the {\zTwo} symmetry. }
\label{tab:ScalarLeptoQuarkscontent}
\end{table}
The new degrees of freedom are: the scalar leptoquarks $\phi$, the gauge singlet scalar $S$, which is assumed to be real, and finally the vector-like triplet fermions $\chi$. The vector-like $SU(2)_L$ triplet can be expressed in $2 \times 2$ matrix notation as
\begin{align}
	\chi &= \begin{pmatrix}
	\frac{\chi^0}{\sqrt{2}} & \chi^+ \\
	\chi^- & -\frac{\chi^0}{\sqrt{2}}
	\end{pmatrix}\,.
\end{align}
Due to the discrete {\zTwo} symmetry the neutral component $\chi^0$ is stable and a suitable DM candidate. In Ref.~\cite{Bauer:2015boy} the hypercharge $Y_\chi$ and the number of generations $N_\chi$ are left as free parameters. In order to boost the ratio of decay widths $\Gamma_{S\to\gamma\gamma}/\Gamma_{S\to gg}$ they claim that the minimal choice is $N_\chi =2$ and $Y_\chi=0$. The model files are implemented with these values.

The Yukawa interactions of the model and the vector-like mass term can be expressed as 
\begin{align} \label{eq:LagYukScalarLeptoquarks}
-\mathcal{L}_Y &= - \mathcal{L}_Y^{\text{SM}} +\lambda^L \,\overline{Q^c} \, L \,\phi^* + \lambda^R \,\overline{u_R^c}\, e_R\, \phi^*+\frac{1}{2} M_\chi\, \overline\chi\, \chi + g_{S\chi} S\,\overline{\chi}\,\chi + \hc\,,
\end{align}
whereas the scalar potential is given by
\begin{align} \label{eq:LagPotScalarLeptoquarks}
V &= V_{\rm SM} + M_\phi^2 |\phi|^2 + \frac{1}{2} M_S^2 S^2 + \frac{1}{3}\lambda_{S_3} S^3 + \frac{1}{4}\lambda_{S_4} S^4 +\frac{1}{2} \lambda_\phi |\phi|^4 \notag \\
&+ g_{SH} S^2 |H|^2 + g_{H\phi} |\phi|^2 |H|^2 + g_{S\phi} S^2|\phi|^2 + \kappa_{S\phi} S |\phi|^2 + \kappa_{SH} S |H|^2 \, .
\end{align}
The Lagrangian in \cref{eq:LagYukScalarLeptoquarks} is not generic. One could add more renormalizable operators involving the leptoquark, like $\overline{d^c}_R u_R \phi$ for instance. An operator of this kind, together with those in \cref{eq:LagYukScalarLeptoquarks}, would lead to rapid proton decay, thus it must be forbidden~\cite{Bauer:2015knc}. This can be achieved by imposing a discrete symmetry (for instance another $\mathbb{Z}_2$ under which the leptoquark and the SM quarks are odd and everything else is even). In the model file we simply omit these dangerous operators. 
The potential in \cref{eq:LagPotScalarLeptoquarks}, on the other hand, is generic. It could be simplified by introducing yet another $\mathbb{Z}_2$ to avoid terms with odd powers of $S$, and we would still have the necessary ingredients to fit the diphoton excess. 

We assume the EWSB proceeds by the usual Higgs VEV in addition to the singlet VEV $v_S$, where the scalar singlet is expressed as 
\begin{align} \label{eq:VEVsScalarLeptoquarks}
S = v_S + \sigma_S \,.
\end{align}

\modelparalabel{Two scalar leptoquark model}{sec:TwoScalarLeptoquark}
\begin{itemize}
\item {\bf Reference:} \cite{Chao:2015nac}
\item {\bf Model name:}  \texttt{Leptoquarks/TwoScalarLeptoquarks}
\end{itemize}

This model includes four additional scalars, two of which, $\Phi$ and $\Omega$ are leptoquarks. The other two are a singlet $S$, which is the \SI{750}{\GeV} resonance, and a scalar $\Theta$ which carries hypercharge. The additional quantum numbers for these new field are shown in \cref{tab:TwoSLQcontent}.
\begin{table}[h]
\centering
\begin{tabular}{ c c c c c}
\toprule
Field & Gen. & $SU(3)_C$ &  $SU(2)_L$ & $U(1)_Y$  \\
\midrule 
$S$ & 1 & \one & \one & $0$ \\
$\Theta$ & 1 & \one & \one & $1$ \\
$\Phi$ & 1 & \threeS & \one & $\frac43$ \\
$\Omega$ & 1 & \threeS & \one & $\frac13$ 
\\ \bottomrule
\end{tabular}
\caption{Additional scalar particles beyond the SM appearing in the two scalar leptoquark model.}
\label{tab:TwoSLQcontent}
\end{table}

The Yukawa interactions consistent with these new fields is given by
\begin{align}
-\mathcal{L}_Y &= \mathcal{L}_Y^{\rm SM} + Y_\Theta L^T i \sigma_2 L \Theta + Y_\Omega Q_L^T i \sigma_2 L \Omega + Y_\Phi e_R^T d_R \Phi + \hc \, ,
\end{align}
where the generation indices have been suppressed. These indices are important so that one obtains non-zero contractions for the first term in the above Yukawa. The scalar potential of the model is given by
\begin{align}
V &=V_{\rm SM} + \sum_i\mu_\phi^2 |\phi_i|^2 +\frac12 \sum_{i,j} \lambda_{ij} |\phi_i|^2 |\phi_j|^2 + \left\{\lambda S \,\Theta \,\Phi^\dagger \Omega + \hc\right\}\,,
\end{align}
where the index $i=\{S,\Theta,\Phi,\Omega\}$ runs over the introduced scalar fields. The potential has an additional $U(1)$ symmetry which can be gauged or global. In the implementation in {\tt SARAH} we take it as global. This symmetry can be used to forbid terms such as $u_R d_R \Omega^\dag$, which could potentially lead to proton decay. 

One can generate neutrino masses at the two-loop level using a combination of the new Yukawa couplings and the VEV of the singlet. Beyond the neutral components of the Higgs doublet, the singlet also obtains a VEV of the form
\begin{align} \label{eq:TwoScalarLeptoquarks:VEVs}
S &=  v_S + \phi_S\,.
\end{align}

\modelparalabel{Georgi-Machacek model}{sec:Georgi-Machacek}
\begin{itemize}
\item {\bf Reference:} \cite{Fabbrichesi:2016alj,Chiang:2016ydx}
\item {\bf Model name:}  \texttt{Georgi-Machacek}
\end{itemize}
The model originally proposed in Ref.~\cite{Georgi:1985nv} extends the SM by one real scalar $SU(2)_L$-triplet $\eta$ with $Y=0$ and one complex scalar $SU(2)_L$-triplet with $\chi$ with $Y=1$, which can be written as
\begin{equation} \label{eq:fieldsGM}
 \eta  = \frac{1}{\sqrt{2}} \, \left( \begin{array}{cc}
\eta^0 & - \sqrt{2} \left(\eta^-\right)^*\\
- \sqrt{2}  \eta^- & - \eta^0  \end{array} \right) \,, \quad 
 \chi  = \frac{1}{\sqrt{2}}  \, \left( \begin{array}{cc}
\chi^- & \sqrt{2} (\chi^0)^* \\
-\sqrt{2} \chi^{--} & -\chi^-  \end{array} \right) \,.
\end{equation}
The CP-even components of the new scalars mix with the neutral SM Higgs. Usually, the lightest state is the $125$ GeV Higgs boson, while the second mass eigenstate is identified with the $750$ GeV resonance. 
The most compact form to write the Lagrangian in a $SU(2)_L \times SU(2)_R$ invariant form is to express the Higgs doublet and the two scalar triplets as
\begin{align}
\Phi =
\left(
\begin{array}{cc}
\phi^{0*}  & \phi^+     \\
 \phi^- & \phi^0      
\end{array}
\right)\,, \qquad
\quad \Delta =
\left(
\begin{array}{ccc}
 \chi^{0*} & \eta^+  & \chi^{++}   \\
 \chi^- & \eta^0  & \chi^+  \\
 \chi^{--} & \eta^{-}  & \chi^0   
\end{array}
\right)\,.
\end{align}
In this form, the scalar potential reads
\begin{align}
V(\Phi, \Delta) & = \mu_2^2 \mathrm{Tr} \Phi^\dagger \Phi + \frac{\mu_3^2}{2} \mathrm{Tr} \Delta^\dagger \Delta  + \lambda_1 \left[  \mathrm{Tr} \Phi^\dagger \Phi \right]^2 + \lambda_2  \mathrm{Tr} \Phi^\dagger \Phi \, \mathrm{Tr} \Delta^\dagger \Delta \nonumber \\
&  + \lambda_3 \mathrm{Tr} \Delta^\dagger \Delta \Delta^\dagger \Delta +   \lambda_4 \left[\mathrm{Tr} \Delta^\dagger \Delta \right]^2   - \lambda_5 \mathrm{Tr} \left( \Phi^\dagger \sigma^a \Phi \sigma^b  \right) \, \mathrm{Tr}  \left(\Delta^\dagger t^a \Delta t^b \right) \nonumber \\
&   - M_1  \mathrm{Tr} \left( \Phi^\dagger \tau^a \Phi \tau^b \right) (U \Delta U^\dagger)_{ab}\nonumber    - M_2 \mathrm{Tr} \left( \Delta^\dagger t^a \Delta t^b \right) (U \Delta U^\dagger)_{ab} \, ,
\end{align}
$\tau^a$ and $t^a$ are the $SU(2)$ generators for the doublet and triplet representations respectively, while $U$ is given for instance in Ref.~\cite{Hartling:2014zca}. Because of the custodial symmetry, the VEVs for the triplets are identical, and there are no tree-level contributions to electroweak precision observables. The compact form of the potential can not be implemented in {\tt SARAH}, so we have translated into the form of \cref{eq:fieldsGM}. For example: 
\begin{align}
 \lambda_2  \mathrm{Tr} \Phi^\dagger \Phi \, \mathrm{Tr} \Delta^\dagger \Delta \quad \to \quad  4 \lambda_2^a |H|^2 \mathrm{Tr}(\chi^\dagger \chi) + 2 \lambda_2^b |H|^2 \mathrm{Tr}(\eta^\dagger \eta) \, .
\end{align}
This introduces more couplings, which have to be identical to preserve the custodial symmetry. A substitution rule to apply these relations is included in the model file. 

The triplets receive VEV as
\begin{equation} \label{eq:GM:VEVs}
\langle \eta \rangle = \frac{1}{\sqrt{2}} \, \left( \begin{array}{cc}
v_\eta & 0\\
0 & -v_\eta  \end{array} \right) \,, \quad 
\langle \chi \rangle =  \, \left( \begin{array}{cc}
0 & v_\chi \\
0 & 0  \end{array} \right) \, ,
\end{equation}
which fulfil $v_\eta=v_\chi$ if the custodial symmetry is preserved.

%% file: tex/ModelCategories/THDmodels.tex
Two-Higgs-doublet models (THDMs) are among the most common candidates
to explain the diphoton excess. In this family of models, the
\SI{750}{\GeV} resonance is typically a heavy Higgs, whose diphoton
rate is enhanced with the addition of new states that contribute to
its production cross-section and/or its decay width into pairs of
photons. Several realisations of this idea can be found in the recent
literature. Here we review a few representative examples.

\paragraph{THDM generalities}\hfill \break
\label{subsec:THDMgeneralities}
We first comment on some general features and conventions used
in THDMs. For a complete review we refer to Ref. \cite{Branco:2011iw}.

\paragraph{Couplings to fermions}\hfill \break
We will consider three types of THDM, depending on the way the two
Higgs doublets $H_1$ and $H_2$ couple to the SM fermions, as shown in
this table:
\begin{center}
\begin{tabular}{ c | c c c }
\toprule
& $u_R$ & $d_R$ & $e_R$ \\
\midrule
Type-I & $H_2$ & $H_2$ & $H_2$ \\
Type-II & $H_2$ & $H_1$ & $H_1$ \\
Type-III & both & both & both \\
\bottomrule
\end{tabular}
\end{center}
We note that, by convention, the right-handed up-type quarks $u_R$
always couple to $H_2$, and that the MSSM can be seen as a particular
example of a type-II THDM. The type-I and type-II THDMs are usually
said to be models with \emph{natural flavour conservation}. This is
because, in contrast to the general type-III THDM, flavour changing
neutral currents are completely absent at tree-level, thus satisfying
flavour constraints more easily. This is achieved by coupling each
fermion species to only one Higgs doublet, which can be enforced using
a discrete symmetry. For example, the type-I THDM couplings to
fermions can be obtained by imposing that $H_1 \to - H_1$ is a
symmetry of the Lagrangian. In our \SARAH implementations we will not
introduce these discrete symmetries explicitly, but simply allow only
for those couplings that characterise the type of THDM under
consideration. We point out that \SARAH includes also other THDM
versions, such as the lepton-specific or flipped ones. These versions
can be easily combined with the extensions presented here for type
I--III to include additional matter in order to explain the diphoton
excess. Finally, we have also not implemented explicitly all
representations for the vector-like states proposed in the literature
so far. For instance, Ref.~\cite{Kang:2015roj} introduces quarks with
a very large hypercharge. However, it is easy to change the considered
quantum numbers by changing the model files accordingly.

\paragraph{Scalar potential}\hfill \break
We assume the following scalar potential for all THDMs considered
below,
\begin{align}
  \label{eq:scalar-pot}
V &= m_{11}^2  |H_1|^2 + m_{22}^2 |H_2|^2 + \left[ m_{12}^2 \, H_1^\dagger H_2 + \, \hc \right] \nonumber \\
  & + \frac{\lambda_1}{2} |H_1|^4 + \frac{\lambda_2}{2}  |H_2|^4
    + \lambda_3 |H_1|^2 |H_2|^2 \nonumber \\
  & + \lambda_4 \left(H_1^\dagger H_2\right)\left(H_2^\dagger H_1\right) 
    + \left[ \frac{\lambda_5}{2} \left(H_1^\dagger H_2\right)^2 
    + \, \hc \right] \, .
\end{align}
In principle, the quartic terms $\lambda_6 \left(H_1^\dagger
H_1\right)\left(H_1^\dagger H_2\right) $ and $\lambda_7
\left(H_2^\dagger H_2\right)\left(H_1^\dagger H_2\right) $ are also
 allowed by the gauge symmetry. However, in our \SARAH
implementation we neglect these terms. This corresponds to 
assuming a global symmetry, which would be
softly broken by the $m_{12}^2 \, H_1^\dagger H_2$ term. 

\paragraph{Symmetry breaking pattern}\hfill \break
The symmetry breaking pattern is the same in all THDM variations,
namely
\begin{equation} \label{eq:THDmodels:VEVs}
\langle H_i \rangle = \left\langle \begin{pmatrix}
H_i^+ \\ H_i^0
\end{pmatrix} \right\rangle = \frac{1}{\sqrt{2}} \begin{pmatrix}
0 \\ v_i
\end{pmatrix} \, ,
\end{equation}
with $i=1,2$ running over the two Higgs-doublets.

In the following, we describe the THDM variations implemented in
\SARAH\ related to the diphoton excess. Broadly speaking, there are two main categories: THDMs with extra
vector-like fermions and THDMs with extra scalar content. In the first
case we have three examples, whereas for the second scenario we have
only a single example.

\modelparalabel{Minimal vector-like THDM}{sec:minTHDM}
\begin{itemize}
\item {\bf Reference:} \cite{Bizot:2015qqo} 
\item {\bf Model name:} \\
\texttt{THDM+VL/min-3}\\
\texttt{THDM+VL/min-8}
\end{itemize}

Ref. \cite{Bizot:2015qqo} considers a type-III THDM extended with two
new vector-like colored fermions, $S$ and $Q^V$, aiming at a simultaneous
explanation of the diphoton excess and of the CMS hint for the Higgs
lepton flavor violating (LFV) decay $h \to \tau \mu$
\cite{Khachatryan:2015kon}. The type-III THDM has already been shown
in several papers to be able to accommodate this LFV signal, see
e.g. \cite{Sierra:2014nqa,Dorsner:2015mja}, but \cite{Bizot:2015qqo}
is the first work that addresses the diphoton excess at the same
time. The representations of the new vector-like fermions under SM gauge group
are shown in \cref{tab:minTHDM38}. 
\begin{table}[h]
\centering
\begin{tabular}{ c c c c c}
\toprule
Field & Gen. & $SU(3)_C$ &  $SU(2)_L$ & $U(1)_Y$  \\
\midrule 
$S$ & 1 & $R_c$ & \one & $Q$ \\
$Q^V$ & 1 & $R_c$ & \two & $Q+\frac12$ \\
\bottomrule
\end{tabular}
\caption{Additional fermion field content for the minimal vector-like THDM.}
\label{tab:minTHDM38}
\end{table}

We choose the explicit realizations $R_c=3$, $Q=2$ ({\tt
  THDM+VL/minTHDM-3}) and $R_c=8$, $Q=2$ ({\tt THDM+VL/minTHDM-8}). In
both cases, the additional interaction terms beyond those in the
standard THDM are
\begin{align}
-\mathcal L_{\text{new}} &= M_Q^V \overline{Q^V_R} Q^V_L + M_S \overline{S_R} S_L 
+ \lambda_{i}^D \overline{S_R} \tilde{H_i} Q^V_L + \lambda_i^S \overline{Q^V_R} H_i S_L + \hc \, ,
\end{align}
where $M_Q^V$ and $M_S$ are masses for the $SU(2)_L$-doublet and
singlet vector-like fermions, respectively, and $\lambda_i^{D,S}$ (with
$i=1,2$) are new Yukawa interactions.

\modelparalabel{THDM with exotic vector-like fermions}{sec:THDMexotic}
\begin{itemize}
\item {\bf References:} \cite{Angelescu:2015uiz,Jung:2015etr}
\item {\bf Model name:}\\
\texttt{THDM+VL/Type-I-VL}\\
\texttt{THDM+VL/Type-II-VL}
\end{itemize}

The THDMs in Refs. \cite{Angelescu:2015uiz,Jung:2015etr} consider a
less minimal framework where three generations of new vector-like
fermions are added to the standard THDM scenario. Both type-I and
type-II THDMs are studied. Furthermore, the vector-like leptons have
exotic hypercharge values shown in \cref{tab:THDMVL12}.

\begin{table}[h]
\centering
\begin{tabular}{ c c c c c c}
\toprule
Field & Gen. & $SU(3)_C$ &  $SU(2)_L$ & $U(1)_Y$ & $\mathbb{Z}_2$  \\
\midrule 
$Q^V$ & 3 & \three & \two & $\frac16$& $-$ \\
$d^V$ & 3 & \three & \one & $-\frac13$ & $-$\\
$u^V$ & 3 & \three & \one & $\frac23$& $-$ \\
$L^V$ & 3 & \one & \two & $-\frac32$& $-$ \\
$e^V$ & 3 & \one & \one & $-2$& $-$ \\
$\nu^V$ & 3 & \one & \one & $-1$& $-$\\
\bottomrule
\end{tabular}
\caption{Additional fermion field content for the THDM with exotic vector-like fermions. All SM fields are even under the imposed $\mathbb{Z}_2$ discrete symmetry.}
\label{tab:THDMVL12}
\end{table}
The vector-like lepton states can be decomposed (or denoted) as
\begin{equation}
L^V = \left( \begin{array}{c}
\ell^{\prime -} \\
\ell^{--}
\end{array} \right) \,, \qquad e^V \equiv \left( e^V \right)^{--} \, , \qquad \nu^V \equiv \left( \nu^V \right)^{-} \, ,
\end{equation}
where we explicitly highlight the presence of doubly-charged
leptons. The additional interaction terms beyond those in the standard
THDMs are
\begin{align}
-\mathcal L_{\text{new}} &= M_Q^V \overline{Q^V_R} Q^V_L + M_L^V \overline{L_R} L_L + 
M_d^V \overline{d_R^V} d_L^V + M_u^V \overline{u_R^V} u_L^V + M_{\nu}^V \overline{\nu_R^V} \nu_L^V + M_{e}^V \overline{e_R^V} e_L^V  \notag\\  
&+ Y_{dL}^V  \overline{d_R^V} \tilde H_1 Q_L^V + Y_{dR}^V  \overline{Q^V_R} H_1 d_L^V
+ Y_{uL}^V  \overline{u_R^V} H_2 Q_L^V + Y_{uR}^V  \overline{Q^V_R} \tilde H_2 u_L^V \label{eq:lagTHDMVL} \notag\\  
&+ Y_{eL}^V  \overline{e_R^V} \tilde H_1 L_L^V + Y_{eR}^V  \overline{L^V_R} H_1 e_L^V
+ Y_{\nu L}^V  \overline{\nu_R^V} H_2 L_L^V + Y_{\nu R}^V  \overline{L^V_R} \tilde H_2 \nu_L^V 
+ \hc \, .
\end{align}
Here $M_i^V$ are the bare vector-like masses and $Y_i^V$ are Yukawa
couplings. For simplicity, we assume that the vector-like states
do not mix with the SM fermions. This can be enforced with a discrete
$\mathbb{Z}_2$ symmetry under which the vector-like states are odd and
the rest of the states even.

\modelparalabel{THDM with SM-like vector-like fermions}{sec:THDMVL}
\begin{itemize}
\item {\bf Reference:} \cite{Badziak:2015zez}
\item {\bf Model name:} \\
\texttt{THDM+VL/Type-I-SM-like-VL} \\
\texttt{THDM+VL/Type-II-SM-like-VL}
\end{itemize}

The THDM proposed in \cite{Badziak:2015zez} introduces vector-like
copies of the SM quarks and leptons, the $SU(2)_L$ doublets $Q^V$ and
$L^V$, as well as the $SU(2)_L$ singlets $d^V$, $u^V$, $\nu^V$ and
$e^V$ with their respective quantum numbers shown in \cref{tab:THDMSMlikeVL}.

\begin{table}[h]
\centering
\begin{tabular}{ c c c c c c}
\toprule
Field & Gen. & $SU(3)_C$ &  $SU(2)_L$ & $U(1)_Y$ & $\mathbb{Z}_2$  \\
\midrule 
$Q^V$ & 1 & \three & \two & $\frac16$ & $-$ \\
$d^V$ & 1 & \three & \one & $-\frac13$& $-$ \\
$u^V$ & 1 & \three & \one & $\frac23$& $-$ \\
$L^V$ & 1 & \one & \two & $-\frac12$& $-$ \\
$e^V$ & 1 & \one & \one & $-1$& $-$ \\
$\nu^V$ & 1 & \one & \one & $0$& $-$\\
\bottomrule
\end{tabular}
\caption{Additional fermion field content for the THDM with SM-like vector-like fermions. All SM fields are even under the imposed $\mathbb{Z}_2$ discrete symmetry.}
\label{tab:THDMSMlikeVL}
\end{table}

In contrast to other THDMs proposed to explain the diphoton excess,
where low values of $\tan \beta$ are taken in order to increase the
heavy Higgs coupling to the top quark and induce a large gluon fusion
cross-section, this paper considers moderate and large values of $\tan
\beta$. In this case, the heavy Higgs production is mainly induced by
the new vector-like colored states. Moreover, the advantage of using
largish $\tan \beta$ values is that the total decay width of the heavy
Higgs is suppressed, thus allowing for a smaller $\Gamma(H \to \gamma
\gamma)$ to explain the excess.

The new interaction terms take the same form as those in
\cref{eq:lagTHDMVL}. Furthermore, as in the previous models, we
assume that the new vector-like states do not mix with the SM
fermions. For this reason, we introduce a $\mathbb{Z}_2$ parity under
which the vector-like states are odd and the rest of the states
even. Finally, Ref.~\cite{Badziak:2015zez} considers two variants of
this scenario in what concerns the couplings to the SM fermions: a
type-I THDM and a type-II THDM.

\modelparalabel{THDM with a real scalar septuplet}{sec:THDMseptuplet}
\begin{itemize}
\item {\bf Reference:} \cite{Han:2015qqj,Moretti:2015pbj}
\item {\bf Model name:} \texttt{THDM/ScalarSeptuplet}
\end{itemize}

Ref. \cite{Han:2015qqj} explores an extension of the type-III THDM
with new scalar $SU(2)_L$ representations, namely an inert complex Higgs
triplet and a real scalar septuplet. A more general analysis can be
found in Ref.~\cite{Moretti:2015pbj}, where a generic real scalar $SU(2)_L$
multiplet is considered. In both papers, the $750$ GeV resonance is
identified with the usual heavy Higgs of the THDM, with the additional
scalars being introduced to push the diphoton rate to higher values.

For the realization of this idea here we focus on the septuplet case,
closely following Ref.~\cite{Han:2015qqj}. The introduction of such large
scalar multiplets has become fairly popular recently in the context of
minimal DM scenarios \cite{Cirelli:2005uq}, although
\cite{Han:2015qqj} does not explore any dark matter implications. It
only takes advantage of the multicharged states contained in the
septuplet which, due to their couplings to the THDM doublets, lead to
a large diphoton rate for the heavy Higgs $H$.

The type-III THDM is extended with the addition of a real scalar which
transforms as a $\seven$ under $SU(2)_L$, see \cref{tab:THDMSsept}.

\begin{table}[h]
\centering
\begin{tabular}{ c c c c c}
\toprule
Field & Gen. & $SU(3)_C$ &  $SU(2)_L$ & $U(1)_Y$   \\
\midrule 
$T$ & 1 & \one & $\mathbf{7}$ & $0$  \\
\bottomrule
\end{tabular}
\caption{Additional fermion field content for the scalar septuplet extended THDM.}
\label{tab:THDMSsept}
\end{table}

It proves convenient to use tensor notation for the
septuplet. This is the usual choice, see
Refs.~\cite{Hisano:2013sn,Alvarado:2014jva}, and the one implemented
in \SARAH. The septuplet $T$ is represented by a symmetric tensor with
six indices, $T^{ijklmn}$, all of which can take values of either $1$ or $2$. The
relation with the vector notation, employed for example in
\cite{Han:2015qqj}, is given by
\begin{equation}
T \equiv \left(
\begin{array}{c}
T^{+++}\\
T^{++}\\
T^{+}\\
T^0\\
T^{-}\\
T^{--}\\
T^{---}
\end{array}
\right) = i\left(
\begin{array}{c}
+T^{111111}\\
+\sqrt{6} \, T^{111112}\\
+\sqrt{15} \, T^{111122}\\
-\sqrt{20} \, T^{111222}\\
-\sqrt{15} \, T^{112222}\\
+\sqrt{6} \, T^{122222}\\
-T^{222222}
\end{array}
\right) \, .
\end{equation}
The prefactor $i$ and the sign for each component are introduced in
order to satisfy $T^c = T$ where $T^c$ denotes charge conjugation of
the field $T$, and $T^0$ is a real scalar. The new potential terms
involving the septuplet $T$ are
\begin{align}
V_T &= M_T^2 \, T^2 + \sum_{i=1}^2 \lambda_i^T \, \left[ T^4 \right]_i + \lambda_3^T \, |H_1|^2 \, T^2 + \lambda_4^T \, |H_2|^2 \, T^2 \, .
\label{eq:VT}
\end{align}
We note that two independent gauge invariant contractions are possible
in case of $T^4$, which is reflected by the second term of
\cref{eq:VT}. Finally, the authors of \cite{Han:2015qqj} assume a
minimum of the scalar potential with $\langle T \rangle = 0$. In this
case, the components of the septuplet do not mix with the scalar
doublets, but only participate in the $H \to \gamma \gamma$ rate due
to the interaction terms $\lambda_{3,4}^T$.

%% file: tex/ModelCategories/U1models.tex
In this section we consider a class of models which extend the SM by a
new $U(1)_X$ gauge group.  One typically introduces, beyond the SM
Higgs doublet, new scalars charged under $U(1)_X$, which serve two
purposes: (i) a linear combination of them results in
the \SI{750}{\GeV} particle, and (ii) via spontaneous symmetry
breaking they give mass to the new gauge boson, the $Z'$ boson.
Typically, one also introduces new fermions charged under the $U(1)_X$
symmetry that can either be singlets under the SM gauge group, hence
forming a dark or hidden sector, or vector-like under the SM.  An
advantage of these models is that, through a suitable choice of charge
assignments under $U(1)_X$, one can avoid flavour constraints present
when allowing the additional quarks to decay. Finally, the presence of
a massive $Z'$ boson can lead to new collider signals, well studied in
the literature, which can serve as a smoking gun for these types of
models. We note that the mixing matrix for the neutral gauge bosons
can be parametrised by two angles, $\Theta$ and $\Theta'$, with
$\Theta'$ highly constrained by LEP data.

We distinguish two cases in the following: models in which the SM
fermions are charged under the new Abelian group, and models in which
they are singlet.

\subsubsection{Models with SM states uncharged under the new $U(1)$}
\modelparalabel{Dark $U(1)'$ extension}{subsubsec:darkU1}
\begin{itemize}
\item {\bf Reference:} \cite{Ko:2016wce}
\item {\bf Model name:} \texttt{U1Extensions/darkU1}
\end{itemize}
This model is based on a gauged $U(1)_{X}$ extension of the SM with a dark sector that includes
three generations of dark $SU(2)_L$-singlet fermions 
and a DM scalar candidate. 

The properties of the new particles introduced in this model are described in Table~\ref{darkU1:tab:1}. 
We have added three right-handed neutrinos $\nu^i_R$, neutral under $U(1)_X$, to the original model, since $N_R$ (their dark partners) were considered here. $\Phi$ is the scalar field responsible for the spontaneous symmetry breaking of $U(1)_X$, while $X$ is the DM candidate. The $U(1)_X$ charges $a,b$ are left arbitrary, with their assignment chosen such that they fulfil the anomaly cancellation conditions.
\begin{table}[h]
\centering
\begin{tabular}{ c c c c c c}
\toprule
Field & Gen. & $SU(3)_C$ &  $SU(2)_L$ & $U(1)_Y$ & $U(1)_X$ \\
\midrule 
$\Phi$ & 1 & \one & \one & $0$ & $a+b$ \\
$X$ & 1 & \one & \one & $0$ & $a$ \\
\midrule
$E_L$ & 3 & \one & \one & $-1$ & $a$ \\
$E_R$ & 3 & \one & \one & $-1$ & $-b$ \\
$N_L$ & 3 & \one & \one & $0$ & $-a$ \\
$N_R$ & 3 & \one & \one & $0$ & $b$ \\
$U_L$ & 3 & \three & \one & $\frac23$ & $-a$ \\
$U_R$ & 3 & \three & \one & $\frac23$ & $b$ \\
$D_L$ & 3 & \three & \one & $-\frac13$ & $a$ \\
$D_R$ & 3 & \three & \one & $-\frac13$ & $-b$ \\
\bottomrule
\end{tabular}
\caption{New fermions and scalar fields of the {\tt darkU1} and their charge assignments under the gauge 
symmetry $SU(3)_C \times SU(2)_L\times U(1)_Y\times U(1)_X$. The scalar and fermion fields are shown in the top and bottom of the table respectively.}
\label{darkU1:tab:1}
\end{table}

The Yukawa interactions and the scalar potential including new fields in the dark sector are described by 
\begin{align}
-\mathcal L_Y^{\rm new} &= Y'_E\, \overline{E_R} \,\Phi^*\, E_L  + Y'_N \,\overline{N_R}\, \Phi\, N_L 
+ Y'_U \,\overline{U_R}\, \Phi\, U_L + Y'_D \,\overline{D_R} \,\Phi^*\, D_L 
+ Y_{XE} \, \overline{e_R} X^*\, E_L\,  \notag\\ &+ Y_{XU} \, \overline{u_R} \,X \,U_L
+ Y_{XD} \,\overline{d_R} \,X^* \,D_L + Y_{XN}\, \overline{\nu_R} \,X \,N_L   + {\rm h.c.}\,, \\
V &= \mu^2 |H|^2 + \lambda |H|^4 + \mu_\Phi^2 |\Phi|^2 
+ \mu_X^2 |X|^2 + \lambda_\Phi |\Phi|^4 + \lambda_X |X|^4 +\lambda_{H\Phi} |H|^2|\Phi|^2 \nonumber \\  
&+ \lambda_{HX} |H|^2|X|^2+ \lambda_{X\Phi} |X|^2|\Phi|^2 \,,
\end{align}
where $H$ denotes the SM Higgs doublet. For $a=b=1$, an extra term $\Phi^\dagger X^2$ would be allowed in the potential, which breaks $U(1)_X$ down to a $Z_2$ subgroup after $\Phi$ develops a non-zero VEV.  
Likewise, for $3a = (a+b)$, there appears an extra term $\Phi^\dagger X^3$, which breaks $U(1)_X$
down to a $Z_3$ subgroup after $\Phi$ develops a nonzero VEV.  These possibilities are not considered here.
 
The gauge symmetry is broken after $H$ and $\Phi$ get non-zero VEVs, $v$ and $v_\Phi$ respectively, while $X$ does not receive a VEV\footnote{We also checked that this condition remains valid at NLO.}. The scalar fields after EWSB can be expressed as
\begin{align}\label{eq:scalarfieldexp}
H &= \frac{1}{\sqrt{2}} \begin{pmatrix}
0 \\
v + \phi_H + i \sigma_H
\end{pmatrix}\,, \qquad
\Phi = \frac{1}{\sqrt{2}} \left(v_\Phi+ \phi_\Phi + i \sigma_\Phi \right)\,,
\end{align}
where $\phi_i$ and $\sigma_i$ are the CP-even and odd components respectively. The \SI{750}{\GeV} candidate is the mixture of the SM Higgs $\phi_H$ and $\phi_\Phi$, which leads to constraints on $\lambda_{H\Phi}$. It is produced in gluon-gluon fusion via loops of the dark quarks, and it decays into diphotons via loops comprised only of charged dark fermions, assuming that the mixing with the SM-like Higgs is negligibly small.

\modelparalabel{Hidden $U(1)$}{sec:hiddenU1}

\begin{itemize}
\item {\bf Reference:} \cite{Das:2015enc}
\item {\bf Model name:} \texttt{U1Extensions/hiddenU1}
\end{itemize}

This is a particularly simple realisation of a gauged $U(1)_X$ extension of the SM.
As previously, a hidden $U(1)_X$ is added to the SM gauge group, under which all SM particles are singlets. A scalar $S_1$ is added for its spontaneous symmetry breaking, and a further scalar $S_2$ is added which is a singlet under the entire gauge symmetry of the model. Here we assume that both $S_1$ and $S_2$ can develop a VEV, in principle.

The \SI{750}{\GeV} candidate is considered to be predominantly composed by $S_2$. To explain the diphoton signal, a vector-like quark is also included, carrying the same charge under $U(1)_X$ as the $S_1$ field. Hence, it couples only to $S_2$ due to the choice of charge assignments.

\begin{table}[h]
\centering
\begin{tabular}{ c c c c c c}
\toprule
Field & Gen. & $SU(3)_C$ &  $SU(2)_L$ & $U(1)_Y$ & $U(1)_X$ \\
\midrule 
$S_1$ & 1 & \one & \one & $0$ & $a$ \\
$S_2$ & 1 & \one & \one & $0$ & $0$ \\
\midrule
$X_L$ & 1 & \three & \one & $Y_X$ & $a$ \\
$X_R$ & 1 & \three & \one & $Y_X$ & $a$ \\
\bottomrule
\end{tabular}
\caption{New fermions and scalar fields of the {\tt hiddenU1} and their charge assignments under the gauge 
symmetry $SU(3)_C \times SU(2)_L\times U(1)_Y\times U(1)_X$. The scalar and fermion fields are shown in the top and bottom of the table respectively.}
\label{hiddenU1:tab:1}
\end{table}

In Table~\ref{hiddenU1:tab:1} the hypercharge of the new vector-like quark $Y_X$ is left arbitrary. The implemented case is the most favourable one in terms of the diphoton excess, $Y_X = 2/3$, which allows a mixing with the up-type quarks. We did not considered the case of adding also a vector-like lepton to the spectrum, which would lead to even larger rates. The $U(1)_X$ charge of the vector-like quark fields $a$ does not affect the diphoton rate, but it has to be the same as for $S_1$, which is relevant for the $Z'$ boson mass. 

The scalar potential, with the usual doublet Higgs field $H$ is given by
\begin{align}
V &= \mu_H^2 |H|^2 + \mu_{S_1}^2 |S_1|^2  + \mu_{S_2}^2 S_2^2  
- \lambda_H |H|^4  -  \lambda_{HS_1} |H|^2|S_1|^2 - \lambda_{S_1} |S_1|^4
\nonumber \\
&-  \lambda_{S_2} S_2^4  -  \lambda_{HS_2} |H|^2 S_2^2 
- \lambda_{S_1 S_2} |S_1|^2 S_2^2 
- \sigma_{1} S_2^3 -\sigma_{2} |H|^2 S_2 -\sigma_{3}|S_1|^2S_2 \,.
\label{eq:hiddenU1:Lpotential}
\end{align}
This potential leads to a mixing between the three physical neutral scalars. The structure of the new scalar fields is given by
\begin{align}
S_1 &= \frac{1}{\sqrt{2}} \left(v_{S_1} + \phi_{S_1} + i \sigma_{S_1} \right)\,,\qquad S_2 = v_{S_2} + \phi_{S_2}\,,
\end{align}
where once again $\phi_i$ and $\sigma_i$ are the CP-even and odd components respectively. As discussed in \cref{sect:motivation:vev} we have allowed all scalar fields to obtain VEVs. 

The Yukawa interactions and fermionic mass terms of the hidden sector read
\begin{align}
-\mathcal{L}^{\rm new} &= M_X \, \overline{X_R} \, X_L + Y_{X_L} \, \overline{u_R} \,S_1^* \, X_L  +  f_{X} \,\overline{X_R} \,S_2  X_L \,\,+{\rm h.c.}\, .
\end{align}
The mixing of the vector-like quark $X$ with SM quarks via the interaction with $S_1$ is kept small, and purely serves the purpose of letting the new quark decay. 

\modelparalabel{Simple $U(1)$}{sec:simpleU1}

\begin{itemize}
\item {\bf Reference:} \cite{Chang:2015bzc}
\item {\bf Model name:} \texttt{U1Extensions/simpleU1}
\end{itemize}

This $U(1)_X$ extension of the SM augments its particle content by a
pair of exotic vector-like quarks, $\chi_1$ and $\chi_2$, doublets
under $SU(2)_L$ and with hypercharge $7/6$, and by one complex scalar
$\Sigma$ responsible for the spontaneous breaking of $U(1)_X$. The
$U(1)_X$-breaking Higgs boson is the \SI{750}{\GeV} candidate. The
particle content beyond the SM is summarized in
Table~\ref{tab:simpleU1:content}.
\begin{table}[h]
\centering
\begin{tabular}{ c c c c c c c }
\toprule
\text{Field} & Gen. & $SU(3)_C$ & $SU(2)_L$ & $U(1)_Y$ & $U(1)_X$ \\
\midrule
$\Sigma$ & 1 & $\bf{1}$ & $\bf{1}$ & $0$ & $1$ \\
\midrule
$\chi_{1}$ & 1 & $\bf{3}$ & $\bf{2}$ & $7/6$ & $1$ \\ 
$\chi_{2}$ & 1 & $\bf{3}$ & $\bf{2}$ & $7/6$ & $2$ \\ 
\bottomrule
\end{tabular}
\caption{Particle content of the {\tt simpleU1} beyond the SM fields. The scalar and fermion fields are shown in the top and bottom of the table respectively. The exotic $\chi_1$ and $\chi_2$ quarks are vector-like.}
\label{tab:simpleU1:content}
\end{table}

The scalar potential of the model is given by
\begin{align}
V = -\mu_H^2 |H|^2 - \mu_{\Sigma}^2 |\Sigma|^2 +\lambda_H |H|^2 + \lambda_{H\Sigma} |H|^2|\Sigma|^2 + \lambda_\Sigma |\Sigma|^4\,,
\label{eq:simpleU1:potential}
\end{align}
while the Yukawa and fermionic mass terms read
\begin{equation}
\mathcal{L} = \mathcal{L}^{\rm SM}_Y - M_1 \, \overline{\chi_{1R}} \,\chi_{1L}  - M_2\, \overline{\chi_{2R}}\, \chi_{2L} - \lambda_1 \, \Sigma\,\overline{\chi_{2R}}\, \chi_{1L} - \lambda_2\, \Sigma^* \overline{\chi_{1R}} \chi_{2L}\,\,.
\end{equation}
Note that the original model proposed in \cite{Chang:2015bzc} contains an effective operator
\begin{align*}
	\mathcal{L} \supset -\frac{1}{\Lambda}\Sigma^* H \chi_1 u^c \,.
\end{align*}
As \SARAH cannot handle effective operators this term is dropped from the model and subsequent constraints pertaining to stable charged particles are ignored. Expanding the new scalar field yields
\begin{align}
\Sigma &= \frac{1}{\sqrt{2}} \left(v_\Sigma + \phi_\Sigma + i \sigma_\Sigma\right)\,,
\end{align}
where once again $\phi_i$ and $\sigma_i$ are the CP-even and odd components respectively.

\modelparalabel{Scotogenic $U(1)$ Model}{sec:scotoU1}

\begin{itemize}
\item {\bf Reference:} \cite{Yu:2016lof}
\item {\bf Model name:} \texttt{U1Extensions/scotoU1}
\end{itemize}
The matter particle content of the Scotogenic $U(1)$ Model is summarized in \cref{tab:scotoU1:content}, where, in addition to the fields charged under $U(1)_D$, we introduce three copies of right-handed neutrinos $\nu_R$ which 
are singlets under the full gauge group. 
\begin{table}[h]
\centering
\begin{tabular}{ c c c c c c c }
\toprule
\text{Field} & Gen. & $SU(3)_C$ & $SU(2)_L$ & $U(1)_Y$ & $U(1)_D$ & $\mathbb{Z}_2$ \\
\midrule
$\Phi$ & 1 & $\bf{1}$ & $\bf{1}$ & $0$ & $2$ & $+$ \\
$H'$ & 1 & $\bf{1}$ & $\bf{2}$ & $\frac{1}{2}$ & $-1$ & $-$ \\
\midrule
$\nu_R$ & 3 & $\bf{1}$ & $\bf{1}$ & 0 & 0 & $+$ \\
$T$ & 1 & $\bf{3}$ & $\bf{1}$ & $\frac{2}{3}$ & $-1$ & $-$ \\
$T'$ & 1 & $\bf{3}$ & $\bf{1}$ & $\frac{2}{3}$ & $1$ & $-$ \\
$N$ & 1 & $\bf{1}$ & $\bf{1}$ & $0$ & $-1$ & $-$ \\
\bottomrule
\end{tabular}
\caption{Matter particle content of the {\tt scotoU1} beyond the SM fields. The scalar and fermion fields are shown in the top and bottom of the table, respectively. The fermions $T,T',N$ are vector-like degrees of freedom (4-component spinors).}
\label{tab:scotoU1:content}
\end{table}
%
Note, that the $U(1)_D$ charge of the $H'$ field has been changed to
$-1$ compared to Ref.~\cite{Yu:2016lof} in order to make the Yukawa
interaction terms gauge invariant. The SM fields are not charged under the new 
$U(1)$ gauge group. The discrete $\mathbb{Z}_2$ symmetry is introduced to stabilize the dark matter candidates $N$.

The scalar potential reads
\begin{align}
    V &=
    -\mu_1^2 |H|^2
    + \mu_2^2 |H'|^2
    - \mu_s^2 |\Phi|^2
    + \lambda_1 |H|^4
    + \lambda_2 |H'|^4
    + \lambda_s |\Phi|^4 \\
    & + \lambda_h |H|^2 |\Phi|^2
    + \lambda_{h'} |H'|^2 |\Phi|^2
    + \lambda_3 |H|^2 |H'|^2
    + \lambda_4 |H^\dagger H'|^2\,, \nonumber
\end{align}    
The term
$\frac{\lambda_5}{2}[(H^\dagger H')^2 + \text{h.c.}]$ proposed in
Ref.~\cite{Yu:2016lof} has been omitted here, because it is not
invariant under $U(1)_D$ gauge transformations.
    
The Yukawa interactions beyond the  SM read
\begin{align}
   \mathcal L &\supset
    + Y_\nu \nu_R H \ell
    + Y_T  \overline{ T_R} H' q
    + y_N \overline{ N_R} H' L
    + m_T (\overline{T} T + \overline{T'} T') \nonumber \\
    &+ M_D \overline{N} N
    + \eta_1 \overline{T_R'} \Phi T_L
    + \eta_2 \overline{T_R} \Phi^* T_L'
    + \eta_3 \overline{N_R^c} \Phi N_L + \text{h.c.}\,,
\end{align}
where $H=(H^+,H^0)$ is the SM Higgs field and $q$ and $\ell$ the SM left-handed quark and lepton doublets.

The $U(1)_D$ symmetry is eventually broken by the VEV of the scalar field $\Phi$ which one can decompose as
\begin{align}
  \langle \Phi \rangle &= \frac{1}{\sqrt{2}} (v_S + S + i A_S)\, , 
\end{align}
whereas the CP-even component $S$ is considered by the authors as the candidate for the 750~GeV resonance.
The spontaneous symmetry breaking leaves the $\mathbb{Z}_2$ parity intact, and $H'$ will therefore not develop any VEV.

\subsubsection{Models with SM states charged under the new $U(1)$}
\modelparalabel{$U(1)_{B-L}$ model with unconventional $B-L$ charges }{sec:BLVL}
\begin{itemize}

\item {\bf Reference:} \cite{Modak:2016ung}
\item {\bf Model name:} \texttt{U1Extensions/BL-VL}
\end{itemize}

This model is based on Refs. \cite{Ma:2014qra,Ma:2015mjd}. It considers a gauged $U(1)_{B-L}$ extension of the SM with an unconventional $B-L$ charge
assignment for the right-handed neutrinos $\nu_R$, and further requires 3 extra Dirac neutrinos $N$.
It was originally proposed to explain the smallness of neutrino masses if neutrinos were Dirac particles. 
Finally, it also contains a scalar DM candidate, stabilised by the residual \zThree\, symmetry from the breaking of the $B-L$ symmetry.
\begin{table}[h]
\centering
\begin{tabular}{ c c c c c c}
\toprule
Field & Gen. & $SU(3)_C$ &  $SU(2)_L$ & $U(1)_Y$ & $U(1)_{B-L}$ \\
\midrule 
$\Phi$ & 1 & \one & \two & $\frac12$ & $0$ \\
$\chi_2$ & 1 & \one & \one & $0$ & $2$ \\
$\chi_3$ & 1 & \one & \one & $0$ & $3$ \\
$\chi_6$ & 1 & \one & \one & $0$ & $-6$ \\ \midrule
$Q_L$ & 3 & \three & \two & $\frac16$ & $\frac13$ \\ 
$u_R$ & 3 & \three & \one & $\frac23$ & $\frac13$ \\
$d_R$ & 3 & \three & \one & $-\frac13$ & $\frac13$ \\
$L_L$ & 3 & \one & \two & $-\frac12$ & $-1$ \\
$l_R$ & 3 & \one & \one & $-1$ & $-1$ \\
$\nu_R^1$ & 1 & \one & \one & $0$ & $5$ \\
$\nu_R^2$ & 1 & \one & \one & $0$ & $-4$ \\
$\nu_R^3$ & 1 & \one & \one & $0$ & $-4$ \\
$N_L$ & 3 & \one & \one & $0$ & $-1$ \\
$N_R$ & 3 & \one & \one & $0$ & $-1$ \\
$X_L$ & 1 & \three & \one & $\frac23$ & $3$ \\
$X_R$ & 1 & \three & \one & $\frac23$ & $0$ \\
$Y_L$ & 1 & \three & \one & $-\frac23$ & $-3$ \\
$Y_R$ & 1 & \three & \one & $-\frac23$ & $0$ \\
\bottomrule
\end{tabular}
\caption{Particle content and charge assignments of the {\tt BL-VL} model. The scalar and fermion fields are shown in the top and bottom of the table respectively.}
  \label{BLVL:tab1}
\end{table}
To fit the diphoton excess, new vector-like quarks are added so that the \SI{750}{\GeV} scalar, a linear combination of the SM Higgs and the $\chi$ fields, can be produced in gluon-gluon fusion. The particle content and the quantum numbers for this model are summarised in Table \ref{BLVL:tab1}. In addition to the SM particle content, the model features three right-handed neutrinos $\nu^i_R$, three pairs of $SU(2)_L$ singlet heavy fermions $N^i_{L,R}$, and two pairs of exotic quarks $X_{L,R}$, $Y_{L,R}$ which carry color and electromagnetic charges but are singlets under $SU(2)_L$. 
 
The scalar potential of the model reads
\begin{align}
V & =  - \mu^2_0 |\Phi|^2 + m_2^2 |\chi_2|^2 - \mu_3^2 |\chi_3|^2 - \mu_6^2 |\chi_6|^2 + \dfrac{1}{2} \lambda_0|\Phi|^4 
+  \dfrac{1}{2} \lambda_2 |\chi_2|^4  +  \dfrac{1}{2} \lambda_3|\chi_3|^4  +  \dfrac{1}{2} \lambda_6 |\chi_6|^4  \nonumber \\
& +  \lambda_{02}|\chi_2|^2|\Phi|^2 + \lambda_{03} |\chi_3|^2|\Phi|^2 + \lambda_{06} |\chi_6|^2|\Phi|^2 
 + \lambda_{23}|\chi_2|^2|\chi_3|^2 + \lambda_{26} |\chi_2|^2|\chi_6|^2\nonumber \\
& + 	 \lambda_{36} |\chi_3|^2|\chi_6|^2 + \left[\dfrac{1}{2} f_{36} (\chi_3^{2}\chi_6) + \dfrac{1}{6} \lambda^{\prime}_{26} (\chi_2^{3}\chi_6)+{\rm h.c.}\right] \,.
\label{BLVL:scapot}
\end{align}
The residual global \zThree\, symmetry protects the singlet scalar $\chi_2$ from acquiring a VEV, i.e. $\langle \chi_2 \rangle = 0$. All leptons carry a charge $\omega = e^{2i\pi/3}$ under \zThree . The CP-even degree of freedom of $\chi_2$ is the DM candidate.
The Yukawa interactions of the new sector read
\begin{align}
-\mathcal{L}_Y^{\rm new} &= Y_{NR}\, \overline{N_R} \,\widetilde{H} \,L_L 
    + f_X\, \overline{X_R} \,\chi_3^*\, X_L 
    + f_Y\, \overline{Y_R} \,\chi_3 \,Y_L \nonumber \\
    &+ f_N\, \overline{\nu_{R_{2,3}}} \, \chi_3^* \, N_L 
    + f_{N6}\, \overline{\nu_{R_{1}}} \,\chi_6^*\, N_L \notag\\
	&+ f_{NL}\, \chi_2 \,\overline{N_L^c} \,N_L 
	+ f_{NR}\, \chi_2 \,\overline{N_R^c}\, N_R
	+ m_N\, \overline{N_R} \,N_L  
	+ {\rm h.c.} \,.
\end{align}
In the set up considered in Ref~\cite{Modak:2016ung}, the \SI{750}{\GeV} scalar is given by the combination
($\chi_6 - \chi_3$), that couples to gluons and photons
via loops of $X$ and $Y$ fermions proportional to the Yukawa couplings $f_X$ and $f_Y$. The fit to the diphoton signal was studied in a simplified scenario, where special relations are imposed to the scalar parameters. As already described in section~\ref{sect:motivation:mixing}, these relations are not protected by symmetries, and hence lead to a large amounts of fine tuning.

\modelparalabel{Sample of $U(1)'$ models based on different charge assignments}{sec:U1sample}

\begin{itemize}
\item {\bf Reference:} \cite{Chao:2015nsm}
\item {\bf Model name:} \texttt{U1Extensions/VLsample}
\end{itemize}

The particle content of this model is shown in Table~\ref{tab:U1sample:content}. 
Similarly to all the previous models, a complex scalar $S$ is added to break the $U(1)_X$ symmetry.
In this paper, the mixing between the scalars is kept small, hence the \SI{750}{\GeV} candidate is predominantly the CP-even part of the $U(1)_X$-breaking field. New doublet and singlet vector-like quarks, charged under $U(1)_X$, are added to fit the diphoton signal strength. In this model,  the SM fermions also carry $X$-charges, which are fixed according to the anomaly cancellation conditions. Only the Higgs doublet is not charged, while the charge of the new scalar field $S$ is double (in absolute magnitude) that of the vector-like quarks.
\begin{table}[h]
\centering
\begin{tabular}{ c c c c c c c }
\toprule
\text{Field} & Gen. & $SU(3)_C$ & $SU(2)_L$ & $U(1)_Y$ & $U(1)_X$ \\
\midrule
$H$ & 1 & \one & \two & $-\frac12$ & $0$ \\ 
$S$ & 1 & \one & \one & $0$ & $-2b$ \\
\midrule
$Q_L$ & 3 & \three & \two & $\frac16$ & $m$ \\ 
$u_R$ & 3 & \three & \one & $\frac23$ & $m$ \\
$d_R$ & 3 & \three & \one & $-\frac13$ & $m$ \\
$L_L$ & 3 & \one & \two & $-\frac12$ & $k$ \\
$e_R$ & 3 & \one & \one & $-1$ & $k$  \\
$\nu_R$ & 3 & \one & \one & $0$ & $k$ \\
$X_L$ & 1 & \three & \two & $a$ & $b$ \\
$X_R$ & 1 & \three & \two & $a$ & $-b$ \\
$y_{1L}$ & 1 &\three & \one & $a+\frac12$ & $-b$ \\
$y_{2L}$ & 1 &\three & \one & $a-\frac12$ & $-b$ \\
$y_{1R}$ & 1 &\three & \one & $a+\frac12$ & $b$ \\
$y_{2R}$ & 1 &\three & \one & $a-\frac12$ & $b$ \\
\bottomrule
\end{tabular}
\caption{Particle content of the {\tt VLsample}. The scalar and fermion fields are shown in the top and bottom of the table respectively.}
\label{tab:U1sample:content}
\end{table}

There are several possibilities to assign the $U(1)_X$ charges in an anomaly-free way, with 
different physical interpretations \cite{Chao:2015nsm}:
\begin{itemize}
\item $X \equiv B-L$: $b=0, ~k=-1, ~m=1/3$
\item $X \equiv  B+L$: $b=-1,~ k=1, ~m=1/3$
\item $X \equiv B$: $b=-1/2, ~k=0, ~m=1/3$
\item $X \equiv L$: $b=-1/2, ~k=1, ~m=0$
\end{itemize}
Note that in all interpretations $a$ remains a free parameter.

The Yukawa interactions of the extra particle content are given by
\begin{align} \label{eq:U1sample:lagY}
\mathcal{L}^{\rm new}_Y &= Y_V^1 \overline{y_{1R}} S^* y_{1L} +  Y_V^2  \overline{y_{2R}} S^* y_{2L} + Y_V^3 \overline{X_R} S X_L \\ \notag &+ \eta_1 \overline{X_R} H y_{1L}  + \eta_2 \overline{X_R} \tilde H y_{2L} + \eta_3 \overline{y_{2R}} H X_{L}  + \eta_4 \overline{y_{1R}} \tilde H X_{L}\, + \hc\,
\end{align}
Finally, the scalar potential is given by
\begin{align}
\label{eq:U1sample:V}
-V  &=  \mu_H^2 |H|^2 + \mu_S^2 |S|^2 - \lambda_H |H|^4 - \lambda_S |S|^4 - \lambda_{HS} |S|^2 |H|^2 \, ,
\end{align}
the symmetry breaking pattern being
\begin{equation} \label{eq:U1sample:VEVs}
\langle H \rangle = \frac{1}{\sqrt{2}} \, \left( \begin{array}{c}
v \\
0  \end{array} \right) \,, \quad 
\langle S \rangle = \frac{1}{\sqrt{2}} \, v_S\,\,,
\end{equation}
while the expansion of the scalar fields in to CP-even and odd components is analogous to \cref{eq:scalarfieldexp}.

\modelparalabel{Model with flavour-nonuniversal quark $U(1)'$ charges}{sec:nonUniU1}

\begin{itemize}
\item {\bf Reference:} \cite{Martinez:2015kmn}
\item {\bf Model name:} \texttt{U1Extensions/nonUniversalU1}

\end{itemize}

In this model the first generation of left-handed SM quarks carries a $U(1)_X$ charge while the second and third generations do not. In this way it is possible to add exotic quarks which are vector-like under the SM gauge group and achieve anomaly cancellation with less than three generations. The scalar sector then needs to be extended with a second Higgs doublet, with a different $U(1)_X$ charge compared to the first ones, in order to have Yukawa interactions for all quark families. A further complex scalar field $S$, which is a singlet under the SM gauge group but charged under $U(1)_X$, is also added for the spontaneous symmetry breaking of the $U(1)_X$ symmetry.

The charge assignments that cancel all anomalies are given in \cite{Martinez:2013qya} and are
summarised in \cref{tab:nonUniU1:content}.\footnote{In an updated version of their paper, the authors of Ref.~\cite{Martinez:2015kmn} have added a further scalar $\sigma$ with the same quantum numbers as $S$ which shall be the dark matter candidate as well as extra fermionic singlets 
in order to allow for a seesaw mechanism in the neutrino sector.}
\begin{table}[h]
\centering
\begin{tabular}{ c c c c c c c }
\toprule
\text{Field} & Gen. & $SU(3)_C$ & $SU(2)_L$ & $U(1)_Y$ & $U(1)_X$ \\
\midrule
$H_1$ & 1 & \one & \two & $-\frac12$ & $-\frac23$ \\ 
$H_2$ & 1 & \one & \two & $-\frac12$ & $-\frac13$ \\ 
$S$ & 1 & \one & \one & $0$ & $-\frac13$ \\
\midrule
$Q_L^1$ & 1 & \three & \two & $\frac16$ & $\frac13$ \\ 
$Q_L^i$ & 2 & \three & \two & $\frac16$ & $0$ \\ 
$u_R$ & 3 & \three & \one & $\frac23$ & $\frac23$ \\
$d_R$ & 3 & \three & \one & $-\frac13$ & $-\frac13$ \\
$L_L$ & 3 & \one & \two & $-\frac12$ & $-\frac13$ \\
$e_R$ & 3 & \one & \one & $-1$ & $-1$  \\
$\nu_R$ & 3 & \one & \one & $0$ & $\frac13$ \\
$T_L$ & 1 & \three & \one & $\frac23$ & $\frac13$ \\
$T_R$ & 1 & \three & \one & $\frac23$ & $\frac23$ \\
$J_L$ & 2 &\three & \one & $-\frac13$ & $0$ \\
$J_R$ & 2 &\three & \one & $-\frac13$ & $-\frac13$ \\
\bottomrule
\end{tabular}
\caption{Particle content of the {\tt nonUniversalU1}. The scalar and fermion fields are shown in the top and bottom of the table respectively.}
\label{tab:nonUniU1:content}
\end{table}

The Yukawa Lagrangian of the model (ignoring here flavour indices) is given by 
\begin{align} \label{eq:nonUniU1:lagY}
\mathcal{L}_Y &= h_{1}^D \overline{d_R} H_1 Q_L^1+ h_1^U \overline{u_R} \tilde H_1 Q_L^i  +  h_2^D  \overline{d_R} H_2 Q^i_L 
              +  h_2^U \overline{u_R} \tilde H_2 Q^1_L 
             + h_1^J  \overline{J_R} H_1 Q_L^1 \notag\\  
			 &+ h_2^J \overline{J_R} H_2 Q_L^i + 
             h_2^T \overline{T_R} \tilde H_2 Q_L^1 + h_1^T \overline{T_R} \tilde H_1 Q_L^i 
          + h_X^U \overline{u_R} S^* T_L+ h_X^T \overline{T_R} S^* T_L  \notag\\ 
             &+ h_X^D \overline{d_R} S J_L + h_X^J \overline{J_R} S J_L 
           + Y_e \overline{e_R} H_1 L_L + Y_v \overline{\nu_R} \tilde H_1 L_L  + {\rm h.c.}\,,
\end{align}
and the scalar potential is
\begin{align}
\label{eq:nonUniU1:V}
V  &=  \mu_{11}^2 |H_1|^2 + \mu_{22}^2 |H_2|^2 + \mu_S^2 |S|^2 
      + \lambda_1 |H_1|^4+ \lambda_2 |H_2|^4 
     + \lambda_3 |H_1|^2 |H_2|^2 - \lambda_{H_1 S} |S|^2 |H_1|^2  \notag\\
	 &- \lambda_S |S|^4+ \lambda_4 (H_2^\dagger H_1) (H_1^\dagger H_2)  
      - \lambda_{H_2 S} |S|^2 |H_2|^2+ \left\{ \kappa_{HS} H_1^\dagger H_2 S+ \hc\right\}\, .
\end{align}
The pattern of EWSB follows
\begin{align}
	H_i &= \frac{1}{\sqrt{2}}\begin{pmatrix} 0 \\ v_i + \phi_i + i \sigma_i \end{pmatrix}\,, \qquad S = \frac{1}{\sqrt{2}}\left(v_S + \phi_S + i \sigma_S\right)\,,
\end{align}
where once again $\phi_i$ and $\sigma_i$ are the CP-even and CP-odd
components respectively. In order to obtain a massive pseudo-scalar
state, $\kappa_{HS}$ needs to be non-zero. Hence, for keeping the
$S-H_i$ mixing small while making the pseudo-scalar massive, either
the condition $v_S \gg v$ must hold, or $\kappa_{HS}$ must be small,
in conjunction with $\frac{v_1}{v_2} \to 0$ or $\infty$.  $\phi_S$,
the CP-even component of $S$, is then identified with
the \SI{750}{\GeV} candidate.  

\modelparalabel{Leptophobic $U(1)$ model}{sec:U1leptophobic}

\begin{itemize}
\item {\bf Reference:} \cite{Ko:2016lai}
\item {\bf Model name:} \texttt{U1Extensions/U1Leptophobic}
\end{itemize}

This model is inspired by an $E_6$ Grand Unified Theory (GUT), but the authors only consider the low energy version where the SM gauge group is augmented by an extra gauged $U(1)_X$ symmetry. This extra $U(1)$ symmetry has zero charges for both left- and right-handed leptons making it entirely lepotophobic. However it is impossible to arrange for this to happen by taking linear combinations of the $U(1)_\chi$ and $U(1)_\psi$ groups that appear from the breakdown of $E_6$. Instead, the charges from this extra $U(1)$ can only be obtained by including gauge kinetic mixing, so that the introduced mixture of $U(1)_Y$ charges exactly cancel the non-zero leptonic charges.  This can be done with the $U(1)_\eta$ gauge symmetry and the charges used in this model correspond exactly to those given in Table I of Ref.~\cite{Buckley:2011mm}. It is these charges rather than the charges of the $U(1)_\eta$ which are set in this model.  Of course one may discard the $E_6$ motivation and treat it as an ad-hoc choice of $U(1)$ charges.

The model contains two Higgs doublets, a complex scalar SM singlet ($\Phi$), charged under $U(1)_X$, plus right handed neutrinos and other new fermions, charged under both the SM and the $U(1)_X$ symmetries. The latter are odd under an imposed \zTwo symmetry, so that the lightest neutral fermion is a DM candidate.

\begin{table}[h]
\centering
\begin{tabular}{ c c c c c c c }
\toprule
\text{Field} & Gen. & $SU(3)_C$ & $SU(2)_L$ & $U(1)_Y$ & $U(1)_X$ & $\mathbb{Z}_2$ \\
\midrule
$H_1$ & 1 & \one & \two & $-\frac12$ & $0$ & $+$ \\
$H_2$ & 1 & \one & \two & $-\frac12$ & $-1$ & $+$ \\
$\Phi$ & 1 & \one & \one & $0$ & $-1$ & $+$ \\
\midrule
$Q_L$ & 3 & \three & \two & $\frac16$ & $-\frac13$ & $+$ \\ 
$u_R$ & 3 & \three & \one & $\frac23$ & $\frac23$ & $+$\\
$d_R$ & 3 & \three & \one & $-\frac13$ & $-\frac13$ & $+$\\
$L_L$ & 3 & \one & \two & $-\frac12$ & $0$& $+$ \\
$e_R$ & 3 & \one & \one & $-1$ & $0$ & $+$ \\
$\nu_R$ & 3 & \one & \one & $0$ & $1$ & $+$ \\
$D_L$ & 3 & \three & \one & $-\frac13$ & $\frac23$ & $-$\\
$D_R$ & 3 & \three & \one & $-\frac13$ & $-\frac13$ & $-$\\
$\tilde{H}_L$ & 3 & \one & \two & $-\frac12$ & $0$ & $-$ \\
$\tilde{H}_R$ & 3 & \one & \two & $-\frac12$ & $-1$ & $-$ \\
$N_L$ & 3 & \one & \one & $0$ & $-1$ & $-$ \\
\bottomrule
\end{tabular}
\caption{Particle content of the {\tt U1Leptophobic}. The scalar and fermion fields are shown in the top and bottom of the table respectively. }
\label{tab:U1leptophobic:content}
\end{table}
The particle content is summarized in \cref{tab:U1leptophobic:content}, while the scalar potential reads
\begin{align}
V_{\rm scalar}&= \tilde{m}_1^2 |H_1|^2
+\tilde{m}_2^2 |H_2|^2
+ \frac{\lambda_1}{2}  |H_1|^4
+ \frac{\lambda_2}{2}  |H_2|^4
+ \lambda_3 |H_1|^2 |H_2|^2
+ \lambda_4 H_1^\dagger H_2 H_2^\dagger H_1
\nonumber\\
&+ \tilde{m}_\Phi^2 |\Phi|^2
+\frac{\lambda_\Phi}{2}  |\Phi|^4
+ \left( \mu_\Phi H_1^\dagger H_2 \Phi + \textrm{h.c.} \right)
+ \tilde{\lambda}_1 |H_1|^2 |\Phi|^2
+ \tilde{\lambda}_2 |H_2|^2 |\Phi|^2\,,
\end{align}
and the Yukawa interactions are given by
\begin{eqnarray}
\mathcal{L}_Y &=& y^u\overline{u_R} H^{\dagger}_1 Q + y^d \overline{d_R} H_2 Q + y^e\overline{e_R}H_2 L + y^n \overline{n_R} H^{\dagger}_1  L\nonumber\\
&&
+ y^D \overline{D_R} \Phi  D_L  +y^H_{ij}   \overline{ \widetilde H_R^{ j}} \Phi  \widetilde H_L^i + y^N_{ij} \overline{N^{c\,i}_L}  H_1^\dagger \widetilde{H}^j_L  
+y'^N_{ij}   \overline{\widetilde{H}_R^{ i}}  H_2 N^j_L +\hc \,.
\end{eqnarray}

We assume the following symmetry breaking pattern
\begin{equation} \label{eq:U1leptophobic:VEVs}
 H_{1/2} = \frac{1}{\sqrt{2}} \, \left( \begin{array}{c}
v_{1/2} + \phi_{1/2} + i \sigma_{1/2} \\
0  \end{array} \right) \,, \qquad 
\Phi  = \frac{1}{\sqrt{2}}\left( v_\Phi+ \phi_\Phi + i \sigma_\Phi\right)\,,
\end{equation}
where once again $\phi_i$ and $\sigma_i$ are the CP-even and CP-odd
components, respectively, and we define
$\tan \beta \equiv \frac{v_2}{v_1}$.

The \SI{750}{\GeV} candidate is taken to be dominantly composed by $\phi_\Phi$, that is, the real CP-even degree of freedom of the $\Phi$ field after $U(1)_X$ symmetry breaking. It couples to photons and gluons via loops of the new fermions.
The model cannot explain the diphoton excess with Yukawa couplings in the perturbative range, but the authors use values between $5$ and $10$. As stressed in \cref{sec:motivation_perturbation}, this renders the perturbative calculation, and hence the results, very questionable.

\modelparalabel{$U(1)'$ extension with a $Z'$ mimicking a scalar resonance}{sec:trickingLY}

\begin{itemize}
\item {\bf Reference:} \cite{Chala:2015cev}
\item {\bf Model name:} {\tt U1Extensions/trickingLY}
\end{itemize}

The idea of this model is that the extra neutral gauge boson decays
into $S \gamma$, whereas the scalar $S$ itself decays into a diphoton
final state. Because of the high boost, the two photons from $S$
appear to be a single photon in the detector.

Ref.~\cite{Chala:2015cev} works in a model realization where the third-generation quarks are charged under $U(1)'$ whereas the first and second generations are not. While that can be viewed as a toy model to make a point, an actual realisation would either need additional Higgs representations or flavour-universal $U(1)'$ charges in order to reproduce the correct CKM matrix. Consequently, we will work with $U(1)'$ charges for all three generations of SM quarks and also have to use three generations of each additional exotic particle for anomaly cancellation.

The particle content of the model is summarized in \cref{tab:trickingLY:content}. For anomaly cancellation, the condition $Q_1-Q_2=-3$ must hold. Furthermore, there are four different valid choices for the hypercharge assignments $Y_i$ \cite{Chala:2015cev}:
\begin{equation}
(Y_1,\,Y_2,\,Y_3) =  (a,a+\frac{1}{2}, a-\frac{1}{2})\,.
\end{equation}
In the model implementation at hand, we choose $(Y_1,\,Y_2,\,Y_3)=(\frac{1}{2},1, 0)$.

\begin{table}[h]
\centering
\begin{tabular}{ c c c c c c}
\toprule
Field & Gen. & $SU(3)_C$ &  $SU(2)_L$ & $U(1)_Y$ & $U(1)'$ \\
\midrule 
$H$ & 1 & \one & \two & $-\frac12$ & $0$ \\
$S$ & 1 & \one & \one & $0$ & $Q_1-Q_2=-3$ \\
 \midrule
$Q_L$ & 3 & \three & \two & $\frac16$ & $1$ \\ 
$u_R$ & 3 & \three & \one & $\frac23$ & $1$ \\
$d_R$ & 3 & \three & \one & $-\frac13$ & $1$ \\
$L_L$ & 3 & \one & \two & $-\frac12$ & $0$ \\
$l_R$ & 3 & \one & \one & $-1$ & $0$ \\
$\nu_R$ & 3 & \one & \one & $0$ & $0$ \\
$X_L$ & 3 & \three & \two & $Y_1$ & $Q_1$ \\
$X_R$ & 3 & \three & \two & $Y_1$ & $Q_2$ \\
$R_L$ & 3 & \one & \one & $Y_2$ & $Q_2$ \\
$R_R$ & 3 & \one & \one & $Y_2$ & $Q_1$ \\
$\xi_L$ & 3 & \one & \one & $Y_3$ & $Q_2$ \\
$\xi_R$ & 3 & \one & \one & $Y_3$ & $Q_1$ \\
\bottomrule
\end{tabular}
\caption{Fermionic and scalar particle content of the {\tt trickingLY}. Here
  $X_L = (x_{1L},x_{2L})$, $X_R = (x_{1R},x_{2R})$ and $H=(H^0,H^-)$.}
  \label{tab:trickingLY:content}
\end{table}

The Yukawa interactions including fields beyond the SM read
\begin{align}
\mathcal L \supset Y_v\, \bar \nu_R\, \tilde H L_L 
           +\eta_1 \,\bar X_R \,S^* \,X_L + \eta_2 \bar R_R\, S\, R_L + \eta_3 \,\bar \xi_R \, S \, \xi_L\,.
\end{align}

The scalar potential is given by
\begin{align}
V = -\mu^2 |H|^2 - \mu_S^2 |S|^2 
            + \lambda_H |H|^4+ \lambda_S |S|^4
            + \lambda_{HS} |S|^2 |H|^2\,,
\end{align}
where the $U(1)'$ symmetry is broken spontaneously as soon as $S$
receives a VEV according to $\langle S \rangle = v_S/\sqrt{2}$.

Unfortunately, the most interesting vertex for this model, $Z' - S - \gamma$, only arises at the loop level. This would require a handling via an effective operator which is currently not supported in the automatized tools advertised. Hence, it is not possible to recast the results of Ref.~\cite{Chala:2015cev} based on this model implementation only.

Note that, in principle, the decay $Z' \to Z S$ is already possible at tree level due to $Z-Z'$ mixing and dominates over the decay into $S\gamma$. Therefore, in order to achieve the desired effect, the mixed gauge interaction term $F^{\mu\nu}F'_{\mu\nu}$ must be forbidden while allowing for $S F^{\mu\nu}F'_{\mu\nu}$ which is hard to justify in general.

%% file: tex/ModelCategories/LRmodels.tex
Left-right (LR) symmetric models can potentially provide an interesting explanation of the diphoton excess through the use of an extended scalar sector that is necessary to spontaneously break the enlarged gauge group $\mathcal{G}_{L-R}\equiv SU(3)_C \times SU(2)_L \times SU(2)_R \times U(1)_{B-L}$ to the SM gauge group. However, due to the large number of fields, these models are often difficult to analyse even at tree-level. Therefore as a starting point we provide model files for four different left-right models that have been proposed in the literature to explain the diphoton excess. These four models are based on the above mentioned gauge group $\mathcal{G}_{L-R}$ with, in two cases, further Abelian gauged or global symmetries. 

\modelparalabel{Left-right symmetric model without bi-doublets}{sec:LRVL}

\begin{itemize}
\item {\bf Reference:} \cite{Dasgupta:2015pbr,Deppisch:2016scs,Dev:2015vjd}
(see also Ref.~\cite{Gu:2010yf})
\item {\bf Model name:} \texttt{LRmodels/LR-VL}
    \end{itemize}

In Ref.~\cite{Dasgupta:2015pbr} the authors explored the possibility
of explaining the observed diphoton excess in the context of the
minimal left-right symmetric model. They show that it is not possible
and that an alternative model is necessary.  Therefore they give up on
standard Yukawa couplings, for which bi-doublets are necessary, and
consider separate $SU(2)_L$- and $SU(2)_R$-Higgs-doublets. In order to
be able to introduce Yukawa interactions, new vectorlike
$SU(2)_i$-singlet fermions are introduced.  After integrating out the
vectorlike fermions, the SM fermion masses are generated through a
universal seesaw mechanism \cite{Brahmachari:2003wv, Davidson:1987mh}.

\begin{table}[h]
\centering
\begin{tabular}{c c c c c c}
\toprule
Field & Generations & $SU(3)_C$ &  $SU(2)_L$ & $SU(2)_R$ & $U(1)_{B-L}$ \\
\midrule
$H_L$ & 1 & $\one$ & $\two$ & $\one$ & $-\frac{1}{2}$ \\
$H_R$ & 1 & $\one$ & $\one$ & $\two$ & $-\frac{1}{2}$ \\
$S$ & 1 & $\one$ & $\one$ & $\one$ & $0$\\
\midrule
$q_{L} = \left ( u_{L,i}, d_{L,i} \right )^T$ & 3 & $\bf{3}$ & $\bf{2}$
  & $\bf{1}$ & $\frac{1}{6}$  \\
$q_{R} = \left ( u_{R,i}, d_{R,i} \right )^T$ & 3 & $\bf{3}$ & $\bf{1}$
  & $\bf{2}$  & $\frac{1}{6}$ \\
$l_{L} = \left ( \nu_{L,i}, e_{L,i} \right )^T$ & 3 & $\bf{1}$ & $\bf{2}$
  & $\bf{1}$ & $-\frac{1}{2}$ \\
$l_{R} = \left ( \nu_{R,i}, e_{R,i} \right )^T$ & 3 & $\bf{1}$ & $\bf{1}$
  & $\bf{2}$ & $-\frac{1}{2}$ \\
$U_L$ & 3 & $\three$ & $\one$ & $\one$ & $\frac{2}{3}$ \\
$U_R$ & 3 & $\three$ & $\one$ & $\one$ & $\frac{2}{3}$ \\
$D_L$ & 3 & $\three$ & $\one$ & $\one$ & $-\frac{1}{3}$ \\
$D_R$ & 3 & $\three$ & $\one$ & $\one$ & $-\frac{1}{3}$ \\
$E_{L/R}$ & 3 & $\one$ & $\one$ & $\one$ & $-1$ \\
$N_{L/R}$ & 3 & $\one$ & $\one$ & $\one$ & $0$ \\
\bottomrule
\end{tabular}
\caption{Matter content and charge assignments for the {\tt LR-VL} model. The
scalar/fermionic fields are shown in the top/bottom of the table respectively.
The generation index $i$ runs over $i=1,2,3$.}
\label{tab:LR-VL}
\end{table}
The particle content for the model is shown in \cref{tab:LR-VL}.  Note that
the authors consider the second-lightest CP-even Higgs, which should be
predominantly singlet-like, as the particle responsible for the observed
resonance.  The scalar potential given the particle content and consistent
with the symmetries ($SU(3)_C\times SU(2)_L \times SU(2)_R \times U(1)_{B-L}$)
is
\begin{align}
V &= M_S^2 S^2 + \left(\mu_L^2 - \mu_{SL} S\right) |H_L|^2 +  \left(\mu_R^2
+ \mu_{SR} S\right) |H_R|^2 - \lambda_S S^4 - \lambda_L |H_L|^4 - \lambda_R |H_R|^4 \notag\\
&- \lambda_{LR}
|H_L|^2 |H_R|^2 - \lambda_{SL} S^2 |H_L|^2 - \lambda_{SR} S^2 |H_R|^2 \,.
\end{align}

The Yukawa interactions can be written as\footnote{Note that our Yukawa
interactions differ from the literature: in Ref.~\cite{Dasgupta:2015pbr}, they
are defined as, e.g.,  $ \overline{q_L} H^\dagger_L U_L$ which contracts to
zero because of the implicit left/right projection operators. Moreover, in
Refs.~\cite{Dasgupta:2015pbr,Deppisch:2016scs} the `conjugate' assignments of
the $H_{L/R}$ need to be exchanged in order to obtain a gauge-invariant
Lagrangian.}
\begin{align}
\mathcal{L}_Y &= Y_U \left(\overline{q_L} H_L U_R+ \overline{q_R} H_R U_L
\right) + Y_D \left(\overline{q_L} \tilde H_L D_R + \overline{q_R} \tilde H_R
D_L\right)+ Y_E \left(\overline{l_L} \tilde H_L E_R + \overline{l_R}
\tilde H_R E_L\right) \notag\\
&+ Y_N \left(\overline{l_L} H_L N_R + \overline{l_R} H_R N_L \right) +
\frac{1}{2} m_{NM} \left(\overline{N_R^c} N_R + \overline{N_L^c}N_L\right)
+ m_{ND} \overline{N_L} N_R \notag\\
&+\left(m_U + \lambda_U S\right) \overline{U_L} U_R + \left(m_D + \lambda_D
S\right) \overline{D_L} D_R + \left(m_E + \lambda_E S\right) \overline{E_L}
E_R + \hc\,,
\end{align}
where $N_L^c$ is the charge conjugate of $N_L$. Note that we have
included both Majorana and Dirac mass terms for the fermionic singlet
$N_{L/R}$.  We assume the following symmetry breaking VEVs
\begin{align}
\langle H_L \rangle &= \frac{1}{\sqrt{2}}\begin{pmatrix} v_L \\ 0
\end{pmatrix}\,, \qquad \langle H_R \rangle = \frac{1}{\sqrt{2}}
\begin{pmatrix} v_R \\ 0 \end{pmatrix}\,, \qquad \langle S \rangle = v_S\,.
\end{align}

\modelparalabel{Left-right symmetric model  with $U(1)_L \times U(1)_R$}{sec:LRLR}

\begin{itemize}
\item {\bf Reference:} \cite{Cao:2015xjz}
\item {\bf Model name:} \texttt{LRmodels/LRLR}
\end{itemize}

This model is based on the gauge group $SU(3)_C \times SU(2)_L \times SU(2)_R
\times U(1)_L \times U(1)_R$.  The inclusion of extra vectorlike
$SU(2)$-singlet fermions allows for the generation of the SM fermion masses
via a see-saw mechanism.  Additionally, a parity symmetry is imposed to ensure
a vanishing $\bar{\theta}$ parameter at tree-level in the QCD Lagrangian
\cite{Babu:1989rb}, in order to solve the strong CP-problem without introducing an
axion.

The particle content of this model and charge assignments under the
gauge symmetries are shown in \cref{tab:LRLR-charges}.  The proposed
candidate for the 750 GeV resonance is taken to be one of the CP-even scalars
associated with the $SU(2)$-singlet Higgs scalars $\sigma_D$, $\sigma_U$
and $\sigma_E$ that are responsible for the breaking $U(1)_L \times U(1)_R
\to U(1)_{B-L}$.  The decays of this state into digluon and diphoton final
states are assumed to proceed via loops involving the $SU(2)$-singlet
fermions.

\begin{table}[h]
\begin{center}
\begin{tabular}{c c c c c c c}
\toprule
Field & $SU(3)_c$ & $SU(2)_L$ & $SU(2)_R$ & $U(1)_L$ & $U(1)_R$ \\
\midrule
$\sigma_U$ & $\bf{1}$ & $\bf{1}$ & $\bf{1}$ & $\frac{2}{3}$ & $-\frac{2}{3}$ \\
$\sigma_D$ & $\bf{1}$ & $\bf{1}$ & $\bf{1}$ & $-\frac{1}{3}$ & $\frac{1}{3}$ \\
$\sigma_E$ & $\bf{1}$ & $\bf{1}$ & $\bf{1}$ & $-1$ & $1$ \\
$\phi_L$ & $\bf{1}$ & $\bf{2}$
  & $\bf{1}$ & $-\frac{1}{2}$ & $0$ \\
$\phi_R$ & $\bf{1}$ & $\bf{1}$
  & $\bf{2}$ & $0$ & $-\frac{1}{2}$ \\
$\Delta_L$ & $\bf{1}$ & $\bf{3}$ & $\bf{1}$ & $1$ & $0$ \\
$\Delta_R$ & $\bf{1}$ & $\bf{1}$ & $\bf{3}$ & $0$ & $1$ \\ \midrule
$q_{L,i} = \left ( u_{L,i}, d_{L,i} \right )^T$ & $\bf{3}$ & $\bf{2}$ & $\bf{1}$
  & $\frac{1}{6}$ & $0$ \\
$q_{R,i} = \left ( u_{R,i}, d_{R,i} \right )^T$ & $\bf{3}$ & $\bf{1}$ & $\bf{2}$
  & $0$ & $\frac{1}{6}$ \\
$l_{L,i} = \left ( \nu_{L,i}, e_{L,i} \right )^T$ & $\bf{1}$ & $\bf{2}$
  & $\bf{1}$ & $-\frac{1}{2}$ & $0$ \\
$l_{R,i} = \left ( \nu_{R,i}, e_{R,i} \right )^T$ & $\bf{1}$ & $\bf{1}$
  & $\bf{2}$ & $0$ & $-\frac{1}{2}$ \\
$U_{L,i}$ & $\bf{3}$ & $\bf{1}$ & $\bf{1}$ & $-\frac{2}{3}$ & $0$ \\
$U_{R,i}$ & $\bf{3}$ & $\bf{1}$ & $\bf{1}$ & $0$ & $-\frac{2}{3}$ \\
$D_{L,i}$ & $\bf{3}$ & $\bf{1}$ & $\bf{1}$ & $\frac{1}{3}$ & $0$ \\
$D_{R,i}$ & $\bf{3}$ & $\bf{1}$ & $\bf{1}$ & $0$ & $\frac{1}{3}$ \\
$E_{L}$ & $\bf{1}$ & $\bf{1}$ & $\bf{1}$ & $1$ & $0$ \\
$E_{R}$ & $\bf{1}$ & $\bf{1}$ & $\bf{1}$ & $0$ & $1$ \\
\bottomrule
\end{tabular}
\end{center}
\caption{The $SU(3)_c \times SU(2)_L \times SU(2)_R \times U(1)_L \times U(1)_R$
charge assignments for the scalar/fermions in the {\tt LRLR} shown in the
top/bottom of the table. The generation index $i$ runs over $i = 1,2,3$.}
\label{tab:LRLR-charges}
\end{table}
The Yukawa interactions consistent with the imposed parity are given by
\begin{align}
-\mathcal{L} &= y_U \left ( \bar{q}_L \phi_L U_L^c + \bar{q}_R \phi_R U_R^c
\right ) + f_U \sigma_U^* \bar{U}_L U_R + y_D \left ( \bar{q}_L \tilde{\phi}_L
D_L^c + \bar{q}_R
\tilde{\phi}_R D_R^c \right ) + f_D \sigma_D^* \bar{D}_L D_R \nonumber \\
& \quad {} + y_E \left ( \bar{l}_L \tilde{\phi}_L E_L^c + \bar{l}_R
\tilde{\phi}_R E_R^c \right ) + f_E \sigma_E^* \bar{E}_L E_R + Y_L \left (
\bar{l}_L^c i \tau_2 \Delta_L l_L
+ \bar{l}_R^c i \tau_2 \Delta_R l_R \right ) + \hc \label{eq:LRLR-yukawas}
\end{align}
The parity symmetry is taken to be softly broken, so that the part of the
scalar potential considered in Ref.~\cite{Cao:2015xjz}  is given by
\begin{align}
V &= \lambda \left ( \sigma_E^* \sigma_D^3 + \hc \right ) + \xi \left (
\sigma_U \sigma_D^2 + \hc \right ) \nonumber \\
& \quad {} + \eta \left ( \sigma_E \sigma_D^* \sigma_U + \hc \right )
+ \mu_{\phi_L}^2 \phi_L^\dagger \phi_L + \mu_{\phi_R}^2 \phi_R^\dagger \phi_R
\nonumber \\
& \quad {} + \mu_{\Delta_L}^2 \mathrm{Tr} \left ( \Delta_L^\dagger \Delta_L
\right ) + \mu_{\Delta_R}^2 \mathrm{Tr} \left ( \Delta_R ^\dagger \Delta_R
\right ) \nonumber \\
& \quad {} + \rho_L \left ( \phi_L^T i \tau_2 \Delta_L \phi_L + \hc \right )
+ \rho_R \left ( \phi_R^T i \tau_2 \Delta_R \phi_R + \hc \right ) .
\label{eq:LRLR-scalar-potnl}
\end{align}
The couplings and masses $\lambda$, $\xi$, $\eta$, $\mu_{\phi_L}^2$,
$\mu_{\phi_R}^2$, $\mu_{\Delta_L}^2$, $\mu_{\Delta_R}^2$, $\rho_L$ and
$\rho_R$ are taken to be real, with $\mu_{\phi_L}^2 \neq \mu_{\phi_R}^2$,
$\mu_{\Delta_L}^2 \neq \mu_{\Delta_R}^2$ and $\rho_L \neq \rho_R$.
Note that Eq.~(\ref{eq:LRLR-yukawas}) and Eq.~(\ref{eq:LRLR-scalar-potnl})
differ from Eq.~(6) and Eq.~(7) in Ref.~\cite{Cao:2015xjz}, which as given
are not gauge invariant.  One may also include a large number of additional
terms that are allowed by the gauge symmetries, given by
\begin{align}
V^\prime &= \kappa \left ( \sigma_D \sigma_E \sigma_U^2 + \hc \right ) +
\mu_D^2 |\sigma_D|^2 + \lambda_{DD}|\sigma_D|^4 + \mu_U^2 |\sigma_U|^2
+ \lambda_{UU} |\sigma_U|^4 + \mu_E^2 |\sigma_E|^2 \nonumber \\
& \quad {} + \lambda_{EE} |\sigma_E|^4 + \lambda_{DU} |\sigma_D|^2 |\sigma_U|^2
+ \lambda_{DE} |\sigma_D|^2 |\sigma_E|^2 + \lambda_{UE} |\sigma_U|^2
|\sigma_E|^2 + \lambda_{LL} |\phi_L|^4 \nonumber \\
& \quad {} + \lambda_{RR} |\phi_R|^4 + \lambda_{LR}
|\phi_L|^2 |\phi_R|^2 + \rho_{R_1} \mathrm{Tr}(\Delta_R^\dagger \Delta_R)
\mathrm{Tr} (\Delta_R^\dagger \Delta_R) + \rho_{R_2} \mathrm{Tr}(\Delta_R
\Delta_R) \mathrm{Tr}(\Delta_R^\dagger \Delta_R^\dagger) \nonumber \\
& \quad {} + \rho_{L_1} \mathrm{Tr} (\Delta_L^\dagger \Delta_L)
\mathrm{Tr}(\Delta_L^\dagger \Delta_L) + \rho_{L_2} \mathrm{Tr}(\Delta_L
\Delta_L) \mathrm{Tr}(\Delta_L^\dagger \Delta_L^\dagger) + \rho_3
\mathrm{Tr}(\Delta_L^\dagger \Delta_L) \mathrm{Tr} (\Delta_R^\dagger
\Delta_R) \nonumber \\
& \quad {} + \eta_{LL} |\phi_L|^2 \mathrm{Tr}(\Delta_L^\dagger \Delta_L)
+ \eta_{RL} |\phi_R|^2 \mathrm{Tr}(\Delta_L^\dagger \Delta_L) + \eta_{LR}
|\phi_L|^2 \mathrm{Tr}(\Delta_R^\dagger \Delta_R) + \eta_{RR} |\phi_R|^2
\mathrm{Tr}(\Delta_R^\dagger \Delta_R) \nonumber \\
& \quad {} + e_{RR_1} \phi_R^\dagger \Delta_R^\dagger \Delta_R \phi_R
- e_{RR_2} \phi_R^\dagger \Delta_R \Delta_R^\dagger \phi_R + e_{LL_1}
\phi_L^\dagger \Delta_L^\dagger \Delta_L \phi_L - e_{LL_2} \phi_L^\dagger
\Delta_L \Delta_L^\dagger \phi_L \nonumber \\
& \quad {} + \sum_{f=U,D,E} |\sigma_f|^2 \left [ \lambda_{fL} |\phi_L|^2
+ \lambda_{fR} |\phi_R|^2 + \tilde{\lambda}_{fL} \mathrm{Tr} \left (
\Delta_L^\dagger \Delta_L \right ) + \lambda_{fR} \mathrm{Tr} \left (
\Delta_R^\dagger \Delta_R \right ) \right ] \,.
\label{eq:LRLR-extra-terms}
\end{align}
The full scalar potential that we consider is then $V + V^\prime$.

The $SU(2)$-singlet Higgs scalars are assumed to acquire VEVs of the form
\begin{equation} \label{eq:LRLR-singlet-vevs}
\langle \sigma_D \rangle = \frac{v_D}{\sqrt{2}} \, , \qquad
\langle \sigma_U \rangle = \frac{v_U}{\sqrt{2}} \, , \qquad
\langle \sigma_E \rangle = \frac{v_E}{\sqrt{2}} .
\end{equation}
resulting in the breaking $U(1)_L \times U(1)_R \to U(1)_{B-L}$.  The
$SU(2)$-doublet Higgs scalars,
\begin{equation} \label{eq:LRLR-doublet-scalars}
\phi_{L,R} = \begin{pmatrix} \phi_{L,R}^0 \\ \phi_{L,R}^- \end{pmatrix} \, ,
\end{equation}
are taken to develop VEVs of the form
\begin{equation} \label{eq:LRLR-doublet-scalar-vevs}
\langle \phi_{L,R} \rangle = \frac{1}{\sqrt{2}}\begin{pmatrix} v_{L,R} \\ 0
\end{pmatrix} \, .
\end{equation}
The non-zero VEV of $\phi_R^0$ leads to the breakdown of
$SU(2)_L \times SU(2)_R \times U(1)_{B-L}$ to $SU(2)_L \times U(1)_Y$,
which is subsequently broken by the VEV of $\phi_L^0$.  As a result, the
triplet scalars
\begin{equation} \label{eq:LRLR-triplet-scalars}
\Delta_{L,R} = \begin{pmatrix} \frac{\delta_{L,R}^+}{\sqrt{2}} &
\delta_{L,R}^{++} \\ \delta_{L,R}^0 & -\frac{\delta_{L,R}^+}{\sqrt{2}}
\end{pmatrix}
\end{equation}
also acquire VEVs of the form
\begin{equation}
\langle \Delta_{L,R} \rangle = \frac{1}{\sqrt{2}}\begin{pmatrix} 0 & 0 \\
u_{L,R} & 0 \end{pmatrix} \, .
\end{equation}

\modelparalabel{Left-right symmetric model with fermionic and scalar triplets}{sec:tripletLR}

\begin{itemize}
\item {\bf Reference:} \cite{Berlin:2016hqw}
\item {\bf Model name:} \texttt{LRmodels/tripletLR}
\end{itemize}
In this model the diphoton signal is produced through a cascade decay, namely,
$pp \to Z^\prime \to X Y \to X X (\delta^0 \to \gamma \gamma)$ where $X$ and
$Y$ are unspecified soft states and $\delta_0$ is the neutral component of the
$SU(2)_R$ scalar triplet. However, in order to sufficiently boost the rate
three $SU(2)_R$ triplet fermion fields are added to the model, $T_1$, $T_2$
and $T_3$. The model is based on a $SU(3)_C \times SU(2)_L \times SU(2)_R
\times U(1)_{B-L}$ gauge group which is broken to the SM gauge group through
the VEV of the triplet field $\Delta_R$ whereby EWSB proceeds through the
bi doublet $\Phi$ VEVs. The entire particle content of the model is illustrated
in \cref{tab:tripletLR}.

\begin{table}[h]
\centering
\begin{tabular}{c c c c c c}
\toprule
Field & Generations & $SU(3)_C$ &  $SU(2)_L$ & $SU(2)_R$ & $U(1)_{B-L}$ \\
\midrule
$\Phi$ & 1 & $\one$ & $\two$ & $\two$ & $0$ \\
$\Delta_R$ & 1 & $\one$ & $\one$ & $\three$ & $1$ \\
\midrule
$Q_{L} = \left ( u_{L,i}, d_{L,i} \right )^T$ & 3 & $\bf{3}$ & $\bf{2}$
  & $\bf{1}$ & $\frac{1}{6}$  \\
$Q_{R} = \left ( u_{R,i}, d_{R,i} \right )^T$ & 3 & $\bf{3}$ & $\bf{1}$
  & $\bf{2}$  & $\frac{1}{6}$ \\
$L_{L} = \left ( \nu_{L,i}, e_{L,i} \right )^T$ & 3 & $\bf{1}$ & $\bf{2}$
  & $\bf{1}$ & $-\frac{1}{2}$ \\
$L_{R} = \left ( \nu_{R,i}, e_{R,i} \right )^T$ & 3 & $\bf{1}$ & $\bf{1}$
  & $\bf{2}$ & $-\frac{1}{2}$ \\
$T_1$ & 1 & $\one$ & $\one$ & $\three$ & $0$ \\
$T_2$ & 1 & $\one$ & $\one$ & $\three$ & $1$ \\
$T_3$ & 1 & $\one$ & $\one$ & $\three$ & $-1$ \\
\bottomrule
\end{tabular}
\caption{Matter content and charge assignments for the {\tt tripletLR} model.
The scalar/fermionic fields are shown in the top/bottom of the table
respectively. The generation index $i$ runs over $i=1,2,3$.}
\label{tab:tripletLR}
\end{table}
The scalar fields of the model can be expressed as
\begin{align}
\Phi &= \begin{pmatrix}
\phi_1^0 & \phi_2^+ \\
\phi_1^- & \phi_2^0 \end{pmatrix}\,, \quad \text{and} \quad \Delta_R =
\begin{pmatrix}
\frac{\delta_R}{\sqrt{2}} & \delta_R^{++} \\
\delta_R^0 & -\frac{\delta_R^+}{\sqrt{2}}
\end{pmatrix}\,,
\end{align}
leading to a scalar potential of the form
\begin{align}
V &= \mu_1^2 \mathrm{Tr} \left(\Phi^\dag \Phi \right) + \mu_2^2 \left[
\mathrm{Tr} (\widetilde{\Phi}\Phi^\dag)+ \mathrm{Tr} (\widetilde{\Phi}^\dag
\Phi)\right] + \mu_3^2 \mathrm{Tr}\left(\Delta_R \Delta_R^\dag\right)
+ \lambda_1 \left[\mathrm{Tr} \left(\Phi^\dag \Phi\right)\right]^2 \notag \\
&+ \lambda_2\left\{\left[\mathrm{Tr} (\widetilde{\Phi}\Phi^\dag)\right]^2
+ \left[\mathrm{Tr} (\widetilde{\Phi}^\dag \Phi)\right]^2\right\}
+\lambda_3 \mathrm{Tr} (\widetilde{\Phi}\Phi^\dag) \mathrm{Tr} (
\widetilde{\Phi}^\dag \Phi) \notag\\
&+ \lambda_4  \mathrm{Tr} (\Phi^\dag \Phi)\left[ \mathrm{Tr} (\widetilde{\Phi}
\Phi^\dag)+ \mathrm{Tr} (\widetilde{\Phi}^\dag \Phi)\right] + \rho_1 \left[
\mathrm{Tr} (\Delta_R \Delta_R^\dag)\right]^2 + \rho_2 \mathrm{Tr} (\Delta_R
\Delta_R) \mathrm{Tr} (\Delta_R^\dag \Delta_R^\dag) \notag\\
&+ \alpha_1 \mathrm{Tr} \left(\Phi^\dag \Phi \right) \mathrm{Tr} (\Delta_R
\Delta_R^\dag) + \left\{ \alpha_2 e^{i \delta}\mathrm{Tr} (
\widetilde{\Phi}^\dag \Phi)\mathrm{Tr} (\Delta_R \Delta_R^\dag) + \hc \right\}
+ \alpha_3 \mathrm{Tr} (\Phi \Phi^\dag \Delta_R \Delta_R^\dag)\,,
\end{align}
where $\tilde{\Phi} \equiv -\sigma_2 \Phi^* \sigma_2$.
The Yukawa interactions of the model are given by
\begin{align}
\mathcal{L}_Y &= Y_1^\alpha \overline{\Psi}_L \Phi \Psi_R +  Y_2^\alpha
\overline{\Psi}_L \widetilde{\Phi} \Psi_R + Y_{DR} L_R^T \mathcal C (i \sigma_2)
\Delta_R \, L_R \notag \\
&+ \frac{1}{2} m_1 \mathrm{Tr} \left( T_1 T_1 \right) + m_{23}\mathrm{Tr}
\left( T_2 T_3 \right)+ \lambda_{T_{13}} \mathrm{Tr}\left( T_1 T_3 \Delta_R
\right) + \lambda_{T_{12}} \mathrm{Tr} (T_1 T_2 \Delta_R^\dag)\,,
\end{align}
where $\alpha$ runs over the quarks and leptons $\alpha=Q,L$ and $\Psi_{L,R}
=(\psi^u_{L,R}, \psi^d_{L,R})$ with $\psi^u=u,\nu$ and $\psi^d=d,\ell$.
$\mathcal C$ is the charge conjugation operator. The VEVs of the scalar fields
in the model take the form
\begin{align}
\langle \Phi \rangle &= \frac{1}{\sqrt{2}}\begin{pmatrix}
\kappa_1 & 0 \\
0 & \kappa_2
\end{pmatrix}\, \quad \text{and} \quad \langle \Delta_R \rangle =
\frac{1}{\sqrt{2}} \begin{pmatrix}
0 & 0 \\
v_R & 0
\end{pmatrix}\,.
\end{align}

\modelparalabel{Dark left-right symmetric model}{sec:darkLR}
\begin{itemize}
\item {\bf Reference:} \cite{Dey:2015bur}
\item {\bf Model name:} \texttt{LRmodels/darkLR}
\end{itemize}

The main idea of this model is to add an additional symmetry in order to
stabilize the DM candidate, namely right-handed neutrinos, so that they cannot
decay via a $W^\prime$ channel. This additional symmetry takes the form of a
global Abelian symmetry labelled as $U(1)_S$. The spontaneous breaking of
$SU(2)_R \times U(1)_S$ is such that the combination $\tilde{L}=S- T_{3R}$,
where $T_{3R}$ is the third component of the right-handed isospin, remains
unbroken. Here $\tilde{L}$ is interpreted as a generalised lepton number.

The full particle content of the model is shown in \cref{tab:darkLR}. Note
that the scalar sector of the model is enlarged to include both left- and
right-handed triplets and doublets as well as the usual bi-doublet.
\begin{table}
\centering
\begin{tabular}{c c c c c c c c}
\toprule
Field & Gen. & $SU(3)_C$ &  $SU(2)_L$ & $SU(2)_R$ & $U(1)_{B-L}$ & $U(1)_S$
  & Lep.  \\
\midrule
$\Phi$ & 1 & $\one$ & $\two$ & $\two$ & $0$ & $\frac{1}{2}$ & $0$ \\
$\Delta_L$ & 1 & $\one$ & $\three$ & $\one$ & $1$ & $-2$ & $0$ \\
$\Delta_R$ & 1 & $\one$ & $\one$ & $\three$ & $1$ & $-1$ & $0$\\
$H_L$ & 1 & $\one$ & $\two$ & $\one$ & $\frac{1}{2}$ & $0$ & $0$\\
$H_R$ & 1 & $\one$ & $\one$ & $\two$ & $\frac{1}{2}$ & $-\frac{1}{2}$ & $0$\\
\midrule
$Q_{L} = \left ( u_{L,i}, d_{L,i} \right )^T$ & 3 & $\bf{3}$ & $\bf{2}$
  & $\bf{1}$ & $\frac{1}{6}$ & $0$ & $(0,0)$ \\
$Q_{R} = \left ( u_{R,i}, d_{R,i} \right )^T$ & 3 & $\bf{3}$ & $\bf{1}$
  & $\bf{2}$  & $\frac{1}{6}$ & $\frac{1}{2}$ & $(0,1)$ \\
$L_{L} = \left ( \nu_{L,i}, e_{L,i} \right )^T$ & 3 & $\bf{1}$ & $\bf{2}$
  & $\bf{1}$ & $-\frac{1}{2}$ & $1$ & $(1,1)$ \\
$L_{R} = \left ( \nu_{R,i}, e_{R,i} \right )^T$ & 3 & $\bf{1}$ & $\bf{1}$
  & $\bf{2}$ & $-\frac{1}{2}$ & $\frac{1}{2}$ & $(0,1)$ \\
$d_R$ & 3 & $\one$ & $\one$ & $\one$ &  $-\frac{1}{3}$ & $0$ & $0$ \\
$x_L$ & 3 & $\one$ & $\one$ & $\one$ & $-\frac{1}{3}$ & $1$ & $1$ \\
\bottomrule
\end{tabular}
\caption{Matter content and charge assignments for the {\tt darkLR} model.
The scalar/fermionic fields are shown in the top/bottom of the table
respectively. The generation index, denoted Gen. in the table, $i$ runs over
$i=1,2,3$. Additionally the model includes a lepton number symmetry where the
quantum numbers are denoted with Lep. above.}
\label{tab:darkLR}
\end{table}
These scalar fields can be expressed as
\begin{align}
\Phi &= \begin{pmatrix}
\phi_1^0 & \phi_2^+ \\
\phi_1^- & \phi_2^0 \end{pmatrix}\,, \quad \Phi_{L,R} = \begin{pmatrix}
H_{L,R}^+ \\ \phi_{L,R}^0
\end{pmatrix}\,, \quad
\Delta_{L,R} = \begin{pmatrix}
\frac{\delta_{L,R}}{\sqrt{2}} & \delta_{L,R}^{++} \\
\delta_{L,R}^0 & -\frac{\delta_{L,R}^+}{\sqrt{2}}
\end{pmatrix}\,.
\end{align}
Subsequently, the proposed candidate for the diphoton excess is
$\phi_R^0$, the neutral component of the $SU(2)_R$ doublet. Running in
the loop will be $W^\prime$-bosons, as well as $\delta_R^+$ and
$\delta_R^{++}$ Higgses.  However, these particles are insufficient to
boost the rate to diphotons, therefore additional quarks $x_L$ and
$d_R$ are introduced.

The scalar potential of the model is
\begin{align}
V &= \mu_1^2 \mathrm{Tr} \left(\Phi^\dag \Phi \right)
+ \mu_{TR}^2 \mathrm{Tr}\left(\Delta_R \Delta_R^\dag\right)
+\mu_{TL}^2 \mathrm{Tr}\left(\Delta_L \Delta_L^\dag\right) \notag\\
&+\mu_{DL}^2 H_L^\dag H_L + \mu_{DR}^2 H_R^\dag H_R
+\lambda_1 \left[\mathrm{Tr} \left(\Phi^\dag \Phi\right)\right]^2 +\lambda_3 \mathrm{Tr} (\widetilde{\Phi}\Phi^\dag) \mathrm{Tr} (\widetilde{\Phi}^\dag \Phi) \notag\\
&+ \rho_1 \left\{\left[ \mathrm{Tr} (\Delta_L \Delta_L^\dag)\right]^2+\left[ \mathrm{Tr} (\Delta_R \Delta_R^\dag)\right]^2 \right\}
+ \beta_2 \left[\mathrm{Tr}(\widetilde{\Phi} \Delta_R \Phi^\dag \Delta_L^\dag) + \mathrm{Tr} (\widetilde{\Phi}^\dag \Delta_L \Phi \Delta_R^\dag)\right] \notag\\
&+\rho_2 \left\{\mathrm{Tr} (\Delta_L \Delta_L) \mathrm{Tr} (\Delta_L^\dag \Delta_L^\dag)+ \mathrm{Tr} (\Delta_R \Delta_R) \mathrm{Tr} (\Delta_R^\dag \Delta_R^\dag) \right\}
+\rho_3  \mathrm{Tr} (\Delta_L \Delta_L^\dag)\mathrm{Tr} (\Delta_R \Delta_R^\dag) \notag\\
&+ \alpha_1 \mathrm{Tr} \left(\Phi^\dag \Phi \right) \left[\mathrm{Tr} (\Delta_L \Delta_L^\dag)+\mathrm{Tr} (\Delta_R \Delta_R^\dag)\right]
+ \alpha_3 \left[ \mathrm{Tr} (\Phi \Phi^\dag \Delta_L \Delta_L^\dag ) + \mathrm{Tr} (\Phi \Phi^\dag\Delta_R \Delta_R^\dag ) \right]\notag\\
&+ \eta_{LL} H_L^\dag H_L \Delta_L \Delta_L^\dag
+\eta_{RL} H_R^\dag H_R \Delta_L \Delta_L^\dag
+\eta_{LR} H_L^\dag H_L \Delta_R \Delta_R^\dag
+\eta_{RR_1} H_R^\dag H_R \Delta_R \Delta_R^\dag \notag \\
&+\eta_{RR_2} H_R^\dag \Delta_R^\dag \Delta_R H_R
+\eta_{RR_3} H_R^\dag \Delta_R \Delta_R^\dag H_R
+\lambda_L |H_L|^4
+\lambda_R |H_R|^4
\lambda_{LR} |H_L|^2 |H_R|^2 \notag\\
&+\beta_L |H_L|^2 \Phi^\dag \Phi
+\beta_R |H_R|^2 \Phi^\dag \Phi \notag\\
&+\left\{ \alpha_4 H_L^\dag \Phi \Delta_R H_R^\dag
+ \xi_R \tilde H_R^\dagger \Delta_R^\dag H_R
+ \xi_{LR} H_R \Phi H_L^\dag + \hc \right\}\,,
\end{align}
where $\tilde{\Phi} \equiv -\sigma_2 \Phi^* \sigma_2$.  Note that there are a
number of extra terms present in this potential compared to \cite{Dey:2015bur},
which are allowed under the symmetries of the model.  Finally the Yukawa
interactions are given by
\begin{align}
-\mathcal{L}_Y &= Y_{L_1} \overline{L}_L \Phi L_R + Y_{Q_1} \overline{Q}_L
\tilde \Phi Q_R + Y_{Q_2} \overline{Q}_L H_L d_R + Y_{Q_3} \overline{x}_L
\tilde H_R Q_R \notag \\
&+ \frac{1}{2}Y_{DL} L_L^T \mathcal{C} (i \sigma_2) \Delta_L L_L  +
\frac{1}{2} Y_{DR} L_R^T \mathcal{C} (i \sigma_2) \Delta_R L_R  + \hc\,, 
\end{align}
where $\mathcal C$ is the charge conjugation operator.  The structure of the
VEVs of the model are
\begin{align}
\langle \Phi \rangle &= \frac{1}{\sqrt{2}}\begin{pmatrix}
\kappa_1 & 0 \\
0 & \kappa_2
\end{pmatrix}\, \quad \langle \Phi_{L,R} \rangle = \frac{1}{\sqrt{2}}
\begin{pmatrix}
0 \\v^D_{L,R}
\end{pmatrix}\,,
\quad \langle \Delta_{L,R} \rangle =\frac{1}{\sqrt{2}} \begin{pmatrix}
0 & 0 \\
v^T_{L,R} & 0
\end{pmatrix}\,.
\end{align}

%% file: tex/ModelCategories/331models.tex
Models based on the {\ThreeThreeOne} gauge symmetry \cite{Singer:1980sw,
  Valle:1983dk,Pisano:1991ee,Frampton:1992wt,Foot:1992rh,Montero:1992jk,
  Pleitez:1992xh}, 331 for short, constitute an extension of the SM that could
explain the number of generations of matter fields.  This is possible as
anomaly cancellation forces the number of generations to be equal to the
number of quark colours.

Regarding the diphoton excess, 331 models automatically
include all the required ingredients to explain the hint.  First, the
usual $SU(2)_L$ Higgs doublet must be promoted to a $SU(2)_L$ triplet,
the new component being a singlet under the standard {\ThreeTwoOne}
symmetry.  Similarly, the group structure requires the introduction of
new coloured fermions to complete the $SU(3)_L$ quark multiplets, these
exotic quarks being {\ThreeTwoOne} vector-like singlets after the
breaking of {\ThreeThreeOne}.  Therefore, {\ThreeThreeOne} models naturally
embed the simple \emph{singlet $+$ vector-like fermions} framework proposed to
explain the diphoton excess.

There are several variants of {\ThreeThreeOne} models.  These are
characterized by their $\beta$ parameter~\footnote{See \cite{Buras:2012dp}
for a complete discussion of 331 models with generic $\beta$.}, which defines
the electric charge operator as~\footnote{\cref{eq:betadef} assumes that the
$SU(3)$ generators are $T_a = \frac{\lambda_a}{2}$, with $\lambda_a$
($a=1,\dots,8$) the Gell-Mann matrices. However, this is not the
convention used in \SARAH, see below.}
\begin{equation} \label{eq:betadef}
Q=T_3 + \beta \, T_8+\mathcal{X}\,.
\end{equation}
First, in \cref{sec:331v1} we consider the model in
Ref.~\cite{Boucenna:2015pav}.  This 331 variant has $\beta = 1/\sqrt{3}$,
which fixes the electric charges of all the states contained in the
$SU(2)_L$ triplets and anti-triplets to the usual $0,\pm 1$ values.  In
\cref{sec:331v2} we consider a 331 model with $\beta =
-\sqrt{3}$, a value leading to exotic electric charges.  This 331
variant has been discussed in the context of the diphoton excess in
\cite{Cao:2015scs,Dong:2015dxw,Hernandez:2015ywg}.  Although the
mechanism to explain the diphoton excess is exactly the same as in
\cite{Boucenna:2015pav}, the presence of the exotic states leads to
slightly different numerical results.

\paragraph{On the $SU(3)$ generators in \SARAH}\hfill \break
\label{sec:SU3gen}
The most common choice for the $SU(3)$ generators is $T_a =
\frac{\lambda_a}{2}$, with $\lambda_a$ ($a=1,\dots,8$) the Gell-Mann
matrices.  However, this is just one of the possible
representations.  In fact, \SARAH uses a different set of matrices,
$T_a^{\text{\SARAH}} = \frac{\Lambda_a}{2}$, following the conventions
of \Susyno \cite{Fonseca:2011sy}.  The relation between the
non-diagonal matrices in the two bases is
\begin{subequations}
\begin{eqnarray}
\lambda_1 &=& \Lambda_1 \, , \\
\lambda_2 &=& \Lambda_4 \, , \\
\lambda_4 &=& - \Lambda_6 \, , \\
\lambda_5 &=& - \Lambda_3 \, , \\
\lambda_6 &=& \Lambda_2 \, , \\
\lambda_7 &=& \Lambda_5 \, .
\end{eqnarray}
\end{subequations}
Concerning the diagonal matrices, the usual $\lambda_{3,8}$
Gell-Mann matrices,
\begin{equation}
\lambda_3 = \left( \begin{array}{ccc}
1 & 0 & 0 \\
0 & -1 & 0 \\
0 & 0 & 0 \end{array} \right) \quad ,
\quad \lambda_8 = \frac{1}{\sqrt{3}} \, \left( \begin{array}{ccc}
1 & 0 & 0 \\
0 & 1 & 0 \\
0 & 0 & -2 \end{array} \right) \, ,
\end{equation}
are replaced by $\Lambda_{7,8}$,
\begin{equation}
\Lambda_7 = \frac{1}{\sqrt{3}} \, \left( \begin{array}{ccc}
2 & 0 & 0 \\
0 & -1 & 0 \\
0 & 0 & -1 \end{array} \right) \quad ,
\quad \Lambda_8 = \left( \begin{array}{ccc}
0 & 0 & 0 \\
0 & 1 & 0 \\
0 & 0 & -1 \end{array} \right) \, .
\end{equation}
The electric charge operator can be written, using the
conventions in \SARAH, as
\begin{equation} \label{eq:betadefSARAH}
Q^{\text{\SARAH}} = -T_8 - \beta \, T_7 + \mathcal{X} \, .
\end{equation}
This in turn implies that the charge assignments in the $SU(3)$
multiplets must be adapted as well.  For example, one can easily check
that the electric charges of the first and third components of a
$SU(3)$ triplet $t$ are exchanged when going from the usual Gell-Mann
representation to the basis choice employed in \SARAH,
\begin{equation} \label{eq:tripletchange}
t = \left( \begin{array}{c}
t_1 \\
t_2 \\
t_3 \end{array} \right) \quad \longrightarrow \quad
t^{\text{\SARAH}} = \left( \begin{array}{c}
t_3 \\
t_2 \\
t_1 \end{array} \right) \, .
\end{equation}
In the following we will use the standard conventions based on the
Gell-Mann matrices in order to keep the discussion as close to the
original works as possible.  However, we emphasize that the
implementation of the 331 models in \SARAH requires this dictionary
between the bases.  It should also be noted that in the current implementation
in \SARAH of the 331 models described below, vertices involving four
vector bosons in the generated model files for \CalcHep cannot yet be
handled correctly.  In order to generate model files that will work with
\CalcHep, one must therefore exclude these vertices from being written out
by \SARAH, as described in \cref{app:calc-relic}. 

\modelparalabel{331 model without exotic charges}{sec:331v1}

\begin{itemize}
\item {\bf Reference:} \cite{Boucenna:2015pav}
\item {\bf Model name:} \texttt{331/v1}
\end{itemize}

\begin{table}[h]
\centering
\begin{tabular}{ c c c c c c c}
\toprule
Field & Gen. & $SU(3)_C$ &  $SU(2)_L$ & $U(1)_{\mathcal{X}}$ & $U(1)_{\mathcal{L}}$ & $\mathbb{Z}_2$ \\
\midrule 
$\Phi_1$ & 1 & \one & \threeS & $\frac23$ & $\frac23$ & $+$ \\
$\Phi_2$ & 1 & \one & \threeS & $-\frac13$ &  $-\frac43$ & $+$ \\
$\Phi_3$ & 1 & \one & \threeS & $-\frac13$ & $\frac23$ & $-$ \\
$\Phi_X$ & 1 & \one & \threeS & $-\frac13$ & $-\frac43$ & $+$ \\
\midrule
$\psi_L$ & 3 & \one & \threeS &  $-\frac13$ & $-\frac13$ & $+$ \\
$e_R$ & 3 & \one & \one & $-1$ & $-1$ & $+$ \\
$s$ & 3 & \one & \one & $0$ & $1$ & $+$ \\
$Q_L^{1,2}$ & 2 & \three & \three & $0$ & $-\frac23$ & $+$ \\
$Q_L^3$ & 1 & \three & \threeS & $\frac13$ & $\frac23$ & $-$ \\
$u_R$ & 3 & \three & \one & $\frac23$ & $0$ & $+$ \\
$T_R$ & 1 & \three & \one & $\frac23$ & $0$ & $-$ \\
$d_R$ & 3 & \three & \one & $-\frac13$ & $0$ & $-$ \\
$D_R,S_R$ & 2 & \three & \one & $-\frac13$ & $0$ & $+$ \\
\bottomrule
\end{tabular}
\caption{Fermionic and scalar particle content of the \texttt{331-v1} model. The scalar and fermion fields are shown in the top and bottom of the table respectively.}
\label{tab:331-v1-particle-content}
\end{table}

The model is based on the {\ThreeThreeOne} gauge symmetry, extended
with a global $U(1)_{\mathcal{L}}$ and an auxiliary {\zTwo} symmetry
to forbid some undesired couplings. The fermionic and scalar particle
content of the model is summarized in \cref{tab:331-v1-particle-content}. In
addition, due to the extended group structure, the model contains $17$
gauge bosons: the usual $8$ gluons; $8$ $W_i$ bosons associated to
$SU(3)_L$ and the $B$ boson associated to $U(1)_\mathcal{X}$.

The fermionic $SU(3)_L$ triplets of the model can be decomposed as
\begin{equation} \label{eq:frepI}
\psi_L = \left( \begin{array}{c}
\ell^- \\
- \nu \\
N^c \end{array} \right)_L^{e,\mu,\tau}\,,\,
Q^1_L = \left( \begin{array}{c}
u \\
d \\
D \end{array} \right)_L\,,\,
Q^2_L = \left( \begin{array}{c}
c \\
s \\
S \end{array} \right)_L \,,\,
Q^3_L = \left( \begin{array}{c}
b \\
- t \\
T \end{array} \right)_L \,.\,
\end{equation}
The notation used for the extra quarks that constitute the third
components of the $SU(3)_L$ triplets $Q^{1,2,3}_L$ is motivated by the
fact that their electric charges are $-1/3$ and $2/3$ for $D$/$S$ and
$T$, respectively. The scalar multiplets can be written as
\begin{equation} \label{eq:srepI}
\Phi_1 = \left( \begin{array}{c}
\phi_1 \\
- \phi_1^- \\
S_1^- \end{array} \right) \,, \quad 
\Phi_2 = \left( \begin{array}{c}
\phi_2^+ \\
- \phi_2 \\
S_2\end{array} \right) \,, \quad
\Phi_3 = \left( \begin{array}{c}
\phi_3^+ \\
- \phi_3 \\
S_3 \end{array} \right) \,, \quad
\Phi_X = \left( \begin{array}{c}
\phi_X^+ \\
- \phi_X \\
X \end{array} \right) \,.
\end{equation}
While $\phi_1^-$, $\phi_{2,3}^+$ and $S_1^-$ are electrically charged
scalars, the components $\phi_{1,2,3,X}$, $S_{2,3}$ and $X$ are
neutral.

The Yukawa Lagrangian of the model can be split as
\begin{equation} \label{eq:lagI}
\mathcal{L}_Y = \mathcal{L}_Y^q + \mathcal{L}_Y^\ell \, ,
\end{equation}
where

\begin{eqnarray}
\mathcal{L}_Y^q &=& 
\bar Q_L^{1,2}\, y^u  u_R \Phi_1^\ast + \bar Q_L^{3}\, \tilde y^d d_R \Phi_1
+ \bar Q_L^{1,2}\, \bar y^d   \hat d_R \Phi_2^\ast  +  \bar Q_L^3\, \bar y^u  T_R \Phi_2   
+\,\bar Q_L^3\, \tilde y^u  u_R \Phi_3 + \bar Q_L^{1,2}\, y^d d_R \Phi_3^\ast  \nonumber\\
&+& \bar Q_L^{1,2}\, \bar y^d_X   \hat d_R \Phi_X+ \bar Q_L^3\, \bar y^u_X  T_R \Phi_X   + 
\, \hc\,,
\label{eq:lagYqI}
\end{eqnarray}
and
\begin{equation}
\mathcal{L}_Y^\ell = y^\ell \bar \psi_L e_R \Phi_1 
+ y^a \overline{\psi_L^c} \psi_L \Phi_1 
+ y^s \bar \psi_L \, s \, \Phi_2 
+ \frac{m_s}{2} \, \overline{s^c} \, s + \hc \, .
\label{eq:lagYlI}
\end{equation}
We defined $\hat d_R \equiv ( D_R, S_R )$. We note that
Eq. \eqref{eq:lagYlI} leads to an inverse seesaw mechanism for neutrino
masses \cite{Mohapatra:1986bd,Boucenna:2015zwa}.  Here, $y^a$ is
anti-symmetric while $m_s$ is symmetric, whereas the rest of Yukawa
couplings are generic matrices, including those in
Eq. \eqref{eq:lagYqI}. An additional term $y^s_X \bar \psi_L s \Phi_X$
could be added to Eq. \eqref{eq:lagYlI}, but given that $\langle \Phi_X
\rangle = 0$, it does not contribute to
neutrino masses and we will drop it for simplicity. Finally, the
scalar potential is given by
\begin{eqnarray}
\label{eq:VI}
V  &=& \sum_i \mu_i^2 |\Phi_i|^2 + \lambda_i |\Phi_i|^4 + \sum_{i \ne j} \lambda_{ij} |\Phi_i|^2|\Phi_j|^2 \nonumber\\
&& + f\, (\Phi_1 \Phi_2 \Phi_3+\hc)  + \frac{\kappa}{2} \left[ (\Phi_2^\dagger \Phi_X)^2 +\hc \right] \, , 
\end{eqnarray}
where $i=1,2,3,X$.  The {\zTwo}-soft-breaking term, $f \Phi_1 \Phi_2
\Phi_3$, is required to break unwanted accidental symmetries in the
scalar potential.

We will assume the following symmetry breaking pattern
\begin{equation} \label{eq:VEVsI}
\langle \Phi_1 \rangle = \frac{1}{\sqrt{2}} \, \left( \begin{array}{c}
k_1 \\
0 \\
0 \end{array} \right) \,, \quad 
\langle \Phi_2 \rangle = \frac{1}{\sqrt{2}} \, \left( \begin{array}{c}
0 \\
0 \\
n \end{array} \right) \,, \quad
\langle \Phi_3 \rangle = \frac{1}{\sqrt{2}} \, \left( \begin{array}{c}
0 \\
k_3 \\
0 \end{array} \right) \,, \quad
\langle \Phi_X \rangle = \left( \begin{array}{c}
0 \\
0 \\
0 \end{array} \right) \,.
\end{equation}

\modelparalabel{331 model with exotic charges}{sec:331v2}

\begin{itemize}
\item {\bf Reference:} \cite{Cao:2015scs} (see also \cite{Dong:2015dxw,Hernandez:2015ywg} for similar constructions)
\item {\bf Model name:} \texttt{331/v2}
\end{itemize}

\begin{table}[h]
\centering
\begin{tabular}{ c c c c c }
\toprule
Field & Gen. & $SU(3)_C$ &  $SU(2)_L$ & $U(1)_{\mathcal{X}}$  \\
\midrule 
$\rho$ & 1 & \one & \three & $1$ \\
$\eta$ & 1 & \one & \three & $0$ \\
$\chi$ & 1 & \one & \three & $-1$ \\
\midrule
$\psi_L$ & 3 & \one & \threeS &  $-1$ \\
$e_R$ & 3 & \one & \one & $-1$  \\
$E_R$ & 3 & \three & \three & $-2$ \\
$Q_L^{1,2}$ & 2 &\three & \three & $\frac23$ \\
$Q_L^3$ & 1 & \three & \threeS & $-\frac13$ \\
$u_R$ & 3 & \three & \one & $\frac23$  \\
$T_R$ & 1 & \three & \one & $-\frac43$ \\
$d_R$ & 3 & \three & \one & $-\frac13$  \\
$D_R,S_R$ & 2 & \three & \one & $\frac53$ \\
\bottomrule
\end{tabular}
\caption{Fermionic and scalar particle content of the 331-v2 model. The scalar and fermion fields are shown in the top and bottom of the table respectively.}
\label{tab:331-v2-particle-content}
\end{table}

Now, we will consider a 331 variant with $\beta = -\sqrt{3}$, as
discussed in the context of the diphoton excess in
\cite{Cao:2015scs}. The fermionic and scalar particle content of the
model is summarized in \cref{tab:331-v2-particle-content}.  In addition, the
model contains $17$ gauge bosons: the usual $8$ gluons; $8$ $W_i$
bosons associated to $SU(3)_L$ and the $B$ boson associated to
$U(1)_\mathcal{X}$.

The fermionic $SU(3)_L$ triplet representations of the model can be
decomposed as
\begin{equation} \label{eq:frepII}
\psi_L = \left( \begin{array}{c}
\ell^- \\
- \nu \\
E^{--} \end{array} \right)_L^{e,\mu,\tau}\,,\,
Q^1_L = \left( \begin{array}{c}
u \\
d \\
D \end{array} \right)_L\,,\,
Q^2_L = \left( \begin{array}{c}
c \\
s \\
S \end{array} \right)_L \,,\,
Q^3_L = \left( \begin{array}{c}
b \\
- t \\
T \end{array} \right)_L \,.\,
\end{equation}
Due to the choice $\beta = - \sqrt{3}$, the electric charges for the
extra quarks that constitute the third components of the $SU(3)_L$
triplets $Q^{1,2,3}_L$ are $5/3$, $5/3$ and $-4/3$, respectively. The
scalar triplets can be written as
\begin{equation} \label{eq:srepII}
\rho = \left( \begin{array}{c}
\rho^+ \\
\rho^0 \\
\rho^{++} \end{array} \right) \,, \quad 
\eta = \left( \begin{array}{c}
\eta^0 \\
\eta_1^- \\
\eta_2^{+} \end{array} \right) \,, \quad
\chi = \left( \begin{array}{c}
\chi^- \\
\chi^{--} \\
\chi^0 \end{array} \right) \,.
\end{equation}
Therefore, the particle spectrum of the model contains the exotic
quarks in Eq. \eqref{eq:frepII}, as well as the doubly-charged fermion
$E^{--}$ and the scalars $\rho^{++}$ and $\chi^{--}$.

The Yukawa Lagrangian of the model can be split as
\begin{equation} \label{eq:lagII}
\mathcal{L}_Y = \mathcal{L}_Y^q + \mathcal{L}_Y^\ell \, ,
\end{equation}
where
\begin{eqnarray}
\mathcal{L}_Y^q &=& 
y^d \, \overline{Q_L^{1,2}} \, \rho \, d_R + \tilde y^d \, \overline{Q_L^3} \, \eta^\ast \, d_R \nonumber \\
&+& y^u \, \overline{Q_L^{1,2}} \, \eta \, u_R + \tilde y^u \, \overline{Q_L^3} \, \rho^\ast \, u_R \nonumber \\
&+& y^J \, \overline{Q_L^{1,2}} \, \chi \, \hat d_R + \tilde y^J \, \overline{Q_L^3} \, \chi^\ast \, T_R
\, + \hc\,,
\label{eq:lagYqII}
\end{eqnarray}
where we have defined $\hat d_R \equiv ( D_R, S_R )$, and
\begin{equation}
\mathcal{L}_Y^\ell = y^\ell \, \overline{\psi_L} \, \eta^\ast \, e_R 
+ y^E \, \overline{\psi_L} \, \chi^\ast \, E_R
+ \hc \, .
\label{eq:lagYlII}
\end{equation}
We note that the exotic fermions $E$, $D$, $S$ and $T$ only couple to
the $\chi$ scalar triplet, and thus only via its vacuum expectation
value (VEV) they will acquire masses. Finally, the scalar potential is
given by
\begin{align}
\label{eq:VII}
V &= \mu_1^2 \, |\rho|^2 + \lambda_1 |\rho|^4 +
 \mu_2^2 \, |\eta|^2 + \lambda_2 |\eta|^4+
 \mu_3^2 \, |\chi|^2 + \lambda_3 |\chi|^4  
 + \lambda_{12}|\rho|^2 |\eta|^2+
\lambda_{13}|\eta|^2|\chi|^2 \nonumber \\
&+\lambda_{23}|\eta|^2|\chi|^2+  
 + {\tilde \lambda}_{12}(\rho^\dagger \eta)(\eta^\dagger \rho)+
{\tilde \lambda}_{13}(\rho^\dagger \chi)(\chi^\dagger \rho)+
{\tilde \lambda}_{23}(\eta^\dagger \chi)(\chi^\dagger \eta )\nonumber  \\
&+ \sqrt{2} \, f \left( \epsilon_{ijk} \, \rho^i \eta^j \chi^k + \, \hc \right) \, .
\end{align}

We will assume the following symmetry breaking pattern
\begin{equation} \label{eq:VEVsII}
\langle \rho \rangle = \frac{1}{\sqrt{2}} \, \left( \begin{array}{c}
0 \\
v_1 \\
0 \end{array} \right) \,, \quad 
\langle \eta \rangle = \frac{1}{\sqrt{2}} \, \left( \begin{array}{c}
v_2 \\
0 \\
0 \end{array} \right) \,, \quad
\langle \chi \rangle = \frac{1}{\sqrt{2}} \, \left( \begin{array}{c}
0 \\
0 \\
v_3 \end{array} \right) \,.
\end{equation}

In this case, the non-zero VEV of $\chi$ is responsible for the breaking
$SU(3)_L \times U(1)_X \to SU(2)_L \times U(1)_Y$.  The requirement that
this occurs at a scale much above the EW scale then imposes a hierarchy
amongst the VEVs, namely that $v_3 \gg v_1, v_2$.  Consequently, one of the
CP-even scalar states is predominantly from the $\chi$ triplet and decouples
from the SM.  This scalar is then identified as the candidate for the 750
GeV resonance in this model.  The decays of this state into two photons
proceed via loops involving the heavy fermions, as well as those involving
the charged scalars and additional charged vector bosons.

%% file: tex/ModelCategories/OtherBSMmodels.tex
\paragraph{Gauged THD model}
\label{sec:gTHDM}
\begin{itemize}
\item {\bf Reference:} \cite{Huang:2015rkj}
\item {\bf Model name:} \texttt{GTHDM}
\end{itemize}

The GTHDM model \cite{Huang:2015wts} comes with an additional gauged
$SU(2)_H$ symmetry and a $U(1)_X$ symmetry, which is either global or
gauged as well. Since the minimal, gauged version suffers from the
fact that two massless vector bosons are present, $U(1)_X$ is treated
as global symmetry.  The scalar and fermion fields are listed in
\cref{tab:GTHDM-matter-content}. 

\begin{table}
\begin{tabular}{c c c c c c c} 
\toprule
Field &  Gen. & $SU(3)_C$ & $SU(2)_L$ & $U(1)_Y$ & $SU(2)_H$ & $U(1)_X$ \\ 
\midrule 
$H = \left(\begin{array}{cc}
       H_2^c & H_1^+ \\
       H_2^0 & H_1^0
      \end{array}\right)$   & 1 & ${\bf 1}$ & ${\bf 2}$  & $\frac{1}{2}$ & ${\bf 2}$ & 1 \\ 
$\Delta_H = \left(\begin{array}{cc}
       \delta^0/\sqrt{2} & (\delta^-)^* \\
       \delta^- & -\delta^0/\sqrt{2}
      \end{array}\right)$ &  1 & ${\bf 1}$ & ${\bf 1}$ & $0$ & ${\bf 3}$ & 0 \\ 
$\Phi = (\phi^c \, \phi^0 )^T$ & 1 & ${\bf 1}$ & ${\bf 1}$ & $0$ & ${\bf 2}$ & $-1$ \\ 
\midrule 
\(q = (u_L \, d_L )^T \) & 3 & ${\bf 3}$ & ${\bf 2}$ & $\frac{1}{6}$ & ${\bf 1}$&  0\\ 
\(l = (\nu_L \, e_L)^T\) &3 & ${\bf 1}$ & ${\bf 2}$ & $-\frac{1}{2}$ & ${\bf 1}$& 0 \\ 
\(d = (d_R^H \, d_R)^T \)  & 3 & ${\bf \overline{3}}$ & ${\bf 1}$ & $\frac{1}{3}$ & ${\bf 2}$ & 1\\ 
\(u = (u_R \, u_R^H)^T\) & 3 & ${\bf \overline{3}}$ & ${\bf 1}$ & $-\frac{2}{3}$ & ${\bf 2}$ & $-1$\\ 
\(\nu = (\nu_R \, \nu_R^H)^T\) & 3 & ${\bf 1}$ & ${\bf 1}$ & $0$ & ${\bf 2}$ &  $-1$\\ 
\(e = (e_R^H \, e_R)^R\) & 3 & ${\bf 1}$ &${\bf 1}$ &$1$ &${\bf 2}$ & 1\\ 
\(\chi_d\) & 3 & ${\bf 3}$&${\bf 1}$&$-\frac{1}{3}$&${\bf 1}$& 0\\ 
\(\chi_u\) & 3 & ${\bf 3}$&${\bf 1}$&$\frac{2}{3}$&${\bf 1}$& 0\\ 
\(\chi_\nu\) & 3 & ${\bf 1}$&${\bf 1}$ & $0$&${\bf 1}$& 0\\ 
\(\chi_e\) & 3 &${\bf 1}$& ${\bf 1}$& $-1$& ${\bf 1}$ &  0\\ 
\bottomrule
\end{tabular}
\caption{Scalar and fermion fields in the {\tt GTHDM}}
\label{tab:GTHDM-matter-content}
\end{table}
The Lagrangian of the GTHDM contains the SM Lagrangian, extended by
the new terms
\begin{align}
  \nonumber \mathcal{L} = &  \left( Y_d\, q\, d\, H^* - Y'_d \, \chi_d\, d\, \Phi  + Y_e\, l\, e\, H^* - Y'_e \,\chi_e\, e\, \Phi \right. \\ \nonumber & \left. + Y_u \,q\, u\, H - Y'_u \,\chi_u\, u\, \Phi^* + Y_\nu\, l\, \nu\, H - Y'_\nu\, \chi_\nu\, \nu\, \Phi^*   + \text{h.c.} \right) \\
  & +\mu^2_{\Delta} |\Delta_H|^2 - \mu^2_{H} |H|^2 - \mu^2_{\Phi} |\Phi|^2 - M_H H^* \Delta_H H + M_\Phi \Phi^* \Delta_H \Phi - {\lambda}_{D} \text{Tr}(\Delta_H^\dagger \Delta_H)^2   \nonumber \\ 
 & -{\lambda}_{H\Delta} |H|^2 \text{Tr}(\Delta_H^\dagger \Delta_H) - {\lambda}_{H} |H|^4 - {\lambda}_{H\Phi} |H|^2 |\Phi|^2  - {\lambda}_{\Phi\Delta} |\Phi|^2 \text{Tr}(\Delta_H^\dagger \Delta_H) - {\lambda}_{\Phi} |\Phi|^4 \, .
 \end{align}
The breaking of $SU(2)_L \times U(1)_Y \times SU(2)_H \to U(1)_{em}$
is triggered by
\begin{equation} \label{eq:GTHDM:VEVs}
\langle \Phi \rangle = \frac{1}{\sqrt{2}} \, \left( \begin{array}{c}
0 \\
v_\phi  \end{array} \right) \,, \quad 
\langle H \rangle = \frac{1}{\sqrt{2}} \, \left( \begin{array}{cc}
0 & 0\\
0 & v  \end{array} \right) \,, \quad 
\langle \Delta_H \rangle = \frac{1}{2}  \, \left( \begin{array}{cc}
v_T & 0 \\
0 & - v_T  \end{array} \right) \,.
\end{equation}
After EWSB, there are three neutral gauge bosons which mix giving rise to the $\gamma,Z, Z'$ mass eigenstates and two charged ones ($W$, $W'$) which do not mix. The neutral component of the SM-singlet $\Phi$, $\phi^0$, is considered to be the candidate for the 750~GeV resonance while its VEV gives mass to the exotic fermions that are needed to run in the loops. As $\phi^0$ will typically mix with the $SU(2)_L$ doublet Higgs, this mixing needs to be suppressed by a specific parameter choice in order to avoid the tight bounds from the dijet, $ZZ$ or dilepton channels.

%% file: tex/ModelCategories/SUSYmodels.tex
There are several ideas to explain the diphoton excess within a
supersymmetric framework. Some of them make use of SUSY models which
already exist in the literature, and for which also \SARAH model files
exist: the MSSM with trilinear $R$-parity
violation \cite{Ding:2015rxx,Allanach:2015ixl}, the simplest models
with Dirac gauginos \cite{Carpenter:2015ucu}, or the model with gauged
$U(1)_L \times U(1)_B$ \cite{Feng:2015wil}. We will not make any
further comment on these models, but concentrate in the following on
the models which are newly implemented. \\

\subsubsection{NMSSM extensions with vector-like multiplets}
\begin{itemize}

\item {\bf Reference:} \cite{Dutta:2016jqn,Tang:2015eko,Hall:2015xds,Wang:2015omi}
\item {\bf Model name:} \texttt{NMSSM+VL}
\end{itemize}

The scalar component of the gauge-singlet superfield $\hat S$ of the Next to Minimal Supersymmetric Standard Model (NMSSM) can explain the 750 GeV resonance, if
one adds vector-like $SU(5)$ multiplets to enhance the diphoton rate. The new multiplets are added in pairs of $(\five , \fiveb)$ and/or $(\ten , \tenb)$ in order to preserve gauge coupling unification. Typically one also imposes a {\zTwo} symmetry to forbid mixing of the new vector-like particles with the MSSM particles. The authors of Ref.~\cite{Wang:2015omi} mention the possibility to interpret the resonance as two nearly degenerate singlet-like bosons, roughly the scalar and pseudoscalar components of the singlet $\hat S$.
There are some differences in the singlet and Higgs superpotential interactions included in the different papers:
\begin{itemize}
 \item in Ref. \cite{Tang:2015eko} the authors assume $\lambda \hat S \hat H_u \hat H_d$ and $\kappa/3 \hat{S}^3$ to be present, but $\mu \hat H_u \hat H_d$ to be absent;
 \item in Ref. \cite{Dutta:2016jqn} $\mu \hat H_u H_d$, $\lambda \hat S \hat H_u \hat H_d$ and $M_S/2 \, \hat{S}^2$ are present;
 \item in Ref. \cite{Hall:2015xds} only $M_S/2 \, \hat{S}^2$ is present. This does not cause a mixing between the singlet and Higgs doublet at the tree level, but such a mixing is unavoidable at the loop level.  
\end{itemize}
The \SARAH implementations use the most general version of the superpotential: all possible interactions are present. The different limits according to the proposals of Refs.~\cite{Dutta:2016jqn,Tang:2015eko,Hall:2015xds} can be obtained by setting the corresponding parameters to zero in numerical studies. In what follows we describe the models with the vector-like multiplets in different representations of $SU(5)$.

\modelparalabel{NMSSM with vectorlike top}{sec:NMSSMVLtop}
\begin{itemize}
\item {\bf Model name:} \texttt{NMSSM+VL/VLtop}
\item {\bf Reference:} \cite{Wang:2015omi}
\end{itemize}
This model is an extension of the NMSSM by a vector-like top. There is a global $\mathbb{Z}_2$ $R$-parity and a $\mathbb{Z}_3$ symmetry, under which all particles transform as $X \to \exp(i \frac{2 \pi}{3}) \, X$. In this way, only terms with three superfields are allowed in the superpotential. The particle content is given in \cref{tab:NMSSMVLtopcontent}. Since this model does not introduce complete multiplets ($\five$ or $\ten$) of $SU(5)$, gauge coupling unification is not achieved.
\begin{table}[h]
\centering
\begin{tabular}{ c c c c c c c}
\toprule
Field & Gen. & $SU(3)_C$ &  $SU(2)_L$ & $U(1)_Y$ & $\mathbb{Z}_2$ & $\mathbb{Z}_3$ \\
\midrule 
$\hat S$ & 1 & $\one$ & $\one$ &  $0$ & $+$ & $\exp(i\frac{2\pi}{3})$ \\ \midrule
$\hat T$ & 1 & $\threeS$ & $\one$ & $-\frac23$ & $-$ & $\exp(i\frac{2\pi}{3})$ \\
$\hat T^\prime$ & 1 & $\three$ & $\one$ & $\frac23$  & $-$ & $\exp(i\frac{2\pi}{3})$ \\
\bottomrule
\end{tabular}
\caption{Superfield content beyond the MSSM superfields, including a singlet and a vector-like top.}
\label{tab:NMSSMVLtopcontent}
\end{table}
The superpotential is given by
\begin{align}
  \nonumber W = &  - Y_d \,\hat{d}\,\hat{q}\hat{H}_d\,- Y_e \,\hat{e}\,\hat{l}\hat{H}_d\,+Y_u\,\hat{u}\,\hat{q}\hat{H}_u  \nonumber\\
  & + \frac{1}{3}\kappa \hat{S}^3 + \lambda \hat S \hat H_u \hat H_d +  Y_t \hat T \hat Q \hat H_u + \lambda_T \hat T \hat T^\prime \hat S + \lambda_{U} \hat U \hat T^\prime \hat S
\label{eq:superpotVLtop}
 \end{align}
Beyond the neutral scalar components of the two Higgs doublets, after EWSB the complex singlet gets a VEV and can be decomposed as
\begin{align} \label{eq:VEVs5SU5}
S &= \frac 1{\sqrt 2} \left( v_S + \phi_S + i \, \sigma_S\right).
\end{align}
The fermionic components of $\hat T,\hat T^\prime$ mix with the up-type quarks, while the scalar components mix with the up-like squarks.

\modelparalabel{Pairs of {\bf 5} of $SU(5)$}{sec:NMSSM5}
\begin{itemize}
\item {\bf Model name:}  \texttt{NMSSM+VL/5plets}
\end{itemize}

The superfields beyond the MSSM are shown in \cref{tab:pair5}.	
\begin{table}[h]
\centering
\begin{tabular}{ c c c c c }
\toprule
Field & Gen. & $SU(3)_C$ &  $SU(2)_L$ & $U(1)_Y$  \\
\midrule 
$\hat S$ & 1 & $\one$ & $\one$ &  $0$ \\ \midrule
$\hat D$ & 1 & $\threeS$ & $\one$ &  $\frac{1}{3}$ \\
$\hat D'$ & 1 & $\three$ & $\one$  &  $-\frac{1}{3}$ \\
$\hat L$ & 1 &  $\one$ & $\two$ & $-\frac{1}{2}$ \\
$\hat L'$ & 1 & $\one$ &  $\two$ &   $\frac{1}{2}$ \\
\bottomrule
\end{tabular}
\caption{Superfield content in the case of a pair of $\mathbf{5}$'s of $SU(5)$.}
\label{tab:pair5}
\end{table}
In the current implementation we only have one copy of $(\five , \fiveb)$ fields, but having at least three copies of them should give a better fit to the diphoton resonance. According to \cite{Dutta:2016jqn} the fit is even better with four copies, however in that case one might hit a Landau pole. 

The superpotential is given by
\begin{align}
  \nonumber W = &  - Y_d \,\hat{d}\,\hat{q}\hat{H}_d\,- Y_e \,\hat{e}\,\hat{l}\hat{H}_d\,+Y_u\,\hat{u}\,\hat{q}\hat{H}_u + \mu \hat H_u \hat H_d \nonumber\\
  & + \frac{1}{3}\kappa \hat{S}^3 + \lambda \hat S \hat H_u \hat H_d + M_S \hat{S}^2 + t_S \hat S \nonumber \\
  & +  \lambda_D \hat S \hat D \hat D' + \lambda_L \hat S \hat L \hat L'
  + M_L \hat L \hat L' + M_D \hat D \hat D' \, .
\label{eq:superpot5SU5}
 \end{align}

Beyond the neutral scalar components of the two Higgs doublets, also the singlet gets a VEV after EWSB and can be decomposed as in \cref{eq:VEVs5SU5}.

\modelparalabel{Pairs of {\bf 10} of $SU(5)$}{sec:NMSSM10}
\begin{itemize}
\item {\bf Model name:}  \texttt{NMSSM+VL/10plets}
\end{itemize}

The superfields beyond the MSSM are shown in \cref{tab:pair10}.
\begin{table}[h]
\centering
\begin{tabular}{ c c c c c }
\toprule
Field & Gen. & $SU(3)_C$ &  $SU(2)_L$ & $U(1)_Y$  \\
\midrule 
$\hat S$ & 1 & $\one$ & $\one$ &  $0$ \\ \midrule
$\hat U$ & 1 & $\threeS$ & $\one$ &  $-\frac{2}{3}$ \\ 
$\hat U'$ & 1 & $\three$ & $\one$  & $\frac{2}{3}$  \\
$\hat Q$ & 1 & $\three$ & $\two$ & $\frac{1}{6}$ \\
$\hat Q'$ & 1 &$\threeS$ & $\two$ & $- \frac{1}{6}$ \\
 $\hat E$ & 1 & $\one$ & $\one$ & $1$\\
 $\hat E'$ & 1 &$\one$ & $\one$ & $-1$ \\
\bottomrule
\end{tabular}
\caption{Superfield content in the case of a pair of $\mathbf{10}$'s of $SU(5)$.}
\label{tab:pair10}
\end{table}
The superpotential is given by
\begin{align}
  \nonumber W = &  - Y_d \,\hat{d}\,\hat{q}\hat{H}_d\,- Y_e \,\hat{e}\,\hat{l}\hat{H}_d\,+Y_u\,\hat{u}\,\hat{q}\hat{H}_u + \mu \hat H_u \hat H_d \nonumber\\
  & + \frac{1}{3} \kappa \hat{S}^3 + \lambda \hat S \hat H_u \hat H_d + M_S \hat{S}^2 + t_S \hat S \nonumber \\
  & + Y_{10} \hat Q \hat U \hat H_u + Y'_{10} \hat Q' \hat U' \hat H_d + \lambda_Q \hat S \hat Q \hat Q' + \lambda_U \hat S \hat U \hat U' + \lambda_E \hat S \hat E \hat E' \nonumber \\
  & + M_U \hat U \hat U' + M_Q \hat Q \hat Q' + M_E \hat E \hat E' \, .
\label{eq:superpot10SU5}
 \end{align}
The symmetry breaking pattern is the same as for the model with 5-plets.

\modelparalabel{Pairs of {\bf 5} and {\bf 10} of $SU(5)$}{sec:NMSSM15}
\begin{itemize}
\item {\bf Model name:}  \texttt{NMSSM+VL/5+10plets}
\end{itemize}

This model combines the previous two setups, adding pairs of
vectorlike {\bf 5} and {\bf 10} representations of $SU(5)$. The
superfields beyond the MSSM are shown in \cref{tab:pair10pair5}.
\begin{table}[h]
\centering
\begin{tabular}{ c c c c c }
\toprule
Field & Gen. & $SU(3)_C$ &  $SU(2)_L$ & $U(1)_Y$  \\
\midrule 
$\hat S$ & 1 & $\one$ & $\one$ &  $0$ \\ \midrule
$\hat D$ & 1 & $\threeS$ & $\one$ &  $\frac{1}{3}$ \\
$\hat D'$ & 1 & $\three$ & $\one$  &  $-\frac{1}{3}$ \\
$\hat L$ &  1 & $\one$ & $\two$ & $-\frac{1}{2}$ \\
$\hat L'$ & 1 & $\one$ &  $\two$ &   $\frac{1}{2}$ \\ \midrule
$\hat U$ & 1 & $\threeS$ & $\one$ &  $-\frac{2}{3}$ \\ 
$\hat U'$ & 1 & $\three$ & $\one$  & $\frac{2}{3}$  \\
$\hat Q$ & 1 & $\three$ & $\two$ & $\frac{1}{6}$ \\
$\hat Q'$ & 1 &$\threeS$ & $\two$ & $- \frac{1}{6}$ \\
 $\hat E$ & 1 & $\one$ & $\one$ & $1$\\
 $\hat E'$ & 1 &$\one$ & $\one$ & $-1$ \\
\bottomrule
\end{tabular}
\caption{Superfield content in the case of a pair of $\mathbf{5}$'s and $\mathbf{10}$'s of $SU(5)$.}
\label{tab:pair10pair5}
\end{table}
The superpotential is given by
\begin{align}
  \nonumber W = &  - Y_d \,\hat{d}\,\hat{q}\hat{H}_d\,- Y_e \,\hat{e}\,\hat{l}\hat{H}_d\,+Y_u\,\hat{u}\,\hat{q}\hat{H}_u + \mu \hat H_u \hat H_d \nonumber\\
  & + \frac{1}{3}\kappa \hat{S}^3 + \lambda \hat S \hat H_u \hat H_d + M_S \hat{S}^2 + t_S \hat S \nonumber \\
  &+ Y'_D \,\hat{D}\,\hat{Q}\hat{H}_d\,+ Y'_E \,\hat{E}\,\hat{L}\hat{H}_d\,+Y'_U\,\hat{U}\,\hat{Q}\hat{H}_u + Y^{''}_D \,\hat{D}'\,\hat{Q}'\hat{H}_u\,+ Y^{''}_E \,\hat{E}'\,\hat{L}'\hat{H}_u\,+Y^{''}_U\,\hat{U}'\,\hat{Q}'\hat{H}_d \nonumber\\
  & +  \lambda_D \hat S \hat D \hat D' + \lambda_L \hat S \hat L \hat L' +  \lambda_Q \hat S \hat Q \hat Q' + \lambda_E \hat S \hat E \hat E' + \lambda_U \hat S \hat U \hat U' \nonumber \\
  & + M_L \hat L \hat L' + M_D \hat D \hat D'  + M_Q \hat Q \hat Q' + M_E \hat E \hat E'  + M_U \hat U \hat U' \, .
\label{eq:superpot15}
 \end{align}
The symmetry breaking pattern is the same as for the model with 5-plets.

\modelparalabel{Pairs of {\bf 5} of $SU(5)$ and $R$-parity violation}{sec:NMSSMRpV}
\begin{itemize}
\item {\bf Model name:}  \texttt{NMSSM+VL/5plets+RpV}
\end{itemize}

One can relax the assumption of the {\zTwo} symmetry that forbids mixing between vector-like fields and MSSM fields, in which case terms like $\kappa_5 \hat S \hat L \hat H_u$ are added to model~\cite{Dutta:2016jqn}. Furthermore, this also breaks $R$-parity.

The superfields beyond the MSSM are shown in \cref{tab:pair5} and the
superpotential is given by
\begin{align}
  \nonumber W = &  - Y_d \,\hat{d}\,\hat{q}\hat{H}_d\,- Y_e \,\hat{e}\,\hat{l}\hat{H}_d\,+Y_u\,\hat{u}\,\hat{q}\hat{H}_u + \mu \hat H_u \hat H_d \nonumber\\
  & + \frac{1}{3}\kappa \hat{S}^3 + \lambda \hat S \hat H_u \hat H_d + M_S \hat{S}^2 + t_S \hat S \nonumber \\
  & + \kappa_{5} \hat S \hat L \hat H_u + \kappa'_{5} \hat S \hat L' \hat H_d + \lambda_D \hat S \hat D \hat D' + \lambda_L \hat S \hat L \hat L'
  + M_L \hat L \hat L' + M_D \hat D \hat D'  \, .
\label{eq:superpot5RPV}
 \end{align}

The inclusion of $R$-parity violating terms triggers VEVs for the neutral components of $\tilde L$ and $\tilde{L}'$,
\begin{equation} \label{eq:VEVs5RPV}
\langle \tilde{L} \rangle = \frac{1}{\sqrt{2}} \, \left( \begin{array}{c}
0 \\
v_L  \end{array} \right) \,, \quad 
\langle \tilde{L}' \rangle = \frac{1}{\sqrt{2}} \, \left( \begin{array}{c}
0 \\
v_{L'}  \end{array} \right) \, ,
\end{equation}
and causes mixing between the vector-like states and the Higgs components. 

\subsubsection{Broken MRSSM}
\label{sec:bMRSSM}
\begin{itemize}

\item {\bf Reference:} \cite{Chakraborty:2015gyj}
\item {\bf Model name:} \texttt{brokenMRSSM}

\end{itemize}
In the minimal $R$-supersymmetric model (MRSSM) the scalar $R_u$ (see
\cref{tab:bMRSSM}) is proposed as an explanation to the 750 GeV
resonance. In order to explain the diphoton excess, it is necessary
to add explicitly an $R$-symmetry breaking term to the Lagrangian
\begin{equation}
\label{eq:Lr}
 L_{\slashed{R}} = T_u H_u \tilde{Q} \tilde{u} \, ,
\end{equation}
where $T_u$ is a dimensionful trilinear coupling. This source of
$R$-symmetry breaking has several consequences, not discussed in
Ref.~\cite{Chakraborty:2015gyj}, which however are taken into account
in the model implementation:
\begin{enumerate}
 \item The term in \cref{eq:Lr} will introduce Majorana gaugino masses via RGE effects
 \item The Majorana gaugino masses will also generate all other trilinear and bilinear soft-terms
 \item This causes $R$-symmetry breaking terms $ R_i H_i$ $i=d,u$ which will trigger VEVs for the R-fields
 \item The neutralinos and gluinos are no longer Dirac particles, but mix to Majorana fermions
 \item There is a mixing between fields of different $R$-charges.
\end{enumerate}
The superfields beyond the MSSM are listed in \cref{tab:bMRSSM}.
\begin{table}[h]
\centering
\begin{tabular}{ c c c c c }
\toprule
Field & Gen. & $SU(3)_C$ &  $SU(2)_L$ & $U(1)_Y$  \\
\midrule 
$\hat S$ & 1 & $\one$ & $\one$ & $0$ \\
$\hat T$ & 1 & $\one$ & $\three$ &  $0$ \\ 
$\hat O$ & 1 & $\eight$ & $\one$  & $0$  \\ \midrule
$\hat R_d$ & 1 & $\one$ & $\two$ &  $+ \frac{1}{2}$ \\
$\hat R_u$ & 1 &$\one$ &$\two$ & $- \frac{1}{2}$ \\
\bottomrule
\end{tabular}
\caption{Superfields of the broken MRSSM beyond the MSSM particle content.}
\label{tab:bMRSSM}
\end{table}
The superpotential, assumed to conserve $R$-symmetry, is given by
\begin{align}
  \nonumber W = &  - Y_d \,\hat{d}\,\hat{q}\hat{H}_d\,- Y_e \,\hat{e}\,\hat{l}\hat{H}_d\,
  +Y_u\,\hat{u}\,\hat{q}\hat{H}_u + \mu_D\,\hat{R}_d \hat{H}_d\,
  \, \\ \nonumber &
  +\mu_U\,\hat{R}_u\hat{H}_u\,+\hat{S}(\lambda_d\,\hat{R}_d\hat{H}_d\,+\lambda_u\,\,\hat{R}_u\hat{H}_u) \\
 &+  \lambda^T_d\,\hat{R}_d \hat{T}\,\hat{H}_d\,+\lambda^T_u\,\hat{R}_u\hat{T}\,\hat{H}_u \,~.~\,
\label{eq:superpot}
\end{align}
As explained above, because the $R$-symmetry is broken in the soft
sector of the model, all possible tri- and bilinear soft-breaking
terms corresponding to the superpotential terms will be generated.

The following VEVs appear after EWSB, beyond those of the neutral scalar
components of the two Higgs doublets
\begin{equation} 
\langle R_d \rangle = \frac{1}{\sqrt{2}} \, \left( \begin{array}{c}
0 \\
v_{R_d}  \end{array} \right) \,, \quad 
\langle R_u \rangle = \frac{1}{\sqrt{2}} \, \left( \begin{array}{c}
0 \\
v_{R_u}  \end{array} \right) \,, \quad 
\langle S \rangle = \frac{v_S}{\sqrt{2}}, \quad
\langle T \rangle = \frac{1}{2}  \, \left( \begin{array}{cc}
v_T & 0 \\
0 & - v_T  \end{array} \right) \,.
\end{equation}

The authors favor to have large stop mixing for a not too large
$R$-symmetry breaking term $T_u$ by considering the limit $v_d \sim
v_u$ and $m_{\tilde t_L} \sim m_{\tilde t_R}$. However, in this limit
the mass of the SM-like Higgs is tiny and often tachyonic: in the
MSSM, the Higgs tree-level mass vanishes for $\tan\beta \to 1$, and
this model has additional negative contributions to the mass from the
new $D$-terms present in models with Dirac gauginos. It is also
questionable if the case with very large $T_u$ is a viable scenario
because these values are highly restricted by charge and colour
breaking minima \cite{Camargo-Molina:2014pwa,Camargo-Molina:2013sta},
which demands careful checks. This is similar to the vacuum stability
issues discussed in \cref{sec:motivation}.

\subsubsection{$U(1)^\prime$-extended MSSM}
\label{sec:U1pMSSM}
\begin{itemize}
\item {\bf Reference:} \cite{Jiang:2015oms,An:2012vp}
\item {\bf Model name:} \texttt{MSSM+U1prime-VL}
\end{itemize}
\begin{table}[h]
\centering
\begin{tabular}{ c c c c c c}
\toprule
Field & Gen. & $SU(3)_C$ &  $SU(2)_L$ & $U(1)_Y$ & $U(1)^\prime$  \\
\midrule
$\hat Q$ & $3$ &$\three$ &$\two$ & $\frac{1}{6}$ &   $\frac{1}{2}$ \\
$\hat d^c$ &$3$ & $\threeS$ & $\one$ & $\frac{1}{3}$ &  $\frac{1}{2}$ \\
 $\hat u^c$ &$3$ & $\threeS$ & $\one$  & $-\frac{2}{3}$  & $\frac{1}{2}$ \\
 $\hat L$ & $3$ & $\one$ & $\two$ & $-\frac{1}{2}$ & $\frac{1}{2}$ \\
 $\hat e^c $ & $3$ &$\one$  &  $\one$ & $1$ & $\frac{1}{2}$ \\
 $\hat \nu^c$ & $3$ &$\one$ & $\one$ & $0$ &$\frac{1}{2}$ \\
 $\hat H_d$ & $1$ &$\one$ &$\two$ &  $-\frac{1}{2}$ &$-1$ \\
 $\hat H_u$ & $1$ & $\one$ &$\two$ & $\frac{1}{2}$ & $-1$\\
\bottomrule
\end{tabular}
\caption{Quantum numbers of the MSSM fields under the full gauge group
  in the {\tt MSSM+U1prime-VL}.}
\label{tab:MSSMU1VL:MSSMsuperfields}
\end{table}
\begin{table}[h]
\centering
\begin{tabular}{ c c c c c c}
\toprule
Field & Gen. & $SU(3)_C$ &  $SU(2)_L$ & $U(1)_Y$ & $U(1)^\prime$  \\
\midrule
$\hat T$ & 2 &$\three$ & $\one$  & $\frac{2}{3}$ & $-1$ \\
$\hat T^c$ & 2 & $\threeS$ &  $\one$  & $-\frac{2}{3}$ & $-1$ \\
$\hat T_3$ & 1 & $\three$ &$\one$  &   $-\frac{1}{3}$  & $-1$ \\
$\hat T^c_3$ & 1 & $\threeS$ & $\one$  & $\frac{1}{3}$ & $-1$ \\
$\hat D $ & 2 & $\one$  & $\two$  & $\frac{1}{2}$ & $-1$ \\
$\hat D^c$ & 2 & $\one$ &$\two$  &  $-\frac{1}{2}$ & $-1$ \\
$\hat X$ & 1 & $\one$  & $\one$  & $1$ & $2$ \\
$\hat X^c$ & 1 & $\one$  &$\one$  & $-1$ & $2$ \\
$\hat N$ & 1 &  $\one$  & $\one$  & $0$ & $2$ \\
$\hat N^c$& 1 & $\one$ & $\one$  & $0$ & $2$ \\
$S$ & 1 & $\one$  & $\one$  & $0$ & $2$ \\
$S^c$ &  1 &$\one$  & $\one$  & $0$ & $-2$ \\
$S_1$ &  1 &$\one$  &$\one$  & $0$ &  $-4$ \\
$S^c_1$ & 1 &  $\one$ & $\one$  & $0$ &  $4$ \\
$S_2$ & 1 &$\one$ &$\one$  & $0$ &  $-2$ \\
\bottomrule
\end{tabular}
\caption{Extra superfield content of the {\tt MSSM+U1prime-VL} and
  their quantum numbers under the full gauge group.}
\label{tab:MSSMU1VL:ExtraSuperfields}
\end{table}
%
In this model all MSSM fields carry a non-zero
$U(1)^\prime$-charge so that anomaly cancellation requires additional
superfields (see \cref{tab:MSSMU1VL:MSSMsuperfields}), which are also responsible for the spontaneous
$U(1)^\prime$ breaking. Furthermore, colour-charged and colour-uncharged
matter superfields which are vector-like with respect to the MSSM
gauge group are introduced. A combination of scalar singlets $S$ and $S_i$ is supposed to give the 750 GeV resonance.

The complete superfield content with all gauge quantum numbers is
given in Tables~\ref{tab:MSSMU1VL:MSSMsuperfields} and
\ref{tab:MSSMU1VL:ExtraSuperfields}. In addition to the usual matter
parity, we impose a {\zTwo} symmetry under which all exotic matter
superfields are odd and all other superfields are even in order to
reduce the number of superpotential terms and hence reduce the
complexity of the model.

The superpotential is given by
\begin{align}
  \nonumber W = &  - Y_d \,\hat{d}^c\,\hat{Q}\,\hat{H}_d\,- Y_e \,\hat{e}^c\,\hat{L}\,\hat{H}_d\,+Y_u\,\hat{u}^c\,\hat{Q}\,\hat{H}_u + Y_\nu\,\hat{\nu}^c\,\hat{L}\,\hat{H}_u + \lambda \,\hat S \,\hat H_u \,\hat H_d \nonumber\\
  & + \lambda_N \,\hat S_1 \,\hat N \,\hat N^c + \lambda_D \,\hat S \,\hat D \,\hat D^c + \lambda_X \,\hat S_1\, \hat X \,\hat X_c 
  + \lambda_T \,\hat S \,\hat T^c\, \hat T + \lambda_{T3}\, \hat S \,\hat T_3^c \,\hat T_3 \\
  & + \mu_S\, \hat S\, \hat S^c + \mu_{1S} \,\hat S_1 \,\hat S_1^c + \mu_{2S}\, \hat S \,\hat S_2 + 
  \kappa_1 \,\hat S \,\hat S \,\hat S_1 + \kappa_2 \,\hat S^c \,\hat S_2 \,\hat S_1^c \nonumber \,.
\label{eq:MSSMU1VL:superpot}
\end{align}
In addition to the neutral components of the two Higgs doublets, the
MSSM singlets get VEVs according to
\begin{equation} \label{eq:MSSMU1VL:VEVs}
\langle S_i^{(c)} \rangle = \frac{v_{S_i}^{(c)}}{\sqrt{2}} \,.
\end{equation}

\subsubsection{$E_6$-inspired SUSY models with extra $U(1)$ model}
\label{sec:E6MSSM}

\begin{itemize}
\item {\bf Reference:} \cite{Chao:2016mtn}
\item {\bf Model name:} \texttt{SUSYmodels/E6SSMalt}
\end{itemize}
$E_6$-inspired SUSY models predict extra SM-gauge singlets and extra
exotic fermions, so they immediately have the ingredients that many
authors have tried to use to fit the diphoton excess.  These models
are often motivated as a solution to the $\mu$-problem of the MSSM,
because the extra $U(1)$ gauge symmetry forbids the $\mu$-term, while
when one of the singlet fields develops a VEV at the TeV scale this
breaks the extra $U(1)$ giving rise to a massive $Z^\prime$ vector
boson and at the same time generates an effective $\mu$ term through
the singlet interaction with the up- and down-type Higgs fields,
$\lambda \hat{S} \hat{H}_u \hat{H}_d$. The matter content of the model
at low energies fills three generations of complete {\bf 27}-plet
representations of $E_6$, which ensures that anomalies automatically
cancel.

A number of models of this nature have been proposed as explanations
of the diphoton excess \cite{Chao:2016mtn, Ma:2015xmf, King:2016wep}.
The example we implement here \cite{Chao:2016mtn} is a variant of the
E$_6$SSM \cite{King:2005jy,King:2005my}.  In this version two singlet
states develop VEVs and the idea is that the $750$ GeV excess is
explained by one of these singlet states with a loop-induced decay
through the exotic states.

In $E_6$ models the extra $U(1)$ which extends the SM gauge group is
given as a linear combination of $U(1)_\psi$ and $U(1)_\chi$ which
appear from the breakdown of the $E_6$ symmetry as $E_6\to
SO(10)\times U(1)_{\psi}$ followed by $SO(10)$ into $SU(5)$,
$SO(10)\to SU(5) \times U(1)_{\chi}$.  In the E$_6$SSM and the variant
implemented here the specific combination is,
\be U(1)_N = \frac{1}{4}U(1)_{\chi} + \frac{\sqrt{15}}{4}U(1)_{\psi}.  \ee 

To allow one-step gauge coupling unification however some incomplete
multiplets must be included in the low energy matter content.  So in
addition to the matter filling complete {\bf 27} representations of
$E_6$ there are also two $SU(2)$ multiplets $\hat{H}^\prime$ and
$\hat{\overline{H}}^\prime$, which are the only components from
additional $\boldsymbol{27'}$ and $\boldsymbol{\overline{27}'}$ that
survive to low energies.  All gauge anomalies cancel between these two
states, so they do not introduce any gauge anomalies. Furthermore,
the low energy matter content of the model beyond the MSSM one
includes three generations of exotic diquarks\footnote{In the original
E$_6$SSM these states could be either diquark or leptoquark in nature,
depending on the choice of a discrete symmetry, but in the model
considered here the allowed superpotential terms for the decay of
these exotic quarks imply they are diquark.},
$\hat{D}_i, \hat{\bar{D}}_i$, three generations of SM singlet
superfields $\hat{S}_i$ and extra Higgs-like states $H^u_{1,2}$ and
$H^d_{1,2}$ that do not get VEVs.

The full set of superfields are given in Table \ref{tab:charges} along with their representations under $SU(3)$ and $SU(2)$ and the charges of the two U(1) gauge groups and the discrete symmetries, which we will now discuss.
\begin{table*}[ht]
\centering
\begin{tabular}{c c c c c c c c}
\toprule 
Field & Gen & $SU(3)_C$ & $SU(2)_L$ & $U(1)_Y$ & $U(1)_N$& $\mathbb{Z}_2^H$ & $\mathbb{Z}_2^L$ \\
 \hline
 $\hat{Q}_i$                 &   3     &    $\bf{3}$                     &             $\bf{2}$                 &               $\frac{1}{6}$       &          $1$      &           -       &      +      \\
 $\hat{u}_i^c$               &   3     &    $\bf{\overline{3}}$          &             $\bf{1}$                 &               $-\frac{2}{3}$      &          $1$      &           -       &      +      \\
 $\hat{d}_i^c$               &   3     &     $\bf{\overline{3}}$         &              $\bf{1}$                &               $\frac{1}{3}$       &          $2$      &           -       &      +      \\
 $\hat{L}_i$                 &   3     &    $\bf{1}$                     &             $\bf{2}$                 &               $-\frac{1}{2}$      &          $2$      &           -       &      -      \\
 $\hat{e}_i^c$               &   3     &    $\bf{1}$                     &             $\bf{1}$                 &               $1$                 &          $1$      &           -       &      -      \\
 $\hat{N}_i^c$               &   3     &    $\bf{1}$                     &             $\bf{1}$                 &               $0$                 &          $0$      &           -       &      -      \\
 $\hat{S}_i$                 &   2     &    $\bf{1}$                     &             $\bf{1}$                 &               $0$                 &          $5$      &           +       &      +      \\
 $\hat{S}_1$                 &   1     &    $\bf{1}$                     &             $\bf{1}$                 &               $0$                 &          $5$      &           -       &      +      \\
 $\hat{H}_u$                 &   1     &    $\bf{1}$                     &             $\bf{2}$                 &               $\frac{1}{2}$       &          $-2$     &           +       &      +      \\
 $\hat{H}_d$                 &   1     &    $\bf{1}$                     &             $\bf{2}$                 &               $-\frac{1}{2}$      &          $-3$     &           +       &      +      \\
 $\hat{H}_{\alpha}^u$        &   2     &    $\bf{1}$                     &             $\bf{2}$                 &               $\frac{1}{2}$       &          $-2$     &           -       &      +      \\
 $\hat{H}_{\alpha}^d$        &   2     &    $\bf{1}$                     &             $\bf{2}$                 &               $-\frac{1}{2}$      &          $-3$     &           -       &      +      \\
 $\hat{D}_i$                 &   3     &     $\bf{3}$                    &              $\bf{1}$                &               $-\frac{1}{3}$      &          $-2$     &           -       &      +      \\
 $\hat{\overline{D}}$        &   3     &    $\bf{\overline{3}}$          &             $\bf{1}$                 &               $\frac{1}{3}$       &          $-3$     &            -      &       +     \\
 $\hat{L}_4$                 &   1     &    $\bf{1}$                     &             $\bf{2}$                 &               $-\frac{1}{2}$      &          $2$      &            -      &       +     \\
 $\hat{\overline{L}}_4$      &   1     &    $\bf{1}$                     &             $\bf{\overline{2}}$      &               $\frac{1}{2}$       &          $-2$     &           -       &      +      \\
\bottomrule
\end{tabular}
\caption{The representations of the chiral superfields under the $SU(3)_C$ and
$SU(2)_L$ gauge groups, and their $U(1)_Y$ and $U(1)_N$ charges without the $E_6$ normalisation. The GUT normalisations are $\sqrt{\frac{5}{3}}$ for $U(1)_Y$ and $\sqrt{40}$ for $U(1)_N$. The transformation properties under the discrete symmetries $\mathbb{Z}_2^H$, $\mathbb{Z}_2^L$ are also shown, where `$+$' indicates the superfield is even under the symmetry and `$-$' indicates that it is odd under the symmetry.}
\label{tab:charges}
\end{table*}

The $\mathbb{Z}_2^L$ symmetry plays a role similar to R-parity in the MSSM,
being imposed to avoid rapid proton decay in the model.  However with
this imposed there are still terms in the superpotential that can lead
to dangerous flavour changing neutral currents (FCNCs).  To forbid
these, an approximate $\mathbb{Z}_2^H$ symmetry is
imposed.  In the original E$_6$SSM model only $\hat{S}_3$, $\hat{H}_d$
and $\hat{H}_u$ were even under the $\mathbb{Z}_2^H$ symmetry, however in this
variant $S_2$ is also even under this approximate symmetry.

The full superpotential before imposing any discrete symmetries is given by
\begin{equation}
W_{E6} = W_0 + W_1 + W_2,
\end{equation} where
\begin{eqnarray}
W_0 &=& \lambda_{ijk} \hat{S}_i \hat{H}^d_{j} \hat{H}^u_{k} + \kappa_{ijk} \hat{S}_i \hat{D}_j \hat{\bar{D}}_k + h^N_{ijk} \hat{N}^c_i \hat{H}^u_{j} \hat{L}_k \nonumber \\
& & + h^U_{ijk} \hat{u}^c_i \hat{H}^u_{j} \hat{Q}_k + h^D_{ijk} \hat{d}^c_i \hat{H}^d_{j} \hat{Q}_k + h^E_{ijk} \hat{e}^c_i \hat{H}^d_{j} \hat{L}_k, \\
W_1 &=& g^Q_{ijk} \hat{D}_i \hat{Q}_j \hat{Q}_k + g^q_{ijk} \hat{\bar{D}}_i \hat{d}^c_j \hat{u}^c_k, \\
W_2 &=& g^N_{ijk} \hat{N}^c_i \hat{D}_j \hat{d}^c_k + g^E_{ijk} \hat{e}^c_i \hat{D}_j \hat{u}^c_k + g^D_{ijk} \hat{Q}_i \hat{L}_j \hat{\bar{D}}_k.
\end{eqnarray}
However, with the discrete symmetries imposed and integrating out the heavy right-handed neutrinos, the superpotential in this specific variant reduces to\footnote{In the paper proposing this variant to explain the excess \cite{Chao:2016mtn}, the terms involving the surviving Higgs states on the second line are omitted from the superpotential.},
\begin{eqnarray}
\label{Eq:E6supvar}
W_{E_6SSM \, \text{variant}} &=&  W_{MSSM}^{(\mu = 0)} + \sum_{\alpha=2}^3\sum_{i=1}^3 \hat{S}^\alpha(\lambda_{\alpha \, i} \hat{H}_u^i \hat{H}_d^i + \kappa_{\alpha \, i} \hat{D}^i \hat{\overline{D}}^i) \nonumber \\ &&+ \mu^\prime \hat{H}\prime \hat{\overline{H}}^\prime + h^E_{4\, j}(\hat{H}_d\hat{H}^\prime) \hat{e}^c_j    \end{eqnarray}
One should remember that the $\mathbb{Z}_2^H$ can only be an approximate symmetry as otherwise the exotic quarks could not decay.  In this variant the exotic quarks decay therough the $\mathbb{Z}_2^H$ violating interactions of $W_1$. 

In the paper it is assumed that the singlet mixing can be negligible
and the numerical calculation was performed under this assumption, neglecting
any mixing between the singlet state which decays to $\gamma\gamma$
via the exotic states and the other CP-even Higgs states from the standard
$SU(2)$ doublets.  However it is clear that there must be some mixing
from the D-terms, and therefore if that is included one important
check would be to test whether other decays are sufficiently
suppressed.  Moreover, the parameters needed to simultaneously get a
$125$ GeV SM-like Higgs state and a $750$ GeV singlet-dominated state
are not given.  In this respect we note that the singlet VEVs appear
both in the diagonal entries of the mass matrix and in the off-diagonal
entries that mix the singlet states with the doublet states.

We finally note that other similar $E_6$ models have also been
proposed in the context of the diphoton excess.  These include a model
by two authors from the original paper \cite{King:2016wep}, a model
with a different $U(1)$ group at low energies \cite{Ma:2016qvn}, and a
model that is still $E_6$-inspired, but where no extra $U(1)$ survives
down to low energies \cite{Karozas:2016hcp}.

%% file: tex/example.tex
\section{Study of a natural SUSY explanation for the diphoton excess}
\label{sec:example}
We show in this section how one can use the described setup to perform easily a detailed study of a new model that aims at explaining the diphoton anomaly. 
\subsection{The model}
We are now going to study a SUSY model which enhances the tree-level Higgs mass due to non-decoupling $D$-terms. The model is based on that proposed in Ref.~\cite{Capdevilla:2015qwa} as a natural SUSY model which allows for light stops compatible with the measured Higgs boson mass, extended by three generations of pairs of vector-like quarks and leptons.  We want to achieve a tree-level enhancement of the SM-like Higgs mass and an explanation of the diphoton excess via the loop-induced decay of a CP-odd scalar. In addition, we will also check whether one can get a broad diphoton resonance in this model.
\begin{table}[h]
\centering
\begin{tabular}{c|c|c|c|cccc} 
\toprule
SF & Spin 0 & Spin \(\frac{1}{2}\) & Generations & $U(1)_Y$ & $SU(2)_L$ & $SU(3)_C$ & $U(1)_X$ \\ 
\midrule
$\hat{q}$ & $\tilde{q}$ & $q$ & 3 & $\frac{1}{6}$ & ${\bf 2}$ & ${\bf 3}$ & $0 $ \\ 
$\hat{l}$ & $\tilde{l}$ & $l$ & 3 & $-\frac{1}{2}$ & ${\bf 2}$ & ${\bf 1}$ & $0 $ \\ 
$\hat{d}$ & $\tilde{d}_R^{*}$ & $d^*_R$ & 3 & $\frac{1}{3}$ & ${\bf 1}$ & ${\bf \overline{3}}$ & $\frac{1}{2} $ \\ 
$\hat{u}$ & $\tilde{u}_R^{*}$ & $u^*_R$ & 3 & $-\frac{2}{3}$ & ${\bf 1}$ & ${\bf \overline{3}}$ & $-\frac{1}{2} $ \\ 
$\hat{e}$ & $\tilde{e}_R^*$ & $e^*_R$ & 3 & $1$ & ${\bf 1}$ & ${\bf 1}$ & $\frac{1}{2} $ \\ 
$\hat{\nu}$ & $\tilde{\nu}_R^*$ & $\nu^*_R$ & 3 & $0$ & ${\bf 1}$ & ${\bf 1}$ & $-\frac{1}{2} $ \\ 
$\hat{U}$ & $\tilde{U}^*$ & $U^*$ & 3 & $-\frac{2}{3}$ & ${\bf 1}$ & ${\bf \overline{3}}$ & $-\frac{1}{2} $ \\ 
$\hat{\bar U}$ & $\tilde{\bar U}$ & $\bar U$ & 3 & $\frac{2}{3}$ & ${\bf 1}$ & ${\bf 3}$ & $\frac{1}{2} $ \\ 
$\hat{E}$ & $\tilde{E}^*$ & $E^*$ & 3 & $1$ & ${\bf 1}$ & ${\bf 1}$ & $\frac{1}{2} $ \\ 
$\hat{\bar E}$ & $\tilde{\bar E}$ & $\bar E$ & 3 & $-1$ & ${\bf 1}$ & ${\bf 1}$ & $-\frac{1}{2} $ \\ 
\midrule
$\hat{H}_d$ & $H_d$ & $\tilde{H}_d$ & 1 & $-\frac{1}{2}$ & ${\bf 2}$ & ${\bf 1}$ & $-\frac{1}{2} $ \\ 
$\hat{H}_u$ & $H_u$ & $\tilde{H}_u$ & 1 & $\frac{1}{2}$ & ${\bf 2}$ & ${\bf 1}$ & $\frac{1}{2} $ \\ 
$\hat{\eta}$ & $\eta$ & $\tilde{\eta}$ & 1 & $0$ & ${\bf 1}$ & ${\bf 1}$ & $-1 $ \\ 
$\hat{\bar{\eta}}$ & $\bar{\eta}$ & $\tilde{\bar{\eta}}$ & 1 & $0$ & ${\bf 1}$ & ${\bf 1}$ & $1 $ \\ 
$\hat{S}$ & $S$ & $\tilde{S}$ & 1 & $0$ & ${\bf 1}$ & ${\bf 1}$ & $0 $ \\ 
\bottomrule
\end{tabular} 
\caption{Scalars and fermions in the $U(1)_X$-extended MSSM}
\label{tab:U1xMSSMparticles}
\end{table}
The matter field content is shown in \cref{tab:U1xMSSMparticles} and the considered superpotential reads:
\begin{align} 
\nonumber W &= - Y_d \,\hat{d}\,\hat{q}\,\hat{H}_d\,- Y_e \,\hat{e}\,\hat{l}\,\hat{H}_d  +Y_u\,\hat{u}\,\hat{q}\,\hat{H}_u +Y_\nu\,\hat{\nu}\,\hat{l}\,\hat{H}_u +Y_x\,\hat{\nu}\,\hat{\bar{\eta}}\,\hat{\nu} +(\mu + {\lambda} \hat{S})\,\hat{H}_u\,\hat{H}_d \nonumber \\ 
&\phantom{={}} + \hat{S} (\xi\, +  {\lambda}_X\,\hat{\eta}\,\hat{\bar{\eta}}) + M_S\,\hat{S}\,\hat{S} +\frac{1}{3} \kappa \,\hat{S}\,\hat{S}\,\hat{S} + \tilde{M}_E \hat e\hat{\bar{E}} + \tilde{M}_U \hat u \hat{\bar{U}} \nonumber \\
&\phantom{={}} + \hat{S} ({\lambda}_{e}\,\,\hat{E}\,\hat{\bar E}+{\lambda}_{u}\,\hat{U}\,\hat{\bar U})  +  M_e\,\hat{E}\,\hat{\bar E}\,+M_u\,\hat{U}\,\hat{\bar U}  +{Y'_e}\,\hat{E}\,\hat{l}\,\hat{H}_d +{Y'_u}\,\hat{U}\,\hat{q}\,\hat{H}_u .
\end{align} 
We will not make the simplifying assumption that mixings between the
MSSM states and the new vector-like fields can be neglected. Of
course, such mixing could have been forbidden by choosing different
$U(1)_X$ charges for the new particles. However, in such case there would be a conserved $\mathbb{Z}_2$ symmetry associated to the vector-like states (under which all vector-like superfields are odd and the rest are even) that would make the lightest of them absolutely stable. This would be a problem unless that state is neutral and colourless, and thus this scenario can only be viable if we also consider additional singlet vector-like states, such as vector-like partners for the right-handed neutrinos, and make them lighter than the other vector-like states. Thus, this setup would predict two stable particles to make the dark matter. Such a scenario could also be studied with the tools presented here. However, we decided not to consider this option in the following.

The other main ingredients of the model are the general soft-SUSY breaking terms, which read 
\begin{align}
\label{eq:Lsoft}
 - \mathcal{L} &= \big(T_d \tilde{d}\tilde{q}{H}_d + (T_e \tilde{e}+{T'_e}\tilde{E})\tilde{l}{H}_d  + (T_u\tilde{u}+{T'_u}\tilde{U})\tilde{q}{H}_u +T_\nu\tilde{\nu}\tilde{l}{H}_u +T_x\tilde{\nu}{\bar{\eta}}\tilde{\nu} +(B_\mu + T_{\lambda}{S}){H}_u{H}_d \nonumber \\ 
&\phantom{={}} + {S} (t_\xi +  T_X{\eta}{\bar{\eta}}) + B_S{S}{S} +\frac{1}{3} T_\kappa {S}{S}{S} + {S} (T_{E}\tilde{E}\tilde{\bar E}+T_{U}\tilde{U}\tilde{\bar U}) \nonumber \\
&\phantom{={}} +   B_E\tilde{E}\tilde{\bar E}+B_U\tilde{U}\tilde{\bar U}  +  \tilde{B}_E \tilde{e} \tilde{\bar E} + \tilde{B}_U \tilde{u} \tilde{\bar U} + \text{h.c.} \big) \nonumber \\
&\phantom{={}} +  \tilde{q}^\dagger m_q^2 \tilde{q} + \tilde{u}^\dagger m_u^2 \tilde{u} + \tilde{d}^\dagger m_d^2 \tilde{d} + \tilde{e}^\dagger m_e^2 \tilde{e} + \tilde{l}^\dagger m_l^2 \tilde{l} + \tilde{U}^\dagger m_{U}^2 \tilde{U}  + \tilde{\bar U}^\dagger m_{\bar U}^2 \tilde{\bar U}   + \tilde{E}^\dagger m_{E}^2 \tilde{E}  + \tilde{\bar E}^\dagger m_{\bar E}^2 \tilde{\bar E} + \nonumber \\
&\phantom{={}} + ( \tilde{U}^\dagger m_{Uu}^2 \tilde{u} +  \tilde{E}^\dagger m_{Ee}^2 \tilde{E} + \text{h.c.}) + m_{H_d}^2 |H_d|^2+ m_{H_u}^2 |H_u|^2 + m_{H_s}^2 |S|^2 + m_{\eta}^2 |\eta|^2 + m_{\bar \eta}^2 |\bar \eta|^2 \nonumber \\
&\phantom{={}} + (M_1 \lambda_B \lambda_B + M_2 \lambda_W \lambda_W + M_3 \lambda_g \lambda_g + M_X \lambda_X \lambda_X + M_{1X} \lambda_B \lambda_X + \text{h.c.})
\end{align}
Note that we have included the gaugino mass term $ M_{1X}$ arising from gauge kinetic mixing. All the terms shown in Eq.~\eqref{eq:Lsoft} are automatically added by \SARAH based on the information provided by the user about the particle content and the superpotential. Several scalar fields acquire VEVs. We decompose them as
\begin{eqnarray} 
& H_d^0 =  \, \frac{1}{\sqrt{2}} \left( \phi_{d}  + v_d  + i  \sigma_{d} \right)\,, \quad 
H_u^0 =  \, \frac{1}{\sqrt{2}} \left(\phi_{u}  + v_u  + i  \sigma_{u} \right) \,, & \\ 
& \eta =  \, \frac{1}{\sqrt{2}} \left(\phi_{\eta}  +  v_{\eta}  + i  \sigma_{\eta}\right)\,, \quad
\bar{\eta} =  \, \frac{1}{\sqrt{2}} \left( \phi_{\bar{\eta}}  + v_{\bar{\eta}}  + i  \sigma_{\bar{\eta}}\right) \,, & \\ 
& S =  \, \frac{1}{\sqrt{2}} \left( {\phi}_{s}  + v_S  + i  {\sigma}_{s} \right) \, . &
\end{eqnarray} 
We define $\tan\beta=\frac{v_u}{v_d}$, $v=\sqrt{v_d^2+v_u^2}$ as well as $\tan\beta_x = \frac{v_{\eta}}{v_{\bar \eta}}$, $x=\sqrt{v_\eta^2+v_{\bar \eta}^2}$.  In addition, the sneutrinos are decomposed with respect to their CP eigenstates,
\begin{eqnarray}
\tilde \nu_{L,i} \to \frac{1}{\sqrt{2}} \left( \phi_{L,i} + i \sigma_{L,i}\right) \,,\quad \tilde \nu_{R,i} \to \frac{1}{\sqrt{2}} \left( \phi_{R,i} + i \sigma_{R,i}\right) \, ,
\end{eqnarray}
which in general have different masses due to the Majorana mass-term $Y_X \langle \bar\eta \rangle$ in the superpotential.
Since $H_d^0$ and $H_u^0$ carry charges under both $U(1)$ gauge groups, there will be non-zero $Z$--$Z'$ mixing even in the limit of vanishing gauge kinetic mixing. The list of particle mixings, which go beyond the usual MSSM mixings reads
\begin{align}
 (B,W_3,B') &\quad \longrightarrow \quad (\gamma,Z, Z') , \\
 (\phi_d,\phi_u,\phi_\eta,\phi_{\bar \eta}, \phi_s) &\quad \longrightarrow \quad h_i, \quad i=1\dots5, \\
 (\sigma_d,\sigma_u,\sigma_\eta,\sigma_{\bar \eta},\sigma_s) &\quad \longrightarrow \quad A^0_i, \quad i=1\dots5, \\
 (\phi_{L,i},\phi_{R,i}) &\quad \longrightarrow \quad \tilde \nu^R_j, \quad i=1\dots3, \, j=1\dots6, \\ 
 (\sigma_{L,i},\sigma_{R,i}) &\quad \longrightarrow \quad \tilde \nu^I_j, \quad i=1\dots3, \, j=1\dots6, \\
 (\tilde B, \tilde W_3, \tilde H_d^0, \tilde H_u^0, \tilde X, \tilde \eta, \tilde{\bar\eta}, \tilde S) &\quad \longrightarrow \quad \tilde\chi^0_i, \quad i=1\dots8, \\
 (e_{L,i},\bar E_i^*)/(e_{R,i} , E_i) & \quad \longrightarrow \quad e_j, \quad i=1\dots3, \, j=1\dots6, \\
 (u_{L,i},\bar U_i^*)/(u_{R,i} , U_i) & \quad \longrightarrow \quad u_i, \quad i=1\dots3, \, j=1\dots6, \\
 (\tilde e_{L,i}, \tilde e_{R,i}, \tilde E_i, \tilde{\bar{E_i}}) &\quad \longrightarrow \quad \tilde e_j, \quad i=1\dots3, \, j=1\dots12, \\
 (\tilde u_{L,i}, \tilde u_{R,i}, \tilde U_i, \tilde{\bar{U_i}}) &\quad \longrightarrow \quad \tilde u_j, \quad i=1\dots3, \, j=1\dots12,
\end{align}
The model files which implement this model in \SARAH are discussed in
\cref{app:modelU1x}. In addition, we provide all files to reproduce
the computations that follow at
\begin{center}
{\tt \url{http://sarah.hepforge.org/U1xMSSM\_example.tar.gz}}
\end{center}

\subsection{Analytical results with {\tt Mathematica}}
Before we perform a numerically precise study of the model, we show how already with just \SARAH and \Mathematica one can gain a lot of information about a new model. 
\subsubsection{Consistency Checks}
The model is initialised after loading it in \SARAH via 
\begin{lstlisting}[style=mathematica]
<<SARAH.m;
Start["U1xMSSM"];
\end{lstlisting}
\SARAH automatically performs some basic consistency checks for the model. For instance, it checks whether the model is free from gauge anomalies:
\begin{lstlisting}[style=mathematica]
Checking for anomalies: 
 {(hypercharge)^3, (left)^3,  (color)^3, (extra)^3,
  (hypercharge)x(gravity)^2,
  (extra)x(gravity)^2,
  (left)^2 x hypercharge,
  (color)^2 x hypercharge,
  (extra)^2 x hypercharge,
  (hypercharge)^2 x extra,
  (left)^2 x extra,
  (color)^2 x extra,
   Witten Anomalyleft}
\end{lstlisting}
One can see that \SARAH tests all different combinations of gauge anomalies and, given that no warning is printed on the screen, confirms that all of them cancel. Similarly, it also checks that all terms in the superpotential are in agreement with all global and local symmetries. More detailed checks can be carried out by running {\tt CheckModel[]} when the initialisation is finished.

After a few seconds, a message is printed telling that the model is loaded.
\begin{lstlisting}[style=mathematica]
All Done. U1xMSSM is ready! 
\end{lstlisting}

\subsubsection{Gauge sector}
Before we discuss the matter sector or the scalar potential, we have a brief look at the gauge bosons. We make use of the mass matrices calculated by \SARAH during the initialisation of the model. 
We find a handy expression for the mass matrix of the neutral gauge bosons in the limit of vanishing gauge kinetic mixing ($g_{X1}=g_{1X}=0$) via
\begin{lstlisting}[style=mathematica]
matV = Simplify[MassMatrix[VZp] /. {gX1 -> 0, g1X -> 0}] 
  //. {vd^2 + vu^2 -> v^2,  x1^2 + x2^2 -> x^2} 
\end{lstlisting}
which reads
\begin{equation}
\left(
\begin{array}{ccc}
 \frac{g_1^2 v^2}{4} & -\frac{1}{4} g_1 g_2 v^2 & \frac{1}{4} g_1 g_X v^2 \\
 -\frac{1}{4} g_1 g_2 v^2 & \frac{g_2^2 v^2}{4} & -\frac{1}{4} g_2 g_X v^2 \\
 \frac{1}{4} g_1 g_X v^2 & -\frac{1}{4} g_2 g_X v^2 & \frac{1}{4} g_X^2 \left(v^2+4 x^2\right) \\
\end{array}
\right) .
\end{equation}
Note, that {\tt MassMatrix[VP]} and {\tt MassMatrix[VZ]} would have given the same result. We can check the eigenvalues of this matrix to first order in $\frac{v^2}{x^2}$ using the {\tt Series} command of \Mathematica 
\begin{lstlisting}[style=mathematica]
Simplify[Normal[Series[Eigenvalues[matV] 
  /. v -> r x, {r, 0, 2}]] /. r -> v/x, {x > 0, gX > 0}] 
\end{lstlisting}
and find 
\begin{align}
\{0, \quad \frac{1}{4} (g_{1}^{2} + g_{2}^{2})v^{2}, \quad \frac{1}{4} g_{X}^{2} (4 x^{2}  + v^{2}) \}
\end{align}
As expected, the first two eigenvalues are just the ones of the SM gauge bosons, while the mass of the new gauge boson is given by
\begin{equation}
M_{Z'} = \frac12 g_X \sqrt{4 x^2 + v^2} \,.
\end{equation}
We will use this relation in the following to replace $x$ by $M_{Z'}$ in all equations. 

\subsubsection{Scalar Sector}
\paragraph{Solving the tadpole equations}
We turn now to the scalar sector of the model. First, we make a list with a few simplifying assumptions which we are going to use in the following
\begin{lstlisting}[style=mathematica]
assumptions = { 
 conj[x_] -> x, RXi[__] -> 0,
 gX1 -> 0, g1X -> 0, 
 x1 -> X/Sqrt[2], x2 -> X/Sqrt[2],
       X -> Sqrt[4 MZp^2 - gX^2 v^2]/(2 gX),
 vd -> v Cos[ArcTan[TB]], vu -> v Sin[ArcTan[TB]], 
 T[kappa] -> 0 , kappa -> 0, 
       T[lambdaH] -> 0, lambdaH -> 0, L[lw] -> 0}; 
\end{lstlisting}
Here we assume all parameters to be real, remove any complex conjugation (\verb"conj") and use the Landau gauge (\verb"RXi[_]->0"), then we turn off again gauge kinetic mixing and take the VEVs of $\eta$ and $\bar\eta$ to be equal. In the fourth line, we parametrise $v_d$ and $v_u$ as usual in terms of $v$ and $\tan\beta$. Finally, we set the parameters $\kappa$, $T_\kappa$, $\lambda$, $T_\lambda$ and $L_\xi$ to zero. We can now solve the tadpole equations, stored by \SARAH in {\tt TadpoleEquations[Eigenstates]}, with respect to the parameters $m_{H_d}^2$, $m_{H_u}^2$, $m_{\eta}^2$, $m_S^2$ and $\xi$ using the aforementioned assumptions:
\begin{lstlisting}[style=mathematica]
sol = Simplify[
 Solve[(TadpoleEquations[EWSB] //. assumptions) == 0, 
   {mHd2, mHu2, mC12, lw, mS2}][[1]]
               ]; 
\end{lstlisting}
We have saved the solution in the variable {\tt sol} for further usage. 

\paragraph{Obtaining a 750~GeV pseudo-scalar}
We use the solution and our assumptions to get simpler expressions for the mass matrix of the CP-even (called {\tt hh}) and CP-odd (called {\tt Ah}) scalars: 
\begin{lstlisting}[style=mathematica]
mH = FullSimplify[MassMatrix[hh] /. sol //. assumptions]
mA = FullSimplify[MassMatrix[Ah] /. sol //. assumptions]
\end{lstlisting}
These matrices can be expressed as
\begin{equation}
m^2_H \simeq  \left( \begin{array}{cc}
     m_H^{2,\text{MSSM}} & m_H^{2,\text{mix}} \\
     (m_H^{2,\text{mix}})^T & m_H^{2,X}
                    \end{array}\right)\,, \quad
m^2_A \simeq  \left( \begin{array}{cc}
     m_A^{2,\text{MSSM}} & 0 \\
     0 & m_A^{2,X}
  \end{array}\right)
\end{equation}
with 
\begin{align}
m_H^{2,\text{MSSM}} &\simeq \left(\begin{array}{cc}
t_\beta B(\mu )+\frac{v^2 \left(g_1^2+g_2^2+g_X^2\right)}{4 \left(t_\beta^2+1\right)} & -B(\mu )-\frac{t_\beta v^2 \left(g_1^2+g_2^2+g_X^2\right)}{4 \left(t_\beta^2+1\right)} \\
 -B(\mu )-\frac{t_\beta v^2 \left(g_1^2+g_2^2+g_X^2\right)}{4 \left(t_\beta^2+1\right)} & \frac{B(\mu )}{t_\beta}+\frac{t_\beta^2 v^2 
\left(g_1^2+g_2^2+g_X^2\right)}{4 \left(t_\beta^2+1\right)}      
               \end{array}\right) \, , \\
m_H^{2,\text{mix}} &\simeq \left(\begin{array}{ccc}
\frac{1}{4} g_X v \sqrt{\frac{4 M_{Z'}^2-g_X^2 v^2}{2 t_\beta^2+2}} & -\frac{1}{4} g_X v \sqrt{\frac{4 M_{Z'}^2-g_X^2 v^2}{2 t_\beta^2+2}} & 0 \\
-\frac{1}{4} g_X t_\beta v \sqrt{\frac{4 M_{Z'}^2-g_X^2 v^2}{2 t_\beta^2+2}} & \frac{1}{4} g_X t_\beta v \sqrt{\frac{4 M_{Z'}^2-g_X^2 v^2}{2 t_\beta^2+2}} & 0 \\                
               \end{array}\right) \, , \\
m_A^{2,\text{MSSM}}  & \simeq   \left(
\begin{array}{cc}
 t_\beta B(\mu ) & B(\mu ) \\
 B(\mu ) & \frac{B(\mu )}{t_\beta}
\end{array}
\right)  \, . 
\end{align}
We omit here the analytical expressions for $m_A^{2,X}$ and $m_H^{2,X}$ because of their length and since they are not needed for the following brief discussion. The mass matrix for the CP-odd states is block-diagonal since the MSSM part is unchanged, while we have mixing in the CP-even sector among all five components.\footnote{The mixing between the MSSM scalars and $S$ is vanishing here only because of our simplifying assumption $\lambda=0$ but is non-zero in general.} The additional $D$-Terms can be found in the MSSM block, $m_H^{2,\text{MSSM}}$. 
This also explains our choice of a pseudo-scalar as the resonance behind the diphoton excess: the tree-level mixing between the scalar singlet and the doublets would cause tree-level decays of a $750$~GeV scalar into all kinds of SM particles. In particular, those into $WW$ and $ZZ$ are constrained and could easily spoil our setup as an explanation of the excess in this model. Of course, we have to check whether it is possible to obtain a pseudo-scalar of the correct mass, and get the corresponding scalar sufficiently heavy so as to escape detection. For that purpose, we calculate the eigenvalues of the lower $3 \times 3$ block of the pseudo-scalar mass matrix, and fix $B_S$ by demanding to have a pseudo-scalar of the correct mass:
\begin{lstlisting}[style=mathematica]
Eigenvalues[Take[mA, {3, 5}, {3,5}]];
sol750 = Solve[%[[2]] == M750^2, B[MS]][[1]]; 
\end{lstlisting}
We now make an arbitrary choice for the numerical values of the remaining parameters, except $m_{\bar \eta}^2$ and $M_{Z'}$,
\begin{lstlisting}[style=mathematica]
num =  {g1 -> 0.36, g2 -> 0.65, v -> 246,
  B[\[Mu]] -> 10^6, \[Mu] -> 1000, 
  TB -> 20, gX -> 0.5, MS -> -100, xS -> 500, 
  T[lambdaC] -> -200, lambdaC -> -0.2, M750 -> 750}
\end{lstlisting}
and calculate all CP-even and CP-odd mass eigenvalues for specific values of $m_{\bar \eta}^2$ and $M_{Z'}$:
\begin{lstlisting}[style=mathematica]
Sqrt /@ Eigenvalues[
  MassMatrix[Ah] //. assumptions //. sol /. sol750  //. assumptions /. 
     num /. mC22 -> 10^6 /. MZp -> 3000 ]
Sqrt /@ Eigenvalues[
  MassMatrix[hh] //. assumptions //. sol /. sol750 //. assumptions /. 
     num /. mC22 -> 10^6 /. MZp -> 3000 ]
\end{lstlisting}
The results are 
\begin{lstlisting}[style=mathematica]
{4477.72, 1792.7, 750., 6.60725*10^-6, 0.}
{4477.74, 3319.15, 2797.53, 822.054, 94.6205}
\end{lstlisting}
Thus, as expected, we have two massless (up to numerical errors) states in the CP-odd sector, which are the neutral Goldstone bosons to be eaten by the $Z$ and $Z^\prime$ gauge bosons, accompanied by a particle with a mass of $750$~GeV. In the  scalar sector we find the lightest state with a mass very close to $M_Z$ and another scalar below $1$~TeV. However, checking the composition of the $750$ and $825$~GeV particles via
\begin{lstlisting}[style=mathematica]
Eigensystem[MassMatrix[hh] //. assumptions //. sol /. sol750 //. assumptions /. 
     num /. mC22 -> 10^6 /. MZp -> 3000 ][[2, -2]]
 {0., 0., -0.704932, -0.704932, -0.0783774}     
Eigensystem[MassMatrix[Ah] //. assumptions //. sol /. sol750 //. assumptions /. 
     num /. mC22 -> 10^6 /. MZp -> 3000 ][[2, -3]]     
 {0., 0., -0.299786, -0.299786, 0.90568}    
\end{lstlisting}
we see that the CP-odd state is, as expected, mainly a singlet while the CP-even one is mainly a $X$-Higgs (composed by $\phi_\eta$ and $\phi_{\bar\eta}$). That looks already very promising. 

\begin{figure}[hbt]
\centering
\includegraphics[width=0.49\linewidth]{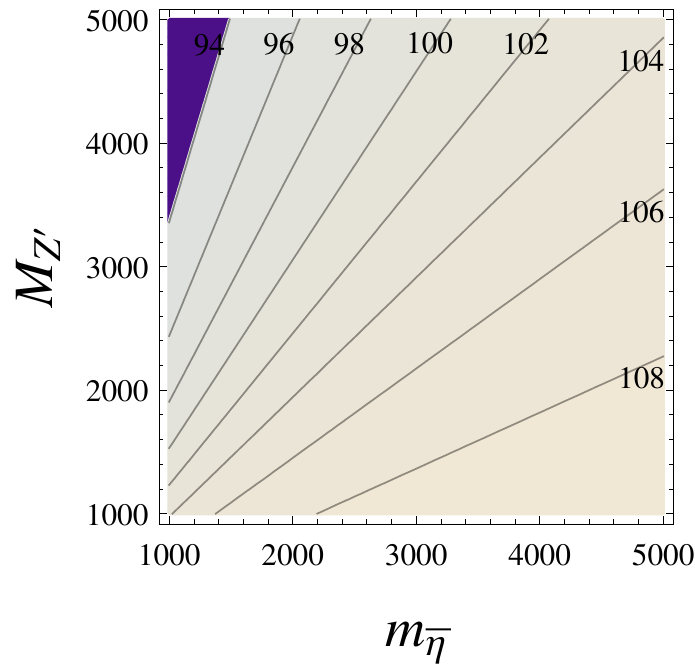}
\caption{Contours of the mass of the lightest CP-even scalar in the model as a function of $M_{Z'}$ and $m_{\bar \eta}$.}
\label{fig:NonDecouplingDTerms}
\end{figure}

\paragraph{Higgs mass enhancement via non-decoupling $D$-terms}
Now, we want to confirm that one gets non-decoupling $D$-terms in this model which cause an enhancement of the tree-level mass of the SM-like scalar. For this purpose, we define a simple function which calculates the lightest CP-even mass for input values of $m_{\bar \eta}$ and $M_{Z'}$,
\begin{lstlisting}[style=mathematica]
TreeMH[msoft_, mzp_] := 
  Sqrt[Eigenvalues[mH /. sol750 /. num  
    /. mC22 -> msoft^2 /. MZp -> mzp ][[-1]]]; 
\end{lstlisting}
and create a contour plot using this function. The result is depicted in \cref{fig:NonDecouplingDTerms}, where one sees that for $m_{\bar \eta} \gg M_{Z'}$ it is indeed possible to find a tree-level mass well above $100$~GeV, while for  $m_{\bar \eta} \ll M_{Z'}$ the tree-level mass approaches $M_Z$.

\paragraph{Is there a second light scalar?}
One can now start to play also with the values we have chosen for {\tt num} to see how the eigenvalues of both matrices change. One finds, for instance, that it is also possible to get a second, relatively light scalar in the model. With the values
\begin{lstlisting}[style=mathematica]
num = {..., lambdaC -> -0.3, MS -> -100, xS -> 3500, 
  TL -> -225,  MZp ->2500, mC22->100}
\end{lstlisting}
we find a tree-level mass of 38~GeV for the lightest CP-even scalar, which is mainly a mixture of $\eta$ and $\bar\eta$. It will be interesting to see if this scenario is still in agreement with all experimental constraints and how important the loop corrections are.  

\paragraph{How to obtain a broad width?}
So far, we have not considered the total decay width of the $750$~GeV scalar. The experimental data shows a slight preference for a rather large width of about $40$~GeV, which is not easy to accommodate in weakly coupled models, typically requiring a large branching ratio into invisible states. Therefore, it would be interesting to see if this can be realised in this model. There are three possibilities for invisible decays: (i) neutralinos, (ii) (heavy) neutrinos, (iii) sneutrinos. We are going to consider the third option here. For this purpose, we have to check two ingredients: can the mass of the sneutrinos be sufficiently light and how can the coupling to the $750$~GeV scalar be maximised? To get a feeling for that, we first consider the mass matrix of the CP-even and CP-odd sneutrinos. We assume that flavour and left-right mixing effects are negligible. In that case, it is sufficient to take a look only at the (4,4) entry of the mass matrices:
\begin{lstlisting}[style=mathematica]
MassMatrix[SvIm][[4, 4]] 
MassMatrix[SvRe][[4, 4]] 
\end{lstlisting}
After some simplifications, we get the following expressions from \SARAH :
\begin{align}
M^2_{\tilde \nu^I,\tilde \nu^R} &= \frac{1}{8} \frac{g_ {X}^{2} v^{2}}{(1 + t_\beta^{2})} -\frac{1}{8} \frac{g_ {X}^{2} t_\beta^{2}  v^{2}}{(1 + t_\beta^{2})} +m_ {\nu,{1 1}}^{2} +\frac{Y_{x,{1 1}}^{2}}{4 g_X^2} (4 M_{Z'}^{2} - g_{X}^{2} v^{2})  \nonumber \\ 
 &\phantom{={}} \pm \left(\frac{v_S Y_{x,{1 1}}}{2\sqrt{2} g_ {X}} {\lambda}_ {C} \sqrt{4 M_{Z'}^2-g_X^2 v^2}  + \frac{1}{2 g_{X}} \sqrt{4 M_{Z'}^2-g_X^2 v^2} T_{x,{1 1}} \right) .
\end{align}
We see that the terms in the second line, $\propto T_x M_{Z'}$ and
$\propto v_S \lambda_C Y_x$, induce a mass splitting between the
CP-even and CP-odd states.  Thus, in order to have the decay $A \to
\tilde{\nu}^I \tilde{\nu}^R$ kinematically allowed, these terms must
be individually small or cancel each other. In addition, one has to
compensate the large terms $\sim M_{Z'}$ in order to get sufficiently
light sneutrinos. This could be done by assuming a negative $m_
{\nu}^{2} = -\frac{1}{4 g_X^2} (4 M_{Z'}^{2} - g_{X}^{2} v^{2}
)Y_{x,{1 1}}^{2}$. Of course, we must check whether this leads to
spontaneous $R$-parity breaking via sneutrino VEVs, and for this
purpose one can use \Vevacious, see below.

We can now check the vertex $A \tilde{\nu}^I \tilde{\nu}^R$ using the
same assumptions:
\begin{lstlisting}[style=mathematica]
Vertex[{Ah, SvIm, SvRe}][[2, 1]] 
\end{lstlisting}
and we obtain after some simplification
\begin{align}
\frac{1}{2} \frac{1}{\sqrt{2} g_{X}} {\lambda}_{C} \sqrt{4 M_{Z'}^2-g_X^2 v^2} Y_{x,{1 1}} Z_{{3 5}}^{A}  + {\lambda}_{C} v_S Y_{x,{1 1}} Z_{{3 3}}^{A}  
- \sqrt{2} Z_{{3 4}}^{A} T_{x,{1 1}} \,.
\end{align}
If the pseudo-scalar is a pure singlet, only the term $\propto Z_{35}^A$ contributes. This term is independent of $v_S$ and $T_x$, i.e.\ we can reduce the mass splitting between the CP-even and CP-odd sneutrinos by adjusting these parameters without having a negative impact on the coupling strength to the $750$~GeV scalar. 

\subsubsection{Vector-like sector}
Before we finish the analytical discussion of the masses,  we briefly discuss the extended matter sector. The mass matrices responsible for the mixing between the SM fermions and the vector-like fermions can be obtained from \SARAH by calling
\begin{lstlisting}[style=mathematica]
MassMatrix[Fe] 
MassMatrix[Fu] 
\end{lstlisting}
which return
\begin{equation}
m_e = \left(
\begin{array}{cc}
 \frac{v_d Y_e}{\sqrt{2}} &  -\frac{v_d Y'_e}{\sqrt{2}} \\
  0 & \frac{\lambda_e v_S}{\sqrt{2}}+M_e \\
\end{array}
\right) \,,\quad 
m_u = \left(
\begin{array}{cccc}
 \frac{v_u Y_u}{\sqrt{2}} &\frac{v_u Y'_u}{\sqrt{2}} \\
 0 & \frac{\lambda_u v_S}{\sqrt{2}}+M_u \\
\end{array}
\right) .
\end{equation}
Thus, for large $\lambda_i$ ($i=u,e$) and $v_S$, there are two important sources for the mass of the vector-like states. The full sfermion matrices containing the new scalars are too lengthy to be shown here. We only check the new mass matrix for one generation of the vector-like selectrons which are the 7th and 10th gauge eigenstates. We can pick the values via 
\begin{lstlisting}[style=mathematica]
({{M[[7, 7]], M[[7, 10]]}, {M[[7, 10]], M[[10, 10]]}} /. 
      M -> (MassMatrix[Se]))
\end{lstlisting}
and obtain by setting all parameters to be diagonal
\begin{equation}
\left(
\begin{array}{cc}
 \frac{ \tilde D+4 \left(t_\beta^2+1\right) \left(\lambda_e^2 v_S^2+2 \sqrt{2} \lambda_e M_e v_S+2 m^2_{E}+2 M_e^2\right)}{8 \left(t_\beta^2+1\right)} & B_E+\lambda_e \left(\frac{\lambda_X M_{Z'}^2}{4 g_X^2}-\frac{\lambda_X v^2}{16}+\xi+\sqrt{2} M_S v_S\right) \\
 B_E+\lambda_e \left(\frac{\lambda_X M_{Z'}^2}{4 g_X^2}-\frac{\lambda_X v^2}{16}+\xi+\sqrt{2} M_S v_S\right) & 
 \frac{4 \left(t_\beta^2+1\right) \left(\lambda_e^2 v_S^2+2 \sqrt{2} \lambda_e M_e v_S+2 m^2_{\bar E}+2   M_e^2\right)-\tilde D}{8 \left(t_\beta^2+1\right)} \\
\end{array}
\right) 
\end{equation}
where we have defined $\tilde D = \left(t_\beta^2-1\right) v^2 \left(2 g_1^2+g_X^2\right)$. There is a potentially dangerous term $\lambda_e \xi$ which rapidly increases for increasing $\lambda_e$. To keep all scalar masses positive, it is necessary to choose a rather large $B_E$ as well. Therefore, we are going to choose always 
\begin{equation}
\label{eq:CondB}
B_E = -  \lambda_e (\xi + \sqrt{2} M_S v_S) \,, \quad B_U = - \lambda_u (\xi + \sqrt{2} M_S v_S)
\end{equation}
in our numerical study to circumvent tachyonic scalars. 

\subsubsection{RGEs and gauge kinetic mixing}
\begin{figure}[hbt]
\centering
\includegraphics[width=0.5\linewidth]{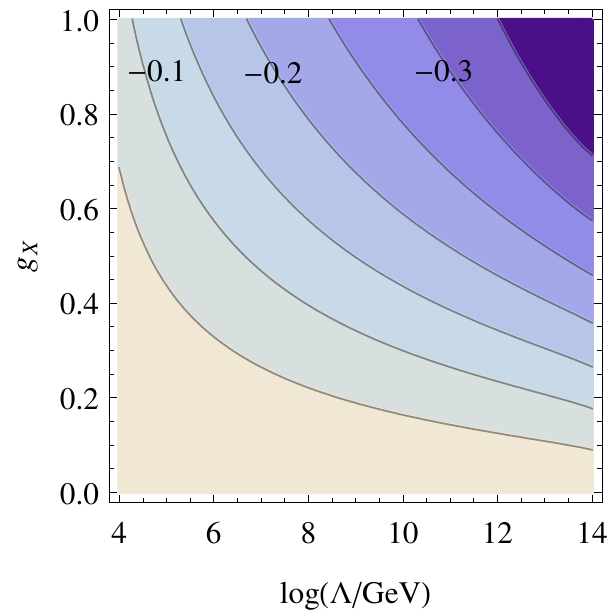}
\caption{Gauge kinetic mixing $\tilde{g}$ at the SUSY scale as a function of the high energy scale $\Lambda$, where it is assumed to vanish, and of the coupling $g_X(\Lambda)$.}
\label{fig:GKM}
\end{figure}

We have so far made the simplifying assumption that gauge kinetic mixing vanishes. However, if the two Abelian gauge groups are not orthogonal, kinetic mixing would be generated via RGE running even if it vanishes at some energy scale. Thus, one of the first checks on the RGEs of the model we can make is whether the two $U(1)$ gauge groups are orthogonal. For this purpose, we first calculate the one-loop RGEs with \SARAH via 
\begin{lstlisting}[style=mathematica]
CalcRGEs[TwoLoop->False];
\end{lstlisting}
We have chosen one-loop RGEs only to save time. Without the \code{TwoLoop->False} flag, the full two-loop RGEs would have been calculated automatically. We can now check the entries in {\tt BetaGauge} and find
\begin{align}
16 \pi^2 \beta_{g_Y} &= 15 g_Y^3 + 15 g_Y \, g_{YX}^2 + 16\sqrt{\frac35} g_Y \, g_{YX} \, g_X
+ 32\sqrt{\frac35} g_Y^2 \, g_{XY} \nonumber \\
 &\quad + 16\sqrt{\frac35} g_{YX}^2 \, g_{XY} + 15 g_{YX} \, g_X \, g_{XY} +  15 g_Y \, g_{XY}^2 , \\
16 \pi^2 \beta_{g_X} &= 15 g_{YX}^2 \, g_X + 15 g_X^3 + 16\sqrt{\frac35} g_Y \, g_X \, g_{XY} + 15 g_X \, g_{XY}^2  \nonumber \\
 &\quad 
+ g_{YX} \left(32\sqrt{\frac35} g_X^2 + 15 g_Y \, g_{XY} + 16\sqrt{\frac35} g_{XY}^2\right) , \\
16 \pi^2 \beta_{g_{XY}} &= 15 g_{YX}^3 + 32\sqrt{\frac35} g_{YX}^2 \, g_X + 15 g_{YX} \, g_X^2 
+ g_Y^2 \left(15 g_{YX} + 16\sqrt{\frac35} g_X\right)  \nonumber \\
 &\quad 
+ g_Y \left(16\sqrt{\frac35} g_{YX} \, g_{XY} + 15 g_X \, g_{XY}\right) , \\
16 \pi^2 \beta_{g_{YX}} &= 15 g_{YX}^3 + 32\sqrt{\frac35} g_{YX}^2 \, g_X + 15 g_{YX} \, g_X^2 
+ g_Y^2 \left(15 g_{YX} + 16\sqrt{\frac35} g_X\right)  \nonumber \\
 &\quad 
+ g_Y \left(16\sqrt{\frac35} g_{YX} \, g_{XY} + 15 g_X \, g_{XY}\right) .
\end{align}
The standard normalisation factor $\sqrt{5/3}$ for the hypercharge has
been included. One can see that the $\beta$-functions for $g_{YX}$ and
$g_{XY}$ are non-zero even in the limit $g_{XY}, g_{YX} \to 0$,
i.e.\ these couplings will be induced radiatively.  Thus, in general
one has not only two couplings $g_1$ and $g_X$ in this model, but a
gauge coupling matrix $G$ defined as
\begin{equation}
G = \left(\begin{array}{cc} g_{YY} & g_{XY} \\ g_{YX} & g_{XX} \end{array} \right)\,.
\end{equation}
In the limit of vanishing kinetic mixing, $g_{YX} = g_{XZ} = 0$, the relations $g_{YY} = g_1$ and $g_{XX} = g_X$ hold. Even if gauge kinetic mixing is present, one has the freedom to perform a change in basis to bring $G$ into a particular form. The most commonly considered cases are the symmetric basis with $g_{XY} = g_{YX}$ and the triangle basis with $g_{YX} = 0$.  The triangle basis has the advantage that the new scalars do not contribute to the electroweak VEV, and the entire impact of gauge kinetic mixing is encoded in one new coupling $\tilde{g}$. The relation between $g_{ij}$ ($i,j=X,Y$) and $g_1$, $g_X$, $\tilde{g}$ are \cite{O'Leary:2011yq} 
\begin{align}
g_1 &= \frac{g_{YY} g_{XX} - g_{XY} g_{YX}}{\sqrt{g_{XX}^2 + g_{XY}^2}}\,, \\
g_X &= \sqrt{g_{XX}^2 + g_{XY}^2}\,, \\
\tilde{g} &= \frac{g_{YX} g_{XX} + g_{YY}  g_{XY}}{\sqrt{g_{XX}^2 + g_{XY}^2}}\,.
\end{align}
It is interesting to see how large $\tilde{g}$ is naturally. With `naturally' we mean under the assumption that the off-diagonal $g_{YX}$ and $g_{XY}$ couplings vanish at some high scale $\Lambda$ and are generated by RGE running down to the SUSY scale. Thus, in this setup, the size of gauge kinetic mixing is a function of $\Lambda$ and $g_X$ at this scale. We can write a simple \Mathematica function to get a feeling for the off-diagonal gauge couplings:
\begin{lstlisting}[style=mathematica]
<< "Output/U1yMSSM/RGEs/RunRGEs.m";
RunningGKM[scale_, gXIN_] := Block[{},
   logS = scale;
   runUp = RunRGEs[{g1 -> 0.45}, 3, logS, TwoLoop -> False][[1]];
   runDown = RunRGEs[{g1 -> (g1[logS] /. runUp), gX -> gXIN}, logS, 3, TwoLoop -> False][[1]];
   g1run = Sqrt[3/5] g1[3] /. runDown;
   gXrun = gX[3] /. runDown;
   g1Xrun = Sqrt[3/5] g1X[3] /. runDown;
   gX1run = gX1[3] /. runDown;
   g1out = (g1run*gXrun - g1Xrun gX1run)/Sqrt[gXrun^2 + gX1run^2];
   gXout = Sqrt[gXrun^2 + gX1run^2];
   g1Xout = (g1Xrun gXrun + g1run gX1run)/Sqrt[gXrun^2 + gX1run^2];
   Return[{g1out, gXout, g1Xout}];
   ];
\end{lstlisting}
In the first line, we load the file written by \SARAH which provides the RGEs in a form which \Mathematica can solve. This file also contains the function {\tt RunRGEs} that can be used to solve the RGEs numerically. As boundary condition, we used $g_1 = 0.45$ at the scale 1~TeV. After the running we rotate the couplings to the basis where $g_{XY}$ vanishes. We can make a contour plot via 
\begin{lstlisting}[style=mathematica]
ContourPlot[RunningGKM[lambda, gx][[3]], 
    {lambda, 4, 17}, {gx, 0, 1},  ContourLabels -> True] 
\end{lstlisting}
and get the result shown in \cref{fig:GKM}.
Thus, we find that at the SUSY scale the gauge mixing coupling $\tilde{g}$ is negative and not much smaller than an ordinary gauge coupling unless $\Lambda$ is assumed to be very small. 

\subsubsection{Boundary conditions and free parameters}
\label{sec:U1xBoundaries}
For the subsequent numerical analysis we are going to assume some simplified boundary conditions applied at the SUSY scale:
\begin{align}
& m_{ q}^2 =  m_{ d}^2 =  m_{ u}^2 =  m_{ l}^2 =  m_{ e}^2 =  m_{ E}^2 =  m_{ \bar E}^2 =  m_{ U}^2 =  m_{ \bar U}^2 \equiv {\bf 1} m^2_{SUSY} \\
& \lambda_e = {\bf 1} \lambda_E\,, \quad \lambda_u \equiv {\bf 1} \lambda_U \\
& M_e = {\bf 1} M_E\,, \quad M_u \equiv {\bf 1} M_U \\
& T_i = A_0 Y_i \; (i=\{u,e,d\})\,,\quad  T'_i = A_0 Y'_i \; (i=\{u,e\})\,,\quad T_i = A_0 \lambda_i \; (i=\{U,E\})\\ 
& T_\lambda = A_\lambda \lambda\,,\quad T_X = A_X \lambda_X \\
& M_1 = M_2 = \frac12 M_3 = M_X \equiv M_\lambda 
\end{align}
In addition, we can set $Y_\nu = 0$ since this parameter is highly
constrained to be small by the small neutrino masses. In addition, we
set the mixing parameters $m_{eE}^2$, $m_{uU}^2$, $\tilde{M}_E$,
$\tilde{M}_U$, $\tilde{B}_E$, $\tilde{B}_U$, $M_{1X}$ to zero and also
assume vanishing $\lambda$, $\kappa$, and $T_\kappa$. However, we
stress that this is just done to keep the following discussion short
and simple. All effects of these parameters can be included without
any additional efforts. Thus, the free parameters mainly considered in
the following are
\begin{eqnarray*}
& m_{SUSY}, M_\lambda, \mu, B_\mu, A_0, \tan\beta, & \\
& g_X, g_{1X}, M_{Z'}, m_{\bar \eta}, \tan\beta_X, \lambda_X, A_X, Y_x, & \\
& M_S, B_S, v_s, A_\lambda, & \\
& \lambda_E, \lambda_U, M_E, M_U, Y'_u, Y'_e. &
\end{eqnarray*}
The tadpole equations are solved for $m_{H_d}^2$, $m_{H_u}^2$, $m_{\eta}^2$, $m_S^2$ and $\xi$, while $B_E$ and $B_U$ are fixed via Eq.~\eqref{eq:CondB}.

\subsection{Analysis of the important loop corrections to the Higgs mass}
We now turn to the numerical analysis of this model. 
In the first step, we have written a {\tt SPheno.m} file for the boundary conditions, see \cref{sec:U1xBoundaries}, and generated the \SPheno code with the \SARAH command 
\begin{lstlisting}[style=mathematica]
MakeSPheno[]; 
\end{lstlisting}
We copy the generated Fortran code to a new sub-directory of {\tt SPheno-3.3.8} and compile it via
\begin{lstlisting}[style=terminal]
$ cd $PATH/SPheno-3.3.8
$ mkdir U1xMSSM
$ cp $PATH/SARAH/Output/U1xMSSM/EWSB/SPheno/* U1xMSSM/
$ make Model=U1xMSSM 
\end{lstlisting}
We now have an executable {\tt SPhenoU1xMSSM} which expects the input
parameters from a file called {\tt LesHouches.in.U1xMSSM}. The \SPheno
code provides many important calculations which would be very
time-consuming to be performed `by hand' for this model, but could be
expected to be relevant. A central point is the calculation of the
pole mass spectrum at the full one-loop (and partially two-loop) level. In
particular, the loop corrections from the vector-like states are known
to be very important. However, the focus in the literature has usually
been only on the impact on the SM-like Higgs. We can automatically go
beyond that and consider the corrections to the 750~GeV state as
well. Moreover, \SPheno calculates all additional two-loop
corrections in the gaugeless limit including all new matter
interactions. Thus, we can check the impact of the vector-like states
even at two-loop level. These effects have not been studied in any of
the SUSY models proposed so far to explain the diphoton excess. Of
course, \SPheno also makes a very precise prediction for the diphoton
and digluon decay rate of all neutral scalars as described in
\cref{sec:diphotoncalc}, and it checks for any potential decay
mode. Thus, it is impossible to miss any important decay as sometimes
has happened in the literature when discussing the diphoton excess.
Finally, there are also other important constraints for this model
like those from flavour observables or Higgs coupling
measurements. As will be shown in the next sections, all of this can
be checked automatically with \SPheno and tools interacting with it.

If not mentioned otherwise, we make the following choice for the input parameters
\begin{eqnarray*}
&m_{SUSY} = 1.5~\TeV\,,M_\lambda=1~\TeV\,,&\\
&\tan\beta = 20\,,\tan\beta_x = 1\,,g_X=0.5\,,M_{Z'} = 3~\TeV, m_{\bar\eta} = 2~\TeV, &\\
&\mu=1~\TeV\,,B_\mu = (1~\TeV)^2\,,v_S=0.5~\TeV\,,M_S=-0.1~\TeV\,,B_S=3.895~\TeV^2, &\\
&\lambda_X = -0.2 \,,A_X = 1~\TeV\,,\lambda_E=\lambda_U=1\,,M_E = 0.4~\TeV\,,M_U=1~\TeV. & 
\end{eqnarray*}

\subsubsection{New loop corrections to the SM-like Higgs}
\begin{figure}[htb]
 \includegraphics[width=0.45\linewidth]{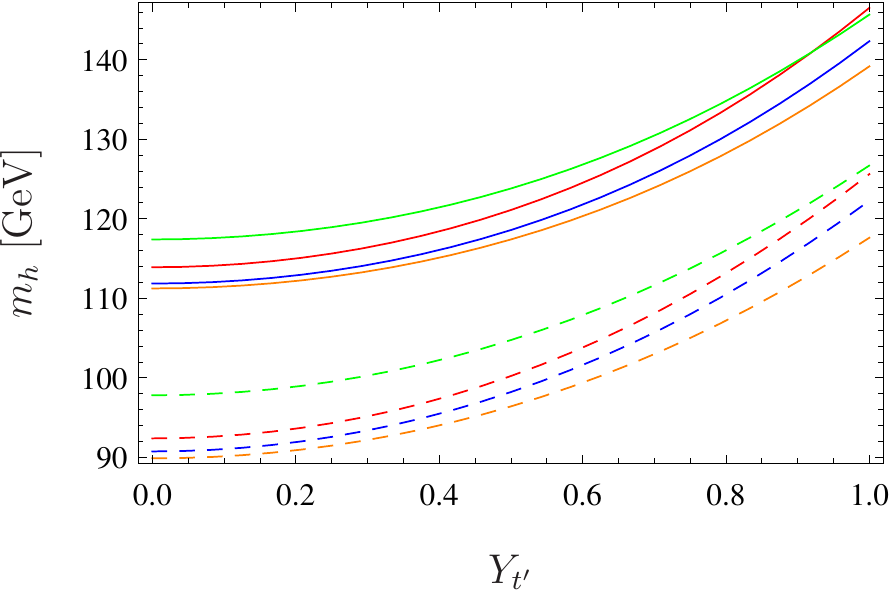}\hfill
 \includegraphics[width=0.45\linewidth]{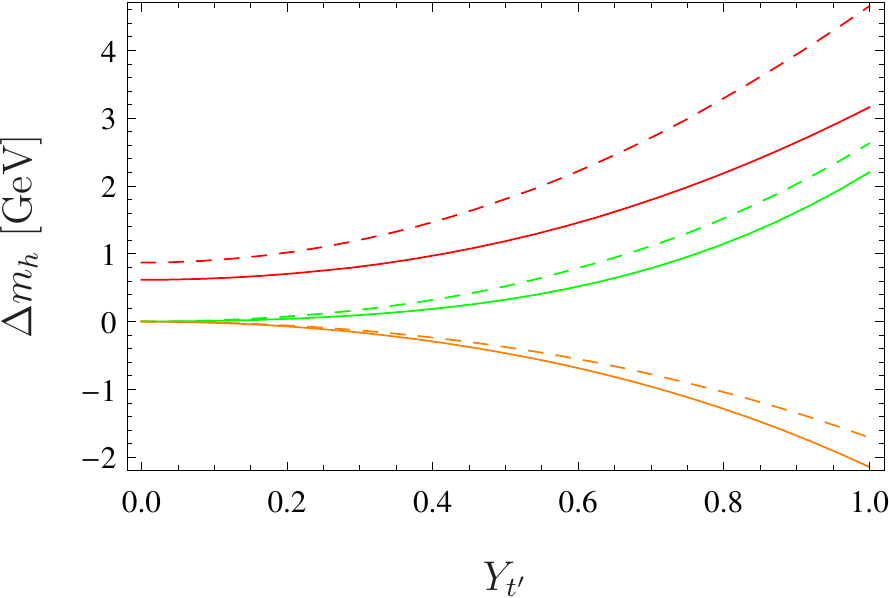} 
  \includegraphics[width=0.45\linewidth]{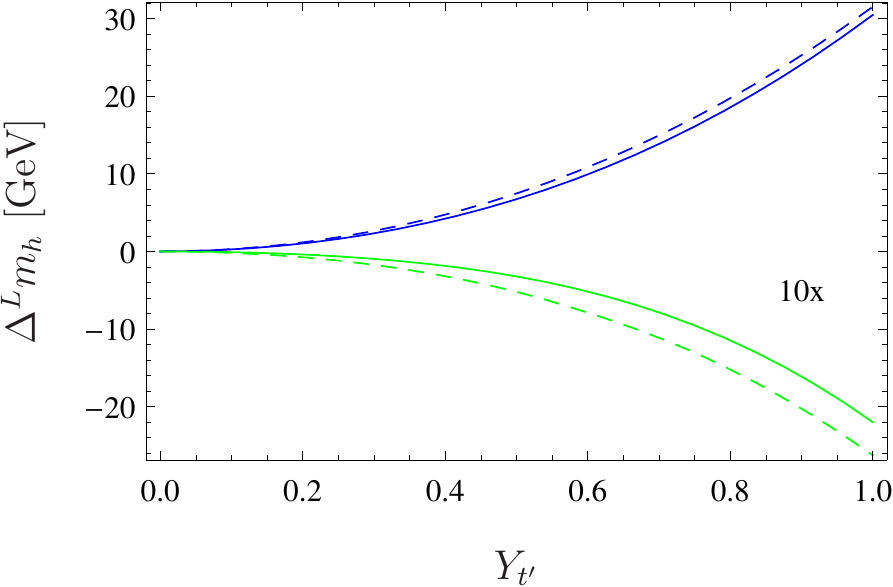}\hfill
  \includegraphics[width=0.45\linewidth]{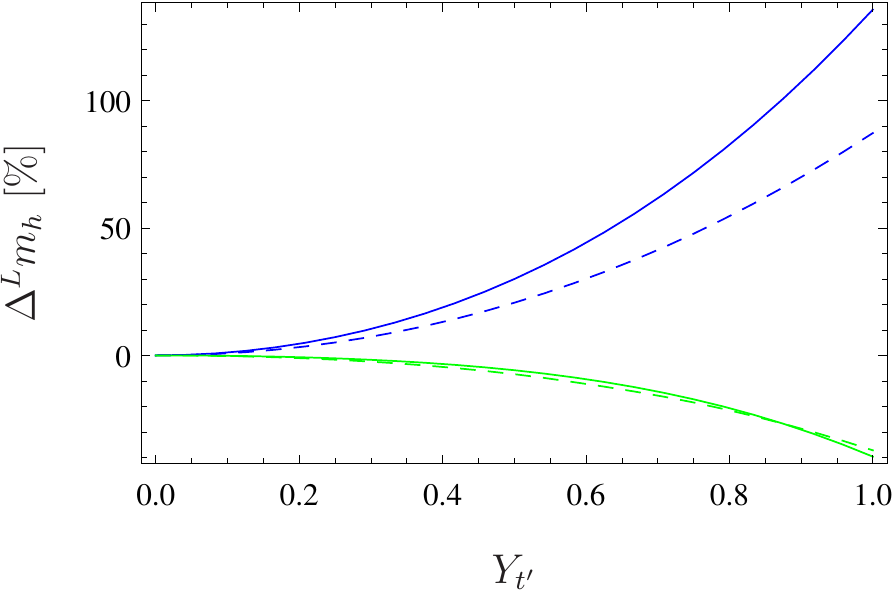} 
 \caption{
Top left: light Higgs mass as a function of $Y_{t'}$ (which corresponds to $Y'_{u,33}$ in this model) with all other entries of $Y'_u$ vanishing. The red line corresponds to the effective potential calculation at one-loop, orange is the one-loop corrections with external momenta but neglecting the new threshold correction stemming from vector-like states,  blue is the full one-loop calculation including the momentum dependence and all thresholds, and green includes the dominant two-loop corrections together with the full one-loop correction. Top right: impact of the threshold corrections (red), the momentum dependence at one-loop (orange) and the two-loop corrections (green), given as the difference $\Delta m_h=m_h - m_h({\text{1L},p^2\neq 0,\text{all thresholds}})$. Bottom left: absolute size of the one- (blue) and two-loop (green) corrections stemming from the vector-like states. For better readability we re-scaled the two-loop corrections by a factor of 10. 
Bottom right: relative importance of the one- (blue) and two-loop (green) corrections normalised to the size of the purely MSSM-like corrections. 
The solid lines are for $\tan\beta=10$ and the dashed ones are for $\tan\beta=2$.  Here, a mass of $1$~TeV for the vector-like quarks was assumed. These plots are taken from Ref.~\cite{Nickel:2015dna}.}
 \label{fig:YtTB}
\end{figure}

In this model we have two new important loop corrections to the SM-like Higgs: (i) the corrections from vector-like states, proportional to $Y'_u$ and $Y'_e$, and (ii) the new corrections from the extended gauge sector. The corrections from vector-like (s)tops up to two-loop have been discussed in detail in Ref.~\cite{Nickel:2015dna} using the \SARAH/\SPheno framework. There are several important effects which are often neglected in studies of vector-like states which only make use of the one-loop effective potential: the momentum effects at one-loop, the two-loop corrections, and the shift of the top-Yukawa coupling. In general, the user does not need to worry about these details because \SARAH/\SPheno take care of them automatically. However, it might be interesting to have an intuitive feeling about the size of the different effects. Since it demands some `hacking' of the code to disentangle the calculation in that way, we are not making this analysis here, but we briefly summarise the main results of Ref.~\cite{Nickel:2015dna} in \cref{fig:YtTB}. 
We see that all these effects can alter the Higgs mass by several GeV. Thus, an estimated uncertainty of about $2$--$3$~GeV when using only the one-loop effective potential approximation is usually over optimistic.

\begin{figure}[hbt]
\includegraphics[width=0.48\linewidth]{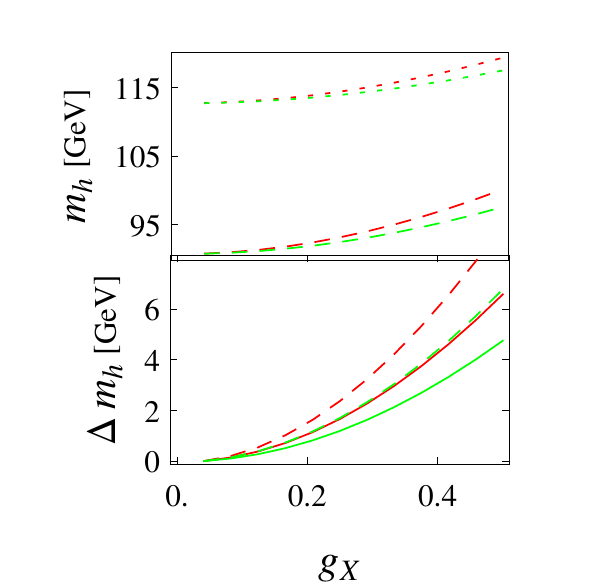} \hfill
\includegraphics[width=0.48\linewidth]{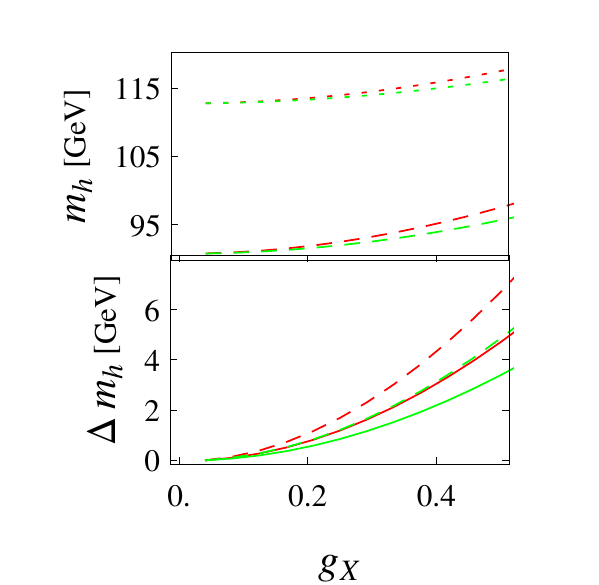}
\caption{Mass of the SM-like Higgs as a function of $g_X$ at tree-level (dashed) and one-loop (full line). The red lines are without gauge-kinetic mixing, for the green ones we set $g_{1X} = \frac15 g_X$. $M_{Z'}$ on the left is 3~TeV and 4~TeV on the right. On the bottom we show the difference $\Delta m_h \equiv m_h(g_X) - m_h(g_X = 0)$ at tree-level (dashed) and including loop corrections (full) for the case with gauge kinetic mixing (green) and without (red). }
\label{fig:MH_gX_tree_loop}
\end{figure}

Furthermore, in models with non-decoupling $D$-terms the new loop corrections are usually neglected in the literature. Therefore, we are going to check whether this is a good approximation or not. For this purpose we show the SM-like Higgs pole mass at tree and one-loop level as a function of $g_X$ for two different values of $M_{Z'}$. Since \SPheno performs the two-loop corrections in the gaugeless limit, additional corrections from the extended gauge sector are not included at two-loop, and we concentrate on the one-loop effects here. For this purpose, we use the different flags in the {\tt Les Houches} input file from \SPheno to turn the corrections at the different loop levels on or off:
\begin{lstlisting}[style=file]
Block SPhenoInput   # SPheno specific input 
 ...
  7  A              # Skip 2-loop Higgs corrections 
 55  B              # Calculate loop corrected masses  
\end{lstlisting}
Here, {\tt A} and {\tt B} are either {\tt 1} or {\tt 0}. With flag {\tt 55} the entire loop-corrections to all masses can be turned on ({\tt 1}) or off ({\tt 0}), while flag {\tt 7} only skips ({\tt 1}) or includes ({\tt 0}) the two-loop corrections in the Higgs sector. The results are shown in \cref{fig:MH_gX_tree_loop}. All scans have been performed using the \Mathematica package {\tt SSP} \cite{Staub:2011dp} for which \SARAH already writes a template input when generating the \SPheno code ({\tt SSP\_Template.m.U1xMSSM}) for a given model. 
We see that the tree-level mass rises quickly with increasing $g_X$. However, for both $M_{Z'}$ values this shift is compensated to some extent when one-loop corrections are included. Thus, the inclusion of non-decoupling $D$-terms only at tree-level would overestimate the positive effect on the SM-like Higgs mass by 20--30\%. In addition, we also see that off-diagonal gauge couplings of a realistic size as consequence of gauge kinetic mixing reduce the positive effect from the non-decoupled $D$-terms on the Higgs mass by a few GeV.

\subsubsection{Loop corrections to the 750~GeV scalar}
\begin{figure}[hbt]
\centering
\includegraphics[width=0.48\linewidth]{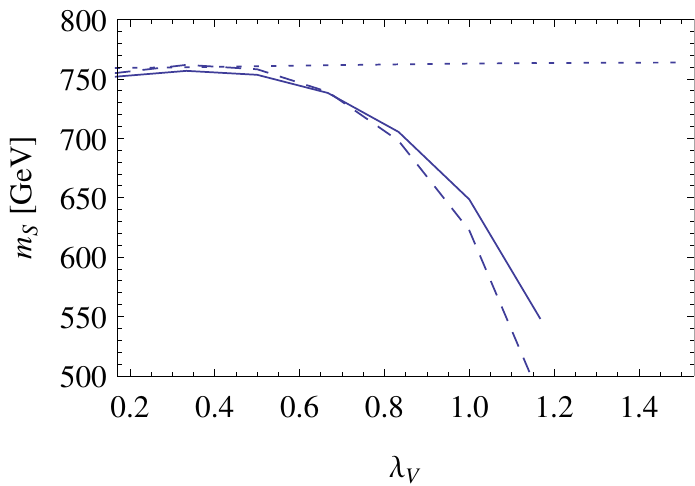} \hfill
\includegraphics[width=0.48\linewidth]{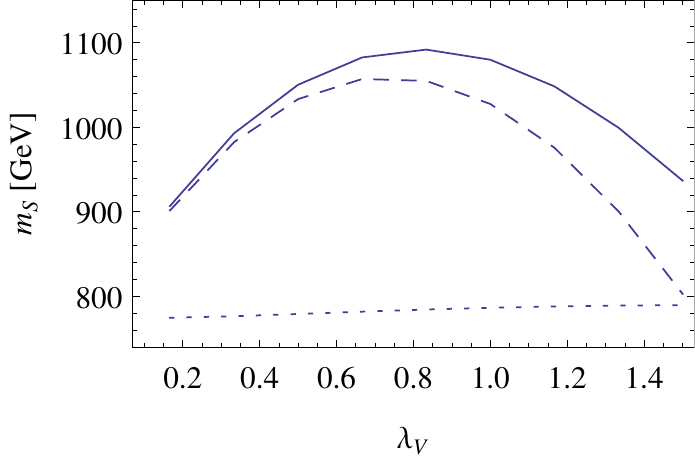}
\caption{Mass of the CP-odd scalar with a tree-level mass of $750$~GeV (dotted), at one-loop (dashed) and two-loop (full) for a variation of  $\lambda_V \equiv \lambda_e=\lambda_u$. On the left we set $m_{SUSY} = 1.5$~TeV, on the right $m_{SUSY}=2.5$~TeV.}
\label{fig:LoopMass750}
\end{figure}

There are also important loop corrections to all other scalars in the model if large Yukawa-like couplings are present. We discuss this briefly for the 750~GeV pseudo-scalar: in \cref{fig:LoopMass750}, the mass at tree, one- and two-loop level for varying $\lambda_V \equiv \lambda_e=\lambda_u$  for two different values of $m_{SUSY}$, $1.5$ and $2.5$~TeV, is given.
For $m_{SUSY}=1.5$~TeV  there is only a moderate difference between tree-level, one- and two-loop for $\lambda_V\to 0$, but for $\lambda_V$ of $O(1)$ the one-loop corrections cause a shift by 100 GeV and more, which is compensated to some extent by the two-loop corrections. For larger $m_{SUSY}$ we see already a large positive shift for small $\lambda_V$, which quickly increases and reaches $300$--$400$~GeV for $\lambda_V \sim 0.8$. For even larger values of $\lambda_V$, the difference between tree-level and the loop corrected mass becomes smaller. Still, the overall shift is more than 100 GeV, and this would be highly overestimated by only including one-loop corrections. As we will see in the next section, one needs $\lambda_V \sim O(1)$ to explain the diphoton signal. For this value, a naive tree-level analysis gives a mass for the lightest CP-odd state which is far off the correct value. Thus, one has to be much more careful with the choice for $B_S$. 

\subsection{Diphoton and digluon rate}
\begin{figure}[hbt]
\includegraphics[width=0.48\linewidth]{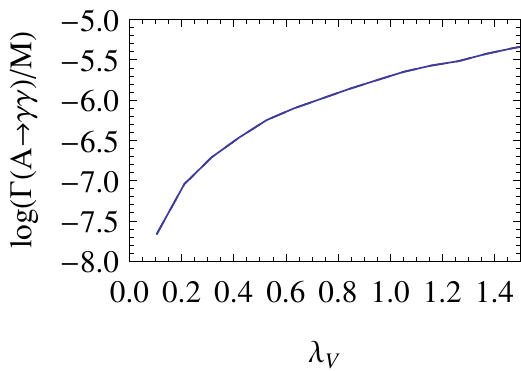} \hfill 
\includegraphics[width=0.48\linewidth]{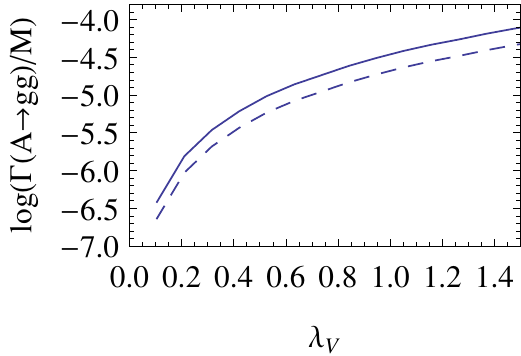} 
\caption{Partial widths into two photons (left) and two gluons (right)
  of the lightest pseudo-scalar, normalised to the mass $M_S$. $B_S$
  was adjusted to keep the mass constant within $(750\pm 5)$~GeV. The
  solid lines were drawn including higher order QCD corrections to
  loop induced decays, the dashed ones at leading order only.}
\label{fig:U1xDiphton}
\end{figure}

We now discuss the diphoton and digluon decay rate of the pseudo-scalar, and its dependence on the new Yukawa-like couplings. As we have just seen, large couplings induce a non-negligible mass shift. Therefore, it is necessary to adjust $B_S$ carefully to get the correct mass, $750$~GeV, after including all loop corrections. This can be done by {\tt SSP}, which can adjust $B_S$ for each point to obtain the correct mass within $5$~GeV uncertainty. The results for the calculated  diphoton and digluon rate at LO and with the higher order corrections discussed in \cref{sec:calculationDecays} are shown in \cref{fig:U1xDiphton}. In order to see the size of the higher order corrections, one can use the flag 521 in \SPheno to turn them on and off
\begin{lstlisting}[style=file]
Block SPhenoInput   # SPheno specific input 
 ...
 521  1             # Diphoton/Digluon widths including higher order
\end{lstlisting}
One finds the expected behaviour: the partial widths rise quadratically with the coupling. For about $\lambda_V \simeq 1.0$ one has $\Gamma(S\to\gamma\gamma)/M_S \sim 10^{-6}$, which is necessary to explain the observed excess. In \cref{fig:U1xDiphton} we also show a comparison between a purely LO calculation and the one including the higher order QCD corrections described in \cref{sec:calculationDecays}. There is no change for the decay into two photons, because its NLO corrections for a pseudo-scalar are non vanishing only for $m_A > 2 M_F$. Instead, the digluon width is enhanced by a factor of 2 when including NLO and NNLO QCD corrections. This also changes the ratio of the digluon-to-diphoton width from about 10 (LO only) to 20 (including higher orders).

\subsection{Constraints on choice of parameters}
\subsubsection{Singlet-doublet mixing}
\begin{figure}[hbt]
\includegraphics[width=0.45\linewidth]{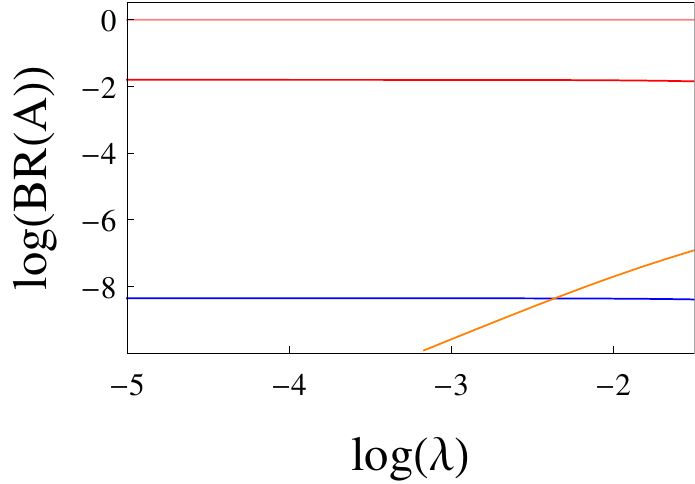} \hfill 
\includegraphics[width=0.45\linewidth]{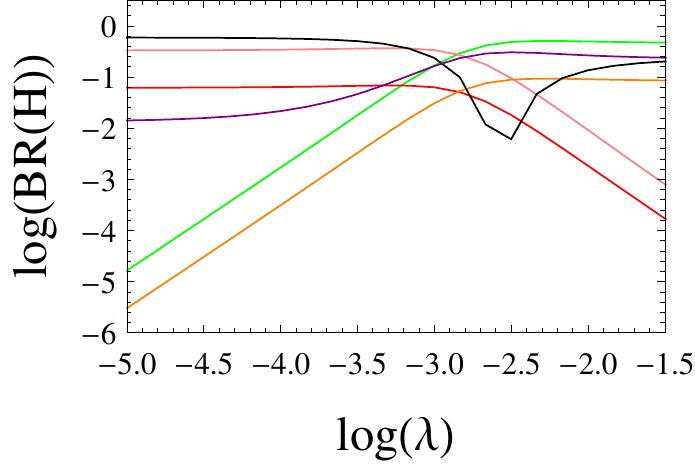} \\ 
\includegraphics[width=0.45\linewidth]{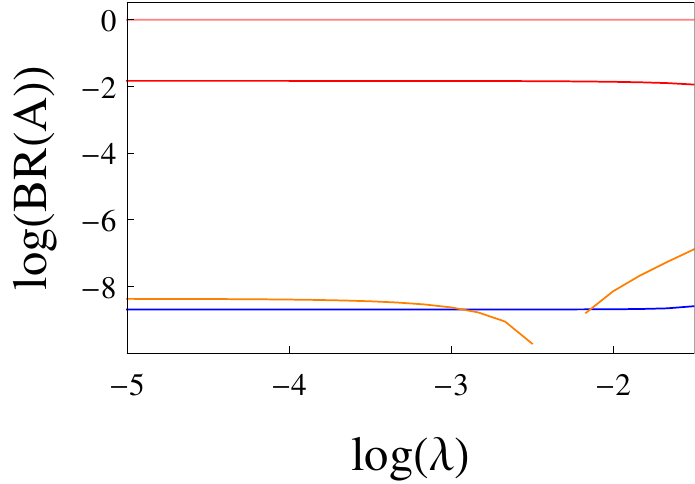} \hfill 
\includegraphics[width=0.45\linewidth]{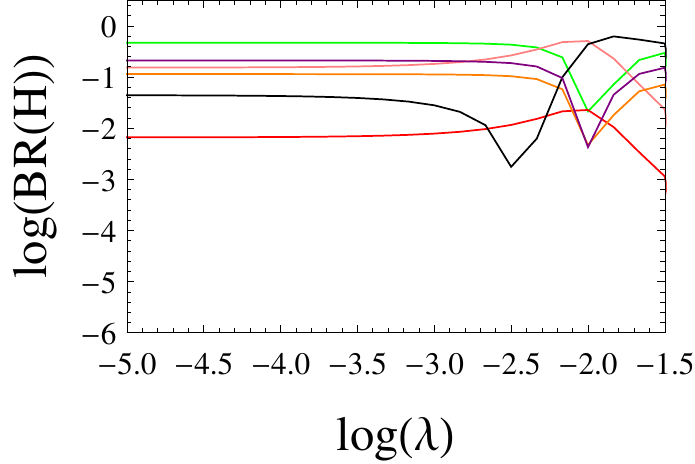} \\
\caption{Branching ratios of the $750$~GeV CP-odd particle (left), and a CP-even scalar (right) close in mass as function of $\lambda$. In the first row the tree-level rotation matrices are used, while in the second row the rotation matrices including loop corrections are used. Here, we set $A_\lambda=1$~TeV. The colour code is as follows:
$\gamma\gamma$ (pink), $gg$ (red), $hZ$ (blue), $t\bar t$ (orange), $hh$ (black), $ZZ$ (purple), $W^+ W^-$ (green). }
\label{fig:BR_H_A_lambda}
\end{figure}

\begin{figure}[hbt]
\centering
\includegraphics[width=0.5\linewidth]{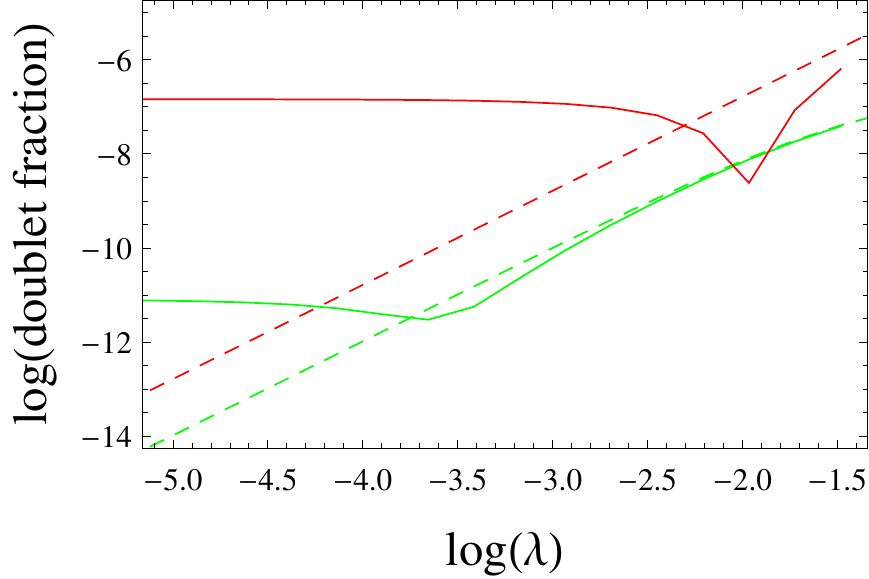} 
\caption{Doublet fraction of the 750~GeV pseudo-scalar (green) and the 800~GeV scalar (red) at tree-level (dashed lines) and including loop corrections (full lines), as function of $\lambda$.}
\label{fig:lambda_SM_mixing}
\end{figure}

So far, we made some strong assumptions about some parameters in this model. In particular, we set the coupling between the singlet and the two Higgs doublets $\lambda=0$. This raises the question how sensitive the results are to this choice. For this purpose, we can test what happens if we slightly deviate from it. The branching ratios of the CP-odd scalar of $750$~GeV mass, which is nearly a pure singlet, as a function of $\lambda$ are shown in \cref{fig:BR_H_A_lambda}. For comparison we also show the branching ratios for the CP-even scalar with a mass around $800$~GeV. This particle is mainly a mixture of $\eta$ and $\bar \eta$ with a small singlet component.  For both particles we depict the branching ratios when calculating only tree-level masses and when including loop-corrections.
At the tree level we find that the impact of $\lambda$ on the branching ratios of $A$
is very small. This does not change much when including the loop
corrections to the pseudo-scalar rotation matrix. On the other hand, for vanishing $\lambda$
 we already have a large
branching ratio of the CP-even scalar into $hh$ even at tree level. Moreover, the decay modes
into two massive vector bosons or $t\bar t$ at tree level increase very quickly with
$\lambda$ and for $\lambda > 0.01$ they already dominate. At one-loop level, the large dependence on $\lambda$ is no longer
visible, because for very small $\lambda$ the branching ratios into massive
SM vector bosons and fermions are already large. This can be seen in
\cref{fig:lambda_SM_mixing} where we compare the doublet fraction of
the two states at tree level and one loop.  In general, the behaviour
shows that a CP-odd scalar might be a much less fine-tuned candidate
for the observed excess.

\subsubsection{Constraints from Higgs coupling measurements}
\begin{figure}[hbt]
 \includegraphics[width=0.49\linewidth]{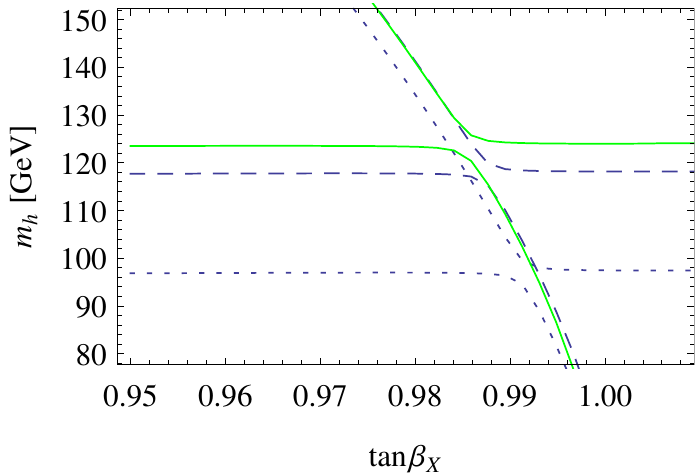}\hfill
 \includegraphics[width=0.49\linewidth]{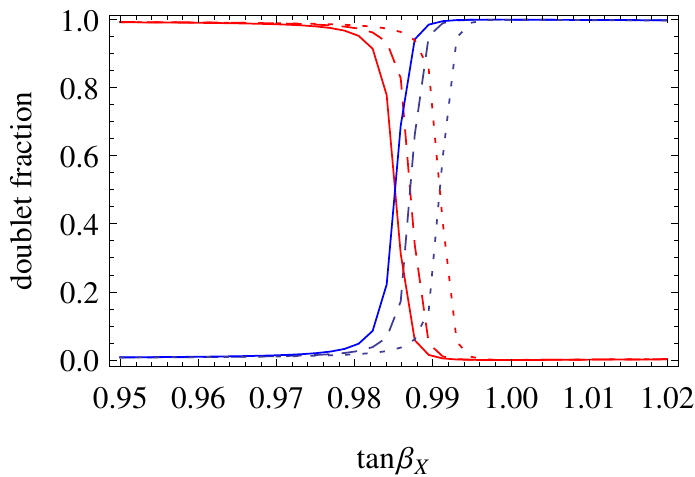}
 \caption{Left: the masses of the two lightest CP-even eigenstates as a function of $\tan\beta_X$ at tree-level (dotted), one-loop (dashed) and two-loop (full green line). 
 Right: the corresponding doublet fraction of the lightest (blue) and second lightest (red) scalar at tree-level (dotted), one-loop (dashed) and two-loop (full) levels. }
\label{fig:U1x_tbx_mh1}
\end{figure}

We have seen in the \Mathematica session that it is possible to obtain
two light scalars at tree-level. One question is: is this also
possible when including all loop contributions? In order to address
this question we change some input parameters to
\begin{eqnarray*}
&m_{SUSY} = 1.75~\TeV\,,\tan\beta = 20\,, m_{\bar\eta} = 1~\TeV\,, M_{Z'} = 2.5~\TeV , &\\
&v_S=3.5~\TeV\,,B_S=45000~\GeV^2, \lambda_X = -0.3 \,,A_X = 750~\GeV .& 
\end{eqnarray*}
The pole masses and the doublet fraction of two lightest CP even
states as a function of $\tan\beta_X$ is shown in
\cref{fig:U1x_tbx_mh1}.
\begin{figure}[hbt]
 \includegraphics[width=0.49\linewidth]{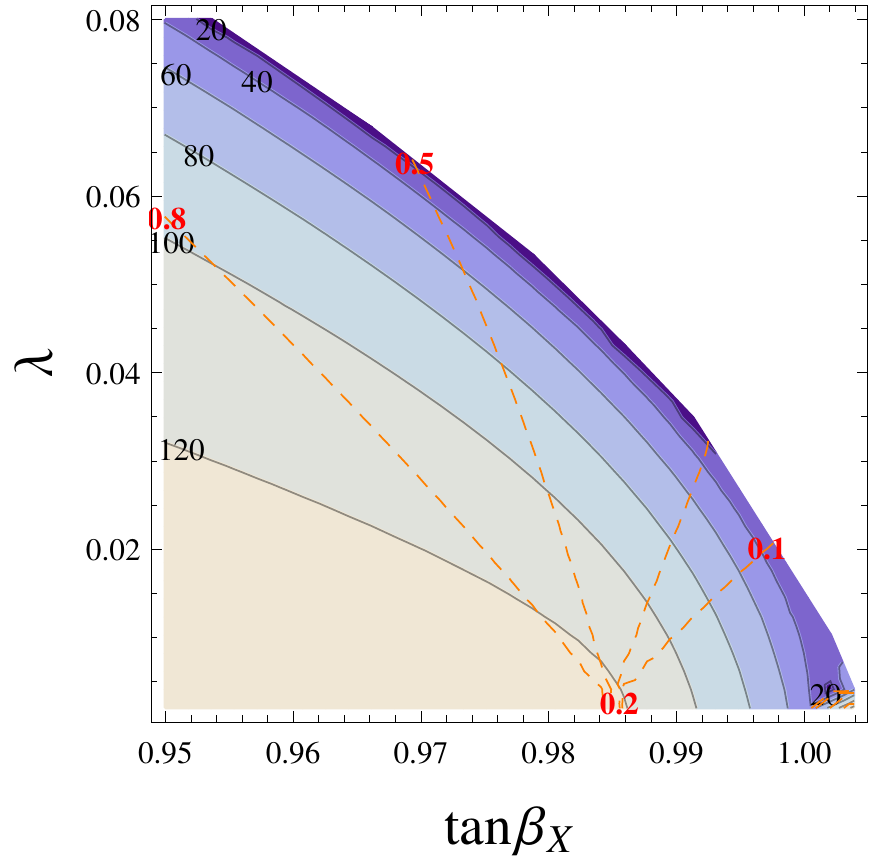}\hfill
 \includegraphics[width=0.49\linewidth]{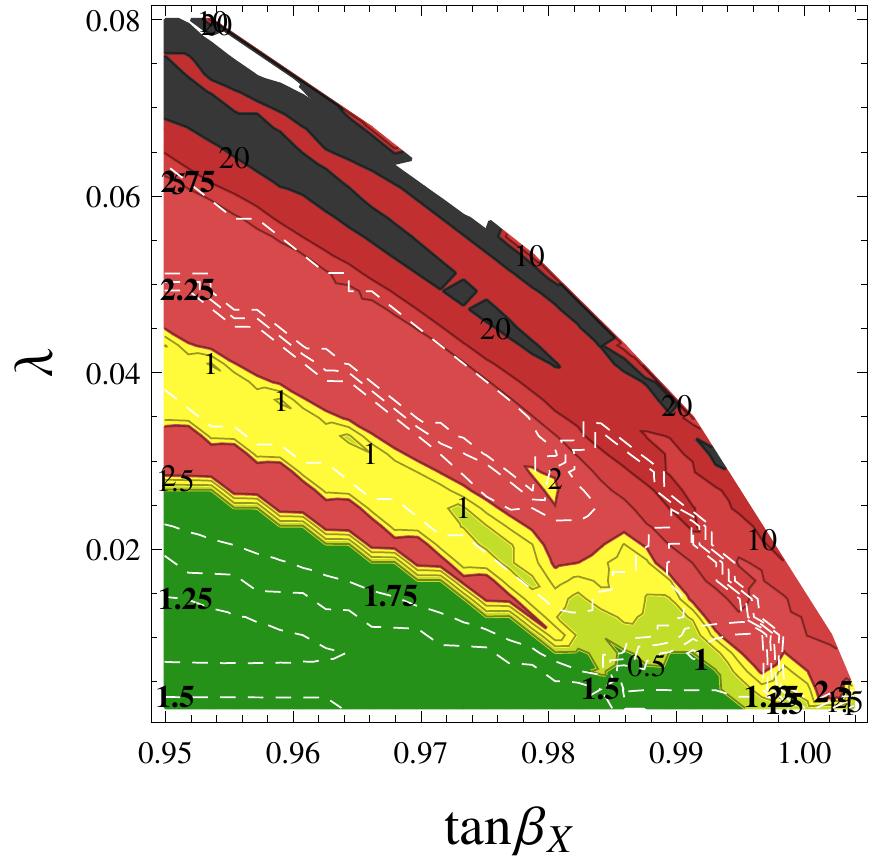} \\
 \includegraphics[width=0.49\linewidth]{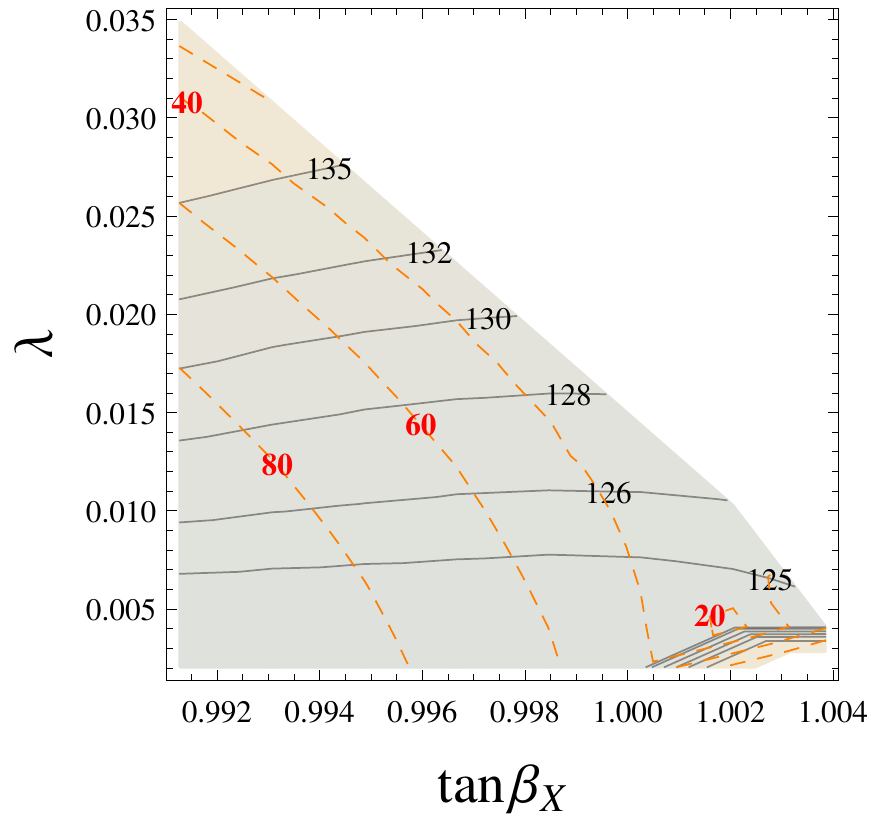}  \hfill 
\includegraphics[width=0.49\linewidth]{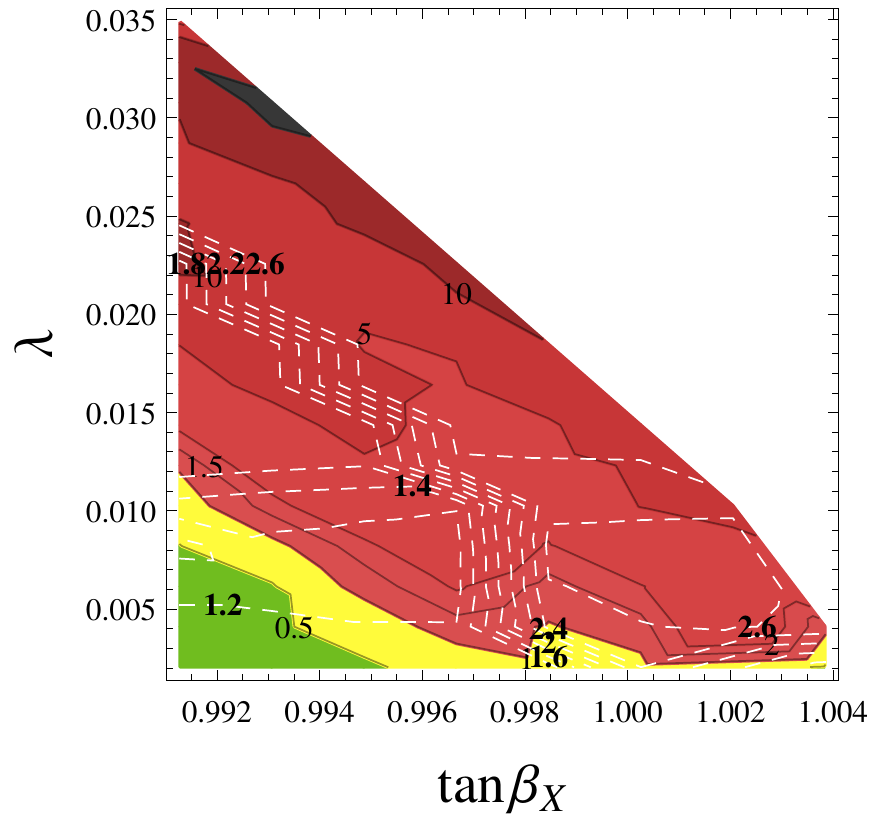} 
 \caption{{\bf First row}: On the left, we show the mass of the lightest CP-even scalar in the $(\tan\beta_X,\lambda)$ plane. On top of this, we give the contour lines for constant values of the doublet fraction of the lightest scalar (orange lines with red labels). On the right, we show the results from \HB and \HS (white contour lines with bold labels for constant $\chi^2$ divided by the number of considered Higgs observables: 81) in the same plane. The red shaded regions are excluded by Higgs searches. {\bf Second row}: Zoom into the region with $\tan\beta_X$ close to 1. On the left, the mass of the two lightest CP-even scalars are shown. The plot on the right provides the same information as the one in the first row. }
 \label{fig:HBHS}
\end{figure}
We find the very strong dependence on $\tan\beta_X$, known in many
$U(1)$ extensions
\cite{O'Leary:2011yq,Hirsch:2011hg,Hirsch:2012kv}. One difference here
is that, due to the mixing with the singlet, the light state in the
extended sector does not become massless for $\tan\beta_X = 1$, but for
small deviations from it. We see in \cref{fig:U1x_tbx_mh1} that the
SM-like Higgs gets a positive mass shift after the level crossing,
while the mass of the lightest state drops very quickly. Of course, it
is important to know if such light Higgs-like particles are compatible
with all limits from Higgs searches at LEP, Tevatron and the LHC. For
this purpose, we can make use of \HB, which checks whether the decay
rates of a scalar into SM states are compatible with the observations
at all experiments performed so far. If any of these rates is above 1
(normalised to the SM expectation), such a parameter point would be
ruled out. Similarly, we can use \HS to obtain a $\chi^2$ estimator
for each parameter point, based on how well the measured Higgs
properties are reproduced. In order to use \HB and \HS, we set the
flag
\begin{lstlisting}[style=file]
Block SPhenoInput   # SPheno specific input 
...
76 1               # Write HiggsBounds file    
\end{lstlisting}
in the input file for \SPheno. In this way, \SPheno writes out all files which are necessary to run \HB and \HS via the effective couplings input mode ({\tt effC}). However, there is one caveat: \SPheno does not automatically write the file {\tt MHall\_uncertainties} which gives an estimate for the theoretical uncertainty in the mass prediction of all scalars. The reason is that \SPheno cannot do such an estimate automatically. However, if this file is missing, \HB and \HS would assume that the uncertainty is zero. Therefore, we add this file by hand and assume a $3$~GeV uncertainty for all masses.
We can now use this setup to make a scan in the ($\tan\beta_X$,$\lambda$) plane, for instance by using the {\tt SSP} option to automatically call \HB and \HS during a parameter scan. The results are shown in  \cref{fig:HBHS}. One can see that both, the discovery potential and the $\chi^2$ value, are very sensitive on small changes in these two parameters. The reason is mainly the large dependence of the masses of the two lightest scalars and their mixing. One sees that the best $\chi^2$ value is found close to the $\tan\beta_X$ range where the SM-like particle is the second lightest CP-even state, and the lightest one is about $80$~GeV. In addition, for a very small stripe close to $\lambda=0$  also points with very light scalars with masses below $40$~GeV pass all constraints, but for slightly larger values of $\lambda$ the mixing already becomes too large and the points are excluded by $e^+ e^-\to (h_1)Z \to (b \bar b)Z$ from LEP searches.

\subsubsection{Large decay width and constraints from vacuum stability}
\begin{figure}[hbt]
\centering
\includegraphics[width=0.32\linewidth]{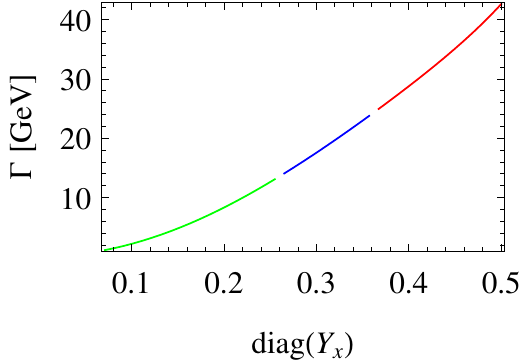}  \hfill
\includegraphics[width=0.32\linewidth]{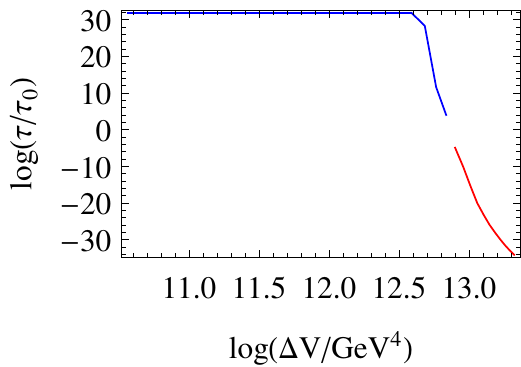}   \hfill 
\includegraphics[width=0.32\linewidth]{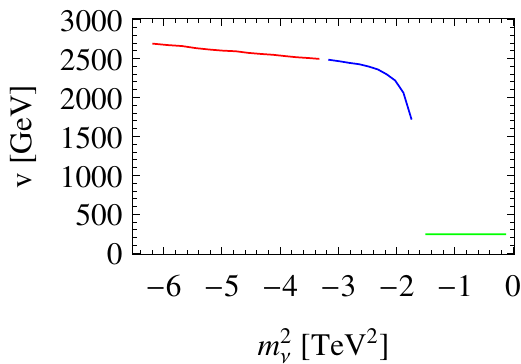}  
\caption{Left: Total width of the 750~GeV particle as function of the diagonal entries in $Y_x$. The stability of the vacuum has been checked with \Vevacious: the green region is absolutely stable, in the blue region the vacuum is unstable but long-lived, while in the red region the EW vacuum would decay too fast. Middle: the life-time $\tau$ in ages of the universe, $\tau_0$, as a function of the potential difference between the electroweak minimum and the global one. Note that the largest value \Vevacious returns is $10^{30}$. Right: the value of the electroweak VEV $v$ at the global minimum of the potential as function of the diagonal entries of $m^2_{\nu}$.  }
\label{fig:U1x_Vacuum}
\end{figure}

We have already considered the possibility to enhance the total decay width of the CP-odd scalar via decays in pairs of right-sneutrinos. In our tree-level analysis with \SARAH we found that one can reduce the mass splitting between the two mass eigenstates by demanding 
\begin{equation}
T_x = - \frac{1}{\sqrt{2}}\lambda_X v_S Y_x .
\end{equation}
In addition, as already discussed above, one has to use a negative soft-mass for the sneutrinos,
\begin{equation}
\label{eq:CondMv2}
m_\nu^2 = - \frac{Y_x^2}{4 g_X^2} \left(4 M_{Z'}^2 - g_X^2 v^2\right) ,
\end{equation}
to get the states light enough. This immediately raises two questions: (i) how large can the total width be for large values of $Y_x$? (ii) Is the electroweak vacuum stable or not? First of all, we notice that a negative $m_\nu^2$ does not necessarily imply spontaneous $R$-parity violation, as shown in Ref.~\cite{CamargoMolina:2012hv}, in contrast to some claims in this direction in the previous literature. However, the danger of disastrous vacuum decays increases, of course, with decreasing $m_\nu^2$. Therefore, we use \Vevacious to check the stability of the potential. For this purpose, we have written a second \SARAH model file where we include the possibility of VEVs for the right sneutrinos. We also added in this new implementation those mixings among states which were forbidden by $R$-parity conservation. This is actually necessary because \Vevacious calculates the one-loop corrections to the effective potential and the full mass matrices are required. 
The \Vevacious model file is generated via 
\begin{lstlisting}[style=mathematica]
MakeVevacious[];
\end{lstlisting}
We can now run a point with \SPheno. If we turn on 
\begin{lstlisting}[style=file]
Block SPhenoInput   # SPheno specific input 
 ...
530 1               # Write Blocks for Vevacious 
\end{lstlisting}
we can pass the \SPheno spectrum file in a second step to \Vevacious,
which finds all minima of the potential with the additional VEV. If
the global minimum is not the local one found by \SPheno with correct
EWSB, \Vevacious uses {\tt CosmoTransitions} \cite{Wainwright:2011kj}
to get the life-time of `our' vacuum. If our survival probability is
found to be below 10\%, we label the points as
short-lived. Metastable points with a longer life-time are called
long-lived. We choose the following set of input
parameters\footnote{This choice might be a bit unlucky but shows the
  dangers of the two-loop effective potential calculation: in the
  gauge-less limit, one of the pseudo-scalars has a tree-level mass
  close to 0. This causes divergencies (`Goldstone catastrophe')
  \cite{Martin:2014bca,Elias-Miro:2014pca} and makes it necessary to
  turn off the 2L corrections in \SPheno via the flag {\tt 7} set to {\tt 1}.}
\begin{eqnarray*}
&m_{SUSY} = 2.5~\TeV\,,\tan\beta = 10\,,\tan\beta_x = 1\,,g_X=0.5\,,M_{Z'} = 2.5~\TeV, m_{\bar\eta} = 1~\TeV, &\\
&v_S=0.5~\TeV\,,B_S=755000~\GeV^2\,,\lambda_X = -0.4 \,,A_X = 0.4~\TeV .& 
\end{eqnarray*}
The final result is summarised in \cref{fig:U1x_Vacuum}.\footnote{For
  this example we had to turn off the thermal corrections to the
  tunnelling by inserting {\tt vcs.ShouldTunnelThermally = False} in
  {\tt Vevacious.py} because {\tt CosmoTransitions} failed otherwise
  to calculate the tunnelling time in the six-dimensional potential.}
To maximise the effect on the total width, we take all sneutrinos to
be degenerate and with the same coupling to the $750$~GeV scalar.

We see that we can get a large total width of the pseudo-scalar for large
diagonal entries in $Y_x$. Up to values of $Y_x$ of $0.25$, which
corresponds to a total width of $15$~GeV, the vacuum is absolutely
stable. One can even reach $Y_x \sim 0.36$ ($\Gamma \sim 30$~GeV)
before the life-time of the correct vacuum becomes too short. The
dependence of the tunnelling time on the value of $m^2_\nu$ is shown
in the middle of \cref{fig:U1x_Vacuum}. One might wonder how
dangerous this vacumm decay is, since spontaneous $R$-parity violation is not a problem
\emph{per se}. However, we show also in the right
plot in \cref{fig:U1x_Vacuum} that the electroweak VEV $v$ changes
dramatically in the global minimum. Therefore, these points are
clearly ruled out.

Even if we cannot reach a width of $45$~GeV with the chosen point, we see that the principle idea to enhance the width is working very well. Thus, with a bit more tuning of the parameters, one might even be able to accommodate this value. However, this is beyond the scope of this example.  We emphasise that, since the large coupling responsible for the large width is a dimensionful parameter, it will not generate a Landau pole. Thus, the large width hypothesis does not necessarily point to a strongly coupled sector close to the observed resonance.

\subsubsection{Dark matter relic density}
We have seen in the last section that light sneutrinos are a good possibility in this model to enhance the width of the 750~GeV particle. Of course, it would be interesting to see if they can also be a dark matter candidate. For this purpose, we can implement the model in {\tt MicrOmegas} to calculate the relic density and to check current limits from direct and indirect  detection experiments. In order to implement the model in {\tt MicrOmegas}, it is sufficient to generate the model files for {\tt CalcHep} with \SARAH via 
\begin{lstlisting}[style=mathematica]
MakeCHep[]
\end{lstlisting}
and copy the generated files into the {\tt work/models} directory of
a new {\tt MicrOmegas} project. \SARAH also writes main files which
can be used to run {\tt MicrOmegas}. For instance, the file {\tt
  CalcOmega.cpp} calculates the dark matter relic density and writes
the result as well as all important annihilation channels to an
external file. This information can then be stored when running a
parameter scan.  The parameters are easily exchanged between {\tt
  MicrOmegas} and a \SARAH-based spectrum generator by copying the
spectrum file into the main directory of the current {\tt MicrOmegas}
project directory.\footnote{If the spectrum file is not called {\tt
    SPheno.spc.\$MODEL}, one can change the file-name by editing the
  fourth line in {\tt func1.mdl} written by \SARAH} However, it is
important to remember that {\tt MicrOmegas} cannot handle complex
parameters. Therefore, one has to make sure, even in the case without
CP violation, that all rotation matrices of Majorana fermions are
real. This can be done by using the following flag for \SPheno:
\begin{lstlisting}[style=file]
Block SPhenoInput   # SPheno specific input 
 ...
50 0                # Majorana phases: use only positive masses
\end{lstlisting}
\begin{figure}[hbt]
\centering
\includegraphics[width=0.66\linewidth]{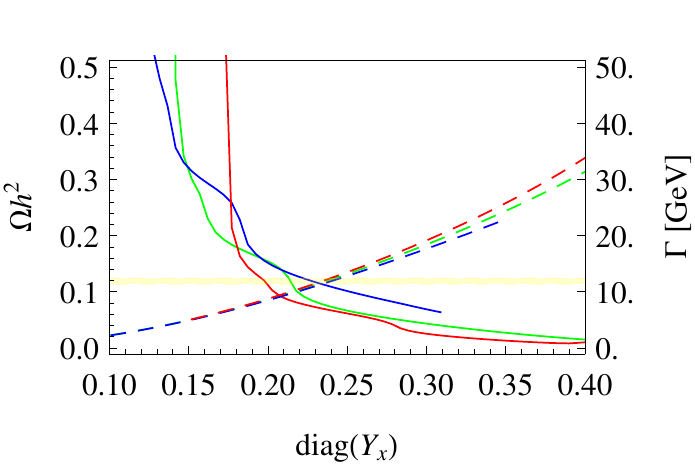}
\caption{Dark matter relic density $\Omega h^2$ (solid lines) of the
  lightest sneutrino and total width of the $750$~GeV scalar (dashed
  lines) as function of $Y_x$. For the green line the relation
  Eq.~\eqref{eq:CondMv2} was used, while the red and blue lines
  deviate from this relation by $\pm 0.4\%$. The yellow shaded region
  is the 3$\sigma$ band of Planck + WP + hihgL + BAO
  \cite{Ade:2013zuv}.}
\label{fig:Yx_DM_width}
\end{figure}
The results from a small scan\footnote{The relic density calculation
  for this model can be very time-consuming, especially for the
  sneutrinos where a large number of co-annihilation channels have to
  be calculated: the first parameter point might take several hours,
  all following points should take no longer than seconds, if no new
  channels are needed.} are shown in \cref{fig:Yx_DM_width}. Here, we
have used again the condition of \cref{eq:CondMv2} as well as very
small deviations from it. One can see that the impact of this small
variation on the total width is marginal, but the relic density is
clearly affected. Thus, with some tuning of the parameters one can
expect that it is possible to explain the dark matter relic density
and the total width by light right-handed sneutrinos. However, also
finding such a point is again beyond the scope of the example here.

Moreover, there are plenty of other dark matter candidates which
mainly correspond to the gauge eigenstates $\tilde S$, $\tilde X$,
$\tilde \eta$, $\tilde{\bar{\eta}}$ beyond the ones from the MSSM. The
properties of all of them could be checked with {\tt MicrOmegas} as
well. A detailed discussion of neutralino and sneutrino dark matter in
$U(1)$ extensions of the MSSM and different mechanism to obtain the
correct abundance was given for instance in Ref.~\cite{Basso:2012gz}.

\subsubsection{Flavour constraints}
\begin{figure}[hbt]
\includegraphics[width=0.49\linewidth]{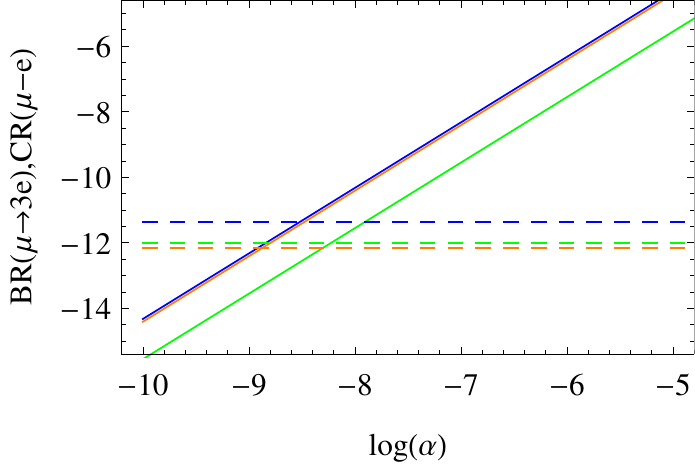} \hfill  
\includegraphics[width=0.49\linewidth]{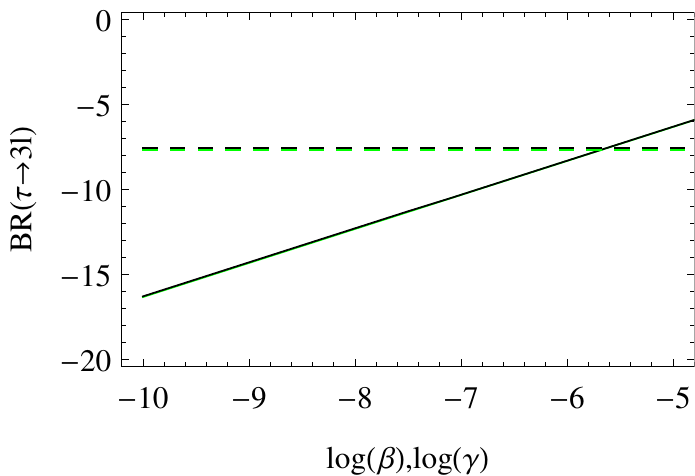}
\caption{Left: Br($\mu \to 3e$) (green), and $\mu$--$e$ conversion in
  Ti (blue) or Au (orange). Right: Br($\tau \to 3\mu$) (green),
  Br($\tau \to 3e$) (black). The dashed lines are the current
  experimental limits
  \cite{Bellgardt:1987du,Hayasaka:2010np,Dohmen:1993mp,Bertl:2006up}. Here,
  we used $X=0.1$.}
\label{fig:LFV}
\end{figure}

As mentioned above, we decided to include in this model mixing terms
between the extra vector-like fermions and the MSSM particles in order
to let the new states decay. In this way, we have a safe solution to
circumvent any potential cosmological problem. If one assumes the new
coupling matrices to have a generic form, i.e.\ large entries of
$O(1)$, including off-diagonal ones as well, they can trigger flavour
violation effects. For instance, let us assume that $Y'_e$ has the
following form
\begin{equation}
 Y'_e = \left(\begin{array}{ccc} 
         X & \alpha & \gamma \\
         \alpha & X & \beta \\
         \gamma & \beta & X
        \end{array}\right),
\end{equation}
with degenerate diagonal entries $X$, and flavour violating entries
$\alpha$, $\beta$, $\gamma$.  We can now check how strong the
constraints on $\alpha$, $\beta$, $\gamma$ would be for given $X$. For
this purpose, we use the results from \SPheno for Br($\mu \to 3e$),
Br($\tau \to 3\mu$), Br($\tau \to 3e$), and $\mu$--$e$ conversion in
Ti and Au, and compare the results with the current experimental
limits, see \cref{fig:LFV}.
We find, for instance for $X=0.1$, that $\alpha$ must be smaller than
$\sim 10^{-9}$, while the limits on $\beta$, $\gamma$, obtained from
$\tau$ decays, can still be as large as $O(10^{-6})$.

If other vector-like states mixing with the left-handed quarks or the
right-handed down-like quarks would be present -- as would be the case for
instance when assuming 5 or 10-plets of $SU(5)$ -- there
would also be stringent constraints on their couplings: they would
cause  tree-level contributions to $\Delta M_{B_s}$. Since
these observables are also calculated by \SPheno, one can easily check
the limits on models featuring those states.


\subsection{$Z'$ mass limits}
\begin{figure}[hbt]
\centering
\includegraphics[width=0.66\linewidth]{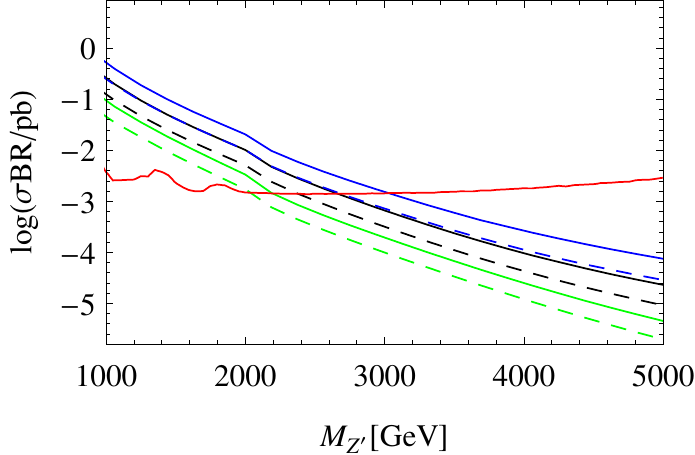}
\caption{Limit on $M_{Z'}$ for three different values of $g_X$: 0.3 (green), 0.5 (black), 0.7 (blue). For the dashed line, we assumed in addition $g_{1X} = -\frac15 g_X$, while for the full lines gauge kinetic mixing has been neglected. The red line shows the exclusion limit from ATLAS \cite{ATLASZp}.}
\label{fig:Limit_MZp}
\end{figure}

So far, we have picked a $Z'$ mass of at least $2.5$~TeV. Of course, we have to check that this is consistent with current exclusion limits. Recent exclusion limits for $pp \to Z' \to e^+ e^-$ have been released by ATLAS using 13~TeV data and 3.2~$\text{fb}^{-1}$ \cite{ATLASZp}. To compare the prediction for our model with these numbers, we can use the  {\tt UFO} model files generated by \SARAH via
\begin{lstlisting}[style=Mathematica]
MakeUFO[];
\end{lstlisting}
and add them to \MG. For this purpose, we copy the \SARAH generated files to a subdirectory {\tt models/U1xMSSM} of the \MG installation. Afterwards, we generate all necessary files to calculate the cross section for the process under consideration by running in \MG
\begin{lstlisting}[style=terminal]
import model U1xMSSM -modelname
generate  p p > Zp > e1 e1bar
output pp_Zp_ee
\end{lstlisting}
Note the option {\tt -modelname} when loading the model. This ensures that \MG is using the names for the particles as defined in our model implementation. Using the default names of \MG causes naming conflicts because of the extended Higgs sector. One can give the spectrum files written by \SPheno as input ({\tt param\_card.dat}) for \MG. One just has to make sure that the blocks written for \HB and \HS are turned off because the {\tt SLHA} parser of \MG is not able to handle them. This can be done by setting the following flag in the Les Houches input file:
\begin{lstlisting}[style=file]
Block SPhenoInput   # SPheno specific input 
 ...
520 0               # Write effective Higgs couplings 
                    # (HiggsBounds blocks)
\end{lstlisting}
In principle, one could also change the mass directly in the {\tt param\_card} without re-running \SPheno for each point. However, the advantage of \SPheno is that it calculates the width 
of the $Z'$ gauge boson including SUSY and non-SUSY states. This usually has some impact on the obtained limits \cite{Krauss:2012ku,Hirsch:2012kv,Basso:2015pka}. 
We can now scan over $M_{Z'}$ for fixed values of $g_X$ and compare the predicted cross section with the exclusion limits. In addition, we can also check the impact of gauge-kinetic mixing: as we have seen, these couplings are negative and can be sizeable. Therefore, we compare the results without gauge kinetic mixing and when setting $g_{1X} = -\frac15 g_X$ at the SUSY scale. The results are summarised in  \cref{fig:Limit_MZp}.
We see that for $g_X = 0.5$ the limit is about 2.8~TeV without gauge-kinetic mixing. Including kinetic mixing, it gets reduced by about 200~GeV. Thus, one sees that kinetic mixing is not necessarily a small effect. This contradicts some claims that sometimes appear in the literature, where it is often argued that kinetic mixing can be ignored. In particular, we emphasise that this is very relevant when discussing a GUT theory with RGE running over many orders of magnitude in energy scale. 

%% file: tex/summary.tex
\section{Summary}
\label{sec:summary}
We have given an overview on weakly-coupled renormalisable models
proposed to explain the excess observed by ATLAS and CMS around
750~GeV in the diphoton channel. We have pointed out that many of the
papers quickly written after the announcement of the excess are based
on assumptions and simplifications which are often unjustified and can
lead to wrong conclusions. A very common mistake is the lack of
inclusion of higher order corrections to the digluon and diphoton
decay rates, which results in underestimating the ratio typically by a
factor of 2.  Several authors assume that the new 750~GeV scalar does
not mix with the SM Higgs, which is often not justified.  Including
such a mixing can give large constraints. These and other problems can
be easily avoided by using \SARAH and related tools which were created
with the purpose of facilitating precision studies of high energy
physics models.  In particular, the link between \SARAH and the
spectrum generators \FS and \SPheno is a powerful approach to obtain
the mass spectrum and all the rotation matrices for any given model
\emph{without neglecting flavour mixing, complex phases or 1st and 2nd
  generation Yukawa couplings}. Optionally, one can also include all
the important radiative corrections up to two loops. In addition, we
have improved the functionality of \FS and \SPheno to calculate the
diphoton and digluon decay widths of neutral scalars, including the
higher order QCD corrections up to N${}^3$LO.  One can now pass on
this information directly to Monte-Carlo tools, like {\tt CalcHep} and
\MG, by using the appropriate model files generated with \SARAH.

In order to study as many models in as much detail as possible, we
have created a database of \SARAH model files for many of the ideas
proposed so far in the literature.  The database is also meant to
provide many examples in the context of the diphoton excess with which
the novel user can try out to familiarise with \SARAH, in order to
build up the level of expertise needed to implement their own models
in the future.

Finally, we have introduced an attractive SUSY model which combines
the idea of non-decoupling $D$-terms with the explanation of the
diphoton excess. We have used this as a new example to show how to use
\SARAH to first understand the model analytically at leading order. As
a second step, we have performed a numerical analysis of the important
loop corrections to the different masses, checked limits from Higgs
searches, neutral gauge bosons searches, and from lepton flavour
violation. We have demonstrated that this model could explain a large
width of the 750~GeV scalar, but in this context limits from
spontaneous $R$-parity violation become important. These limits can be
checked by using the interface to \Vevacious.

%% file: tex/appendix_sarah.tex
\section{How to use \SARAH and related tools}
We briefly summarise here the important commands and steps to use many powerful features of \SARAH. For pedagogical introductions see \cite{Staub:2015kfa,Vicente:2015zba}.
\tocless\subsection{How do I install \SARAH?}
The installation of \SARAH is very simple: the package can be downloaded from 
\begin{center}
{\tt \url{http://sarah.hepforge.org} }
\end{center}
After copying the {\tt tar} file to the directory {\tt \$PATH}, it can be extracted 
\begin{lstlisting}[style=terminal]
$ cp [Download-Directory]/SARAH-X.Y.Z.tar.gz $PATH/
$ cd $PATH
$ tar -xf SARAH-X.Y.Z.tar.gz
$ ln -s SARAH-X.Y.Z SARAH
\end{lstlisting}
{\tt X.Y.Z} must be replaced by the version which was downloaded. In the last line a symbolic link \SARAH to the directory {\tt SARAH-X.Y.Z} is created. There is no compilation necessary, \SARAH can directly be used with any \Mathematica version between 7 and 10.\\

\tocless\subsection{How do I load a model in \SARAH?}
To load an existing model (called {\tt \$MODEL} in the following) run in \Mathematica 
\begin{lstlisting}[style=mathematica]
<<[$PATH]/SARAH/SARAH.m;
Start["$MODEL"];
\end{lstlisting}
After some time, depending on the complexity of the model,
\begin{lstlisting}[style=mathematica]
All Done. $MODEL is ready! 
\end{lstlisting}
should appear and no error messages or warnings should show up during the evaluation. 

\tocless\subsection{How do I get analytical expressions for masses, vertices and tadpoles?}
\begin{itemize}
 \item {\bf Mass matrices} :
 The mass matrix for a particle ({\tt Particle})  is returned by
\begin{verbatim}
 MassMatrix[Particle]
\end{verbatim}
\item {\bf Tadpole equations} :
The tadpole equation corresponding to a scalar ({\tt Scalar}) is printed by using
\begin{verbatim}
 TadpoleEquation[Scalar]
\end{verbatim}
\item {\bf Vertices} :
To calculate the vertices for a list of external states ({\tt Particles}) use
\begin{verbatim}
 Vertex[{Particles},Options];
\end{verbatim}
The options define the considered eigenstates ({\tt Eigenstates -> EWSB/GaugeES}) as well as the treatment of 
dependences among parameters ({\tt UseDependences -> False/True}). \\
All vertices for a set of eigenstates are calculated at once by
\begin{verbatim}
MakeVertexList[$EIGENSTATES, Options];
\end{verbatim}
Here, first the eigenstates ({\tt \$EIGENSTATES: GaugeES, EWSB}) are defined and as an option it can be defined whether only specific generic 
classes should be considered (e.g. {\tt GenericClasses -> FFS}). 
\end{itemize}
\tocless\subsection{How do I get the renormalisation group equations for a model?}
The calculation of the RGEs at the one- and two-loop level can be performed after the initialization of a model via
\begin{verbatim}
CalcRGEs[Options];
\end{verbatim}
The results are saved in three-dimensional arrays: the first entry is the name of  the considered parameter, the second entry is the one-loop \(\beta\)-function ($\times 16 \pi^2$) and the third one is the two-loop \(\beta\)-function ($\times (16 \pi^2)^2$). 
For non-SUSY models, the RGEs of the different parameters are saved in 
 \begin{itemize}
  \item  \verb"Gij": Anomalous dimensions of all fermions and scalars
  \item \verb"BetaGauge": Beta functions of all gauge couplings
  \item \verb"BetaLijkl": Beta functions of all  quartic scalar couplings
  \item \verb"BetaYijk": Beta functions of all  interactions between two fermions and one scalar
  \item \verb"BetaTijk": Beta functions of all  cubic scalar interactions
  \item \verb"BetaMuij": Beta functions of all  bilinear fermion terms
  \item \verb"BetaBij": Beta functions of all  bilinear scalar terms
  \item \verb"BetaVEV": Beta functions of all  VEVs
\end{itemize}
The output for SUSY models is 
saved in the following arrays:
\begin{itemize}
\item \verb"Gij": Anomalous dimensions of all chiral superfields
\item \verb"BetaWijkl": Quartic superpotential parameters
\item \verb"BetaYijk": Trilinear superpotential parameters
\item \verb"BetaMuij": Bilinear superpotential parameters
\item \verb"BetaLi": Linear superpotential parameters
\item \verb"BetaQijkl": Quartic soft-breaking parameters
\item \verb"BetaTijk": Trilinear soft-breaking parameters
\item \verb"BetaBij": Bilinear soft-breaking parameters
\item \verb"BetaSLi": Linear soft-breaking parameters
\item \verb"Betam2ij": Scalar squared masses
\item \verb"BetaMi": Majorana Gaugino masses
\item \verb"BetaGauge": Gauge couplings
\item \verb"BetaVEV": VEVs
\item \verb"BetaDGi": Dirac gaugino mass terms
\item \verb"BetaFIi" Fayet-Iliopoulos terms
\end{itemize}
\tocless\subsection{How can I run the RGEs in Mathematica?}
\SARAH writes a file to run the RGEs numerically within \Mathematica. One can load this file in any \Mathematica session and use the provided function {\tt RunRGEs} to solve the RGEs numerically. For instance,
\begin{verbatim}
<< [$PATH]/SARAH/Output/$MODEL/RGEs/RunRGEs.m;
sol = RunRGEs[{g1 -> 0.46, g2 -> 0.63, g3 -> 1.05}, 3, 17][[1]];
Plot[{g1[x], g2[x], g3[x]} /. sol, {x, 3, 17}, Frame -> True, Axes -> False];
\end{verbatim}
First, the file is loaded and then the RGEs are evaluated from $10^3$ to $10^{17}$ GeV. The initial conditions at 1 TeV are $g_1=0.46$, $g_2=0.63$, $g_3=1.05$. The interpolation functions were saved in the variable {\tt sol} which can then be used for plotting.

\tocless\subsection{How do I get analytical expressions for the one-loop corrections?}
Loop corrections are calculated via
\begin{verbatim}
CalcLoopCorrections[$EIGENSTATES]; 
\end{verbatim}
As argument only the considered eigenstates (e.g. {\tt \$EIGENSTATES=EWSB}) have to be defined. The results are saved in
the variables {\tt Tadpoles1LoopSums[\$EIGENSTATES]} and {\tt SelfEnergy1LoopSum[\$EIGENSTATES]}
as sums of all contributions, or as list of the different contributions  in 
{\tt Tadpoles1LoopList[\$EIGENSTATES]} and  \linebreak {\tt SelfEnergy1LoopList[\$EIGENSTATES]}.

\tocless\subsection{How do I get a spectrum generator based on \SPheno for a new model?}
To obtain the \SPheno output (after \SARAH has been loaded in \Mathematica and the model initialised), type the command
\begin{lstlisting}[style=mathematica]
MakeSPheno[];
\end{lstlisting}
When executing {\tt MakeSPheno}, \SARAH first calculates all the information it needs, i.e. it is not necessary to run the calculation of vertices or RGEs before. When \SARAH is done, the source code for \SPheno is stored in \linebreak
{\tt \$PATH/SARAH/Output/\$MODEL/EWSB/SPheno/}. \\
The compilation of this code is done as follows: enter the directory of the \SPheno installation, 
create a new sub-directory (named {\tt \$MODEL}) and copy the code into this directory.
\begin{lstlisting}[style=terminal]
$ cd $PATH/SPHENO
$ mkdir $MODEL
$ cp $PATH/SARAH/Output/$MODEL/EWSB/SPheno/* $MODEL/
\end{lstlisting}
Afterwards, the code is compiled via 
\begin{lstlisting}[style=terminal]
$ make Model=$MODEL 
\end{lstlisting}
and a new executable {\tt SPheno\$MODEL} is available in the {\tt bin} subfolder. \\
An input file, by default called {\tt LesHouches.in.\$MODEL}, is needed to run {\tt SPheno\$MODEL}. \SARAH writes a template 
for that file which has been copied to the {\tt \$MODEL} subdirectory of \SPheno together with the \Fortran code. You can move it to the root
directory of \SPheno 
\begin{lstlisting}[style=terminal]
$ cp $MODEL/LesHouches.in.$MODEL . 
\end{lstlisting}
By doing this we can work now from the \SPheno main directory and we do not have to give the file as argument when 
running \SPheno. Thus, \SPheno can be called via 
\begin{lstlisting}[style=terminal]
$ ./bin/SPheno$MODEL
\end{lstlisting}
Alternatively, one can keep the Les Houches file in the {\tt \$MODEL} and work with it via
\begin{lstlisting}[style=terminal]
$ ./bin/SPheno$MODEL $MODEL/LesHouches.in.$MODEL
\end{lstlisting}
{\bf Note:} If no default values are given in the model file, the
corresponding parameters are set to zero in the template input
file. The user should enter a set of suitable values before running
\SPheno.

\tocless\subsection{How to I get a spectrum generator with \FS?}

Download \FS\ (at least version 1.4.0) from
\begin{center}
{\tt \url{https://flexiblesusy.hepforge.org}}
\end{center}
and unpack it:
\begin{lstlisting}[style=terminal]
$ wget \
    https://www.hepforge.org/archive/flexiblesusy/
    FlexibleSUSY-1.4.0.tar.gz
$ tar -xf FlexibleSUSY-1.4.0.tar.gz
$ cd FlexibleSUSY-1.4.0/
\end{lstlisting}
The \FS\ model file \code{FlexibleSUSY.m.in} for the model under
consideration should be put into a sub-directory of the
\code{model_files/} directory with the name of the model:
\begin{lstlisting}[style=terminal]
$ mkdir model_files/$MODEL/
$ cp FlexibleSUSY.m.in model_files/$MODEL/
\end{lstlisting}
Afterwards, the spectrum generator for the model can be generated
as\footnote{\FS\ assumes that \SARAH\ can be loaded from within
  Mathematica using the \code{Needs["SARAH`"]} command.  Please refer
  to the installation instructions within the \code{README} file in
  the \FS/ directory for installation instructions.}
\begin{lstlisting}[style=terminal]
$ ./createmodel --name=$MODEL
$ ./configure --with-models=$MODEL
$ make
\end{lstlisting}
To run the spectrum generator, an SLHA input file has to be provided.
The full path to the SLHA input file must be specified using the
\code{--slha-input-file=} argument, for example:
\begin{lstlisting}[style=terminal]
$ ./models/$MODEL/run_$MODEL.x \
    --slha-input-file=LesHouches.in.$MODEL
    --slha-output-file=LesHouches.out.$MODEL
\end{lstlisting}
Please run \code{./models/$MODEL/run_$MODEL.x --help} for more options
or refer to the \code{README} file.

\tocless\subsection{How can I check points with \HB and \HS?}
In the same directory in which the \SPheno spectrum file is located, all other input files for \HB and \HS are saved by \SPheno. The (relative) path to this directory has to be given as the last argument to \HB when executing it. Thus, working from the directory {\tt \$PATH}, \HB is started via:
\begin{lstlisting}[style=terminal]
$ ./HiggsBounds/HiggsBounds LandH effC #neutral #charged SPHENO/
\end{lstlisting}
From other directories, one can use absolute paths:
\begin{lstlisting}[style=terminal]
$ $PATH/HiggsBounds/HiggsBounds LandH effC #neutral #charged $PATH/SPHENO/
\end{lstlisting}
{\tt \#neutral} and {\tt \#charged} are the number of (physical) neutral and charged Higgs bosons in the model. \\
\HS is the complement to \HB and checks how  well a point reproduces the Higgs mass and rate measurements. The syntax is very similar to \HB:
\begin{lstlisting}[style=terminal]
$ ./HiggsSignals/HiggsSignals latestresults peak 2 effC #neutral #charged SPHENO/ 
\end{lstlisting}

\tocless\subsection{How do I implement a model in \MG and link it to \SPheno/\FS?}
A new model in \MG can be implemented via the  \UFO format which is also supported by other tools like \Herwig or \Sherpa. The command to generate the \UFO files is
\begin{lstlisting}[style=mathematica] 
MakeUFO[]
\end{lstlisting}
All files written by SARAH have to be copied to a new sub-directory in \MG's model directory:
\begin{lstlisting}[style=terminal]
cd $PATH/MADGRAPH/
mkdir models/$MODEL
cp $PATH/SARAH/Output/$MODEL/EWSB/UFO/* models/$MODEL
\end{lstlisting}
The model can be loaded in \MG via 
\begin{lstlisting}[style=terminal]
import model $MODEL -modelname
\end{lstlisting}
The option {\tt -modelname} is used to keep the names of the particles as given in the model files. This prevents conflicts with internal \MG conventions. \\
The spectrum files written by \SPheno and \FS can be given as a parameter card to \MG ({\tt param\_card.dat}). One must only make sure that the \HB blocks 
are not included\footnote{The output of these blocks is suppressed via flag 520 in \SPheno.}, because \MG cannot parse them and would consider the file to be corrupted.

\tocless\subsection{How do I implement a model in \WHIZARD/\OMEGA and link it to \SPheno?}
The model files for \WHIZARD/\OMEGA are obtained by
\begin{lstlisting}[style=mathematica]
MakeWHIZARD[]
\end{lstlisting}
After the interface has completed, the generated files are stored in the directory
\begin{verbatim}
$PATH/SARAH/Output/$MODEL/EWSB/WHIZARD_Omega/ 
\end{verbatim}
In order to use the model with \WHIZARD and \OMEGA, the generated code must be compiled and installed.
In most cases this is done by
\begin{lstlisting}[style=terminal] 
$ cd $PATH/SARAH/Output/$MODEL/EWSB/WHIZARD_Omega
$ ./configure --prefix=$PATH/WHIZARD/ WO_CONFIG=$PATH/WHIZARD/bin/
$ make
$ make install
\end{lstlisting}
If \WHIZARD has not been installed globally in the home directory of the current user, \WHIZARD will not be able to find the binaries. Thus, the {\tt WO\_CONFIG} environment variable is used to point explicitly to the binaries.  By default, the {\tt configure} script would install the compiled model into \verb".whizard" in the home directory of the user. If the user wants to have several \WHIZARD installations or install \WHIZARD locally, it might be better to provide a model just for one installation. For these cases the installation path has been defined via the \verb"--prefix" option of the {\tt configure} script. More information on the available options is shown with the command
\begin{lstlisting}[style=terminal]
./configure  --help
\end{lstlisting}
To link \WHIZARD and \SPheno, all \SPheno modules created by \SARAH write the information about the parameters and masses into an additional file. This file is written in the \WHIZARD specific format and can be directly read by \WHIZARD.   One just has to make sure that the corresponding flag is turned on in the Les Houches input for \SPheno to get this output:
\begin{lstlisting}[style=file,numbers=none,title=\hspace{11cm}LesHouches.in.\$MODEL] 
Block SPhenoInput   # SPheno specific input 
...
75 1                # Write WHIZARD files 
\end{lstlisting}
The parameter file can then be included in the {\tt Sindarin} input file for \WHIZARD  via
\begin{lstlisting}[style=file,numbers=none,title=\hspace{13cm}Example.sin] 
model = $model_sarah
...
include("$PATH/SPHENO/WHIZARD.par.$MODEL") 
\end{lstlisting}

\tocless\subsection{How do I implement a model in \CalcHep/\MO and link it to \SPheno?}
\label{app:calc-relic}
Model files for \CalcHep can be obtained by running 
\begin{lstlisting}[style=mathematica]
MakeCHep[Options];
\end{lstlisting}
The options provided may be used to configure exactly what is to be included
in the generated model files.  For example, one might prevent vertices
involving four vector bosons from being included in the model files
by specifying {\tt Exclude -> \{VVVV\}}.  When \SARAH is finished with
{\tt MakeCHep}, the \CalcHep model files are located in the directory
\begin{verbatim}
$PATH/SARAH/Output/$MODEL/EWSB/CHep/ 
\end{verbatim}
To implement the new model in \CalcHep, it is sufficient to use the internal ``import model'' routine from the GUI menu, and give the above absolute path.

The model files for \CalcHep are also suitable for \MO, since the latter uses \CalcHep to obtain the cross section and all necessary decay widths to evaluate the dark matter abundance. 
To implement the model in \MO, a new project has to be created and the files have to be copied in the working directory of this project:
\begin{lstlisting}[style=terminal]
$ cd $PATH/MICROMEGAS
$ ./newProject MODEL
$ cd $MODEL
$ cp $PATH/SARAH/Output/$MODEL/EWSB/CHep/*  work/models
\end{lstlisting}
{\tt c++} files written by \SARAH to run \MO for the given model were copied together with all model files into the working directory of the current project. You can move them to the main project directory and compile them
\begin{lstlisting}[style=terminal]
$ mv work/models/CalcOmega_with_DDetection.cpp .
$ make main=CalcOmega_with_DDetection.cpp
\end{lstlisting}
A new binary {\tt CalcOmega\_with\_DDetection} is now available. 

The only missing piece are the input parameters. 
Providing the numerical parameters is simple because \CalcHep/\MO can read the SLHA files written by \SPheno or \FS\footnote{It might be just necessary to adjust the name of the spectrum file at the beginning of {\tt func1.mdl}.}. However, the user must make sure that no complex rotation matrices show up in the spectrum file: in the case of Majorana matrices and no CP violation, there are two equivalent outputs: (i) all Majorana masses are positive, but some entries of the corresponding rotation matrices are complex; (ii) all mixing matrices are real, but some masses are negative. \CalcHep can only handle the second case with real matrices. Hence, one has to use the flag 
\begin{lstlisting}[style=file,numbers=none,title=\hspace{11cm}LesHouches.in.\$MODEL]
Block SPhenoInput   # SPheno specific input 
 ...
 50 0               # Majorana phases: use only positive masses  
\end{lstlisting}
to get the spectrum according to that convention. Afterwards, the spectrum file must be moved to the same directory as the \CalcHep numerical session (typically the folder \verb"results") for \CalcHep, or where {\tt CalcOmega\_with\_DDetection} is located for \MO, and start the calculation.\\
For \CalcHep
\begin{lstlisting}[style=terminal]
$ cp $PATH/SPHENO/SPheno.spc.$MODEL /$CALCHEP/results/
$ ./n_calchep
\end{lstlisting}
For \MO
\begin{lstlisting}[style=terminal]
$ cp $PATH/SPHENO/SPheno.spc.$MODEL /$MICROMEGAS/$MODEL
$ ./CalcOmega_with_DDetection
\end{lstlisting}

\tocless\subsection{How do I implement a new model in \Vevacious?}
The model files for \Vevacious are generated by \SARAH  via
\begin{lstlisting}[style=mathematica]
MakeVevacious[];
\end{lstlisting}
As soon as the model file is created, it is convenient to copy them to the model directory of the local \Vevacious installation. In addition, one can also generate a new subdirectory which contains the  \SPheno spectrum files for the \verb"$MODEL" used as input  for \Vevacious, as well as the output written by \Vevacious
\begin{lstlisting}[style=terminal]
$ cd $PATH/VEVACIOUS
$ mkdir $MODEL/
$ cp $PATH/SARAH/Output/$MODEL/Vevacious/$MODEL.vin models/
$ cp $PATH/SPHENO/SPheno.spc.$MODEL $MODEL/
\end{lstlisting}
These steps are optional. The user can provide other locations of the model and spectrum files within the initialisation file used by \Vevacious. Independent of the location of the files, one has to write this initialisation file for a new study. The easiest way is to start with the file included in the \Vevacious package in the subdirectory {\tt bin} and edit it
\begin{lstlisting}[style=terminal]
$ cd $PATH/VEVACIOUS/bin
$ cp VevaciousInitialization.xml VevaciousInitialization_$MODEL.xml 
\end{lstlisting}

\tocless\subsection{How do I get the model files for \FeynArts/\FormCalc?}
Run
\begin{verbatim}
MakeFeynArts[]
\end{verbatim}
and copy the files to 
\begin{lstlisting}[style=terminal]
/home/user/.Mathematica/Applications/FeynArts/Models/
\end{lstlisting}
The model can then be chosen as option for the {\tt InsertFields} command of \FeynArts. 

\tocless\subsection{How do I get all expression in \LaTeX\ format?}
Use
\begin{lstlisting}[style=mathematica]
MakeTeX[];
\end{lstlisting}
When \SARAH is done with the output, the {\tt .tex} file are stored in
\begin{lstlisting}[style=terminal]
$PATH/SARAH/Output/$MODEL/EWSB/TeX 
\end{lstlisting}
The main file which can be compiled with {\tt pdflatex} is {\tt \$MODEL\_EWSB.tex}. \SARAH usually also generates Feynman diagrams using the \LaTeX\ package {\tt feynmf} \cite{Ohl:1995kr}.  In order to compile all Feynman diagrams and the pdf file at once, a shell script {\tt MakePDF.sh} is generated by \SARAH.

\section{What is necessary to implement a model in \SARAH?}
All information about the model is saved in three different files: \verb"$MODEL.m", \verb"parameters.m" and \verb"particles.m" must be located in the subdirectory \verb"$MODEL" in the {\tt Models} directory of \SARAH. 
Only the first file, {\tt MODEL.m}, is absolutely necessary and contains all physical information about the model: the symmetries, particle content, (super)potential and mixings. In \verb"parameters.m" properties of all parameters can be defined, e.g. \LaTeX\ name, Les Houches block and number, relations among parameters, real/complex, etc. In \verb"particles.m" additional information about particles are set: mass, width, electric charge, PDG, \LaTeX\ name, output name, and so on. The optional information in \verb"parameters.m" and \verb"particles.m" might be needed for the different outputs of \SARAH. We give here two examples to show that also more complicated  SUSY and non-SUSY models can be defined in \SARAH in a rather short form. Detailed information about the meaning and syntax are given in Refs.~\cite{Staub:2015kfa}

\tocless\subsection{Definition of a non-SUSY model}
As an example how to define a non-SUSY model in \SARAH we picked the model with two scalar leptoquarks discussed in \cref{sec:TwoScalarLeptoquark}. The different pieces of the model file are: \vspace{0.25cm}\\
{\bf Gauge Sector}
\begin{lstlisting}[style=file]
Gauge[[1]]={B,   U[1], hypercharge, g1,False};
Gauge[[2]]={WB, SU[2], left,        g2,True};
Gauge[[3]]={G,  SU[3], color,       g3,False};
\end{lstlisting}
{\bf Matter Content}
\begin{lstlisting}[style=file]
FermionFields[[1]] = {q, 3, {uL, dL},     1/6, 2,  3};  
FermionFields[[2]] = {l, 3, {vL, eL},    -1/2, 2,  1};
FermionFields[[3]] = {d, 3, conj[dR],     1/3, 1, -3};
FermionFields[[4]] = {u, 3, conj[uR],    -2/3, 1, -3};
FermionFields[[5]] = {e, 3, conj[eR],       1, 1,  1};

ScalarFields[[1]] =  {H, 1, {Hp, H0},     1/2, 2,  1};
ScalarFields[[2]] =  {Phi, 1, PhiLQ,     -4/3, 1,  3};
ScalarFields[[3]] =  {Omega, 1, OmegaLQ, -1/3, 1,  3};
ScalarFields[[4]] =  {The, 1, theta,        1, 1,  1};
ScalarFields[[5]] =  {s, 1, sing,           0, 1,  1};
\end{lstlisting}
{\bf Potential}        
\begin{lstlisting}[style=file]
DEFINITION[GaugeES][LagrangianInput]= {
	{LagHC, {AddHC->True}},
	{LagNoHC,{AddHC->False}}};
	
LagNoHC =  -(muH2 conj[H].H 
+ muS2 conj[s].s + muP2 conj[Phi].Phi 
+ muO2 conj[Omega].Omega + muT2 conj[The].The 
+ LambdaH/2 conj[H].H.conj[H].H 
+ LambdaS conj[s].s.conj[s].s 
+ LambdaOS conj[Omega].Omega.conj[s].s 
+ LambdaPS conj[Phi].Phi.conj[s].s 
+ LambdaTS conj[The].The.conj[s].s 
+ LambdaHS conj[H].H.conj[s].s 
+ LambdaHO conj[H].H.conj[Omega].Omega 
+ LambdaHT conj[H].H.conj[The].The 
+ LambdaHP conj[H].H.conj[Phi].Phi 
+ LambdaOP Delta[col1,col2] Delta[col3,col4] \
                *conj[Omega].Omega.conj[Phi].Phi 
+ LambdaOT conj[Omega].Omega.conj[The].The 
+ LambdaPT conj[Phi].Phi.conj[The].The 
+ LambdaO Delta[col1,col2] Delta[col3,col4] \
               *conj[Omega].Omega.conj[Omega].Omega 
+ LambdaT conj[The].The.conj[The].The 
+ LambdaP Delta[col1,col2] Delta[col3,col4] 
               *conj[Phi].Phi.conj[Phi].Phi);
    
LagHC =  -(Yd conj[H].d.q + Ye conj[H].e.l + Yu H.u.q 
+ YT l.l.The 
+ YO q.l.conj[Omega] 
+ YP e.d.Phi 
+ Sqrt[2] LambdaHat s.The.Phi.conj[Omega] );
\end{lstlisting}
{\bf Rotations in gauge sector}
\begin{lstlisting}[style=file]
DEFINITION[EWSB][GaugeSector] =
{ 
  {{VB,VWB[3]},{VP,VZ},ZZ},
  {{VWB[1],VWB[2]},{VWp,conj[VWp]},ZW}
};     
\end{lstlisting}
{\bf VEVs}        
\begin{lstlisting}[style=file]
DEFINITION[EWSB][VEVs]= {
{H0,{v,1/Sqrt[2]},{sigmaH,I/Sqrt[2]},{phiH,1/Sqrt[2]}},
{sing,{vS,1/Sqrt[2]},{sigmaS,I/Sqrt[2]},{phiS,1/Sqrt[2]}}
 };
\end{lstlisting}
{\bf Rotations in matter sector}
\begin{lstlisting}[style=file]
DEFINITION[EWSB][MatterSector]=   
    {{{phiH,phiS},{hh,ZH}},
     {{sigmaH,sigmaS},{Ah,ZA}},
     {{{dL}, {conj[dR]}}, {{DL,Vd}, {DR,Ud}}},
     {{{uL}, {conj[uR]}}, {{UL,Vu}, {UR,Uu}}},
     {{{eL}, {conj[eR]}}, {{EL,Ve}, {ER,Ue}}}};  
\end{lstlisting}
{\bf Dirac spinors}
\begin{lstlisting}[style=file]
DEFINITION[EWSB][DiracSpinors]={
 Fd ->{  DL, conj[DR]},
 Fe ->{  EL, conj[ER]},
 Fu ->{  UL, conj[UR]},
 Fv ->{  vL, 0}
 };
\end{lstlisting} 

\tocless\subsection{Definition of a SUSY model}
\label{app:modelU1x}
As an example how to define a SUSY model in \SARAH we show the model file for the model discussed in \cref{sec:example}.\\
{\bf Global symmetries}
\begin{lstlisting}[style=file]
Global[[1]] = {Z[2],MParity};
MpM = {-1,-1,1};
MpP = {1,1,-1};
\end{lstlisting}
{\bf Gauge symmetries}
\begin{lstlisting}[style=file]
Gauge[[1]]={B,   U[1], hypercharge, g1,  False, MpM};
Gauge[[2]]={WB, SU[2], left,        g2,  True,  MpM};
Gauge[[3]]={G,  SU[3], color,       g3,  False, MpM};
Gauge[[4]]={BX,  U[1], extra,     gX, False, MpM};
\end{lstlisting}
{\bf Chiral superfields}
\begin{lstlisting}[style=file]
SuperFields[[1]] = {q, 3, {uL,  dL},  1/6, 2, 3,  0, MpM};  
SuperFields[[2]] = {l, 3, {vL,  eL}, -1/2, 2, 1,  0, MpM};
SuperFields[[3]] = {Hd,1, {Hd0, Hdm},-1/2, 2, 1,-1/2, MpP};
SuperFields[[4]] = {Hu,1, {Hup, Hu0}, 1/2, 2, 1, 1/2, MpP};

SuperFields[[5]] = {d, 3, conj[dR], 1/3, 1, -3,  1/2, MpM};
SuperFields[[6]] = {u, 3, conj[uR],-2/3, 1, -3, -1/2, MpM};
SuperFields[[7]] = {e, 3, conj[eR],  1, 1,  1,  1/2, MpM};
SuperFields[[8]] = {vR,3, conj[vR],  0, 1,  1, -1/2, MpM};

SuperFields[[9]]  = {C1, 1, C10,  0, 1, 1, -1, MpP};
SuperFields[[10]] = {C2, 1, C20,  0, 1, 1,  1, MpP};
SuperFields[[11]] = {S, 1, sing,  0, 1, 1,  0, MpP};

SuperFields[[12]] = {UX, 3,conj[uRx],-2/3, 1,-3,-1/2, MpM};
SuperFields[[13]] = {UXp,3,uRxp,      2/3, 1, 3, 1/2, MpM};
SuperFields[[14]] = {EX, 3,conj[eRx], 1, 1,  1,  1/2, MpM};
SuperFields[[15]] = {EXp,3,eRxp,     -1, 1,  1, -1/2, MpM};
\end{lstlisting}
{\bf Superpotential}
\begin{lstlisting}[style=file]
SuperPotential = Yu u.q.Hu - Yd d.q.Hd - Ye e.l.Hd  \
 + \[Mu] Hu.Hd + lambdaH S.Hu.Hd + Yv vR.l.Hu \
 + lambdaC S.C1.C2 + lw S + Yn vR.C2.vR 
 + MS S.S + kappa/3 S.S.S + MtE e.EXp + MtU u.UXp \
 + lambdaE S.EX.EXp + MVE EX.EXp + lambdaU S.UX.UXp \
 + MVU UX.UXp + Yep EX.l.Hd + Yup UX.q.Hu; 
\end{lstlisting}
{\bf Rotations in gauge sector}
\begin{lstlisting}[style=file]
DEFINITION[EWSB][GaugeSector] =
{ 
  {{VB,VWB[3],VBX},{VP,VZ,VZp},ZZ},
  {{VWB[1],VWB[2]},{VWm,conj[VWm]},ZW},
  {{fWB[1],fWB[2],fWB[3]},{fWm,fWp,fW0},ZfW}
};
\end{lstlisting}
{\bf VEVs}
\begin{lstlisting}[style=file]
DEFINITION[EWSB][VEVs]= {
 {SHd0,{vd,1/Sqrt[2]},{sigmad,I/Sqrt[2]},{phid,1/Sqrt[2]}},
 {SHu0,{vu,1/Sqrt[2]},{sigmau,I/Sqrt[2]},{phiu,1/Sqrt[2]}},
 {SC10,{x1,1/Sqrt[2]},{sigma1,I/Sqrt[2]},{phi1,1/Sqrt[2]}},
 {SC20,{x2,1/Sqrt[2]},{sigma2,I/Sqrt[2]},{phi2,1/Sqrt[2]}},
 {Ssing,{xS,1/Sqrt[2]},{sigmaS,I/Sqrt[2]},{phiS,1/Sqrt[2]}},
 {SvL,{0,0},{sigmaL,I/Sqrt[2]},{phiL,1/Sqrt[2]}},
 {SvR,{0,0},{sigmaR,I/Sqrt[2]},{phiR,1/Sqrt[2]}}
};
\end{lstlisting}
{\bf Rotations in matter sector}
\begin{lstlisting}[style=file]
DEFINITION[EWSB][MatterSector]= 
{ {{SdL, SdR}, {Sd, ZD}},
  {{SuL, SuR, SuRx, SuRxp}, {Su, ZU}},
  {{SeL, SeR, SeRx, SeRxp}, {Se, ZE}},
  {{sigmaL,sigmaR}, {SvIm, ZVI}},
  {{phiL,phiR}, {SvRe, ZVR}}, 
  {{phid, phiu,phi1, phi2,phiS}, {hh, ZH}}, 
  {{sigmad, sigmau,sigma1,sigma2,sigmaS}, {Ah, ZA}},
  {{SHdm,conj[SHup]},{Hpm,ZP}},
  {{fB, fW0, FHd0, FHu0,fBX,FC10,FC20,Fsing}, {L0, ZN}}, 
  {{{fWm, FHdm}, {fWp, FHup}}, {{Lm,UM}, {Lp,UP}}},
  {{FvL,conj[FvR]},{Fvm,UV}},
  {{{FeL,FeRxp},{conj[FeR],conj[FeRx]}},{{FEL,ZEL},{FER,ZER}}},
  {{{FdL},{conj[FdR]}},{{FDL,ZDL},{FDR,ZDR}}},
  {{{FuL,FuRxp},{conj[FuR],conj[FuRx]}},{{FUL,ZUL},{FUR,ZUR}}} 
       }; 
\end{lstlisting}
{\bf Phases}
\begin{lstlisting}[style=file]
DEFINITION[EWSB][Phases]= 
{    {fG, PhaseGlu}
    };        
\end{lstlisting}
{\bf Dirac Spinors}
\begin{lstlisting}[style=file]
DEFINITION[EWSB][DiracSpinors]={
 Fd ->{  FDL, conj[FDR]},
 Fe ->{  FEL, conj[FER]},
 Fu ->{  FUL, conj[FUR]},
 Fv ->{  Fvm, conj[Fvm]},
 Chi ->{ L0, conj[L0]},
 Cha ->{ Lm, conj[Lp]},
 Glu ->{ fG, conj[fG]}
};	
\end{lstlisting}

\section{How can I define the features of a \SPheno or \FS version?}
Before we can use \FS or \SPheno for a model, it is necessary to
provide an additional input file which defines the basic setup.
In general, there are two different kinds of input versions the user
can create which need a different amount of input:
\begin{itemize}
\item {\bf High-scale version}: In a high-scale version of \SPheno or
  \FS a RGE running between the electroweak scale, an intermediate
  renormalisation scale and a high (or GUT) scale is supported. The
  user can define appropriate boundary conditions at each of these
  three scales. Furthermore, threshold effects by including
  additional scales where heavy particles are integrated out can
  optionally be included. Finally, the user can specify a condition
  which defines the high-energy scale. The most common choice is the
  unification scale of gauge couplings, but other choices such as
  Yukawa unification are possible. In addition, these high-scale
  versions also  include the possibility to define the entire input at
  the intermediate renormalisation scale and skip the RGE running to
  the GUT scale. The high-scale version is usually the appropriate option for SUSY
  models, models with heavy mass spectra, or models which should be
  studied at very high scales.
\item {\bf Low-scale version}: In a low-scale version usually no RGE
  running is included.  The \FS or \SPheno\ low-scale versions expect
  all free parameters to be given at the low-energy or the
  renormalisation scale. This version is usually used for non-SUSY
  models or models with light mass spectra, which should not be
  studied at very high scales.
\end{itemize}
The corresponding files are called {\tt SPheno.m} or {\tt
  FlexibleSUSY.m} and we give here two examples for them.

\tocless\subsection{{\tt FlexibleSUSY.m} for a high-scale version}
{\bf Model information}
\begin{lstlisting}[style=file]
FSModelName = "@CLASSNAME@";
FSDefaultSARAHModel = MSSM;
FSEigenstates = SARAH`EWSB;
\end{lstlisting}
{\bf Input parameters}
\begin{lstlisting}[style=file]
MINPAR = {
    {1, m0},
    {2, m12},
    {3, TanBeta},
    {4, Sign[\[Mu]]},
    {5, Azero}
};
\end{lstlisting}
{\bf Parameters fixed by the electroweak symmetry breaking conditions}
\begin{lstlisting}[style=file]
EWSBOutputParameters = {\[Mu], B[\[Mu]]};
\end{lstlisting}
{\bf Definition of the renormalisation scale, at which the pole masses should be calculated}
\begin{lstlisting}[style=file]
SUSYScaleFirstGuess = Sqrt[m0^2 + 4 m12^2];
SUSYScale = Sqrt[Product[M[Su[i]]^(Abs[ZU[i,3]]^2 + Abs[ZU[i,6]]^2), {i,6}]];
\end{lstlisting}
{\bf Condition defining the GUT scale}
\begin{lstlisting}[style=file]
HighScaleFirstGuess = 2.0 10^16;
HighScale = g1 == g2;
\end{lstlisting}
{\bf Condition defining the Standard Model matching scale}
\begin{lstlisting}[style=file]
LowScaleFirstGuess = LowEnergyConstant[MZ];
LowScale = LowEnergyConstant[MZ];
\end{lstlisting}
{\bf Boundary conditions}
\begin{lstlisting}[style=file]
HighScaleInput = {
    {T[Ye], Azero*Ye},
    {T[Yd], Azero*Yd},
    {T[Yu], Azero*Yu},
    {mq2, UNITMATRIX[3] m0^2},
    {ml2, UNITMATRIX[3] m0^2},
    {md2, UNITMATRIX[3] m0^2},
    {mu2, UNITMATRIX[3] m0^2},
    {me2, UNITMATRIX[3] m0^2},
    {mHd2, m0^2},
    {mHu2, m0^2},
    {MassB, m12},
    {MassWB,m12},
    {MassG, m12}
};

SUSYScaleInput = {
    (* solve EWSB conditions for \[Mu] and B[\[Mu]]
       at this scale *)
    FSSolveEWSBFor[EWSBOutputParameters]
};

LowScaleInput = {
    {vd, 2 MZDRbar Cos[ArcTan[TanBeta]] /
         Sqrt[GUTNormalization[g1]^2 g1^2 + g2^2]},
    {vu, 2 MZDRbar Sin[ArcTan[TanBeta]] /
         Sqrt[GUTNormalization[g1]^2 g1^2 + g2^2]},
    {Yu, Automatic},
    {Yd, Automatic},
    {Ye, Automatic}
};
\end{lstlisting}
{\bf Initial parameter guess}
\begin{lstlisting}[style=file]
InitialGuessAtLowScale = {
   {vd, LowEnergyConstant[vev] Cos[ArcTan[TanBeta]]},
   {vu, LowEnergyConstant[vev] Sin[ArcTan[TanBeta]]},
   {Yu, Automatic},
   {Yd, Automatic},
   {Ye, Automatic}
};

InitialGuessAtHighScale = {
   {\[Mu]   , 1.0},
   {B[\[Mu]], 0.0}
};
\end{lstlisting}
{\bf MSSM-specific options}
\begin{lstlisting}[style=file]
(* use 2L Higgs self-energy contributions *)
UseHiggs2LoopMSSM = True;
EffectiveMu = \[Mu];

(* use 3L MSSM RGEs *)
UseMSSM3LoopRGEs = True;
\end{lstlisting}
{\bf Definition of additional SLHA output blocks}
\begin{lstlisting}[style=file]
(* add FlexibleSUSYOutput block containing the scales *)
ExtraSLHAOutputBlocks = {
   {FlexibleSUSYOutput,
           {{0, Hold[HighScale]},
            {1, Hold[SUSYScale]},
            {2, Hold[LowScale] }}}
};
\end{lstlisting}

\tocless\subsection{{\tt FlexibleSUSY.m} for a low-scale version}
{\bf Model information}
\begin{lstlisting}[style=file]
FSModelName = "@CLASSNAME@";
FSDefaultSARAHModel = SMgaugegroup/TwoScalarLeptoquarks;
FSEigenstates = SARAH`EWSB;
FSRGELoopOrder = 0; (* do not generate RGEs *)
\end{lstlisting}
{\bf Flag to choose a low-scale \FS\ variant without high-scale boundary condition}
\begin{lstlisting}[style=file]
OnlyLowEnergyFlexibleSUSY = True;
\end{lstlisting}
{\bf Input parameters}
\begin{lstlisting}[style=file]
MINPAR = {
    {1, mhIN},
    {2, msIN},
    {3, LambdaTinput},
    {4, LambdaPinput},
    {5, LambdaOinput},
    {6, LambdaHSinput},
    {7, LambdaHTinput},
    {8, LambdaHPinput},
    {9, LambdaHOinput},
    {10, LambdaSTinput},
    {11, LambdaSPinput},
    {12, LambdaSOinput},
    {13, LambdaTPinput},
    {14, LambdaTOinput},
    {15, LambdaPOinput},
    {16, LambdaHATinput},
    {21, mTinput},
    {22, mPinput},
    {23, mOinput},
    {25, vSinput}
};
\end{lstlisting}
{\bf Parameters fixed by the electroweak symmetry breaking conditions}
\begin{lstlisting}[style=file]
EWSBOutputParameters = {muH2, muS2};
\end{lstlisting}
{\bf Definition of the renormalisation scale, at which the pole masses should be calculated}
\begin{lstlisting}[style=file]
SUSYScaleFirstGuess = LowScaleFirstGuess;
SUSYScale = LowScale;
\end{lstlisting}
{\bf Condition defining the Standard Model matching scale}
\begin{lstlisting}[style=file]
LowScaleFirstGuess = LowEnergyConstant[MZ];
LowScale = LowEnergyConstant[MZ];
\end{lstlisting}
{\bf Boundary conditions}
\begin{lstlisting}[style=file]
SUSYScaleInput = {};

LowScaleInput = {
    {vS, vSinput},
    {LambdaT, LambdaTinput},
    {LambdaP, LambdaPinput},
    {LambdaO, LambdaOinput},
    {LambdaHS, LambdaHSinput},
    {LambdaH, (mhIN^2*v^2 + msIN^2*v^2 -
               Sqrt[mhIN^4*v^4 - 2*mhIN^2*msIN^2*v^4 +
                    msIN^4*v^4 - 4*LambdaHS^2*v^6*vS^2]
              )/(2*v^4)},
    {LambdaS, (mhIN^2 + msIN^2 +
               Sqrt[v^4*(mhIN^4 - 2*mhIN^2*msIN^2 + msIN^4
               - 4*LambdaHS^2*v^2*vS^2)]/v^2)/(4*vS^2)},
    {LambdaHT, LambdaHTinput},
    {LambdaHP, LambdaHPinput},
    {LambdaHO, LambdaHOinput},
    {LambdaTS, LambdaSTinput},
    {LambdaPS, LambdaSPinput},
    {LambdaOS, LambdaSOinput},
    {LambdaPT, LambdaTPinput},
    {LambdaOT, LambdaTOinput},
    {LambdaOP, LambdaPOinput},
    {LambdaHat, LambdaHATinput},
    {muT2, (2*mTinput^2 - LambdaHT*v^2 - LambdaTS*vS^2)/2},
    {muP2, (2*mPinput^2 - LambdaHP*v^2 - LambdaPS*vS^2)/2},
    {muO2, (2*mOinput^2 - LambdaHO*v^2 - LambdaOS*vS^2)/2},
    {YT, LHInput[YT]},
    {YO, LHInput[YO]},
    {YP, LHInput[YP]},
    {v, 2 MZDRbar /
        Sqrt[GUTNormalization[g1]^2 g1^2 + g2^2]},
    {Ye, Automatic},
    {Yd, Automatic},
    {Yu, Automatic}
};
\end{lstlisting}
{\bf Initial parameter guess}
\begin{lstlisting}[style=file]
InitialGuessAtLowScale = {
    {vS, vSinput},
    {LambdaT, LambdaTinput},
    {LambdaP, LambdaPinput},
    {LambdaO, LambdaOinput},
    {LambdaHS, LambdaHSinput},
    {LambdaH, (mhIN^2*v^2 + msIN^2*v^2 -
               Sqrt[mhIN^4*v^4 - 2*mhIN^2*msIN^2*v^4 +
                    msIN^4*v^4 - 4*LambdaHS^2*v^6*vS^2]
              )/(2*v^4)},
    {LambdaS, (mhIN^2 + msIN^2 +
               Sqrt[v^4*(mhIN^4 - 2*mhIN^2*msIN^2 + msIN^4
               - 4*LambdaHS^2*v^2*vS^2)]/v^2)/(4*vS^2)},
    {LambdaHT, LambdaHTinput},
    {LambdaHP, LambdaHPinput},
    {LambdaHO, LambdaHOinput},
    {LambdaTS, LambdaSTinput},
    {LambdaPS, LambdaSPinput},
    {LambdaOS, LambdaSOinput},
    {LambdaPT, LambdaTPinput},
    {LambdaOT, LambdaTOinput},
    {LambdaOP, LambdaPOinput},
    {LambdaHat, LambdaHATinput},
    {YT, LHInput[YT]},
    {YO, LHInput[YO]},
    {YP, LHInput[YP]},
    {v , LowEnergyConstant[vev]},
    {Yu, Automatic},
    {Yd, Automatic},
    {Ye, Automatic}
};
\end{lstlisting}

\tocless\subsection{{\tt SPheno.m} for a high-scale version}
{\bf Expected Input parameters}
\begin{lstlisting}[style=file]
MINPAR={{1,m0},
        {2,m12},
        {3,TanBeta},
        {4,SignumMu},
        {5,Azero}};
        
RealParameters = {TanBeta, m0};        
\end{lstlisting}
{\bf Parameters obtained from tadpole equations}
\begin{lstlisting}[style=file]    
ParametersToSolveTadpoles = {\[Mu],B[\[Mu]]};
\end{lstlisting}
{\bf Definition of Renormalisation scale}
\begin{lstlisting}[style=file]    
RenormalizationScaleFirstGuess = m0^2 + 4 m12^2;
RenormalizationScale = Sqrt[(mq2[3, 3] 
 + (vu^2*conj[Yu[3, 3]]*Yu[3, 3])/2)*(mu2[3, 3] 
 + (vu^2*conj[Yu[3, 3]]*Yu[3, 3])/2)
 -((vd*\[Mu]*conj[Yu[3, 3]] - vu*conj[T[Yu][3, 3]])
 *(vd*conj[\[Mu]]*Yu[3, 3] - vu*T[Yu][3, 3]))/2];
\end{lstlisting}
{\bf Condition for GUT scale}
\begin{lstlisting}[style=file]    
ConditionGUTscale = g1 == g2;
\end{lstlisting}
{\bf Boundary conditions}
\begin{lstlisting}[style=file]    
BoundaryHighScale={
{T[Ye], Azero*Ye},
{T[Yd], Azero*Yd},
{T[Yu], Azero*Yu},
{mq2, DIAGONAL m0^2},
{ml2, DIAGONAL m0^2},
{md2, DIAGONAL m0^2},
{mu2, DIAGONAL m0^2},
{me2, DIAGONAL m0^2},
{mHd2, m0^2},
{mHu2, m0^2},
{MassB, m12},
{MassWB,m12},
{MassG,m12}
};
BoundarySUSYScale={};
BoundaryEWSBScale={};
\end{lstlisting}
{\bf List of particles for which the decays shall be calculated by \SPheno}
\begin{lstlisting}[style=file]    
ListDecayParticles = Automatic;
ListDecayParticles3B = Automatic;
\end{lstlisting}

\tocless\subsection{{\tt SPheno.m} for a low-scale version}
{\bf Flag to choose a low-scale \SPheno version}
\begin{lstlisting}[style=file]    
OnlyLowEnergySPheno = True;
\end{lstlisting}
{\bf Expected Input parameters}
\begin{lstlisting}[style=file]    
MINPAR={
 {1, mhIN},
 {2, msIN},
 {3, LambdaTinput},
 {4, LambdaPinput},
 {5, LambdaOinput},
 {6, LambdaHSinput},
 {7, LambdaHTinput},
 {8, LambdaHPinput},
 {9, LambdaHOinput},
 {10, LambdaSTinput},
 {11, LambdaSPinput},
 {12, LambdaSOinput}, 
 {13, LambdaTPinput},
 {14, LambdaTOinput},
 {15, LambdaPOinput},
 {16, LambdaHATinput},
 
 {21, mTinput},
 {22, mPinput},
 {23, mOinput},
 
 {25, vSinput}
};
\end{lstlisting}
{\bf Parameters obtained from tadpole equations}
\begin{lstlisting}[style=file]    
ParametersToSolveTadpoles = {muH2, muS2};
\end{lstlisting}
{\bf Boundary conditions}
\begin{lstlisting}[style=file]    
BoundaryLowScaleInput={
 {v, vSM}, 
 {Ye, YeSM},
 {Yd, YdSM},
 {Yu, YuSM},
 {g1, g1SM},
 {g2, g2SM},
 {g3, g3SM},
 {vS,          vSinput},
 {LambdaT,           LambdaTinput},
 {LambdaP,           LambdaPinput},
 {LambdaO,           LambdaOinput},
 {LambdaHS,           LambdaHSinput},
 {LambdaH,  (mhIN^2*v^2 + msIN^2*v^2 - 
    Sqrt[mhIN^4*v^4 - 2*mhIN^2*msIN^2*v^4 + msIN^4*v^4 
    - 4*LambdaHS^2*v^6*vS^2])/(2*v^4)},
 {LambdaS,  (mhIN^2 + msIN^2 + 
    Sqrt[v^4*(mhIN^4 - 2*mhIN^2*msIN^2 + msIN^4 
    - 4*LambdaHS^2*v^2*vS^2)]/v^2)/(4*vS^2)},
 {LambdaHT,           LambdaHTinput},
 {LambdaHP,           LambdaHPinput},
 {LambdaHO,           LambdaHOinput},
 {LambdaTS,           LambdaSTinput},
 {LambdaPS,           LambdaSPinput},
 {LambdaOS,           LambdaSOinput}, 
 {LambdaPT,           LambdaTPinput},
 {LambdaOT,           LambdaTOinput},
 {LambdaOP,           LambdaPOinput},
 {LambdaHat,           LambdaHATinput},
             
 {muT2,  (2*mTinput^2 - LambdaHT*v^2 - LambdaTS*vS^2)/2},
 {muP2,  (2*mPinput^2 - LambdaHP*v^2 - LambdaPS*vS^2)/2},
 {muO2,  (2*mOinput^2 - LambdaHO*v^2 - LambdaOS*vS^2)/2},

 {YT, LHInput[YT]},
 {YO, LHInput[YO]},
 {YP, LHInput[YP]}
};
\end{lstlisting}
{\bf List of particles for which the decays shall be calculated by \SPheno}
\begin{lstlisting}[style=file]    
ListDecayParticles = {Fu,Fe,Fd,hh,Ah,PhiLQ,theta,OmegaLQ};
ListDecayParticles3B = {{Fu,"Fu.f90"},{Fe,"Fe.f90"},{Fd,"Fd.f90"}};
\end{lstlisting}